\documentclass[11pt]{article}
\usepackage[letterpaper,margin=.83in]{geometry}
\usepackage[T1]{fontenc}
\usepackage{lmodern,microtype}
\usepackage{amsmath,amssymb,amsthm,mathtools,bm,mathrsfs}
\usepackage{graphicx,booktabs,array,tabularx,longtable,enumitem,float,needspace,caption,placeins,tocloft}
\usepackage{xcolor,hyperref}
\hypersetup{pdftitle={The opposing informational effects of AI trading},pdfauthor={Hajime Shimao},colorlinks=true,urlcolor=blue!45!black,citecolor=black,linkcolor=black}
\newcommand{\E}{\mathbb E}
\newcommand{\V}{\operatorname{Var}}
\newcommand{\Cov}{\operatorname{Cov}}
\newcommand{\R}{\mathbb R}
\newcommand{\MI}{\operatorname{I}}
\newcommand{\CE}{\operatorname{CE}}

\newtheorem{assumption}{Assumption}[section]
\newtheorem{definition}{Definition}[section]
\newtheorem{lemma}{Lemma}[section]
\newtheorem{proposition}{Proposition}[section]

\newtheorem{corollary}{Corollary}[section]
\newtheorem{theorem}{Theorem}
\begin{document}

\begin{center}
{\LARGE\bfseries The opposing informational effects of AI trading}\par\vspace{5pt}
{\normalsize Hajime Shimao}\par\vspace{2pt}
{\small Penn State, Great Valley School of Graduate Professional Studies}\par
{\small \href{mailto:hjs5825@psu.edu}{hjs5825@psu.edu}}\par
\end{center}
\vspace{3pt}
\begin{abstract}\noindent
Better artificial intelligence (AI) can improve individual forecasts without making prices better at revealing new information. In a competitive market with resource-constrained forecasting, more capable algorithms can substitute analysis of public records for private observations, reducing the private information aggregated into prices. The same trades can convey sophisticated public-data analysis to investors unable to perform it themselves. We establish conditions under which these effects coexist and the private benefit of further computation falls. Numerical equilibrium calculations and finite-data neural-market experiments exhibit corresponding patterns. Out-of-sample earnings forecasts and controlled language-model tasks show related patterns of information substitution. In the neural experiment, wider adoption increases the predictive value of price for limited observers, reduces it for advanced observers, and lowers the benefit of upgrading by 44\%. Retraining by investors who did not upgrade can amplify the loss of private-information precision. The informational consequences of AI trading depend on what prices reveal and who can use them.
\end{abstract}

\section*{Will smarter trading make prices smarter?}
How much can a market price tell us? From Adam Smith's account of decentralized economic coordination to Friedrich Hayek's argument that prices communicate scattered knowledge, economists have long debated how markets bring information together \cite{smith1776,hayek1945}. Paul Samuelson's analysis of anticipated asset prices and Eugene Fama's work on market efficiency gave that question a sharper form in modern finance \cite{samuelson1965,fama1970}. Prices need not be perfectly efficient to reflect some of what investors know and help direct capital toward better opportunities \cite{grossman1980,bai2016}.

Artificial intelligence (AI) appears to offer a powerful way to improve that process. Algorithms can search large collections of corporate reports, economic data, and past prices for patterns that would escape less capable forecasters. Machine learning has demonstrated predictive value in asset-pricing research \cite{gukelly2020,murray2024}, raising hopes of more efficient markets alongside concerns about instability \cite{imf2024,imfblog2024}. If more accurate forecasts inform more trades, surely prices should become more informative. But \emph{more informative about what?}

Consider two investors. One visits a supplier and discovers unexpectedly strong orders---a clue that is not yet public. The other uses a powerful AI model to identify a subtle pattern in reports that anyone can download. Both may trade profitably, and both may move the same share price. Yet only the first contributes an observation that the published record did not already contain. The second contributes a better \emph{interpretation} of that record. Improving the second investor's forecast does not necessarily improve the market's ability to discover new facts.

Trading algorithms also change the data on which future forecasts are based. When many investors adopt a new algorithm, their orders can change market prices. Those prices enter the public record, so even investors who keep their old algorithms encounter a different history of price movements. They may then retrain their forecasts on the changed market, producing further changes in trading \cite{perdomo2020}.

We show that more capable trading AI can make prices less revealing about newly discovered facts while making them more useful to people who cannot analyze the public record as thoroughly. We establish how both effects can occur in a competitive market and examine them in numerical calculations and markets with trained neural forecasting models. The same analysis shows how prices can reduce the private benefit of further computation. We also examine information substitution using U.S. market data and controlled language-model forecasts.

\section*{How learning algorithms become market prices}
Our model follows investors who trade an asset according to their forecasts of its future payoff. All have access to the same public record: past prices, corporate payouts, and published economic indicators. Each also possesses an imperfect private clue about current conditions. What distinguishes more capable AI is not exclusive access to public data, but an ability to extract more useful patterns from them. A smaller algorithm may capture a few regularities; a more powerful one can build a richer summary of the same history and use it to forecast both future payoffs and uncertainty.

\begin{figure}[H]
\centering\includegraphics[width=.98\linewidth]{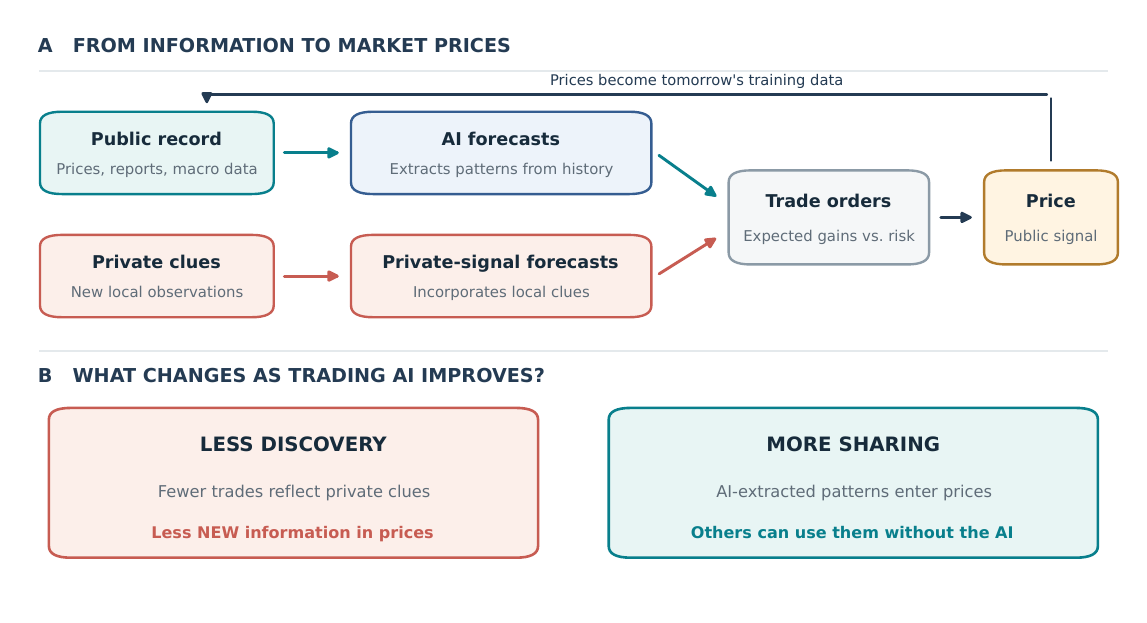}
\caption{\textbf{From private clues and public records to prices.} (A) Investors combine their own private information with forecasts drawn from published data; trading on both inputs determines the market price. Prices then enter the histories used in subsequent AI training. (B) More capable AI can reduce trading on private clues (less discovery) while conveying more information extracted from public data to other price users (more sharing). }
\label{fig:feedback}
\end{figure}

Investors choose positions using their fitted forecasts of payoff and risk. Their orders meet in a competitive market and jointly determine the price. Private clues and AI-generated forecasts therefore affect prices through the same familiar force: people's willingness to buy or sell (Fig.~\ref{fig:feedback}A). The model allows algorithms to learn nonlinear patterns in public data. In a tractable version, we can also calculate forecasts and market prices exactly.

We distinguish three stages. First, a more capable algorithm, trained on the previously prevailing market, changes orders at its first auction. Second, even with every algorithm's parameters fixed, the resulting prices enter subsequent public histories and change forecasts and orders. Third, investors may refit their algorithms to the changed market. The second adjustment requires no retraining. We keep these stages separate when comparing initial deployment, frozen-policy markets, and retraining.

\section*{Better forecasts can conceal new information}
A better public-data forecast may leave less for a private observation to explain. If an AI can already infer likely sales from published reports, a supplier's clue changes its forecast by less. Trading then responds less to the clue, and less privately observed information reaches the price.

The mechanism can be seen in the strength of an investor's response to a private signal. In our analytically tractable benchmark,
\begin{equation}
\boxed{
\underbrace{\frac{\partial h_i}{\partial s_i}}_{\substack{\text{trading response}\\\text{to a private clue}}}
=
\underbrace{\frac{\beta_i}{\gamma_i v_i}}_{\substack{\text{predictive weight}\\\text{adjusted for risk}}}.
}
\label{eq:signalresponse}
\end{equation}
Here $h_i$ is the position held and $s_i$ the private clue. Holding the candidate price fixed, the coefficient $\beta_i$ measures how much the clue changes the expected payoff; $v_i$ is the remaining forecast uncertainty, and $\gamma_i$ describes how cautious the investor is about risk. As an AI explains more of the payoff using public data, the clue contributes less to its forecast, reducing $\beta_i$. A better forecast also lowers uncertainty $v_i$, which can encourage larger positions. Our first theorem accounts for both changes. When private clues concern the fundamental driving the payoff, an upgrade that improves the fitted public-data forecast reduces the net response of trades to those clues at initial deployment. The result permits nonlinear public-data processing under the forecasting assumptions given in the Supplementary Materials.

\begin{figure}[H]
\centering\includegraphics[width=.98\linewidth]{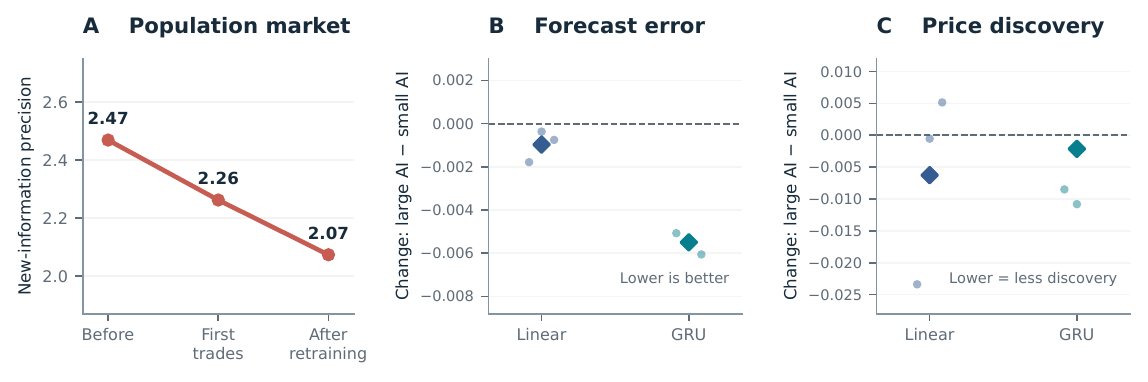}
\caption{\textbf{Better forecasts need not produce better discovery.} (A) In a population-optimal competitive-market example, prices carry progressively less new private information after advanced AI is deployed and after investors retrain. (B,C) In separate finite-data experiments, increasing the size of linear state-space and gated recurrent models reduces forecast error on average while also reducing price sensitivity to new private information on average. Small dots show three training runs; diamonds show their means. The fitted-model experiments are separate from the population calculation.}
\label{fig:forecast}
\end{figure}

When investors trade less on their private clues, their actions leave a weaker trace of that information in prices. Random buying and selling then makes it harder to infer what investors have learned privately. We measure this using the \emph{precision of new information}: how clearly prices distinguish private news from unrelated trading. In a competitive-market model, this precision falls from 2.47 before advanced AI enters to 2.26 at first deployment and 2.07 after retraining (Fig.~\ref{fig:forecast}A)---a decline of about 16\%. The decrease arises from investors' trading decisions even as their forecasts improve. Experiments with trained linear state-space and recurrent neural-network algorithms also find improved forecasts alongside weaker discovery on average (Fig.~\ref{fig:forecast}B,C).

\Needspace{9\baselineskip}
\section*{Prices can also share what AI has learned}
But revealing less \emph{new} information is not the same as being less useful to everyone. A more powerful trading algorithm can recognize patterns in published data that a less capable investor cannot practically extract. When the algorithm trades on those patterns, part of its analysis becomes visible in the market price. Watching the price may therefore give others the benefit of computations they never performed.

Consider two people looking at the same share price. One can analyze years of reports and market data; the other can perform only a smaller set of calculations. The second may learn more from the price because the first can already work out much of what it reveals.

We can describe this difference mathematically. Let $\theta$ be the economic condition investors are trying to infer and $p$ the price. Write $\mathcal F$ for the information a modest algorithm can extract from published data and $\mathcal U\supseteq\mathcal F$ for what a more capable algorithm can extract from the same data. Then
\begin{equation}
\boxed{\begin{aligned}
I(\theta;p\mid\mathcal F)=\;&\underbrace{I(\theta;p\mid\mathcal U)}_{\text{New information beyond public analysis}}\\[-1pt]
&+\underbrace{\left[I(\theta;\mathcal U\mid\mathcal F)
-I(\theta;\mathcal U\mid\mathcal F,p)\right]}_{\text{Public information shared through price}}.
\end{aligned}}
\label{eq:twochannels}
\end{equation}
Here $I$ measures information as a reduction in uncertainty. Imagine first that you could carry out all the public-data analysis performed by the powerful algorithm. The price may still reveal information beyond that feasible analysis, including traders' private observations: this is the first term. Now imagine that you see the same publicly available reports but have only the resources for simpler analysis. The price can pass along some of the sophisticated conclusions that you could not produce yourself: this is the second term. The identity holds for nested information sets; the second term is signed and need not be positive. Our economic model supplies conditions under which sharing increases while discovery declines.

We prove that both effects can occur together even when investors' competitive trading determines prices. In a numerical market, as advanced AI spreads, the additional information that prices reveal beyond advanced public-data analysis falls by about 13\%. At the same time, the information that a less capable observer gains from prices rises by about 36\% (Fig.~\ref{fig:dual}A,B). These opposite changes occur in the same market: prices reveal less that is new while giving more information to people with limited computing resources.

A similar pattern appears when neural networks learn from market data and then trade. We examine how much observing the current price improves dividend forecasts for two outside observers---one with limited computing capacity and one with more advanced tools. As AI adoption rises in the market, the forecasting benefit of seeing the price increases by 12.4\% for the limited observer but decreases by 6.8\% for the advanced observer (Fig.~\ref{fig:dual}D,E). The experiment measures these gains in predictions from trained models rather than from a theoretical information formula.

\section*{Who still needs to invest in more AI?}
If investors can learn from prices what others have already calculated, why pay to repeat the work? But competition could also lead to an arms race in which every investor keeps buying more powerful AI. Which force determines the value of the next upgrade?

\begin{samepage}
In a benchmark with normally distributed signals and correctly fitted forecasts, the benefit of greater computing power has a simple expression. Let $M$ represent the additional public history an advanced algorithm can process beyond the basic information $\mathcal F$; the investor also observes the current price $p$ and a private signal $s$. With optimal trading, the gain from additional computation is
\begin{equation}
\boxed{\text{Value of extra computation}=\frac{1}{\gamma}
I(\text{asset payoff};M\mid p,s,\mathcal F).}
\label{eq:computev}
\end{equation}
\end{samepage}
The formula says that extra computing power is valuable to an investor only to the extent that it reveals something not already learned from the price and available clues. If competitors' AI has already put much of the relevant analysis into the price, doing the same analysis independently brings a smaller advantage. Under the conditions we establish, wider adoption can therefore reduce the private benefit against which an investor weighs the cost of additional computation.

\begin{figure}[H]
\centering\includegraphics[width=.98\linewidth]{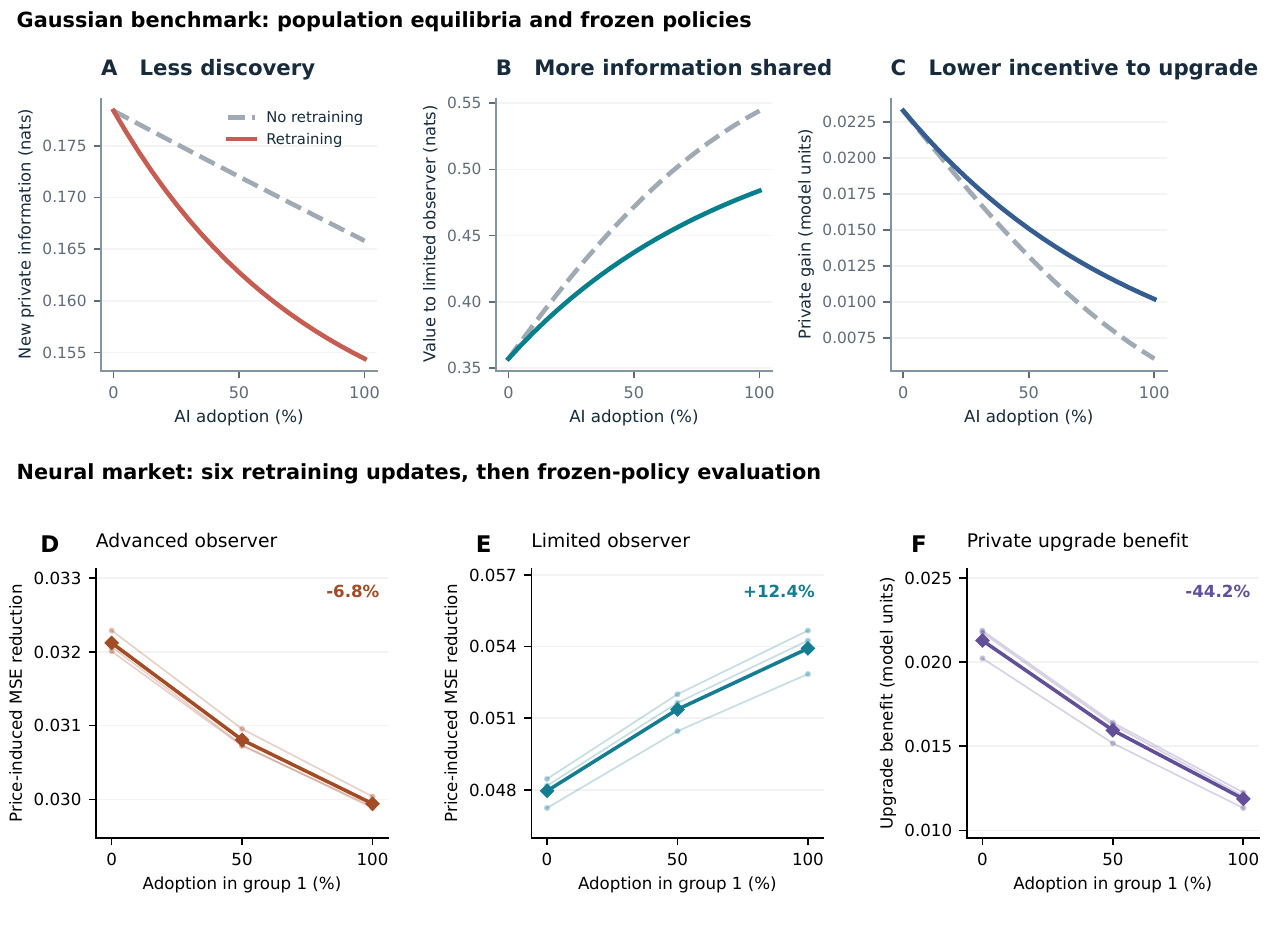}
\caption{\textbf{Opposing price-information effects and a falling benefit of computation.} (A--C) Gaussian benchmark: price information beyond advanced analysis falls, information to a limited observer rises, and the upgrade benefit falls. Solid curves allow population retraining; dashed curves freeze incumbent policies while histories adjust. (D--F) Neural markets after six joint retraining updates: price-induced dividend-MSE reductions fall for advanced observers and rise for limited observers; directly evaluated upgrade benefits also fall. Diamonds are three-seed means; faint lines show individual seeds. Lines connect the three tested adoption shares. Panels D,E exclude own signals; both strategies in F receive the same price and own signal. Adoption is within group 1 (55\% of the neural market). Units and vertical scales differ; neural profiles are not certified retraining equilibria.}
\label{fig:dual}
\end{figure}

In our example, the payoff-equivalent benefit of an AI upgrade falls by more than half, from 0.0233 to 0.0102 in the model's monetary units, as advanced trading AI becomes widespread (Fig.~\ref{fig:dual}C). We establish conditions for this decline and show that it can accompany the simultaneous loss of new information and gain in shared analysis. Earlier financial research established the importance of incentives to acquire and interpret information \cite{grossman1980,han2016,mondria2022,dugast2025,chenma2026}. The additional mechanism here is that competing algorithms process \emph{the same publicly accessible material} and pass the results of that work into a common price.

In the same neural markets, we evaluate the low- and high-capacity models' trading strategies under the true payoff distribution, giving both the same price and private signal. The certainty-equivalent benefit of upgrading falls from 0.0213 to 0.0119 model units, a 44.2\% decline (Fig.~\ref{fig:dual}F). The extra computing power still has value, but its advantage becomes considerably smaller as competitors adopt stronger AI.

\section*{When the new prices become new training data}
The market can change again after the first AI upgrade. A newly deployed algorithm alters prices through its trades; those prices then enter the data used for its next training round. Even investors who did not upgrade may change their behavior because they now learn from a different market history.

In our equilibrium calculations, retraining makes the loss of new private information larger, while reducing some of the gain that limited observers receive from prices (Fig.~\ref{fig:dual}, solid versus dashed curves). In one example, much of the additional loss of discovery comes from investors \emph{who did not buy the stronger AI} but updated their forecasts in response to prices changed by others (Supplementary Materials, Section S7). Other experiments distinguish the improvement in forecasting expected payoffs from the improvement in forecasting risk: the first can reduce trading on private clues, while the second can make investors trade more aggressively (Supplementary Materials, Section S7). The neural experiment supports the same mechanisms: retraining the non-upgrading group accounts for much of the additional reduction in private-information precision, while post-fit substitutions show that changes in predicted risk partly offset changes in predicted means (Supplementary Materials).

The neural results persist with larger training budgets, new training samples, and alternative ways of turning learned information into forecasts. Separate experiments also find the opposing information effects and a falling upgrade benefit when payoffs depend on factors not directly observed by private clues, or when payoff risk changes with public economic conditions (Supplementary Materials).

We also examined whether repeated neural retraining settles into a stable pattern. With the same market disturbances reused across updates, several model coefficients stabilize; with fresh training data, they continue to drift. The neural results therefore describe markets after a specified number of learning rounds, alongside our separate analytical and population-equilibrium results.

\section*{Information substitution beyond simulated markets}
\begin{figure}[H]
\centering\includegraphics[width=.93\linewidth]{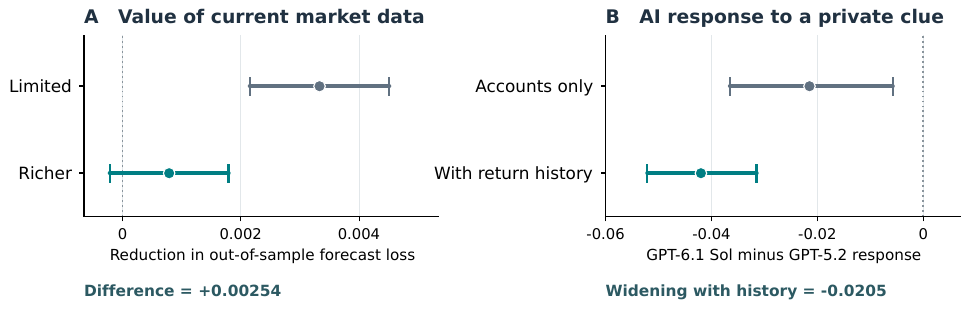}
\caption{\textbf{The value of market information depends on the observer's forecasting tools.} (A) Adding current market information improves a limited predictor's forecasts about 4.2 times as much as a richer predictor's forecasts (58,807 U.S. firm-quarter observations). Dots show mean reductions in out-of-sample forecast loss. (B) With historical returns, GPT-6.1 Sol responds about 6\% less than GPT-5.2 to the same hypothetical private research estimate; the gap between the models is nearly twice as large as without the returns (50 company-quarter cases). Dots show means and 95\% confidence intervals (two-way clustering in A; resampling cases in B). Precise differences are 0.00254 in A and $-0.0205$ for the change in the model gap in B.}
\label{fig:empirical}
\end{figure}

Do the same differences appear outside simulated markets? We first use historical U.S. financial data to predict companies' quarterly earnings relative to analysts' earlier forecasts. We compare a limited statistical model using a short public-data history with a richer model that processes longer histories and more complex patterns. We then give both the same additional current market information: share price, market capitalization, and the latest monthly stock return. Across 58,807 firm-quarter observations outside the training sample, these variables reduce mean squared forecast error by 0.60\% for the limited model, compared with 0.14\% for the richer model (Fig.~\ref{fig:empirical}A). The difference remains when we use more recent quarterly financial statements (Supplementary Materials).

We next examine how actual language models respond to a hypothetical private research estimate of a company's next earnings surprise. We give GPT-5.2 and GPT-6.1 Sol identical anonymous financial information, then change only the private estimate and measure how much each forecast moves. Across 50 company-quarter cases, the newer model responds about 6\% less to the clue when historical market returns are available. The gap between the models is nearly twice as large with that history as without it (Fig.~\ref{fig:empirical}B). Both models respond more to the clue when returns are supplied, but the increase is smaller for the newer model. Together, the market-data and language-model experiments show how the value and use of information depend on what can already be learned from public records.

\section*{What does a smarter market mean for society?}
Market prices matter far beyond the people who trade. Investors use them to judge opportunities, while companies and lenders can use them to assess changing economic conditions \cite{bond2012,goldstein2017}. As AI becomes more widely used in trading, prices may become better summaries of information already available in public records while revealing less about developments that investors have discovered independently. The same price can therefore be more useful for one decision and less useful for another.

These changes can also reshape the incentives to invest in AI. When prices convey some of the conclusions reached by better-equipped traders, others can benefit without performing the same calculations. As advanced trading systems spread, the reward to buying an additional forecasting system can fall rather than rise indefinitely. Meanwhile, their trades can leave a weaker trace of private observations in prices, and those altered prices become the data from which other investors learn.

The familiar promise of markets is that prices bring dispersed knowledge together. AI adds a new dimension: prices can also carry conclusions that require substantial computation to extract from public data. Better individual forecasts need not produce better collective discovery. To understand whether smarter trading produces a smarter market, we must ask not only how well AI predicts, but \emph{what prices still discover and whose analysis they share}.

\clearpage
\newgeometry{margin=24mm}
\setlength{\parindent}{1.2em}
\setlength{\parskip}{2.5pt}
\setlist{topsep=3pt,itemsep=2pt}
\setlength{\cftsecnumwidth}{2.8em}
\setcounter{section}{0}
\setcounter{equation}{0}
\setcounter{figure}{0}
\setcounter{table}{0}
\renewcommand{\thesection}{S\arabic{section}}
\renewcommand{\theHsection}{S.\arabic{section}}
\numberwithin{equation}{section}
\renewcommand{\thetable}{S\arabic{table}}
\renewcommand{\theHtable}{S.\arabic{table}}
\setcounter{tocdepth}{1}
\begin{center}
{\LARGE\bfseries Supplementary Information}\\[5pt]
{\large The opposing informational effects of AI trading}\\[6pt]
{\normalsize Hajime Shimao}\\[2pt]
{\small Penn State, Great Valley School of Graduate Professional Studies}\\[2pt]
{\small \href{mailto:hjs5825@psu.edu}{hjs5825@psu.edu}}
\end{center}
\medskip
\tableofcontents
\clearpage
\section{Economic environment and timing}\label{sec:S1}

This section defines the common economic environment underlying the paper's theoretical and computational results. We distinguish three objects at the outset: (i) the \emph{economic process} that generates fundamental payoffs and publicly released data; (ii) the \emph{information and technology} available to market participants when making forecasts; and (iii) the \emph{price process} that results when they trade. The underlying economic state determines both cash flows and the clues about them. Each trader combines a private clue with an AI-generated summary of the publicly available record to choose a holding. The holdings together determine the price. An improvement in the AI changes the processing of existing data and, through the resulting orders, the price; it does not mechanically improve the underlying fundamentals. Learning the new price process may then change subsequent trading decisions. These distinctions are maintained throughout the Supplementary Information (SI).

We first describe the environment conditional on a fixed profile of deployed forecasting systems. The population training problem and capacity restrictions are developed in S2; the forecast-based optimal demand schedules, clearing map, and endogenous retraining equilibrium are derived in S3. S4--S6 state and prove the three main analytical results using the primitives and timing established here. Additional stochastic-volatility, architecture-specific, and long-duration specifications used in robustness experiments are identified explicitly where introduced, rather than silently being treated as assumptions of the core theoretical results.

\subsection{A stochastic economy with a finite-dimensional latent state}\label{subsec:S1-state}

Trading dates are indexed by $t\in\mathbb Z$. The latent economic state $f_t\in\mathbb R^m$, where $m<\infty$ is arbitrary, follows
\begin{equation}\label{eq:S1-state}
  f_{t+1}=Ff_t+G\eta_{t+1},\qquad
  \eta_t\overset{\mathrm{iid}}{\sim}\mathcal N(0,I_q),\qquad
  \rho(F)<1.
\end{equation}
Here $F$ is an $m\times m$ transition matrix, $G$ is $m\times q$, and $\rho(\cdot)$ denotes spectral radius. The economy is taken in its stationary solution,
\begin{equation}\label{eq:S1-stationary}
  f_t=\sum_{j=0}^{\infty}F^jG\eta_{t-j},\qquad
  \Sigma_f\equiv\operatorname{Var}(f_t)
  =F\Sigma_fF^\top+GG^\top.
\end{equation}
Stability ensures existence and uniqueness of the finite positive-semidefinite stationary covariance $\Sigma_f$. We require positive variance for the scalar fundamental introduced below, without requiring $\Sigma_f$ to be nonsingular on economically irrelevant directions. Full-rank conditions needed for individual linear projections or for a smooth implicit-function argument will be added in the corresponding theorem.

A scalar fundamental about which traders possess dispersed information is
\begin{equation}\label{eq:S1-theta}
  \theta_t=a^\top f_t,\qquad
  P_\theta\equiv\operatorname{Var}(\theta_t)=a^\top\Sigma_f a>0,
\end{equation}
where $a\in\mathbb R^m$ is fixed. Thus $f_t$ is the full economic state, whereas $\theta_t$ is the particular scalar about which traders receive individual clues. The state may contain many other components, including persistent or transitory macroeconomic drivers. The number $m$ describes the economy, not the memory or computational power of an AI. This definition does not itself provide a trader with $\theta_t$. Traders instead infer it from their signals and the disclosed record; a special disclosure technology may happen to reveal it exactly. The individual clue $s_{ci,t}$ is introduced in \eqref{eq:S1-private}, and the AI's public-data summary $z_{c,t}$ in \eqref{eq:S1-adapted}. In particular, a high-dimensional economy need not require a high-dimensional predictor for every task.

Let $m_t\in\mathbb R^r$ denote the vector of public economic releases available before the date-$t$ auction. The unindexed $m$ denotes the latent state dimension, whereas the time-indexed $m_t$ denotes the public release (possibly a vector). The baseline linear-Gaussian observation specification is
\begin{equation}\label{eq:S1-public-release}
  m_t=Hf_t+\omega_t,\qquad
  \omega_t\overset{\mathrm{iid}}{\sim}\mathcal N(0,\Sigma_\omega),
\end{equation}
where $H$ is $r\times m$, $\Sigma_\omega\succeq0$, and $\{\omega_t\}$ is independent of primitive fundamental and trading shocks. The degenerate case $\Sigma_\omega=0$ permits exact disclosure of selected coordinates of $f_t$. The general market mechanism does not depend on \eqref{eq:S1-public-release} being the only possible form of public disclosure: S2 allows algorithms to process any pre-auction, commonly accessible release stream. The linear-Gaussian branch needed for the exact information and certainty-equivalent identities in S5--S6 retains \eqref{eq:S1-public-release} (or a jointly Gaussian specialization).

\subsection{Traded claims, payoffs, and random supply}\label{subsec:S1-assets}

At each date $t$, traders can buy or sell a risky claim at an ex-dividend price $p_t$. This claim pays $D_{t+1}$ at the beginning of date $t+1$ and then expires. The next auction trades a new one-period claim generated by the same stationary economic structure. The baseline payoff is
\begin{equation}\label{eq:S1-payoff}
  D_{t+1}=\bar d+\ell^\top f_t+\varepsilon_{t+1},\qquad
  \varepsilon_t\overset{\mathrm{iid}}{\sim}\mathcal N(0,\sigma_\varepsilon^2),
  \qquad \sigma_\varepsilon^2>0,
\end{equation}
where $\ell\in\mathbb R^m$. The payoff shock is independent of $\{\eta_t,\omega_t\}$, the supply shocks, and all investors' idiosyncratic signals. The time index is deliberate: $D_t$ is known at the date-$t$ auction, whereas $D_{t+1}$ is not yet realized. The claim's one-period duration is a restriction on the \emph{holding horizon}, not on the length of the data history an AI may analyze.

A risk-free account is supplied elastically at gross return $R_f>0$ (we use $R_f>1$ in the numerical examples). Holding $h$ shares of the risky claim and financing the remainder at the risk-free rate produces risky excess payoff
\begin{equation}\label{eq:S1-excess}
  X_{t+1}(p_t)\equiv D_{t+1}-R_fp_t
\end{equation}
per share. In particular, at wealth $W_{ci,t}$ before trading, an investor with risky holding $h_{ci,t}$ has next-date wealth
\begin{equation}\label{eq:S1-wealth}
  W_{ci,t+1}=R_fW_{ci,t}+h_{ci,t}\big(D_{t+1}-R_fp_t\big).
\end{equation}
The primitive payoff-shock variance $\sigma_\varepsilon^2$ should not be confused with a forecasting system's fitted conditional payoff variance $v_{c,t}$. The latter also reflects uncertainty about $f_t$ that the particular algorithm fails to resolve. Accordingly, $v_{c,t}$ may depend on its processed public state even when $\varepsilon_{t+1}$ is homoskedastic. Both quantities are defined and related explicitly in S2--S3.

There is no exogenous resale-price term in \eqref{eq:S1-payoff} or \eqref{eq:S1-wealth}. Allowing resale of a long-lived security, with payoff $D_{t+1}+\delta p_{t+1}$, is an explicitly separate extension; it is not needed for Theorems 1--3.

The net quantity of risky claims to be absorbed by computational traders at date $t$ is
\begin{equation}\label{eq:S1-supply}
  S_t=\bar S+\zeta_t,\qquad
  \zeta_t\overset{\mathrm{iid}}{\sim}\mathcal N(0,\sigma_\zeta^2),
  \qquad \sigma_\zeta^2>0.
\end{equation}
The disturbance $\zeta_t$ represents exogenous liquidity or hedging orders and is not separately disclosed to traders before the auction. It is independent of fundamental, payoff, and private-signal shocks. It enters the aggregate demand-clearing equation and limits what an observer can infer about fundamentals from the price. Because a Gaussian has unbounded support, $S_t$ is interpreted as a \emph{signed net supply}, not a technologically constrained number of physical shares; unconstrained long and short positions are allowed in the analytical benchmark. In numerical experiments with nonzero $\bar S$, the mean net supply can create a level or risk premium, but it does not change the definition of informational precision after removal of predictable components.

The primitive exogenous laws in \eqref{eq:S1-state}--\eqref{eq:S1-supply} are held fixed in computational-adoption comparisons unless a stated robustness exercise changes a primitive. A change in a price time series therefore comes from changed forecasts and trading demands, not from automatically better or worse exogenous fundamentals.

\subsection{The aligned-payoff restriction and the unrestricted payoff}\label{subsec:S1-alignment}

For the paper's information-crowd-out theorem it is useful to isolate an economy in which private clues and the traded payoff concern the same fundamental. We call the payoff \emph{aligned} with the privately signaled scalar if
\begin{equation}\label{eq:S1-aligned}
  \ell=\lambda a,\quad \lambda>0,
  \qquad\Longleftrightarrow\qquad
  D_{t+1}=\bar d+\lambda\theta_t+\varepsilon_{t+1}.
\end{equation}
Alignment says that \emph{all of the current state's effect on the payoff mean} passes through the scalar $\theta_t$. For example, if $f_t=(u_t,v_t)^\top$, $\theta_t=u_t$, and $D_{t+1}=\bar d+2u_t+\varepsilon_{t+1}$, the payoff is aligned even though $v_t$ may help predict future states. Alignment does \emph{not} mean that traders observe $\theta_t$ without error: each sees a different noisy signal of it. Nor does it mean the realized payoff is perfectly correlated with $\theta_t$, because $\varepsilon_{t+1}$ remains. This is a specialization of \eqref{eq:S1-payoff}, not a separate market model. It is the domain of the monotonic private-signal response theorem in S4 and the conditional-variance identities in S5.

The unrestricted payoff also accommodates a distinct publicly predictable component. To make the distinction precise, define
\begin{equation}\label{eq:S1-orthogonal}
  \lambda_0\equiv\frac{\ell^\top\Sigma_f a}{a^\top\Sigma_f a},\quad
  b\equiv\ell-\lambda_0a,\quad
  r_t\equiv b^\top f_t.
\end{equation}
Then $\ell^\top f_t=\lambda_0\theta_t+r_t$ and $\operatorname{Cov}(r_t,\theta_t)=0$ under the stationary fundamental law. This is a contemporaneous \emph{linear projection} decomposition, not independence between $r_t$ and the entire history of $\theta$. For example, if $\theta_t=u_t$, $u_t$ and $v_t$ are contemporaneously uncorrelated, and the payoff is $D_{t+1}=\bar d+2u_t+3v_t+\varepsilon_{t+1}$, then $r_t=3v_t$. The additional term matters for the payoff although private signals continue to concern $u_t$. More generally, when $F$ couples components, $r_t$ can be correlated with lagged or future fundamentals. The difference between aligned and nonaligned payoffs is therefore \emph{not} whether the payoff--fundamental correlation equals one; it is whether $\theta_t$ spans the full current-state component of the expected payoff. In the aligned branch, $b=0$ and $\lambda_0=\lambda>0$. Outside this branch, improved public-data computation may also reduce payoff risks that are not captured by the private signal, so Theorem 1 imposes no universal sign.

\subsection{Trader populations, private observations, and preferences}\label{subsec:S1-traders}

There are finitely many trader types $c\in\mathcal C=\{1,\ldots,C\}$ with population masses $\pi_c>0$ summing to one. Write $\mathcal I_c$ for the continuum of type-$c$ accounts and $\mu_c$ for its normalized population measure, so that aggregate demand weights type-$c$ averages by $\pi_c$. Type $c$ has absolute risk aversion $\gamma_c>0$, noisy private-signal variance $\nu_c>0$, and a computational resource budget $B_c$ specified in S2. Its members share the deployed forecasting technology and readout coefficients but differ in their individual private observations. For ease of interpretation, types may describe groups of accounts served by different forecasting organizations: the organization trains and deploys an algorithm, while each atomistic account chooses its own risky holding. An organization-wide technology adoption may affect the market because it is used by a positive mass of accounts; any individual account remains a price taker. We write $x\in[0,1]$ for the \emph{share of the relevant group of investors} using the higher-capacity system. When only part of type $c$ upgrades, the high- and low-capacity masses are $x\pi_c$ and $(1-x)\pi_c$. In experiments with several adopting groups, $x$ may be replaced by a vector of group-specific shares. It is a population fraction, not an individual holding or an order.

A type-$c$ investor $i$ receives, before submitting demand at $t$, a scalar private signal
\begin{equation}\label{eq:S1-private}
  s_{ci,t}=\theta_t+\xi_{ci,t},\qquad
  \xi_{ci,t}\sim\mathcal N(0,\nu_c),\quad \nu_c>0.
\end{equation}
The errors are independent across individuals and dates, independent of all aggregate primitive shocks, and independent of the public record conditional on the fundamental. Different accounts therefore receive different signals about the \emph{same} $\theta_t$; the signals are private because each account's realization of $\xi_{ci,t}$ is not publicly announced. This need not mean private knowledge is perfectly accurate. As investors trade on these different clues, the price can aggregate information about $\theta_t$ even when no single trader has an error-free observation of it. Signal quality can differ across types without changing the aggregate shock structure. As a continuum-economy idealization, each within-type normalized measure $\mu_c$ obeys the exact aggregation property
\begin{equation}\label{eq:S1-aggregation}
  \int_{\mathcal I_c}\xi_{ci,t}\,d\mu_c(i)=0,
  \qquad
  \int_{\mathcal I_c}s_{ci,t}\,d\mu_c(i)=\theta_t
  \qquad \text{almost surely}.
\end{equation}
Equation \eqref{eq:S1-aggregation} is imposed as an exact law-of-large-numbers condition on the continuum and can be formalized on a suitable Fubini extension; it is not deduced from countably many independent random variables. Individual signal noise affects learning and personal risk exposures but contributes no separate aggregate shock after integration. With affine private-signal forecast heads, this assumption provides a simple aggregate price loading on $\theta_t$ (S3). It does not amount to a trader individually observing $\theta_t$.

Investor preferences over wealth are constant-absolute-risk-aversion (CARA):
\begin{equation}\label{eq:S1-CARA}
  u_c(W)=-\exp(-\gamma_c W),\qquad \gamma_c>0.
\end{equation}
Investors take the candidate clearing price as given. They will select a risky position based on the mean and risk of the payoff under their \emph{own learned forecast}. In S3, CARA utility and a subjective Gaussian forecasting distribution give an exact price-contingent demand function. The rationality requirement is optimal portfolio choice \emph{conditional on the fitted belief}; an arbitrary nonlinear, constrained forecast need not equal the objective conditional distribution of the payoff. Correct Gaussian Bayesian forecasts are invoked separately in the sharper analytical branch.

Computational upgrades are \emph{type-preserving}: a specified mass of investors changes its feasible forecasting technology, while its private-signal precision, risk preferences, the exogenous state law, and the supply process are held fixed. Increasing $x$ shifts some accounts from low to high computational capacity without changing the fraction receiving good or bad private news. Unless explicitly modeled otherwise, the adoption share is varied as an economic counterfactual, not chosen strategically within the auction. The marginal private value of an upgrade is measured for a negligible account, which does not internalize the market-wide price effect of upgrading a positive mass of other accounts.

\subsection{Public records, pre-price information, and the call auction}\label{subsec:S1-filtrations}

The public information record immediately \emph{before} auction $t$ is the $\sigma$-field
\begin{equation}\label{eq:S1-filtration}
  \mathscr H_t^-
  =\sigma\big(m_\tau: \tau\leq t;\;
               p_\tau: \tau\leq t-1;\;
               D_\tau: \tau\leq t\big).
\end{equation}
Equivalently, the newest public input is $o_t=(m_t,p_{t-1},D_t)$ and $\mathscr H_t^-=\sigma(o_t,o_{t-1},\ldots)$. The history contains current publicly released macro observations, \emph{past} clearing prices, and all payouts that have already matured. It excludes the current price $p_t$, the future dividend $D_{t+1}$, and the unrevealed supply shock $\zeta_t$. The post-auction public information field is
\begin{equation}\label{eq:S1-postfield}
  \mathscr H_t^+=\mathscr H_t^-\vee\sigma(p_t).
\end{equation}
At the start of date $t+1$, the newest price $p_t$ and realized payoff $D_{t+1}$ become available in $\mathscr H_{t+1}^-$. This indexing prevents the algorithm from using the price generated by its own current action as a \emph{pre-auction historical input}.

All types can access the same raw public disclosures $\mathscr H_t^-$. The capacity difference is what they can \emph{calculate from} those disclosures, not which public files they are legally allowed to see. An AI of type $c$ produces a state or feature vector $z_{c,t}$ from the pre-auction sequence of public observations $o_t,o_{t-1},\ldots$. Schematically one may write $z_{c,t}=g_{\phi_c}(o_t,o_{t-1},\ldots)$, with the operational form of $g_{\phi_c}$ restricted by the computational budget. The formal timing requirement is
\begin{equation}\label{eq:S1-adapted}
 z_{c,t}\ \text{is }\mathscr H_t^-\text{-measurable}.
\end{equation}
For example, a simple algorithm might retain only the current macro release, $z_{\mathrm{low},t}=m_t$. A more capable algorithm might process $m_t$ together with recent prices, matured dividends, and older macro releases; a recurrent neural network could compress an even longer history into a finite-dimensional state. Importantly, $z_{c,t}$ is constructed \emph{only from public history}, not from an investor's private $s_{ci,t}$ or its past values. Private clues enter the individual forecast separately. S2 defines the admissible encoders, memory and operation budgets, and optimization of the forecasting model. A small state need not consist of the latest raw prices: it may encode persistent patterns in older records. Equally, access to a long public record does not imply that a resource-limited algorithm can extract all its predictive information.

After the public encoder state $z_{c,t}$ is formed, each investor observes her own private signal $s_{ci,t}$ and submits a \emph{demand schedule} $h_{ci,t}(p)$: the number of shares she wishes to hold at each possible price $p$. Thus $h$ denotes the investor's desired \emph{holding}, whereas $x$ denotes the \emph{population share} using a higher-capacity algorithm. The desired holding can depend on all three inputs---$s_{ci,t}$, $z_{c,t}$, and the candidate price $p$---and on the fitted forecasting model. For a fixed deployed forecasting model, we can therefore write $h_{ci,t}(p)=h_c(s_{ci,t},z_{c,t},p)$, with learned model parameters suppressed; S3 derives this function from utility maximization. The candidate price can enter the forecast and demand even though the realized clearing price is not yet known. The clearing price is selected to satisfy
\begin{equation}\label{eq:S1-clearing-condition}
  \sum_{c\in\mathcal C}\pi_c
     \int_{\mathcal I_c}h_{ci,t}(p_t)\,d\mu_c(i)
       =\bar S+\zeta_t.
\end{equation}
The price-taking assumption applies to individual accounts, not to a positive-mass algorithmic technology group. S3 derives the optimal demand schedules, conditions for a strictly downward-sloping aggregate demand curve (equivalently a positive clearing denominator), and the resulting unique temporary clearing price.

\begin{table}[htbp]
\centering\small
\caption{The state, signals, computed features, holdings, and adoption in the information-to-price mechanism.}
\label{tab:S1-information-map}
\begin{tabularx}{\linewidth}{@{}p{25mm}X X@{}}
\toprule
\textbf{Object} & \textbf{Economic role} & \textbf{Who observes or chooses it?}\tabularnewline
\midrule
$f_t$, $\theta_t$ & Latent economy and the scalar target of private clues. & Latent primitives, inferred from private and public observations.\tabularnewline
$s_{ci,t}$ & Account $i$'s noisy clue about $\theta_t$. & Privately observed by account $i$.\tabularnewline
$z_{c,t}$ & Algorithm $c$'s processed summary of past public observations. & Computed from data accessible to everyone; the summary depends on capacity.\tabularnewline
$h_{ci,t}(p)$ & Desired number of risky shares at candidate price $p$. & Chosen by investor $i$ using $s_{ci,t}$, $z_{c,t}$ and $p$.\tabularnewline
$p_t$, $x$ & Clearing price and high-capacity adoption fraction. & $p_t$ emerges from clearing; $x$ specifies a counterfactual technology mix.\tabularnewline
\bottomrule
\end{tabularx}
\end{table}

A candidate price is an \emph{argument} of the demand schedule and the forecast readout; it is not a member of $\mathscr H_t^-$. Once the auction settles, the actual random price $p_t$ is added to the observed data. An individual account's signal noise has no first-order effect on aggregate price under \eqref{eq:S1-aggregation}; its signal and the clearing price can nevertheless be jointly informative about $\theta_t$. The supply shock is not an additional trader's private signal about fundamentals: its role is to prevent the price from perfectly revealing the cross-sectional aggregate private observation.

\subsection{One-period timing and the two clocks}\label{subsec:S1-timing}

The order of events within a trading date and across training epochs matters for every theoretical comparison. Table \ref{tab:S1-timing} records the chronology at a fixed deployed profile of forecasting algorithms.

\begin{table}[htbp]
\centering\small
\caption{Chronology of one date in the fast market.}\label{tab:S1-timing}
\begin{tabularx}{\linewidth}{@{}p{29mm}X@{}}
\toprule
\textbf{Stage} & \textbf{Events and information available}\tabularnewline
\midrule
0. State and releases & $f_t$ is realized; previously traded claim pays $D_t$; the public sees $m_t$, $D_t$, and the prior price $p_{t-1}$, updating $\mathscr H_t^-$. The new payoff $D_{t+1}$ is not known.\tabularnewline
1. Pre-auction computation & Every deployed forecasting algorithm updates its internal state from $\mathscr H_t^-$. This step cannot use $p_t$ or $\zeta_t$.\tabularnewline
2. Private clues & Individual investors observe $s_{ci,t}$ and choose their full price-contingent schedules $h_{ci,t}(p)$.\tabularnewline
3. Clearing & The exogenous net supply $\bar S+\zeta_t$ is presented to the auction; $p_t$ is the price at which integrated demand equals supply.\tabularnewline
4. Public price & The newly determined $p_t$ enters $\mathscr H_t^+$. Forecasting rules are not refitted within this auction.\tabularnewline
5. Next date & $D_{t+1}$ matures and is disclosed before the next auction, entering $\mathscr H_{t+1}^-$. The process repeats with the next one-period claim.\tabularnewline
\bottomrule
\end{tabularx}
\end{table}

There are two separate time indices. The \emph{fast clock} $t$ indexes the trading dates just described, during which forecasting parameters are fixed. The \emph{slow clock} $n=0,1,2,\ldots$ indexes model deployment and retraining epochs. At the beginning of epoch $n$, a profile of learned forecasting systems $\mathbf M_n=(M_{c,n})_{c\in\mathcal C}$ is deployed. Its induced fast market, if it admits an invariant law, generates a stationary distribution of history, price, private signals, and realized payoff. Call this law $Q(\mathbf M_n)$, suppressing type-specific marginals. When population retraining is undertaken, agents' next forecasting systems are chosen using this \emph{endogenously generated} data law. S2 states the training objective; S3 formally defines the fixed point and analyzes the law-to-model map.

We do \emph{not} assert that every admissible nonlinear forecasting algorithm automatically induces a stable price process. A stable latent $F$ in \eqref{eq:S1-state} and a stable encoder considered in isolation do not guarantee fast-market stationarity when their trading decisions feed back through past prices. Existence of an invariant $Q(\mathbf M_n)$, sufficient integrability, uniqueness where invoked, and regularity of the learning map are explicit conditions in later equilibrium claims or are checked numerically for a specified calibration. Likewise, stationarity of the within-epoch price process is logically distinct from convergence of the cross-epoch retraining sequence.

\subsection{Three deployment comparisons and counterfactual values}\label{subsec:S1-experiments}

A statement about the consequence of a stronger forecasting algorithm is incomplete unless the response of market prices and other agents' learning is specified. We use three distinct protocols.

\begin{description}[leftmargin=6mm,style=nextline,itemsep=5pt]
\item[Inherited-law or immediate-deployment comparison.]
Start from a fixed reference market law $Q_0$. Low- and high-capacity candidate forecasting systems are both trained on $Q_0$, including the \emph{inherited} current-price observations in the historical training sample. Holding the inherited pre-auction history fixed, replace a specified positive mass of deployed policies and clear the next auction. This is the first-deployment comparison underlying Theorem 1; it does not require the new trading rule to be optimal for the subsequently altered price distribution.

\item[Frozen-policy stationary comparison.]
Keep all selected incumbent forecasting parameters fixed after changing the adoption mix, but run the associated fast market until its new stationary law $Q_x^{\mathrm F}$ is obtained, when it exists. The resulting price history is endogenous even though the algorithms' parameters do not change. The private value of a negligible \emph{candidate} upgrade evaluated at $Q_x^{\mathrm F}$ is a separate counterfactual: both candidate capacities are optimally evaluated under that law. It is not the realized wealth change of a candidate whose old coefficients were literally frozen.

\item[Population-retrained stationary comparison.]
For a specified capacity distribution indexed by adoption $x$, refit the forecasting systems under the market law they themselves induce. A stationary computational-learning equilibrium is a law $Q_x^*$ and a profile of admissible population-optimal policies that reproduce one another through training and clearing (formalized in S3). Unless specified otherwise, all extant types can retrain, including types that did not adopt a stronger algorithm. The candidate's private incremental value is again evaluated with both feasible forecasting alternatives fitted under the \emph{same} $Q_x^*$.
\end{description}

For the first protocol we avoid labeling a single cleared price observation a new \emph{stationary} distribution. For the other two we do not equate an invariant fast-market law with a stable fixed point of the slow retraining map. Each price-information and investment-value result is tagged by its protocol. These distinctions are needed in particular because the nonlinear state-dependent forecast specification can exhibit a frozen-policy distribution-adjustment effect even before any retraining occurs.

\subsection{Pre-price computational information and the scope of Gaussian statements}\label{subsec:S1-infosets}

We next distinguish the common raw record from the \emph{information retained by a particular feasible encoder}. The notation $\sigma(z)$ means the \emph{$\sigma$-algebra generated by the random variable or vector $z$}: in plain terms, everything that could be inferred from observing that computed summary. This $\sigma(\cdot)$ is an information-set operator, \emph{not} a standard deviation (contrast $\sigma_\varepsilon$ and $\sigma_\zeta$). Let $L$ and $H$ label particular low- and high-capacity public summaries. Under a market law indexed by the high-capacity adoption fraction $x$, suppose their processed information is nested: \emph{all information available from the low-capacity summary can also be recovered from the high-capacity one}. Then
\begin{equation}\label{eq:S1-FU}
  \mathcal F_{L,x,t}=\sigma(z_{L,x,t}),\qquad
  \mathcal U_{H,x,t}=\sigma(z_{H,x,t}),\qquad
  \mathcal F_{L,x,t}\subseteq\mathcal U_{H,x,t}
                       \subseteq\mathscr H_t^-.
\end{equation}
The notation $z_{L,x,t}$ and $z_{H,x,t}$ keeps track of the economic environment as well as the algorithms. When $x$ changes, even an unchanged encoder receives a different \emph{price history}, because the market's trading and prices change. The $x$ subscript does not denote a different legal right to see public releases, and the low and high summaries need not be numerical vectors of different length. A concrete nested example is $z_L=m_t$ and $z_H=(m_t,m_{t-1},p_{t-1},D_t)$. The latter retains all information in the former and can additionally use recent public history. This nesting is imposed on these \emph{selected summaries} only when invoked in S5.

The current clearing price $p_t$ and investor-specific private signal $s_{ci,t}$ are \emph{added after} the pre-price summaries are defined and are included in both forecast alternatives. Thus the individual inference fields are $\mathcal F_{L,x,t}\vee\sigma(p_t,s_{ci,t})$ and $\mathcal U_{H,x,t}\vee\sigma(p_t,s_{ci,t})$. The symbol $\vee$ denotes the smallest information field containing both sets of observations. This structure makes clear why upgrading an AI does not alter the individual's private-signal quality: it changes the public-history computation while keeping her signal and current-price observation available.

Crucially, nested \emph{feasible algorithm classes} do not imply \eqref{eq:S1-FU} for the \emph{outputs independently selected} by two global optimizers. Theorem 1 compares nested \emph{feasible classes} on a fixed training law and does not require their optimal feature maps to be nested. Theorem 2 instead compares the forecast value of an explicitly nested \emph{pair of feasible pre-price information sets}. These are different mathematical claims. If the high-capacity learner's summary does not actually contain the low-capacity learner's summary, one can enlarge it to their joint statistic when the budget permits, or restrict Theorem 2 to a nested finite-information benchmark; one cannot assume containment merely because the high model has more parameters or lower average forecasting loss.

For the experiments, an important finite-history comparison uses
\begin{align}\label{eq:S1-eight-fields}
 \mathcal F_{L,x,t}
   &=\sigma(m_t),\\[-2pt]
 \mathcal U_{H,x,t}
   &=\sigma\!\left(m_t,
       \big\{m_{t-j},p_{t-j},D_{t+1-j}\big\}_{j=1}^{3}\right).
 \nonumber
\end{align}
These are pre-price fields under the same law $Q_x$, and the second field uses exactly three lagged records. In particular, $D_{t+1-j}$ is already realized for every $j\ge1$. Infinite-history or 40-lag oracles can be useful for a \emph{separate} observer-accessibility calculation, but they must not be substituted for $\mathcal U_{H,x,t}$ when stating the finite-$K=3$ private value of additional computation. In the two-factor existence construction of S6, the fields specialize to $\mathcal F_L$ that does not process a publicly released scalar macro signal $m_t$ and $\mathcal U_H=\mathcal F_L\vee\sigma(m_t)$.

Conditional mutual information such as $I(\theta_t;p_t\mid\mathcal F_{L,x,t})$ measures the reduction in uncertainty supplied by observing the \emph{current price} after carrying out the low-capacity analysis. It is not the same mathematical object as likelihood precision of an aggregate private-information loading. Along a correctly specified jointly Gaussian equilibrium price law, the mutual information can be calculated from conditional covariance matrices; in the general nonlinear encoder economy, the price process need not be jointly Gaussian even when $f_t$ and primitive shocks are. S5 states precisely when conditional-variance identities and information-to-certainty-equivalent conversions are valid.

For clarity, \emph{aligned payoff} is a primitive restriction on \eqref{eq:S1-payoff}, whereas \emph{correct joint Gaussianity of prices and forecasts} is a restriction on the equilibrium price map and informational architecture. The former does not imply the latter once nonlinear forecasting and nonlinear price feedback are admitted. State-dependent mean and forecast variance are legitimate components of the general market in S2 and S3 but do not automatically inherit all the analytical sign results of the constant-affine Gaussian submodel.

\subsection{How the three analytical theorems specialize the same environment}\label{subsec:S1-domain}

The paper's results have different degrees of generality. Table \ref{tab:S1-theorems} identifies their maintained economic primitives, the distinct informational restrictions that generate sharp analytical predictions, and the comparison to which each applies. The detailed statements are given in S4--S6, not assumed here.

\begin{table}[htbp]
\centering\small
\caption{Theorem-domain map within one economic model.}\label{tab:S1-theorems}
\begin{tabularx}{\linewidth}{@{}p{17mm}X X@{}}
\toprule
\textbf{Result} & \textbf{Informational and statistical specialization} & \textbf{Claimed comparison}\tabularnewline
\midrule
Theorem 1 (S4) & Aligned payoff; optimally fitted constant-affine Gaussian head; arbitrary feasible nonlinear public encoders with nested complete algorithm classes; fixed inherited law. & Upgrade changes the aggregate private-signal loading and fresh-price likelihood precision in the first deployed auction, with other policies fixed.\tabularnewline
Theorem 2 (S5) & Correct joint Gaussianity; aligned payoff; the same explicitly nested finite-capacity public information fields, with current price and own private signal added to both. & Information accounting and comparative-statics conditions for a negligible investor's private capacity premium along stationary laws $Q_x$.\tabularnewline
Theorem 3 (S6) & A two-factor, publicly released macro signal and a private fundamental innovation; population Gaussian forecasts and CARA competitive clearing; aligned payoff. & Strict local coexistence of negative new-information discovery, positive low-capacity price information, and lower private capacity premium, with both frozen policies and population refitting.\tabularnewline
\bottomrule
\end{tabularx}
\end{table}

None of the three theorems gives an unrestricted sign for every state-dependent-variance forecasting algorithm, every nonaligned payoff, or every stationary retraining equilibrium. The population experiments, finite-data linear state-space and GRU exercises, and nonaligned/stochastic-risk cases are therefore reported with separate statistical assumptions and evaluation metrics in S7--S9. They do not silently strengthen Theorem 1 into a universal welfare or mutual-information monotonicity result.

\subsection{A compact assumptions and notation ledger}\label{subsec:S1-ledger}

Table \ref{tab:S1-assumptions} collects assumptions that are maintained in the common analytical core, distinguishing them from the restrictions introduced only to prove an individual theorem. Where a parameter is varied, the corresponding computational experiment says so explicitly.

\begin{table}[H]
\centering\small
\caption{Shared economic assumptions, with locations of refinements.}\label{tab:S1-assumptions}
\begin{tabularx}{\linewidth}{@{}p{18mm}X X@{}}
\toprule
\textbf{Label} & \textbf{Maintained in core economic environment} & \textbf{Added, rather than presumed, later}\tabularnewline
\midrule
ENV-1 & Finite-dimensional stable exogenous economic state, stationary baseline covariance. & Reachability/observability or full-rank projections where required.\tabularnewline
ENV-2 & One-period claim and risk-free account; payoff noise $\sigma_\varepsilon^2>0$. & A long-lived resale-price recursion or state-dependent structural payoff volatility in robustness studies.\tabularnewline
ENV-3 & Continuum of atomistic traders, Gaussian private signals and exact within-type aggregation. & Alternative cross-sectional signal structures, if used in a separate experiment.\tabularnewline
ENV-4 & Independent noisy net supply, $\sigma_\zeta^2>0$, and unconstrained positions. & Positive clearing denominator (downward-sloping aggregate demand) and invariant market-price law (S3).\tabularnewline
ENV-5 & Common raw pre-price public record with only matured payouts; capacity-specific processed summaries. & Global existence of an optimal learned forecast and nested information fields for a particular theorem (S2, S5).\tabularnewline
ENV-6 & CARA portfolio choice against an individually fitted forecast and competitive price taking. & Subjective Gaussian head for S3 closed form; correct joint Gaussianity for S5--S6.\tabularnewline
\bottomrule
\end{tabularx}
\end{table}

\smallskip
\begingroup\small
\begin{longtable}{@{}p{30mm}p{110mm}@{}}
\caption{Notation established in S1 and used later in the SI.}\label{tab:S1-notation}\\
\toprule
\textbf{Symbol} & \textbf{Definition}\tabularnewline\midrule\endfirsthead
\toprule\textbf{Symbol} & \textbf{Definition (continued)}\tabularnewline\midrule\endhead
$t,n$ & Fast trading date and slow deployment/retraining epoch, respectively.\\
$f_t,F,G,\Sigma_f$ & Latent state, transition, innovation loading, stationary state covariance.\\
$m$ & Dimension of the latent state $f_t$ (not computational capacity).\\
$m_t,H,\omega_t$ & Public macro-release vector, observation loading, and optional release noise. The scalar $m_t$ in S6 is a public macro factor.\\
$\theta_t,a$ & Privately signaled scalar fundamental and its latent-state loading.\\
$D_{t+1},\bar d,\ell,\varepsilon_{t+1}$ & One-period payoff, intercept, factor loading, independent payoff shock.\\
$\lambda,r_t$ & Aligned payoff loading (when imposed), and other predictable payoff component in the unrestricted case.\\
$p_t,R_f,X_{t+1}$ & Date-$t$ price, risk-free gross return, and risky excess payoff per unit.\\
$h_{ci,t}(p)$ & Investor $i$'s price-contingent desired number of risky shares; not a probability or the adoption fraction.\\
$S_t,\bar S,\zeta_t,\sigma_\zeta^2$ & Signed net risky-asset supply, mean, unobserved noise and its variance.\\
$c,\pi_c,\gamma_c,\nu_c$ & Investor type, mass, CARA coefficient and idiosyncratic signal variance.\\
$x\in[0,1]$ & Fraction of a specified investor group using high-capacity AI; adoption counterfactual.\\
$s_{ci,t},\xi_{ci,t},\mu_c$ & Individual signal, its noise, normalized within-type measure.\\
$B_c,\mathbf M_n,z_{c,t}$ & Computational budget, epoch-$n$ deployed model profile, feasible public-history state.\\
$\mathscr H_t^-,\mathscr H_t^+$ & Raw pre- and post-auction public information fields.\\
$\mathcal F_{L,x,t},\mathcal U_{H,x,t}$ & Pre-price processed low-/high-capacity information fields, nested only where imposed.\\
$\sigma(z)$, $\vee$ & Information generated by observing $z$, and the join of two information fields; distinct from a standard deviation.\\
$Q_0,Q_x^{\mathrm F},Q_x^*$ & Inherited market law, frozen-policy stationary law, and population-retrained stationary law.\\
$\bar S+\zeta_t$ & Auction net supply; its shock is not separately announced to traders.\\
\bottomrule
\end{longtable}
\endgroup

\subsection{Interpretive boundaries}\label{subsec:S1-boundaries}

Five distinctions remain relevant in the later proofs. \emph{First}, the observable financial price can be useful by conveying either newly aggregated private observations or public information that another agent could not efficiently process; informational value depends on the observer's conditioning information and need not have a single sign. \emph{Second}, a primitive Gaussian state process does not force market prices under nonlinear trading algorithms to be jointly Gaussian. \emph{Third}, a stronger computational budget means a larger set of feasible algorithms, not that every trained algorithm is better in finite samples or that two selected internal representations must be nested. \emph{Fourth}, a strict stationary fast-market condition does not imply convergence of model updates across epochs. \emph{Fifth}, informational improvement is not, by itself, a proof of an improvement in investors' certainty equivalents or total social welfare. These distinctions are formalized where relevant rather than substituted for their proof-specific conditions.

The core environment thus leaves the adoption-response signs \emph{to be derived}. It specifies a source of dispersed private information, a common public history that can be processed unequally, optimizing investors, and a competitive market whose prices are generated by their orders. S2 and S3 turn these primitives into a constrained-learning competitive market; S4--S6 then identify conditions under which improved computation crowds out fresh price discovery, disseminates processed public information, and changes the incentive to compute again.

\clearpage\section{Resource-constrained learning}\label{sec:S2}

Investors in S1 receive individual private signals and observe a common public record. They trade on forecasts of the asset's next payout, and their combined orders determine the current market price. This section specifies how an AI builds those forecasts. It also explains why deploying a more capable algorithm can change the distribution of the \emph{data} that even an unchanged algorithm receives.

Each forecasting system has two parts. Before the auction, an \emph{encoder} turns the public record into an internal state $z_{c,t}$. At the auction, a \emph{forecasting head} combines $z_{c,t}$, the investor's private signal $s_{ci,t}$, and a candidate price $p$ to predict the next payout $D_{t+1}$ and its uncertainty. The general head permits state-dependent mean and variance forecasts; the restricted head used for Theorem~1 has constant affine coefficients. We also distinguish two meanings of higher computational ability: being \emph{able} to run a richer forecasting program, and actually \emph{retaining} more information after training. They are not equivalent.

\subsection{The objects entering the forecast}\label{subsec:S2-recap}

The underlying economy has a latent, possibly multidimensional state $f_t\in\R^m$. Its scalar component $\theta_t=a^\top f_t$ is the object about which investors receive private clues. The next-period risky payout is
\begin{equation}\label{eq:S2-payoff}
 D_{t+1}=\bar d+\ell^\top f_t+\varepsilon_{t+1}.
\end{equation}
An investor $i$ of type $c$ privately observes $s_{ci,t}=\theta_t+\xi_{ci,t}$, where $\xi_{ci,t}$ is idiosyncratic noise with variance $\nu_c>0$. The \emph{aligned-payoff} specialization, used in the main crowd-out proof, sets $\ell=\lambda a$ for $\lambda>0$. In that case, the current state affects the expected payout only through $\theta_t$: $D_{t+1}=\bar d+\lambda\theta_t+\varepsilon_{t+1}$. This does not give investors perfect knowledge of $\theta_t$: their private signals remain noisy. The general model also permits other components of $f_t$ to affect the payout directly (S1).

A date-$t$ public observation and the corresponding record available \emph{before} the auction are
\begin{equation}\label{eq:S2-input}
 o_t=(m_t,p_{t-1},D_t),\qquad
 \mathscr H_t^-=\sigma(o_t,o_{t-1},\ldots).
\end{equation}
Here $m_t$ denotes macroeconomic data released by the start of date $t$, $p_{t-1}$ is the previous clearing price, and $D_t$ is the payout that has just matured. By contrast, the \emph{current} clearing price $p_t$ is still to be determined, and the next payout $D_{t+1}$ has not occurred. Individual private signals $s_{ci,t}$ are observed by their owners but are not part of the common record. The symbol $\sigma(\cdot)$ means the information generated by its arguments (a $\sigma$-algebra), not a standard deviation. All types start from these same public records but may compute different summaries.

\begin{table}[htbp]
\centering\small
\caption{What an investor observes and what the algorithm produces. The final two columns prevent a common confusion between private signals, public computations, and the current price.}\label{tab:S2-inputs}
\begin{tabularx}{\linewidth}{@{}p{24mm}X p{39mm}@{}}
\toprule
\textbf{Object} & \textbf{Meaning} & \textbf{Available when?}\\\midrule
$f_t,\theta_t$ & Economic state and the common fundamental; generally latent, not supplied to an investor. & Not directly observed.\\
$s_{ci,t}$ & Investor-specific noisy observation of $\theta_t$; its realization is private. & Before submitting demand.\\
$o_t,\mathscr H_t^-$ & Released macro data, past prices, and matured payouts; common raw record. & Before the auction.\\
$z_{c,t}$ & The algorithm's computed representation of that record. & Computed before the auction.\\
$p$ & Candidate \emph{current} clearing price in a demand schedule. & Varied at the auction.\\
$D_{t+1}$ & Target payout for forecasting and for the subsequent payoff realization. & Revealed only next period.\\
\bottomrule
\end{tabularx}
\end{table}

The index $c$ denotes a forecasting-technology type, while $i$ identifies an individual investor. People of the same type use the same public algorithm but can trade differently because they observe different $s_{ci,t}$. Their \emph{demand schedule} $h_{ci,t}(p)$, derived in S3, specifies the holding each would choose if the current price were $p$. The adoption fraction $x\in[0,1]$ is different: it measures the share of a specified investor population using the higher-capacity algorithm. Thus $h$ is an individual \emph{holding as a function of price}, whereas $x$ describes the \emph{mix of forecasting technologies} in the market.

\subsection{An encoder of the public record}\label{subsec:S2-encoder}

An encoder may carry information forward from one public observation to the next, or it may recompute its representation from a finite window. A generic recurrent form is
\begin{equation}\label{eq:S2-encoder}
 z_{c,t}=T_{\phi_c}(z_{c,t-1},o_t),
 \qquad z_{c,t}\in\mathcal Z_{\phi_c},
\end{equation}
Here $T_{\phi_c}$ is a state-update rule, and $\phi_c$ specifies its parameters and initialization convention. The vector $z_{c,t}$ is what the system keeps from the public history; it is not an additional economic signal arriving from outside the market. The update rule can be linear or nonlinear. A finite-window alternative computes $z_{c,t}=E_{\phi_c}(o_t,\ldots,o_{t-L})$ directly. Thus a linear state-space filter, a recurrent neural network, or a finite-context transformer can all be modeled as ways of constructing $z$. The representation may have one or many coordinates, without having the same dimension as the latent economic state.

The fundamental requirement is pre-auction measurability:
\begin{equation}\label{eq:S2-adapted}
 z_{c,t}\text{ is }\mathscr H_t^-\text{-measurable}.
\end{equation}
The encoder cannot use the current price $p_t$, which has not yet cleared, or the investor's private signal $s_{ci,t}$. The private signal enters later, at the forecasting head. This separation is important for the economic comparison: we ask what happens when algorithms become better at processing information that was already \emph{public}, while the quality of the investor's \emph{private} clue remains unchanged.

For example, a low-budget system might keep only today's released macro vector, $z_{L,t}=m_t$. A higher-budget system might retain $z_{H,t}=(m_t,m_{t-1},D_t,p_{t-1})$ or use those observations to compute more complex features. Both see the same underlying public releases. The difference is how much useful information their algorithms extract and retain. Neither state includes the unobserved $\theta_t$ or an individual's $s_{ci,t}$ as a directly supplied input.

\paragraph{A fixed algorithm can receive different data.}
An increase in the adoption share $x$ changes the mix of orders submitted to the market. It can therefore change the current clearing price \emph{before anyone retrains an algorithm}. If the fitted encoders $T_{\phi_c}$ and forecasting heads are then held fixed, the newly generated prices still enter later public observations $o_{t+1},o_{t+2},\ldots$. As a result, the distribution of each encoder's input---and hence the distribution of its output $z_{c,t}$---can change. For example, a low-budget algorithm that retains yesterday's price may see different inputs tomorrow simply because more traders use the high-budget algorithm today. Its forecast \emph{rule} is unchanged, but the forecasts and trades it produces need not be.

This gives two distinct subsequent effects. First, with all fitted algorithms frozen, the changed price history can alter future holdings and prices until the market reaches a new stationary distribution, if one exists. Second, investors may \emph{retrain} the algorithms using observations generated in that new market, changing the rules $T_{\phi_c}$ and the forecasting heads themselves. Retraining therefore adds a feedback channel; it is not necessary for the preceding change in the distribution of $z$. Section S3 defines these stages of adjustment precisely.

For an encoder run on an unbounded recurrent history, population comparisons use its causal stationary state when such a state exists. A finite-window implementation instead specifies the window and its initialization rule.

\subsection{Resource constraints: memory and operations}\label{subsec:S2-budget}

\begin{definition}[Feasible inference algorithm]\label{def:S2-feasible}
An inference algorithm comprises the pre-auction public-data encoder, any stored state and parameters needed to run it, and the forecasting readout that combines its output with $s_{ci,t}$ and a candidate $p$. Let $C^{\mathrm{mem}}(M)$ be a specified bound on its stored memory and $C^{\mathrm{ops}}(M)$ a bound on inference operations per observation and forecast, under a common implementation and precision convention. The resource budget of type $c$ is $B_c=(B_c^{\mathrm{mem}},B_c^{\mathrm{ops}})$, and
\begin{equation}\label{eq:S2-class}
 \mathcal M_c(B_c)=\bigl\{M:\ C^{\mathrm{mem}}(M)\le B_c^{\mathrm{mem}},\quad C^{\mathrm{ops}}(M)\le B_c^{\mathrm{ops}},\quad M\text{ obeys }\eqref{eq:S2-adapted}\bigr\}.
\end{equation}
\end{definition}

The budget limits computation \emph{when the trained algorithm is used}; the cost of training and the amount of training data are tracked separately. A finite-window predictor pays to store and process its window. A recurrent predictor pays to update its saved state and produce a forecast. To compare budgets consistently, we use the same precision and resource-accounting conventions for parameter storage, state memory, and operations. Because the forecasting head is affine in candidate price, it can report the coefficients of an entire price-contingent demand schedule without repeatedly running the encoder.

We write $B_c'\succeq B_c$ when each component of $B_c'$ is at least the corresponding component of $B_c$. Under the same resource accounting and the ability to reuse an existing program unchanged,
\begin{equation}\label{eq:S2-nestedclass}
 B_c'\succeq B_c\quad\Longrightarrow\quad
 \mathcal M_c(B_c)\subseteq\mathcal M_c(B_c').
\end{equation}
This means that the higher-budget learner \emph{can run} any lower-budget program. It does not mean that two programs trained independently will store the same information. Theorem~1 also requires the feasible classes to permit the full constant-affine forecasting specification: increasing the budget may not silently exclude old coefficient choices, public predictors, or admissible constant forecast variances. By contrast, a GRU and a transformer trained separately need not form nested model classes simply because one has more parameters.

For theory, the feasible class may be an appropriate closure or continuous-parameter relaxation of finite-precision programs; S4 states the attainability assumptions needed to use a global optimum. In experiments, memory and operations are measured for finite implementations. Hidden-state dimension and filter order are therefore useful \emph{architecture-specific measures}, not universal definitions of computational ability.

\begin{table}[htbp]
\centering\small
\caption{Examples of admissible public-data encoders. None is imposed as the universal model of computational capacity.}\label{tab:S2-architectures}
\begin{tabularx}{\linewidth}{@{}p{29mm}X X@{}}
\toprule
\textbf{Implementation} & \textbf{How public history is represented} & \textbf{Resources to measure}\\\midrule
Finite impulse response / linear state space & A fixed lag window or a linear recurrent state; includes the analytically solvable Gaussian benchmark. & State memory, coefficients, matrix operations.\\
GRU / recurrent network & Nonlinear recurrent update of a learned hidden representation. & Weights, stored hidden state, update/forecast operations.\\
Finite-context transformer & History processed by attention over a bounded token/window sequence or retained cache. & Context/cache memory, parameter storage, attention/readout operations.\\
Finite-history feature program & A prescribed collection of lags, thresholds, and nonlinear transformations of public data. & Stored history, instructions, output/head costs.\\
\bottomrule
\end{tabularx}
\end{table}

\subsection{Forecasting distributions: the general head and the analytical head}\label{subsec:S2-heads}

After computing the public state $z_{c,t}$, the algorithm combines it with the investor's private signal and a candidate price. Its output is a predictive distribution for the next payout; the investor will choose a holding using that forecast in S3. In the general \emph{Model B}, both the expected payout and its reported uncertainty may depend on $z_{c,t}$:
\begin{align}\label{eq:S2-headB}
 D_{t+1}\mid(z_{c,t},s_{ci,t},p)&\ \overset{M_c}{\sim}\
 \mathcal N\bigl(\mu_{ci,t}^{B}(p),v_c(z_{c,t})\bigr),\\
 \mu_{ci,t}^{B}(p)&=a_c(z_{c,t})+\beta_c(z_{c,t})s_{ci,t}+\chi_c(z_{c,t})p,
 \qquad v_c(z)>0.\nonumber
\end{align}
The superscript $M_c$ over the distribution sign means \emph{according to the learned model} $M_c$, not necessarily according to the true distribution. The learned functions $a_c(z),\beta_c(z),\chi_c(z)$ produce the forecast mean; $v_c(z)$ is the forecast variance. The function $a_c(z)$ is an intercept, not the fixed vector $a$ that defines $\theta_t=a^\top f_t$. The learned $v_c(z)$ includes uncertainty remaining about the latent state and is therefore different from the primitive payout-noise variance $\sigma_\varepsilon^2$.

For a fixed pre-auction state $z$, the forecast mean is linear in $s$ and the candidate price $p$, up to an intercept. The variance depends on $z$, not on the candidate price or the realization of $s$. These restrictions preserve a linear \emph{demand schedule in the current price}, even though the public-state update and the state-dependent coefficients can be nonlinear. In particular, the encoder is \emph{not} run again at each candidate price. S3 derives market clearing under these restrictions.

For the first-deployment theorem we restrict the same forecasting system to \emph{Model A}, the \emph{constant-affine head}:
\begin{align}\label{eq:S2-headA}
 \mu_{ci,t}^{A}(p)&=\alpha_c+b_c^\top z_{c,t}+\beta_cs_{ci,t}+\chi_cp,\\
 v_{ci,t}^{A}&=v_c>0,\nonumber
\end{align}
Here $\alpha_c,b_c,\beta_c,\chi_c,v_c$ are fitted constants. The public predictor $q_{c,t}=b_c^\top z_{c,t}$ can still be a nonlinear function of the raw public history, because $z_{c,t}$ may come from a nonlinear encoder. Model A restricts the final coefficients on private signal and price, and uses one forecast variance per type; it does \emph{not} require a linear public-history encoder. Model A is contained in Model B whenever the relevant constant and affine readout functions are feasible.

\begin{table}[htbp]
\centering\small
\caption{The two nested specifications of the forecasting head. The encoder can be nonlinear in either model.}\label{tab:S2-heads}
\begin{tabularx}{\linewidth}{@{}p{30mm}X X@{}}
\toprule
 & \textbf{Model A: constant affine head} & \textbf{Model B: state-dependent affine head}\\\midrule
Predicted mean & $\alpha_c+b_c^\top z+\beta_cs+\chi_cp$ & $a_c(z)+\beta_c(z)s+\chi_c(z)p$\\
Predicted variance & One fitted number $v_c>0$ per type & Fitted function $v_c(z)>0$\\
Public-data computation & Arbitrary admissible pre-auction nonlinear encoder & Arbitrary admissible pre-auction nonlinear encoder\\
Principal role & General first-deployment sign theorem (S4) & Richer learning and risk experiments (S8--S9)\\
\bottomrule
\end{tabularx}
\end{table}

This distinction matters for the sign of the market effect. In Model A, the learned variance is one number that falls when the globally optimal squared prediction error falls. In Model B, the predicted variance and the effect of a private signal can change differently across states. A lower average forecast loss consequently need not mean that every investor reduces trading on private information. The crowd-out theorem in S4 is proved for Model A, not asserted for the whole Model-B family.

\subsection{Training under a specified market law}\label{subsec:S2-objective}

To understand training, distinguish two prices. In a historical observation, the date-$t$ market cleared at a realized price $p_t$, which is available to the researcher or training algorithm later. The public encoder for that observation must still use only $m_t,p_{t-1},D_t$ and earlier releases. The realized $p_t$ enters the \emph{forecasting head}, and the realized payout $D_{t+1}$ is its later training label. When the fitted model is used in a new auction, the head is evaluated at proposed prices $p$ to form the demand schedule. That day's clearing price is not fed back into its encoder until future dates.

Let $Q$ be the stationary probability law generated by an \emph{existing} set of market policies, or an inherited law specified for a first-deployment comparison. Both a low-capacity and a high-capacity candidate can be fitted \emph{on this same law}. Write $p_t^Q$ for the price that actually cleared in a training observation drawn from $Q$; its superscript identifies the market generating the observation, rather than denoting a power. The historical data available for training the type-$c$ predictor have joint law
\begin{equation}\label{eq:S2-trainingtuple}
 ((o_t,o_{t-1},\ldots),\ s_{ci,t},\ p_t^Q,\ D_{t+1})\sim Q_c.
\end{equation}
Here $Q_c$ combines the market's observable history and payoff with the private signal of a representative type-$c$ investor. A single atomistic investor's signal noise does not affect the market price, but it matters for that investor's forecast. The sequence $(o_t,o_{t-1},\ldots)$ in this equation is a random \emph{history}; $\mathscr H_t^-$ in \eqref{eq:S2-input} is the $\sigma$-algebra it generates. The training head may use the observed historical $p_t^Q$, whereas the pre-auction encoder for that date may not.

For $M_c\in\mathcal M_c(B_c)$ write the output of its encoder as $z_{c,t}(M_c)$, its predicted mean at a realized training price as $\mu_{M_c,t}(p_t^Q)$, and its predicted positive variance as $v_{M_c,t}$. The population Gaussian negative log score is
\begin{equation}\label{eq:S2-loss}
 \mathcal L_c(M_c;Q)=\frac12\E_{Q_c}\!\left[
  \log\bigl(2\pi v_{M_c,t}\bigr)
  +\frac{\bigl(D_{t+1}-\mu_{M_c,t}(p_t^Q)\bigr)^2}{v_{M_c,t}}
 \right].
\end{equation}
We define the population-optimal learned system by
\begin{equation}\label{eq:S2-argmin}
 M_c^*(Q;B_c)\in\underset{M\in\mathcal M_c(B_c)}{\operatorname{argmin}}\ \mathcal L_c(M;Q),
 \qquad \mathcal L_c^*(Q;B_c)=\inf_{M\in\mathcal M_c(B_c)}\mathcal L_c(M;Q).
\end{equation}
The learner chooses the encoder, the forecast-mean coefficients, and the forecast variance jointly, subject to its resource budget. The superscript $*$ denotes the global optimum for a given $Q$, rather than the output of any particular numerical optimizer. Whenever a theorem uses an attained optimum, existence and a selection rule are assumed; different optimal models need not have identical internal states. Holding one part of the forecasting head fixed while retraining another defines a separate, restricted experiment discussed below.

The Gaussian negative log score rewards both an accurate predicted mean and well-calibrated uncertainty. It also yields a simple exact result for Model A: the best constant reported variance equals the forecast's mean-square error. A logarithmic score is proper over unrestricted predictive densities. Within a restricted Gaussian family, however, minimizing it picks the best \emph{admissible working belief}; it need not recover the true conditional payout distribution.

\subsection{What a constant Gaussian variance learns}\label{subsec:S2-constant-profile}

For a fixed feasible mean predictor $\mu_{ci,t}$ in Model A, let
\begin{equation}\label{eq:S2-mse}
 R_c(\mu;Q)=\E_{Q_c}\!\left[\bigl(D_{t+1}-\mu_{ci,t}(p_t^Q)\bigr)^2\right],\qquad 0<R_c(\mu;Q)<\infty.
\end{equation}
The part of the loss depending on the \emph{constant} variance is
\begin{equation}\label{eq:S2-const-profile}
 \mathcal L_c(\mu,v;Q)=\tfrac12\log(2\pi v)+\frac{R_c(\mu;Q)}{2v}.
\end{equation}
Differentiation gives $\partial\mathcal L/\partial v=(v-R_c)/(2v^2)$, so the unique positive optimum is
\begin{equation}\label{eq:S2-vopt}
 v_c^*(\mu;Q)=R_c(\mu;Q),\qquad
 \min_{v>0}\mathcal L_c(\mu,v;Q)=\tfrac12\log\bigl(2\pi e\,R_c(\mu;Q)\bigr).
\end{equation}
Thus Model A's optimal constant reported variance is exactly its mean-squared forecast error. Because the profiled score in \eqref{eq:S2-vopt} increases with that error, minimizing Gaussian log score over Model A is equivalent to minimizing MSE. Nested feasible classes therefore order \emph{best-achievable prediction error}. Theorem~1 uses the aligned-payoff specification and an additional projection lemma (S4) to translate this prediction improvement into a smaller private-signal trading response and, ultimately, a lower fresh-price signal precision. The risk identity alone makes no claim about those market outcomes.

The fit $v_c^*$ is the overall population residual mean-square error of the particular forecast. If residual risk varies across states, Model A deliberately reports its unconditional average rather than a state-dependent risk estimate. The upper or lower bound $v_{\min}\le v\le v_{\max}$ sometimes imposed for numerical stability is innocuous for \eqref{eq:S2-vopt} only when the unconstrained optimum lies strictly within those bounds. Theorem 1 explicitly assumes an interior solution; a binding risk-head restriction would change its profile step.

\subsection{What a state-dependent Gaussian variance learns}\label{subsec:S2-conditional-profile}

Model B permits the forecast variance and the two information slopes to depend on the public state $z$. To see what its learning objective would select in an idealized case, hold the encoder fixed and suppose the head's coefficients can be chosen separately for every $z$, without a binding parameter-sharing or resource constraint. We call this a \emph{conditionally saturated head}. It is an analytical benchmark, not an assumption about a finite neural network. Define
\begin{equation}\label{eq:S2-Rvec}
 u_{ci,t}=\begin{pmatrix}1\\s_{ci,t}\\p_t^Q\end{pmatrix},\quad
 \Omega_c(z)=\E_{Q_c}[u_{ci,t}u_{ci,t}^\top\mid z_{c,t}=z],\quad
 k_c(z)=\E_{Q_c}[u_{ci,t}D_{t+1}\mid z_{c,t}=z].
\end{equation}
The vector $u_{ci,t}$ collects the regressors entering the head, not the history passed to its encoder. The conditional moment matrix $\Omega_c(z)$ is assumed nonsingular almost everywhere. Its conditional least-squares coefficient vector is
\begin{equation}\label{eq:S2-cond-head}
 \begin{pmatrix}a_c^*(z)\\\beta_c^*(z)\\\chi_c^*(z)\end{pmatrix}
 =\Omega_c(z)^{-1}k_c(z),\qquad
 v_c^*(z)=\E_{Q_c}\!\left[
 \bigl(D_{t+1}-{a_c^*(z)}-{\beta_c^*(z)s_{ci,t}}-{\chi_c^*(z)p_t^Q}\bigr)^2
 \ \middle|\ z_{c,t}=z\right].
\end{equation}
The first formula is the best affine prediction of payout from private signal and realized training price, \emph{within a given public state}. The second reports the remaining squared prediction error averaged over signals and prices in that state. Thus $v_c^*(z)$ need not equal the true conditional variance $\V(D_{t+1}\mid z,s,p)$ for each $(s,p)$; it is a $z$-conditional \emph{residual MSE}. After optimizing both quantities, the expected log score is
\begin{equation}\label{eq:S2-condscore}
 \mathcal L_{c,B}^{\mathrm{head},*}(Q;z_{c,\cdot})=\frac12\E_{Q_c}\!\left[\log\bigl(2\pi e\,v_c^*(z_{c,t})\bigr)\right].
\end{equation}
The argument $z_{c,\cdot}$ denotes the \emph{entire fixed encoder-output process}; this optimized population score averages over its states, rather than describing the score at one numerical value of $z$. The superscript $*$ means the best \emph{head at this fixed encoder}, not that the encoder itself has been globally optimized. A finite neural head may share parameters across different $z$, preventing it from attaining these state-by-state optima. Moreover, a lower optimized \emph{log score} under state-dependent risk need not correspond to a lower \emph{unconditional MSE}.

For example, if a publicly computed state distinguishes a high-volatility regime from a low-volatility regime, the two regimes can receive different predicted payout variances. A trader using the same private signal may then take different positions, even if the effect of that signal on the \emph{predicted mean} is unchanged. With a state-dependent coefficient $\beta_c(z)$ as well, computation changes both the statistical usefulness of the private clue and the willingness to bear risk when acting on it. Consequently, ordering overall log scores does not order private-signal responses \emph{state by state}. This is the reason S4 treats Model B's sign questions separately rather than asserting Theorem 1 for it.

\subsection{Two meanings of a more capable learner}\label{subsec:S2-nesting}

Increasing a budget expands what an algorithm \emph{could} compute. It need not expand the information actually retained by the trained output. Fix a forecasting type $c$ and compare its low and high budget options, suppressing the type label on the selected models. Let $M_L^*(Q)$ and $M_H^*(Q)$ be population optima of $\mathcal M_c(B_L)\subseteq\mathcal M_c(B_H)$. Their public representations can summarize different aspects of the same history, so
\begin{equation}\label{eq:S2-nonnestedoutputs}
 \sigma\bigl(z_L^*(Q)\bigr)\not\subseteq\sigma\bigl(z_H^*(Q)\bigr)
 \quad\text{is possible even though}\quad
 \mathcal L_c^*(Q;B_H)\le\mathcal L_c^*(Q;B_L).
\end{equation}
For example, the low-budget model might retain today's macro release, while the high-budget optimum instead retains a lagged price innovation that predicts the payout better. The high-budget model \emph{could} have copied the small model, but its chosen output does not necessarily preserve every fact that the small model retained.

Theorem~1 uses the inclusion of feasible \emph{programs} at a fixed training law, together with the special head restrictions from S4. Theorem~2 instead values a specific addition to the information a trader can \emph{use}; for that comparison we explicitly require the selected outputs to retain nested information fields:
\begin{equation}\label{eq:S2-nestedfields}
 \mathcal F_{L,x,t}=\sigma(z_{L,x,t})
 \subseteq\mathcal U_{H,x,t}=\sigma(z_{H,x,t})
 \subseteq\mathscr H_t^-,
\end{equation}
Both hypothetical forecasts then observe \emph{the same} current price $p_t$ and the same individual's private signal $s_{ci,t}$, and both are evaluated \emph{under the same market law $Q_x$}. The subscript $x$ makes explicit that these information fields contain prices generated at a particular adoption share. Their inclusion is an \emph{additional modeling or implementation condition}, not a conclusion from the budget ordering.

One concrete nested finite-history construction used later is
\begin{align}\label{eq:S2-K3}
 \mathcal F_{L,x,t}&=\sigma(m_t),\\
 \mathcal U_{H,x,t}&=\sigma\!\left(m_t,\left\{m_{t-j},p_{t-j},D_{t+1-j}\right\}_{j=1}^{3}\right).\nonumber
\end{align}
For $j=1$, the payout in the high-budget field is $D_t$, already known before auction $t$; $D_{t+1}$ is never an input. The three-lag example specifies what information a \emph{particular} high-budget benchmark can process, not what all powerful AI systems process. A Gaussian population forecast with a free affine head can use the statistical information in this entire finite vector. A finite neural system may learn a less informative summary, and an unlimited-history observer is a different benchmark again.

\subsection{Correct Gaussian forecasting, working Gaussian beliefs, and portfolio choice}\label{subsec:S2-correct}

The word \emph{Gaussian} plays two different roles. In S1 the primitive state and payoff shocks are Gaussian. In S2 the \emph{forecaster} reports a Gaussian distribution, whether or not it is correct. Nonlinear processing and market feedback can make the joint law of payout, public state, private signal, and price non-Gaussian even if the primitive shocks are Gaussian. By contrast, in a jointly Gaussian market with a linear public representation, the correct conditional mean is affine and the correct conditional variance is constant, so Model A can reproduce the true conditional forecast when its feasible class contains it.

In the general \emph{working-belief} interpretation, an investor trains a Gaussian forecast by \eqref{eq:S2-argmin} and then maximizes CARA expected utility \emph{under that forecast}. The resulting demand is optimal for the belief the investor actually uses, even if the belief differs from the objective conditional distribution. S3 derives this demand. We do not equate constrained prediction with a full-information Bayesian forecast.

In the \emph{correctly specified joint-Gaussian} subeconomy, the forecast supplied by each feasible information field is the true conditional normal distribution and the conditional payoff variance is constant across realizations. This additional property is required for Theorem 2's identity between an ex-ante CARA certainty-equivalent premium and conditional mutual information. For that identity, the high and low feasible information fields must be evaluated using their own \emph{correct} population conditional distributions under the same market law and with the same $p_t,s_{ci,t}$. In a nonlinear working-belief market a reduction in forecast log score is not, by itself, a theorem that the investor's true utility or social welfare rises.

The paper deliberately separates the two optimizations. \emph{Prediction-oriented training} minimizes the statistical loss \eqref{eq:S2-loss}; \emph{portfolio choice} then maximizes utility given the learned forecast. Training directly for trading profits or utility could select a different model when computational resources bind. Similarly, the risk-aversion parameter $\gamma_c$ belongs to portfolio choice, not to the baseline Gaussian log-score training objective.

\subsection{From deployment to a new market law and retraining}\label{subsec:S2-laws}

The learning map of type $c$ is the possibly set-valued correspondence
\begin{equation}\label{eq:S2-learningmap}
 \mathcal B_c(Q;B_c)=\underset{M\in\mathcal M_c(B_c)}{\operatorname{argmin}}\,\mathcal L_c(M;Q).
\end{equation}
A tie-breaking rule selects a fitted model $M_c^*(Q;B_c)$ when the argmin contains multiple systems. To define the market feedback, let $\bm M$ collect the fitted algorithms used by all types. For a given adoption mix $x$, suppose these policies generate a stationary market law $\mathcal G(x,\bm M)$. The population learning-and-market cycle then takes the form
\begin{equation}\label{eq:S2-map}
 \bm M_n\ \xrightarrow{\quad\mathcal G(x,\cdot)\quad}\ Q_n
 \xrightarrow{\quad\mathcal B(\,\cdot\,;\bm B)\quad}\ \bm M_{n+1}.
\end{equation}
At each step, existing fitted rules generate market data; population training on that data produces the next set of rules. The index $n$ counts \emph{training epochs}, whereas $t$ counts \emph{trading dates}. S3 specifies the market map, its stationarity conditions, and the associated fixed points. Even when every training epoch reaches its population optimum, the repeated map need not converge.

These definitions distinguish three objects that the results compare. \emph{First deployment} changes the composition of orders submitted in the next auction, using candidate algorithms trained on a common inherited law $Q_0$. A \emph{frozen-policy stationary market} keeps those fitted algorithms fixed while allowing price histories, encoder outputs, and the market distribution to adjust to the new adoption mix. A \emph{population-retrained equilibrium} repeatedly refits all participating types using their induced market law until a fixed point is reached, if one exists. Thus ``frozen'' means \emph{fixed parameters}, not fixed forecasts or fixed price realizations. When evaluating an atomistic investor's hypothetical upgrade value at either market law, both low- and high-budget candidate forecasts are optimally fitted to \emph{that same law}; this separate valuation does not itself change the prevailing market.

\subsection{Finite training and architecture comparisons in the experiments}\label{subsec:S2-simulation}

In the computational experiments, the target is still the one-period payout and the input timing is still \eqref{eq:S2-input}. Learning is performed with finite time series rather than the population expectation \eqref{eq:S2-loss}. For a training window of $T$ observations, the empirical counterpart is
\begin{equation}\label{eq:S2-sampleloss}
 \widehat{\mathcal L}_{c,T}(M)=\frac1{2T}\sum_{t=1}^{T}
 \left[\log(2\pi v_{M,t})+
 \frac{(D_{t+1}-\mu_{M,t}(p_t))^2}{v_{M,t}}\right].
\end{equation}
As in population training, the encoder for date $t$ must use only data available before its auction. The realized historical $p_t$ enters the forecasting head and loss, not that day's pre-auction encoder. Training, model selection, and evaluation use separate time blocks or shock paths. More training data may improve the approximation to population risk under suitable regularity, but finite samples and nonconvex optimizers do not guarantee the global solution in \eqref{eq:S2-argmin}.

The studies in S8 compare finite-history linear state-space and GRU predictors at controlled state sizes and resource budgets, with optional state-dependent heads. They report the measured forecast negative log score, mean-square error, memory and operation counts, private-signal response, market-clearing residual, and the price-information measure appropriate to each experiment. Transformer architectures are a feasible extension but should not be labeled an executed result unless fitted and evaluated. Separate training-seed and market-shock replications quantify optimizer and simulation variability. Theoretical nesting of \emph{all feasible resource-bounded programs} does not license treating two separately optimized neural architectures as nested without an explicit embedding construction.

To separate mean learning from risk learning, some experiments fit one part of the head while holding the other fixed. In a \emph{mean-only} fit the old variance is held fixed; in a \emph{variance-only} fit the old mean is held fixed; in a \emph{joint} fit both are optimized. These exercises differ from simply swapping fitted coefficients after training. The difference can be seen from a population identity: for information fields $\mathcal F\subseteq\mathcal U$, let $\mu_L=\E[D\mid\mathcal F]$ and $\mu_H=\E[D\mid\mathcal U]$. Then
\begin{equation}\label{eq:S2-variance-only}
 v_{\mathrm{risk\ only}}^*(\mathcal U)
 =\E[(D-\mu_L)^2\mid\mathcal U]
 =\V(D\mid\mathcal U)+(\mu_H-\mu_L)^2.
\end{equation}
This variance includes both the uncertainty remaining after the richer information and the squared error that results from retaining the old mean forecast. Equation~\eqref{eq:S2-variance-only} is the ideal optimum when variance may condition on all of $\mathcal U$. Model B is more restrictive: its variance depends only on the pre-price public state $z$, not on current $p$ or private $s$. At a fixed mean, its unconstrained state-conditional optimum would be $v^*(z)=\E[(D-\mu_L)^2\mid z]$, if implementable. Finite neural heads need not attain either benchmark. S8--S9 report the actual optimization protocols and head constraints.

\subsection{Assumptions used in the analytical results}\label{subsec:S2-proofhandoff}

The general forecasting environment requires integrability and strictly positive fitted variances sufficient to make the population log score in \eqref{eq:S2-loss} well defined, pre-auction adaptedness, and an explicitly defined feasible resource class. To use its output in competitive clearing S3 will require a positive aggregate downward demand slope and an invariant fast-market law whenever a population training distribution is discussed. None of these conditions is automatically implied by $\rho(F)<1$ for the exogenous latent state.

The additional assumptions for each analytical theorem are summarized in Table~\ref{tab:S2-lock}; S4--S6 state them again where they are needed.

\begin{table}[htbp]
\centering\small
\caption{Learning restrictions needed for the main analytical results.}\label{tab:S2-lock}
\begin{tabularx}{\linewidth}{@{}p{21mm}X X@{}}
\toprule
\textbf{Result} & \textbf{Training and head restrictions} & \textbf{Informational comparison}\\\midrule
Theorem 1 (S4) & Model A; aligned payoff; complete nested feasible algorithm classes; free optimal constant affine readout and variance; attained global population minima. & Same inherited $Q$ and its historical $p_t^Q$; a first-auction price-signal precision comparison.\\
Theorem 2 (S5) & Correct joint Gaussian market law; candidate low and high conditional forecasts are both fitted correctly. & Same $Q_x$, current price and private signal for both, and explicitly nested \emph{feasible finite-capacity} public information fields.\\
Theorem 3 (S6) & Linear-Gaussian two-factor specialization; Gaussian-optimal forecasts; CARA-optimal holdings and clearing. & Locally endogenous adoption path, frozen incumbents or all-type population refitting; signs proved near zero adoption.\\
\bottomrule
\end{tabularx}
\end{table}

\begin{longtable}{@{}p{50mm}p{101mm}@{}}
\caption{Notation defined in S2.}\label{tab:S2-notation}\\
\toprule
\textbf{Symbol} & \textbf{Definition}\\\midrule\endfirsthead
\toprule\textbf{Symbol} & \textbf{Definition (continued)}\\\midrule\endhead
$M_c,\phi_c,T_{\phi_c},z_{c,t}$ & Learned forecasting system, encoder parameters, public-state update, resulting pre-auction state.\\
$B_c,C^{\rm mem},C^{\rm ops},\mathcal M_c(B_c)$ & Inference-resource budget, memory and operation costs, feasible full forecasting systems.\\
$\mu_{ci,t}(p),v_c(z)$ & Fitted payout mean and variance evaluated at a candidate current price.\\
$\alpha_c,b_c,\beta_c,\chi_c$ & Model-A affine head intercept, public-state loading, private-signal loading, price loading.\\
$a_c(z),\beta_c(z),\chi_c(z)$ & Model-B state-dependent mean readout functions.\\
$Q,Q_c,p_t^Q$ & Inherited/induced stationary training law, type-$c$ marginal, realized current price in training samples.\\
$\mathcal L_c,\mathcal L_c^*,M_c^*,\mathcal B_c$ & Population log score, optimized score, selected population-optimal predictor, learning best-response correspondence.\\
$\mathcal F_{L,x,t},\mathcal U_{H,x,t}$ & Selected, explicitly nested low-/high-budget information fields when used for Theorem 2.\\
\bottomrule
\end{longtable}

S3 takes the fitted forecasting heads defined here and derives investor demand, competitive clearing, and the market laws compared under adoption and retraining. S4 proves the first-deployment crowd-out result using the constant-variance identity \eqref{eq:S2-vopt}. S5 relates correct Gaussian forecasts and finite, nested public information fields to the private value of computation. S6 constructs a competitive equilibrium where both informational effects coexist. These results retain their own assumptions; the general Model-B head is not claimed to inherit every sign result proved for Model A.

\clearpage\section{Trading equilibrium and endogenous retraining}\label{sec:S3}

The previous two sections establish the economy and the forecasting algorithms. Here we connect them. At the start of a trading date, an investor has an AI-generated summary of previously released information and an individual private signal. She then decides how many shares to hold at each possible current price. The orders of all investors determine that price. The new price becomes part of tomorrow's public record, even if nobody changes an algorithm. When algorithms are subsequently refitted to this new record, a further feedback appears. These are different economic operations, and we give each an explicit mathematical definition.

The principal point is that the model does not assign a price response to AI trading by assumption. We derive it from optimal portfolio choice and competitive market clearing. The resulting clearing price remains explicit even when the forecasting algorithm is nonlinear and its estimated risk varies with its computed public state. What does \emph{not} follow automatically is a stationary market law or convergence of retraining; these require separate conditions.

\subsection{Information and timing at an auction}\label{subsec:S3-timing}

Recall the one-period claim from S1. At date $t$ it trades at price $p_t$, pays $D_{t+1}$ at the start of $t+1$, and then expires. The gross risk-free return is $R_f>0$. Its economic primitives are
\begin{align}\label{eq:S3-primitives}
 f_{t+1}&=Ff_t+G\eta_{t+1}, & \theta_t&=a^\top f_t,\\
 D_{t+1}&=\bar d+\ell^\top f_t+\varepsilon_{t+1}, & s_{ci,t}&=\theta_t+\xi_{ci,t},\nonumber\\
 S_t&=\bar S+\zeta_t.\nonumber
\end{align}
The latent $f_t\in\R^m$ can have any finite dimension. The scalar $\theta_t$ is the target of investors' individual private clues $s_{ci,t}$. Type $c$ has risk aversion $\gamma_c>0$ and idiosyncratic signal-noise variance $\nu_c>0$. The supply disturbance $\zeta_t$ is independently distributed $N(0,\sigma_\zeta^2)$, is not announced to investors before the auction, and has $\sigma_\zeta^2>0$. The errors $\xi_{ci,t}$ aggregate to zero across the continuum of each type. In the aligned-payoff specialization used by the central analytical theorems, $\ell=\lambda a$ with $\lambda>0$, so $D_{t+1}=\bar d+\lambda\theta_t+\varepsilon_{t+1}$. S3's demand and clearing results do not require alignment.

Before auction $t$, the commonly accessible history is $\mathscr H_t^-=\sigma(m_t,p_{t-1},D_t,m_{t-1},p_{t-2},D_{t-1},\ldots)$. Here $m_t$ denotes publicly released economic observations, not the dimension $m$ of $f_t$. The \emph{encoder} of each forecasting type $c$ processes that history to form $z_{c,t}$; in S2, a generic implementation is $z_{c,t}=T_{\phi_c}(z_{c,t-1},o_t)$ for $o_t=(m_t,p_{t-1},D_t)$. Thus $z_{c,t}$ may depend on earlier prices, but it does \emph{not} depend on the current price $p_t$ or an individual investor's signal $s_{ci,t}$. The latter is separately available to the investor.

Investors submit a \emph{demand schedule} $h_{ci,t}(p)$, specifying desired shares at each hypothetical current price $p$. The auction selects the price $p_t$ at which these schedules collectively absorb the net supply $S_t$. We use $h$ for an individual shareholding, whereas $x$ will denote the fraction of a selected investor group using higher computational capacity. The conditional \emph{forecast} can depend on the proposed $p$ although the pre-auction encoder cannot. The timeline is summarized in Table~\ref{tab:S3-timing}.

\begin{table}[htbp]\centering\small
\caption{Two prices and the within-date sequence. The realized $p_t$ is an input to later training, not to date-$t$ encoding.}\label{tab:S3-timing}
\begin{tabularx}{\linewidth}{@{}p{30mm}X p{38mm}@{}}\toprule
\textbf{Stage} & \textbf{What happens} & \textbf{Information available}\\\midrule
Before auction & Public reports, $p_{t-1}$ and matured payout $D_t$ arrive; the AI constructs $z_{c,t}$; investor $i$ observes $s_{ci,t}$. & $\mathscr H_t^-$ and own $s_{ci,t}$.\\
Orders and clearing & The trained head evaluates each candidate $p$; the investor submits $h_{ci,t}(p)$; all schedules clear at $p_t$. & Candidate prices in schedules; no individual takes $p_t$ as personally controllable.\\
After clearing & The realized $p_t$ becomes public. At the next date, $D_{t+1}$ is paid and revealed. & $\mathscr H_t^+=\mathscr H_t^-\vee\sigma(p_t)$.\\
Slow learning epoch & A sequence of trades generates training observations $(z,s,p,D)$; population or sample learning may refit the forecasts. & Realized historical prices and matured payout labels.\\\bottomrule
\end{tabularx}\end{table}

\subsection{Gaussian working beliefs and the CARA portfolio problem}\label{subsec:S3-demand}

A deployed forecasting model produces a \emph{subjective} Gaussian distribution for the next payout. Under S2's general, state-dependent affine head (Model B), its mean and variance evaluated at candidate price $p$ are
\begin{align}\label{eq:S3-headB}
 \mu_{ci,t}(p)&=a_c(z_{c,t})+\beta_c(z_{c,t})s_{ci,t}+\chi_c(z_{c,t})p,\\
 v_{ci,t}&=v_c(z_{c,t})>0.\nonumber
\end{align}
The functions and the encoder have already been fitted. An investor treats these outputs as her beliefs at each quoted price; the Gaussian approximation need not equal the objective conditional distribution in a nonlinear market. In particular, $v_c(z)$ is the algorithm's learned predictive \emph{variance}, not the primitive dividend-shock variance $\sigma_\varepsilon^2$. The price slope $\chi_c(z)$ reflects what the investor expects a quoted price to reveal about the payout.

For an initial wealth $W_{ci,t}$, a risky position $h$ at price $p$ produces next-date wealth
\begin{equation}\label{eq:S3-wealth}
 W_{ci,t+1}=R_fW_{ci,t}+h\bigl(D_{t+1}-R_fp\bigr).
\end{equation}
Utility is $u_c(W)=-e^{-\gamma_c W}$. Under the subjective normal forecast, the conditional expected utility of a position is
\begin{equation}\label{eq:S3-EU}
 \E^{M_c}_{ci,t,p}[u_c(W_{ci,t+1})]
 =-\exp\!\left[-\gamma_c R_fW_{ci,t}-\gamma_c h(\mu_{ci,t}(p)-R_fp)
 +\frac{\gamma_c^2v_c(z_{c,t})h^2}{2}\right].
\end{equation}
The superscript $M_c$ emphasizes that the expectation is taken under the \emph{learned} belief, conditional on $(z_{c,t},s_{ci,t},p)$. Maximizing \eqref{eq:S3-EU} is equivalent to maximizing its quadratic certainty equivalent in $h$,
\begin{equation}\label{eq:S3-CE-quadratic}
 \CE_{ci,t}(h;p)=R_fW_{ci,t}+h\bigl(\mu_{ci,t}(p)-R_fp\bigr)
 -\tfrac12\gamma_cv_c(z_{c,t})h^2.
\end{equation}
\begin{proposition}[Optimal demand schedule under the learned Gaussian belief]\label{prop:S3-demand}
For $\gamma_c>0$ and $v_c(z_{c,t})>0$, the investor's unique optimal risky holding at candidate price $p$ is
\begin{equation}\label{eq:S3-demand}
 \boxed{\quad h_{ci,t}(p)=\frac{a_c(z_{c,t})+\beta_c(z_{c,t})s_{ci,t}-[R_f-\chi_c(z_{c,t})]p}{\gamma_cv_c(z_{c,t})}.\quad}
\end{equation}
Holding the pre-auction information fixed, its price slope and private-signal slope are
\begin{equation}\label{eq:S3-slopes}
 \frac{\partial h_{ci,t}}{\partial p}=-\frac{R_f-\chi_c(z_{c,t})}{\gamma_cv_c(z_{c,t})},\qquad
 \frac{\partial h_{ci,t}}{\partial s_{ci,t}}=\frac{\beta_c(z_{c,t})}{\gamma_cv_c(z_{c,t})}.
\end{equation}
At this optimum, the conditional certainty equivalent is
\begin{equation}\label{eq:S3-optimalCE}
 \CE_{ci,t}^*(p)=R_fW_{ci,t}+\frac{[\mu_{ci,t}(p)-R_fp]^2}{2\gamma_cv_c(z_{c,t})}.
\end{equation}
\end{proposition}
\begin{proof}
The first derivative of \eqref{eq:S3-CE-quadratic} with respect to $h$ is $\mu_{ci,t}(p)-R_fp-\gamma_cv_c(z_{c,t})h$; the second derivative is $-\gamma_cv_c(z_{c,t})<0$. Solve the first-order condition, substitute the mean in \eqref{eq:S3-headB}, and differentiate in $p$ and $s$. Substitution of the optimum into \eqref{eq:S3-CE-quadratic} yields the reported certainty equivalent.
\end{proof}

The second slope in \eqref{eq:S3-slopes} is the direct bridge to the paper's private-information crowd-out result: it measures how strongly an investor's desired position responds to a slightly more favorable \emph{private} observation. Both the private-signal forecast coefficient $\beta_c$ and the fitted risk $v_c$ matter. Better forecasts may reduce $\beta_c$ and $v_c$ at the same time; the sign of their \emph{ratio} is the object proved in S4 under the appropriate restrictions. No sign is imposed for a general state-dependent head.

The individual schedule slopes downward if $R_f-\chi_c(z_{c,t})>0$. A unique market price requires only that the \emph{weighted aggregate} slope be downward, a weaker condition derived next. The optimality in Proposition~\ref{prop:S3-demand} is conditional on an AI's working belief and individual price taking. It is not a claim that the fitted Gaussian model is a correct posterior or that training by log score maximizes the investor's true ex-ante utility.

\subsection{The exact competitive clearing price with nonlinear AI}\label{subsec:S3-clearing}

There are finitely many types $c\in\mathcal C$ with masses $\pi_c>0$, $\sum_c\pi_c=1$. Write $\mathcal I_c$ for type $c$'s continuum of accounts and $\mu_c$ for its normalized population measure. All accounts of a type share $z_{c,t}$ and the learned head, but have distinct private noises. S1 assumes exact within-type aggregation: $\int_{\mathcal I_c}s_{ci,t}\,d\mu_c(i)=\theta_t$. It follows that only the common component of their private signals survives in aggregate demand.

Define the pre-auction, \emph{state-dependent} risk-adjusted weights and the two price coefficients
\begin{align}\label{eq:S3-weights}
 w_{c,t}&\equiv\frac{\pi_c}{\gamma_cv_c(z_{c,t})},\\
 A_t&\equiv\sum_cw_{c,t}\beta_c(z_{c,t}),\qquad
 \Delta_t\equiv\sum_cw_{c,t}[R_f-\chi_c(z_{c,t})],\nonumber\\
 K_t&\equiv\sum_cw_{c,t}a_c(z_{c,t}).\nonumber
\end{align}
For a given policy profile, the three quantities $A_t,\Delta_t,K_t$ are determined by pre-auction public states. $K_t$ measures the public-history contribution to aggregate desired holdings; $A_t$ measures the contribution of the common fundamental carried by private signals; and $\Delta_t$ is the magnitude of the slope of aggregate demand with respect to the quoted price. For example, increasing $\Delta_t$ makes the market more responsive in \emph{quantities} to a given price change, which in turn makes the \emph{price} less sensitive to a given supply disturbance: $\partial p_t/\partial\zeta_t=-1/\Delta_t$ once the fitted coefficients and pre-auction states are fixed. These quantities need not be directly observed by every market participant (for instance, a firm's forecasting weights may be proprietary). Their essential feature is temporal: no $p_t$ enters $K_t$, $A_t$, or $\Delta_t$ contemporaneously, even if the encoder is a neural network.

\begin{proposition}[Competitive clearing under Model B]\label{prop:S3-clearing}
Suppose $0<v_c(z_{c,t})<\infty$, the weights in \eqref{eq:S3-weights} are integrable, and
\begin{equation}\label{eq:S3-positive-delta}
 \boxed{\quad\Delta_t>0\quad\text{almost surely}.\quad}
\end{equation}
Then competitive market clearing has a unique finite price satisfying
\begin{equation}\label{eq:S3-price}
 \boxed{\quad p_t=\frac{K_t+A_t\theta_t-\bar S-\zeta_t}{\Delta_t}.\quad}
\end{equation}
It solves the original continuum clearing condition with the CARA-optimal individual schedules \eqref{eq:S3-demand}. The formula remains valid for an arbitrary admissible nonlinear public encoder and state-dependent affine-mean, state-dependent-variance head.
\end{proposition}
\begin{proof}
Integrating \eqref{eq:S3-demand} over investors of each type and using exact aggregation gives
\begin{equation}\label{eq:S3-agg-demand}
 \sum_c\pi_c\!\int_{\mathcal I_c} h_{ci,t}(p)\,d\mu_c(i)
 =\sum_c\frac{\pi_c}{\gamma_cv_c(z_{c,t})}\bigl[a_c(z_{c,t})+\beta_c(z_{c,t})\theta_t-(R_f-\chi_c(z_{c,t}))p\bigr]
 =K_t+A_t\theta_t-\Delta_tp.
\end{equation}
Equate this expression to $S_t=\bar S+\zeta_t$ and solve for $p$. Since the slope of aggregate demand is $-\Delta_t<0$, the solution is unique and finite.
\end{proof}

Condition \eqref{eq:S3-positive-delta} is an economic regularity condition, not a consequence of having more accurate AI. A convenient sufficient restriction is $R_f-\chi_c(z)>0$ for every active type and state. In a calibration one may instead check that the aggregate $\Delta_t$ remains bounded below by some $\underline\Delta>0$ over states encountered. This matters particularly for Model B, where $v_c(z)$ and $\chi_c(z)$ can change from day to day. A state-dependent price formula is still explicit as long as the head variance and its coefficients do not themselves depend on the \emph{candidate} price. If $v_c$ depended on candidate $p$ or the head were nonlinear in $p$, the schedule would generally cease to be affine and a separate clearing-existence argument would be needed.

\paragraph{A numerical illustration of clearing.}
Consider two types with masses $(\pi_1,\pi_2)=(0.6,0.4)$, risk aversions $(\gamma_1,\gamma_2)=(1,2)$, forecast variances $(v_1,v_2)=(1.5,2)$, private-signal slopes $(\beta_1,\beta_2)=(0.3,0.5)$, and price slopes $(\chi_1,\chi_2)=(0.10,0)$ at $R_f=1.02$. Suppose the two public mean terms are $a_1(z_1)=0.2$ and $a_2(z_2)=0$, with $\bar S=\zeta_t=0$. These values imply $w_1=0.4$, $w_2=0.1$, $A=0.17$, $\Delta=0.47$, and $K=0.08$. When $\theta_t=1$, equation~\eqref{eq:S3-price} gives $p_t=0.25/0.47\simeq0.5319$. The two within-type average demands are approximately $0.00709$ and $-0.01064$ shares, respectively; their population-weighted sum is zero. The exercise illustrates the clearing identity, not a model calibration used in the results.

\subsection{Separating newly aggregated information from public-data forecasts}\label{subsec:S3-info}

Equation~\eqref{eq:S3-price} separates the two channels behind the paper. $A_t\theta_t$ is the part of aggregate demand driven by the common fundamental contained in individuals' private signals. $K_t$ comes from the public record as interpreted by the deployed algorithms. The current supply shock $\zeta_t$ obscures both. A more capable AI can change both $A_t$ and $K_t$---there is no general reason for them to move in the same direction.

The two components affect the usefulness of prices to different observers. Someone who can process the complete pre-auction public record, and who knows the deployed policies, can already reconstruct $K_t$: it is not a new piece of information to that observer. What the observer cannot directly see is the private fundamental $\theta_t$, which enters prices through the aggregate signal loading $A_t$. To isolate this \emph{new} information, we subtract the already predictable public component from the clearing price. Conditional on $\mathscr H_t^-$ and the deployed policies, $A_t,K_t,\Delta_t$ are known, so the rearrangement yields the signal experiment
\begin{equation}\label{eq:S3-signal-experiment}
 \underbrace{\Delta_tp_t-K_t+\bar S}_{\text{price after removing computed public component}}
 =A_t\theta_t-\zeta_t.
\end{equation}
The subtraction does not discard private information: it removes precisely the part a fully informed public-history observer could have calculated without seeing today's price. If desired, subtracting the further known term $A_t\E[\theta_t\mid\mathscr H_t^-]$ isolates $A_t(\theta_t-\E[\theta_t\mid\mathscr H_t^-])-\zeta_t$, the signal about the \emph{innovation} in the fundamental. Conditional on the record, the supply noise is independent of $\theta_t$. If $A_t\ne0$, divide by $A_t$ to obtain a noisy measurement $\theta_t-\zeta_t/A_t$. Its \emph{likelihood precision per unit of $\theta_t$}, in the sense of the inverse observation-noise variance, is
\begin{equation}\label{eq:S3-J}
 \boxed{\quad J_t=\frac{A_t^2}{\sigma_\zeta^2}.\quad}
\end{equation}
At $A_t=0$ the signal carries no incremental information about $\theta_t$ beyond the conditioned public record and we set $J_t=0$. The scaling $\Delta_t$ cancels because it multiplies both the fundamental and supply-noise components of price.

This measure is distinct from a correlation or an unconditional measure of price informativeness. Under \emph{correct conditional Gaussianity}, with $P_t=\V(\theta_t\mid\mathscr H_t^-)$, the incremental conditional mutual information equals
\begin{equation}\label{eq:S3-MI}
 I(\theta_t;p_t\mid\mathscr H_t^-)
 =\frac12\E\!\left[\log(1+J_tP_t)\right].
\end{equation}
When $P_t$ is constant, as in a stationary jointly Gaussian linear price economy with a specified Gaussian public field, the right side simplifies to $\frac12\log(1+JP)$. This conversion is \emph{not} valid merely because the fundamental shocks are Gaussian: a nonlinear feedback market can make the conditional distribution of $\theta_t$ given its own price history non-Gaussian. In that setting $J_t$ remains a valid \emph{price likelihood precision} but need not equal a one-to-one transform of mutual information.

An additional subtlety concerns what an observer can decode. Equation~\eqref{eq:S3-signal-experiment} lets an observer who knows the full public record and the deployed public forecast component remove $K_t$. An observer who can process only a small part of that record may \emph{not} recover $K_t$ cheaply or accurately. To this observer, $p_t$ can be more useful because it transmits information that better-equipped traders already extracted from earlier public data, even while the price's \emph{fresh} likelihood precision $J_t$ falls. S5 formalizes the dual information effects using explicitly defined low- and high-capacity information fields.

\paragraph{The constant-affine analytical case.}
For S2's restricted Model A, the forecast head is
\begin{equation}\label{eq:S3-headA}
 \mu_{ci,t}(p)=\alpha_c+b_c^\top z_{c,t}+\beta_cs_{ci,t}+\chi_cp,
 \qquad v_{ci,t}=v_c>0.
\end{equation}
Given a fixed model profile and adoption mix, the price coefficients simplify to
\begin{equation}\label{eq:S3-modelA}
 w_c=\frac{\pi_c}{\gamma_cv_c},\qquad
 A=\sum_cw_c\beta_c,\qquad
 \Delta=\sum_cw_c(R_f-\chi_c),\qquad
 K_t=\sum_cw_c(\alpha_c+b_c^\top z_{c,t}).
\end{equation}
Thus $A$ and $\Delta$ are \emph{constants} for a fixed profile, even if the encoder is nonlinear and $K_t$ responds endogenously to changing price histories. In particular, its signal precision $J=A^2/\sigma_\zeta^2$ is independent of the induced \emph{distribution} of $z$ as long as the head coefficients are frozen. This does not mean that the entire price distribution or the mutual information computed using a limited public field stays unchanged when the public history changes. For the state-dependent Model B, by contrast, $A_t,\Delta_t$ also vary with $z_{c,t}$, so even the \emph{average} structural signal precision can change under a frozen model. This is why the three terms in \eqref{eq:S3-telescope} are conceptually distinct although the middle term happens to vanish for one particular Model-A precision measure.

\subsection{Computational adoption and the first clearing-price comparison}\label{subsec:S3-adoption}

The index $c$ has so far described an \emph{active forecasting type}, which may differ from others in risk aversion, signal precision, and computational capacity. To study adoption while holding investors' economic characteristics fixed, we now separate an underlying \emph{investor group} $g$ from the capacity $k\in\{L,H\}$ of the algorithm it uses. Thus an active type can be written $c=(g,k)$. Members of group $g$ share risk aversion $\gamma_g$ and private-signal variance $\nu_g$, whether they use low- or high-budget forecasting. Heterogeneity in risk aversion was already allowed in S1; the new index simply allows us to vary \emph{computational capacity within the same economic group}.

Suppose a share $x_g\in[0,1]$ of group $g$'s mass $\pi_g$ uses a high-budget forecasting system, while share $1-x_g$ uses a low-budget system. For clearing purposes these are separate active types with masses $\pi_gx_g$ and $\pi_g(1-x_g)$. An adoption change alters this \emph{mix}, not the underlying $f_t$, individuals' signal quality, or supply process.

For one adopting group, it is helpful to write the contribution of that group to the information loading explicitly:
\begin{equation}\label{eq:S3-ax}
 A_t(x)=A_{-g,t}+\pi_g\!\left[(1-x)\frac{\beta_{gL}(z_{gL,t})}{\gamma_gv_{gL}(z_{gL,t})}
 +x\frac{\beta_{gH}(z_{gH,t})}{\gamma_gv_{gH}(z_{gH,t})}\right],
\end{equation}
where $A_{-g,t}$ aggregates other groups. Analogous expressions with $R_f-\chi$ and $a(z)$ replace $A_t$ in $\Delta_t$ and $K_t$. Equation~\eqref{eq:S3-ax} makes the distinction between adoption $x$ and individual demand $h(p)$ explicit: an individual's position depends on her forecast, while $x$ determines the population weights assigned to different forecast technologies.

\begin{definition}[Immediate first-deployment experiment]\label{def:S3-impact}
Fix an inherited stationary market law $Q_0$ at adoption mix $x_0$. Fit the low- and high-capacity \emph{candidate} systems under that same law, obtaining a common set of forecast policies $\mathbf M_0^{\mathrm{cand}}$. Retain the same date-$t$ pre-auction economic state, public history, and individual private signals. Change the specified mass of active trading accounts from low to high capacity and clear that date's auction using \eqref{eq:S3-price}. The resulting counterfactual price is the \emph{initial-deployment} price. Its policy coefficients are evaluated before any new market histories have been generated or any algorithm has been retrained on them.
\end{definition}

Intuitively, we take the market just before the upgrade, replace some investors' forecasting algorithms with more capable algorithms already fitted to that \emph{same} historical market, and clear the next auction. The underlying economic state, realized private signals, and past public record are unchanged. The comparison therefore identifies the immediate effect of who uses which algorithm, before the changed prices can alter future histories or induce any retraining.

Under Model A, let $\iota_{gk}=\beta_{gk}/(\gamma_gv_{gk})$ denote the privately informed demand slope of capacity $k\in\{L,H\}$ trained on $Q_0$. Then
\begin{equation}\label{eq:S3-impact-derivative}
 \frac{\partial A(x)}{\partial x}=\pi_g(\iota_{gH}-\iota_{gL}),\qquad
 \frac{\partial J(x)}{\partial x}=\frac{2A(x)\pi_g}{\sigma_\zeta^2}(\iota_{gH}-\iota_{gL}).
\end{equation}
These formulas are accounting identities for the first auction; no optimal-learning sign has been assumed. Theorem 1 (S4) establishes $\iota_{gH}\le\iota_{gL}$ under nested learning, aligned payoffs, and constant head coefficients. In addition, the sign of $J'(x)$ in \eqref{eq:S3-impact-derivative} follows as nonpositive when the aggregate $A(x)$ is nonnegative (as in the sign-restricted aligned case). Otherwise a fall in $A$ need not imply a fall in $A^2$. The S4 proof therefore states the loading-sign conditions it needs. This is a comparison of the first auction's price-signal likelihood precision, \emph{not} an unconditional statement about every observer's mutual information or a stationary retrained equilibrium.

\subsection{The stationary price law of a fixed trading policy}\label{subsec:S3-fastlaw}

The immediate comparison in S3.5 changes only the first auction. The next question is what happens over many trading dates if the newly deployed algorithms are \emph{never refitted}. Today's clearing price enters tomorrow's public history; the encoders therefore receive different inputs and can generate different forecasts even though all their weights remain fixed. We call the resulting probability distribution of repeated trading outcomes the \emph{fixed-policy stationary market law}, provided such a distribution exists.

Let $\mathbf M$ collect the deployed trained forecast systems and let $x$ specify their population shares. Combine the exogenous fundamental, public-release and payoff processes with the encoder updates and the clearing formula into a stochastic recursion. Denote its stationary law, \emph{when it exists}, by
\begin{equation}\label{eq:S3-G-map}
 Q_x^{\mathbf M}\equiv\mathcal G(x,\mathbf M).
\end{equation}
The law includes public histories, prices, payoffs, and representative individual signals needed for later population training. The superscript $\mathbf M$ emphasizes that the forecasting algorithms are held fixed. This is a law of the \emph{fast} trading economy, not a fixed point of any learning algorithm.

\paragraph{A sufficient assumption for a stationary market.}
We now introduce an additional assumption that ensures that $\mathcal G(x,\mathbf M)$ is well defined. For nonlinear encoders, let $Y_t$ collect all forecast states, required lagged prices, and other \emph{endogenous} lags; let $E_t$ collect the exogenous shocks and observations that drive the market. The joint forecast--demand--clearing system can be represented as
\begin{equation}\label{eq:S3-fast-recursion}
 Y_t=\mathcal R_{x,\mathbf M}(Y_{t-1};E_t),
\end{equation}
where $x$ and $\mathbf M$ do not change over dates. We require the effects of different initial endogenous states to diminish when both economies experience the same sequence of exogenous disturbances. The following uniform contraction condition is convenient and stronger than necessary.
\begin{assumption}[A sufficient contraction condition for the fixed-policy market]\label{ass:S3-fast}
For a specified $(x,\mathbf M)$, $\mathcal R_{x,\mathbf M}(\cdot;e)$ is jointly measurable and uniformly Lipschitz in $Y$, with some $\kappa<1$ such that
\begin{equation}\label{eq:S3-contraction}
 \|\mathcal R_{x,\mathbf M}(y;e)-\mathcal R_{x,\mathbf M}(y';e)\|
 \le\kappa\|y-y'\|\quad\text{for almost every }e.
\end{equation}
Furthermore, $\E\|\mathcal R_{x,\mathbf M}(0;E_t)\|^2<\infty$, and the driving input sequence is jointly stationary and ergodic.
\end{assumption}

\paragraph{Why an additional assumption is needed.}
Unique clearing at \emph{each} auction does not guarantee that the resulting \emph{sequence} of clearing prices settles into a stable probability law. Even the exogenous economic state may be stable, $\rho(F)<1$, while endogenous price feedback is too strong. For example, take a fitted Model-A forecast whose public component is $b p_{t-1}$ and suppress other predictable public terms. The clearing equation becomes
\begin{equation}\label{eq:S3-instability-example}
 p_t=\frac{b}{\Delta}\,p_{t-1}+\text{a stationary exogenous forcing process}.
\end{equation}
With nondegenerate disturbances, $|b/\Delta|\ge1$ rules out the usual causal finite-variance stationary solution. An encoder that simply stores yesterday's price is not itself unstable. The instability is generated by the \emph{closed loop}: yesterday's price affects today's forecast, today's forecast affects demand and clearing, and today's price becomes tomorrow's input. Thus the relevant stability condition concerns the \emph{entire market with fixed algorithms}, not the encoder or the exogenous fundamentals taken separately. Assumption~\ref{ass:S3-fast} gives a tractable way to rule out such amplification, including with nonlinear forecasting rules.

\begin{proposition}[A sufficient condition for an invariant fast-market law]\label{prop:S3-invariant}
Under Assumption~\ref{ass:S3-fast}, \eqref{eq:S3-fast-recursion} has a unique square-integrable causal stationary solution determined by the infinite past of $E_t$. The associated stationary distribution defines $\mathcal G(x,\mathbf M)$; it is reached in mean square from alternative square-integrable initial endogenous states driven by the same exogenous history.
\end{proposition}
\begin{proof}
Iterate the recursion backward from time $t-n$ using a common driving sequence. The distance between paths starting from distinct states at that date is bounded by $\kappa^n$ times the initial distance. The square-integrable bound at zero and $\kappa<1$ show that the backward iterates form a Cauchy family in $L^2$; their limit is causal, stationary under shifts of the exogenous sequence, and solves \eqref{eq:S3-fast-recursion}. Coupling any other square-integrable causal solution to the same forcing and iterating the Lipschitz inequality forces its distance from the constructed one to vanish. Starting from a separate initial state at a finite date yields the same geometric mean-square forgetting estimate.
\end{proof}

The proposition establishes a stable probability law for \emph{repeated trading with the algorithms frozen}; it does not claim that algorithms will converge when retrained. The condition may be imposed in a weighted norm or over several periods, which is useful for lag-shift registers; a Markov specification may instead use drift and minorization conditions. Theorem 1's \emph{immediate} price comparison does not require this assumption, whereas stationary and retraining comparisons require a well-defined invariant law by some means. For nonlinear numerical experiments, we diagnose stationarity empirically instead of inferring it from Gaussian exogenous shocks.

\subsection{An exact linear-Gaussian stationary benchmark}\label{subsec:S3-linear}

The contraction condition above covers nonlinear forecasting but does not give a convenient way to calculate moments. For the population experiments we therefore also use a linear-Gaussian specialization, in which stationarity can be checked by inspecting a matrix and the full stationary covariance can be computed exactly. The following calculation is a \emph{special case} of the fixed-policy market, not a new price-formation rule. The analytical and population experiments specialize Model A to affine linear encoders as follows. Stack all active types' public states into $z_t\in\R^k$, and write their update in the common linear form
\begin{equation}\label{eq:S3-linear-encoder}
 z_t=Tz_{t-1}+U_m m_t+U_pp_{t-1}+U_DD_t+u_0.
\end{equation}
These matrices are fixed within a training epoch; finite lag windows are included by augmenting the state. Stack the risk-weighted public forecast slopes into a row vector $b^\top$ and their weighted intercepts into a scalar $k_0$, so $K_t=k_0+b^\top z_t$. In Model A, $A,\Delta$ are constants as in \eqref{eq:S3-modelA}. The resulting competitive price is
\begin{equation}\label{eq:S3-linear-price}
 p_t=\frac{k_0+b^\top z_t+A a^\top f_t-\bar S-\zeta_t}{\Delta}.
\end{equation}
Substitute S1's observations $m_t=Hf_t+\omega_t$, matured payout $D_t=\bar d+\ell^\top f_{t-1}+\varepsilon_t$, and state transition $f_t=Ff_{t-1}+G\eta_t$. Setting $Y_t=(f_t^\top,z_t^\top,p_t)^\top$ gives an affine Gaussian vector autoregression
\begin{equation}\label{eq:S3-matrix-recursion}
 Y_t=d_{x,\mathbf M}+\mathsf K_{x,\mathbf M}Y_{t-1}
 +\mathsf L_{x,\mathbf M}\begin{pmatrix}\eta_t\\\omega_t\\\varepsilon_t\\\zeta_t\end{pmatrix},
\end{equation}
whose transition matrix is \emph{derived} by substitution:
\begin{equation}\label{eq:S3-K}
 \boxed{\quad
 \mathsf K_{x,\mathbf M}=\begin{pmatrix}
 F&0&0\\
 U_mHF+U_D\ell^\top&T&U_p\\
 \dfrac{b^\top(U_mHF+U_D\ell^\top)+Aa^\top F}{\Delta}
 &\dfrac{b^\top T}{\Delta}&\dfrac{b^\top U_p}{\Delta}
 \end{pmatrix}.\quad}
\end{equation}
All dimensions are conformable: $F$ is $m\times m$, $T$ is $k\times k$, $U_p$ is $k\times1$, and $b^\top$ is $1\times k$. The noise matrix $\mathsf L$ follows from the same substitutions and is recorded in the accompanying reproducibility code. Explicitly, its blocks for $(\eta_t,\omega_t,\varepsilon_t,\zeta_t)$ are
\begin{equation}\label{eq:S3-L}
\mathsf L=\begin{pmatrix}
 G&0&0&0\\
 U_mHG&U_m&U_D&0\\
 \dfrac{b^\top U_mHG+Aa^\top G}{\Delta}&\dfrac{b^\top U_m}{\Delta}&\dfrac{b^\top U_D}{\Delta}&-\dfrac1\Delta
\end{pmatrix}.
\end{equation}
The primitive noise covariance is $\Omega=\operatorname{diag}(I_q,\Sigma_\omega,\sigma_\varepsilon^2,\sigma_\zeta^2)$.
\begin{proposition}[Linear-Gaussian fast-market stationarity]\label{prop:S3-lyap}
If $\rho(\mathsf K_{x,\mathbf M})<1$, then \eqref{eq:S3-matrix-recursion} has a unique causal square-integrable stationary solution. Its mean and covariance are
\begin{equation}\label{eq:S3-lyapunov}
 \bar Y=(I-\mathsf K)^{-1}d_{x,\mathbf M},\qquad
 \Sigma_Y=\mathsf K\Sigma_Y\mathsf K^\top+\mathsf L\Omega\mathsf L^\top.
\end{equation}
The stationary covariance is uniquely determined by the discrete Lyapunov equation. It yields the population moments used to fit each Gaussian forecasting head under the stationary market law.
\end{proposition}
\begin{proof}
The stable linear recursion has the convergent moving-average representation $Y_t=\bar Y+\sum_{j\ge0}\mathsf K^j\mathsf L u_{t-j}$, with $u_t=(\eta_t,\omega_t,\varepsilon_t,\zeta_t)^\top$. The noise sequence is independent across dates with finite covariance $\Omega$, hence summing the covariance series yields \eqref{eq:S3-lyapunov}. Uniqueness follows from $\rho(\mathsf K)<1$.
\end{proof}

\paragraph{What this benchmark does and does not cover.}
The Lyapunov calculation gives a population law without sample noise or stochastic gradient optimization. A \emph{linear public encoder} with a \emph{constant affine head} has a jointly Gaussian invariant state and price whenever \eqref{eq:S3-lyapunov} is stable. A nonlinear encoder and state-dependent risk generally do not preserve joint Gaussianity. Moreover, the stability condition for the fixed-policy $\mathsf K$ says nothing by itself about convergence of the \emph{across-epoch} learning map. The two eigenvalue tests refer to different mathematical operators.

\subsection{Frozen-policy markets, refitting, and retrained equilibria}\label{subsec:S3-retraining}

The preceding two subsections answer a \emph{fixed-algorithm} question: for a given adoption mix and set of forecasting rules, what distribution of prices and other market data is generated over time? We now allow the algorithms to learn from that distribution. Doing so requires keeping two operations distinct. The \emph{market operator} $\mathcal G(x,\mathbf M)$ maps a fixed collection of deployed models and their population shares into the stationary data law they generate. The \emph{learning operator} $\mathcal B_x(Q)$ takes a data law $Q$ and returns models that are optimally fitted to it within the capacity classes used at adoption mix $x$.

More precisely, S2 defines the type-$c$ population log-score minimization problem $\mathcal B_c(Q;B_c)$ and a selected minimizer $M_c^*(Q;B_c)$ when multiple solutions exist. Let $\mathbf B(x)$ denote the capacities present at adoption mix $x$, and collect the selected best-fitting models for those capacities in $\mathcal B_x(Q)$. In the most general formulation, minimizers and stationary market laws may be nonunique; a particular simulated retraining path then requires explicit selection rules. When the relevant models and invariant fast-market laws are unique, the two operations are ordinary functions.

The sequence of experiments is as follows. Immediately after adoption, the market has \emph{new trading rules} but inherits \emph{old histories}. It next generates \emph{new histories without changing those rules}. Only after this stationary adjustment do we ask what happens when investors \emph{refit} their rules using the new data. The next two definitions formalize the last two stages.

\begin{definition}[Frozen-policy stationary law]\label{def:S3-frozen}
Start from a baseline trained model profile $\mathbf M_0$ and change the adoption mix to $x_1$. If an upgraded capacity was inactive in the baseline, its candidate forecast is first fitted on the same inherited $Q_0$. Call $\mathbf M_1$ the resulting profile \emph{deployed immediately after adoption}: it retains the old policies for non-upgraders and includes the newly fitted high-capacity candidates for upgraders. With every parameter in $\mathbf M_1$ held fixed, let the market evolve according to the fast recursion. Denote its invariant law, if it exists, by
\begin{equation}\label{eq:S3-frozen}
 Q_{x_1}^{\mathrm F}=\mathcal G(x_1,\mathbf M_1).
\end{equation}
The encoder states and forecasts are recomputed each trading date from newly generated public histories, but the encoder weights and head parameters are not refitted. This is the \emph{distributional adjustment without retraining}.
\end{definition}

At the frozen-policy stage, an investor using an unchanged rule $z_{L,t+1}=p_t$ will nevertheless compute a different state whenever the newly formed market generates a different price. The forecast can change, but the coefficients of the rule do not. \emph{Retraining} is a subsequent operation: the coefficients themselves are optimized using the new market history. We allow every active type, including investors who did not upgrade, to do so. For a fixed adoption mix $x$, the \emph{population retraining iteration} is
\begin{equation}\label{eq:S3-iterate}
 \boxed{\quad Q_n=\mathcal G(x,\mathbf M_n),\qquad
 \mathbf M_{n+1}=\mathcal B_x(Q_n).\quad}
\end{equation}
This is the formal counterpart of the feedback described in S2. A non-upgrading type may still change its forecasts because the prices it uses in training have changed. The model-selection map and the market-clearing map are separate; the first evaluates prediction quality under an inherited market law, whereas the second constructs the law created by the selected predictors.

\begin{definition}[Population-retrained computational-market equilibrium]\label{def:S3-equilibrium}
For a fixed adoption mix $x$, a pair $(\mathbf M_x^*,Q_x^*)$ is a population-retrained stationary equilibrium if
\begin{equation}\label{eq:S3-fixedpoint}
 \boxed{\qquad Q_x^*=\mathcal G(x,\mathbf M_x^*),\qquad
 \mathbf M_x^*\in\mathcal B_x(Q_x^*).\qquad}
\end{equation}
Every active type's forecast is population-optimal \emph{within its feasible computational class} for the same stationary law its joint trading decisions generate. This is not the assertion that every forecast is objectively correctly specified, nor that the equilibrium is dynamically attracting under the iteration \eqref{eq:S3-iterate}.
\end{definition}

A retrained equilibrium therefore satisfies two conditions at once: investors use rules fitted to the market's prevailing data distribution, and their trades reproduce that same distribution. A stationary law of the fixed-policy market alone meets only the second condition for a \emph{given} set of rules; it need not make those rules optimal for the data it generates. This equilibrium notion is a \emph{statistical fixed point coupled to a competitive financial market}. It differs from a single-period rational-expectations equilibrium in that the family of feasible predictive models is computationally restricted and refitted to the stationary data generated by its deployment. The model does not assume that an individual atomistic account takes account of the effect of its own infinitesimal forecast change on the market law. The market-wide data feedback instead arises when a positive mass adopts or retrains its technology. An additional ex-ante equilibrium in \emph{adoption shares} can be defined by combining the computed private benefit of an upgrade with a cost schedule, but is not part of definition \eqref{eq:S3-fixedpoint}.

\subsection{Keeping deployment, data adjustment, and retraining separate}\label{subsec:S3-decomposition}

S3.8 defined the markets that exist immediately after adoption, after market histories have adjusted with the algorithms fixed, and after the algorithms have been refitted. This subsection asks a different question: \emph{how much of a change in a specified market statistic is associated with each of those successive stages?} We answer it by evaluating the \emph{same} statistic at four precisely defined counterfactual states. This is an accounting decomposition, not another equilibrium concept.

Take, for example, the mean price-signal likelihood precision under Model B. Write $\mathfrak J(x,\mathbf M;Q)\equiv\E_Q[A_t(x,\mathbf M)^2/\sigma_\zeta^2]$, where the data law $Q$ supplies the pre-auction histories on which the frozen policies operate. We consider the baseline, the first changed auction, the frozen-policy stationary market, and the retrained equilibrium in that order:
\begin{align}\label{eq:S3-stages}
 \mathfrak J_0&=\mathfrak J(x_0,\mathbf M_0;Q_0),& Q_0&=\mathcal G(x_0,\mathbf M_0),\\
 \mathfrak J_{\mathrm{imp}}&=\mathfrak J(x_1,\mathbf M_1;Q_0),& \mathfrak J_{\mathrm F}&=\mathfrak J(x_1,\mathbf M_1;Q_{x_1}^{\mathrm F}),\nonumber\\
 \mathfrak J_{\mathrm R}&=\mathfrak J(x_1,\mathbf M_{x_1}^*;Q_{x_1}^*),& Q_{x_1}^{\mathrm F}&=\mathcal G(x_1,\mathbf M_1).\nonumber
\end{align}
Here $\mathbf M_1$ retains the baseline policies for non-upgraders and includes high-budget policies fitted under the inherited law for newly adopting accounts. The key difference between the middle two evaluations is the distribution of \emph{inputs}, not the algorithms: $\mathfrak J_{\mathrm{imp}}$ uses histories drawn from $Q_0$, whereas $\mathfrak J_{\mathrm F}$ uses those generated by the newly deployed but fixed policies. The final $\mathfrak J_{\mathrm R}$ also replaces the policies with those fitted at the retrained equilibrium. Adding and subtracting the intermediate evaluations gives the exact telescoping identity
\begin{align}\label{eq:S3-telescope}
 \mathfrak J_{\mathrm R}-\mathfrak J_0
 &=(\mathfrak J_{\mathrm{imp}}-\mathfrak J_0)
 +(\mathfrak J_{\mathrm F}-\mathfrak J_{\mathrm{imp}})
 +(\mathfrak J_{\mathrm R}-\mathfrak J_{\mathrm F})\\
 &\equiv\underbrace{\Delta_{\mathrm{impact}}}_{\text{policy mix}}
 +\underbrace{\Delta_{\mathrm{frozen}}}_{\text{distributional adjustment}}
 +\underbrace{\Delta_{\mathrm{retrain}}}_{\text{model refitting}}.\nonumber
\end{align}
This separates effects without assuming that they are additive \emph{structural} causal channels under arbitrary interventions: the terms are an ordered set of well-defined counterfactual changes. In particular, the result depends on using the specified order for the decomposition, and the intermediate outcomes need not be independent.

For \emph{Model A with all head coefficients frozen}, $\mathfrak J(x_1,\mathbf M_1;Q)$ is independent of $Q$, so $\Delta_{\mathrm{frozen}}=0$ for this specific \emph{likelihood precision} measure. It does \emph{not} follow that the frozen-policy market produces unchanged prices, forecasts, learning inputs, or conditional mutual information. For Model B, $A_t$ can change as fixed nonlinear encoders process new histories, and the frozen-policy contribution can therefore be nonzero even for $\mathfrak J$.

For an information measure such as $I(\theta_t;p_t\mid\mathcal F_{L,x,t})$, each stage instead requires the corresponding joint counterfactual law of fundamentals, public signals and the actual current clearing price. It is incorrect to substitute the \emph{structural} $J$ into the mutual-information formula without ensuring the same conditioning field and correct Gaussianity. The main text's frozen/retrained information curves are backed by separate calculations on the appropriate stationary laws.

\subsection{Local stability and comparative statics of retraining}\label{subsec:S3-local}

The previous subsection compared \emph{levels} of a market outcome across four stages. We now ask a complementary question: when the adoption rate changes slightly, how much does the \emph{retrained equilibrium} change, and is the retraining process itself stable? Both questions concern the slow sequence of model updates, rather than the within-epoch market recursion studied in S3.6--S3.7.

For a finite-dimensional, differentiable representation of the market law, let $q\in\R^d$ be a vector of stationary moments or price coefficients and let $\Phi(q,x)$ be the law-to-law retraining map: fit the optimal algorithms to the law represented by $q$, deploy them at adoption share $x$, and calculate the moments of the new stationary market. Then
\begin{equation}\label{eq:S3-qmap}
 q_{n+1}=\Phi(q_n,x),\qquad q^*(x)=\Phi(q^*(x),x).
\end{equation}
This representation is a convenient exact coordinate chart in a Gaussian population submodel; it need not be finite-dimensional in an arbitrary nonlinear market. A fixed point is a \emph{separate} mathematical condition, not a consequence of the existence of $\mathcal G$.

\begin{proposition}[Sufficient conditions for a retraining fixed point]\label{prop:S3-slowexist}
Suppose the law-to-law map $\Phi(\cdot,x)$ is a continuous self-map of a nonempty compact convex set $\mathcal Q\subset\R^d$ of admissible stationary-law coordinates. Then it has at least one fixed point. If in addition $\Phi$ is a contraction on $\mathcal Q$ in some norm, the fixed point is unique within $\mathcal Q$ and the iteration $q_{n+1}=\Phi(q_n,x)$ converges to it from any $q_0\in\mathcal Q$.
\end{proposition}
\begin{proof}
Existence follows from Brouwer's fixed-point theorem. Under contraction, the Banach fixed-point theorem yields uniqueness and convergence.
\end{proof}
The conditions are stated for a finite-dimensional chart with feasible trained policies and invariant fast-market laws; they do not assert that arbitrary neural-network retraining always generates such a compact continuous self-map.

Whenever differentiability and a nonsingular $I-D_q\Phi$ hold at a fixed point, the implicit-function theorem gives
\begin{equation}\label{eq:S3-IFT}
 \boxed{\quad \frac{dq^*}{dx}=(I-D_q\Phi)^{-1}\partial_x\Phi.\quad}
\end{equation}
For any differentiable equilibrium informational outcome $W(q,x)$, its total adoption derivative is
\begin{equation}\label{eq:S3-total}
 \frac{dW(q^*(x),x)}{dx}
 =\underbrace{\partial_xW}_{\text{direct composition effect at fixed law coordinates}}
 +\underbrace{D_qW\,(I-D_q\Phi)^{-1}\partial_x\Phi}_{\text{equilibrium learning and market feedback}}.
\end{equation}
The second term need not share the sign of the first. In particular, even if the initial first-deployment result in S4 gives a negative direct effect, the long-run response need not be universally negative. S5--S6 analyze other informational outcomes using the same stationary-law map, with separate assumptions.

If $\rho(D_q\Phi(q^*,x))<1$, the fixed point is locally attracting for the \emph{slow learning iteration} under the usual smoothness conditions. If an eigenvalue crosses $-1$, a local period-doubling (flip) mechanism may arise, but a stable two-cycle must be established by additional nonlinear analysis or direct verification; eigenvalue crossing alone is insufficient. In the computational population experiments, one can evaluate the Jacobian by differentiating the stationary Lyapunov equation and the population least-squares head, rather than by differentiating through a finite stochastic training sample.

For clarity about the source of this derivative, suppose a locally unique population-optimal model has parameters $\psi^*(q)$ and satisfies $\nabla_\psi\mathcal L(\psi^*(q);q)=0$, with nonsingular objective Hessian $\mathsf H=\nabla^2_{\psi\psi}\mathcal L$. Local differentiation yields
\begin{equation}\label{eq:S3-learning-IFT}
 \frac{d\psi^*(q)}{dq}=-\mathsf H^{-1}\nabla^2_{\psi q}\mathcal L.
\end{equation}
Combining \eqref{eq:S3-learning-IFT} with the derivative of the market-law map gives $D_q\Phi$. For neural learning with nonunique minima, changing active constraints, or stochastic early stopping, these derivatives need not characterize the implemented training process. The fast-market condition $\rho(\mathsf K)<1$ in the linear case and the slow-learning condition $\rho(D_q\Phi)<1$ address different clocks.

\subsection{Interpretation and links to the analytical results}\label{subsec:S3-handoff}

Three separations matter for every result that follows. First, the structural price loading $J=A^2/\sigma_\zeta^2$ describes fresh private-information precision, not the total information available to an observer whose public-data processing is limited. Second, training minimizes a statistical Gaussian log score, whereas demand solves a CARA utility problem \emph{using the trained belief}; equality with correct Bayesian behavior requires additional assumptions. Third, an immediate price response, a new stationary law with fixed algorithms, and a stationary retraining fixed point are distinct counterfactual objects. Conflating them would turn a narrow first-deployment sign result into an unwarranted claim about all long-run informational effects.

\begin{table}[htbp]\centering\small
\caption{Market-clearing and informational inputs to the main analytical results.}\label{tab:S3-handoff}
\begin{tabularx}{\linewidth}{@{}p{26mm}X X@{}}\toprule
\textbf{SI proof} & \textbf{S3 input} & \textbf{What must still be established there}\\\midrule
S4 / Theorem 1 & Demand slope $\beta/(\gamma v)$; inherited-law $A(x)$; likelihood precision $J=A^2/\sigma_\zeta^2$. & Nested optimal learning, aligned payoff and the correct sign of private-signal response.\\
S5 / Theorem 2 & Common price and pre-price low/high information fields; stationarity and, where used, jointly Gaussian law. & Information identity, Gaussian posterior variances, ex-ante capacity value and its derivative.\\
S6 / Theorem 3 & Price formed by optimal CARA demands and actual clearing; locally differentiable equilibrium equations. & A two-factor primitive economy with strict coexistence signs and an open neighborhood.\\\bottomrule
\end{tabularx}\end{table}

The methods in S3 also underlie later experiments. The stationary Lyapunov benchmark supplies the population covariance moments used to calculate the figures, whereas the general nonlinear clearing equation permits resource-constrained SSM and GRU forecasting policies to trade in the same economic environment. Numerical fast-market diagnostics and repeated-training runs do not substitute for population existence proofs unless the stated regularity conditions are verified.

S4 will derive the main crowd-out sign theorem under nested computational classes and the inherited-law comparison. S5 proves how the price can transmit previously processed public information and how this affects the private value of computation. S6 shows that these channels coexist in a genuine competitive market with population Gaussian learning. Their shared economic inputs are precisely the objects defined above; no distinct ad hoc pricing equation is introduced for the individual analytical results.

\begin{longtable}{@{}p{42mm}p{106mm}@{}}
\caption{S3 notation and the corresponding economic object.}\label{tab:S3-notation}\\
\toprule\textbf{Symbol} & \textbf{Meaning}\\\midrule\endfirsthead
\toprule\textbf{Symbol} & \textbf{Meaning (continued)}\\\midrule\endhead
$h_{ci,t}(p)$ & Individual CARA-optimal holding at candidate clearing price $p$.\\
$\mu_{ci,t}(p),v_c(z)$ & Learned subjective Gaussian conditional mean and state-dependent predictive variance.\\
$w_{c,t}$ & Population mass divided by risk aversion and fitted predictive variance.\\
$K_t$ & Aggregate public-history component of demand, before subtracting the supply level.\\
$A_t$ & Aggregate loading on the privately signaled scalar fundamental $\theta_t$.\\
$\Delta_t$ & Positive aggregate price slope magnitude; $-\Delta_t$ is the aggregate demand derivative.\\
$J_t$ & Fresh-price likelihood precision $A_t^2/\sigma_\zeta^2$; not conditional mutual information.\\
$x_g$ & Fraction of investor group $g$ deploying the high-budget rather than low-budget system.\\
$\mathcal G(x,\mathbf M)$ & Stationary law of the fast market when fitted policies are fixed.\\
$Q_x^{\mathrm F}$ & Fast-market stationary law after adopting new policies but before retraining them.\\
$\mathcal B_x(Q)$ & Typewise population log-score refitting correspondence at adoption composition $x$.\\
$(\mathbf M_x^*,Q_x^*)$ & Population-retrained stationary computational-market equilibrium.\\
$q,\Phi(q,x)$ & Finite-dimensional population-law coordinate and retraining law-to-law map where available.\\
$\mathsf K,\mathsf L,\Sigma_Y$ & Fast linear-Gaussian market transition, noise loading and stationary Lyapunov covariance.\\
\bottomrule
\end{longtable}

\clearpage
\setcounter{table}{0}\renewcommand{\thetable}{S\arabic{section}.\arabic{table}}
\renewcommand{\theHtable}{S.\arabic{section}.\arabic{table}}
\section{Inherited-law computational crowd-out}\label{sec:S4}

An investor who can predict more of an asset's payoff from public records may put less weight on a private clue. But this alone does not show that she trades less on that clue. Better prediction also reduces her estimated risk, which encourages larger positions. This section establishes when the first effect dominates the second. The result connects an improvement in the best feasible forecast to a weaker private-signal response in demand and, through competitive clearing, to a less precise price signal about the privately observed fundamental.

To isolate the immediate effect of a capacity upgrade, we first fit the low- and high-capacity algorithms to the same inherited market data law. We then replace the forecasting rules of some investors with their higher-capacity alternatives, keeping the pre-auction public history, private signals and market primitives fixed. Their new demand schedules determine a new clearing price. This comparison, formalized in S3.5, captures the change in privately informed trading before new prices enter subsequent public histories or anyone retrains on the changed market. Theorem~\ref{thm:one} establishes when this change makes the price a less precise signal of the privately observed fundamental. The price can also convey computations on public records, whose value to different observers is studied in S5.

\subsection{The learning and deployment comparison}\label{subsec:S4-comparison}

Maintain the aligned-payoff economy of S1,
\begin{equation}\label{eq:S4-primitives}
 D_{t+1}=\bar d+\lambda\theta_t+\varepsilon_{t+1},\qquad
 s_{ci,t}=\theta_t+\xi_{ci,t},\qquad \lambda>0,
\end{equation}
where $\V(\varepsilon_{t+1})=\sigma_\varepsilon^2>0$ and $\V(\xi_{ci,t})=\nu_c>0$. The payoff disturbance is independent of the fundamental and of the information available at the auction. The individual signal error is independent of the aggregate history, fundamental and price, under the atomistic aggregation assumption in S1. The supply disturbance has unchanged distribution $\zeta_t\sim N(0,\sigma_\zeta^2)$, with $\sigma_\zeta^2>0$, and is independent of the inherited pre-auction history and current fundamental. Risk aversion is $\gamma_c>0$.

An admissible public encoder constructs $z_{c,t}$ from the pre-auction record $\mathscr H_t^-$. It may be nonlinear, recurrent or based on a finite history. The forecasting system uses S2's Model-A final head,
\begin{equation}\label{eq:S4-head}
 \mu_{ci,t}(p)=\alpha_c+b_c^\top z_{c,t}+\beta_cs_{ci,t}+\chi_cp,
 \qquad v_c>0,
\end{equation}
with constant fitted coefficients and one constant fitted variance per type. Nonlinear computation is allowed in $z_{c,t}$, while the private-signal coefficient and reported risk remain constant across public states. The resource constraint applies to the complete encoder and forecasting head, using the common accounting convention of S2.3.

Write $Q$ for the inherited stationary data law and $p_t^Q$ for the realized price in a historical observation drawn from that law. In training, the mean in \eqref{eq:S4-head} is evaluated at $p_t^Q$. At deployment, the same fitted coefficients are evaluated at a candidate price $p$ to form a demand schedule. Neither price is an input to that date's public encoder: historical date-$t$ encoding uses only information available before the date-$t$ auction. This distinction permits us to fit a model under one law and use its schedule to clear a changed market.

\begin{assumption}[Conditions for inherited-law crowd-out]\label{ass:S4-T1}
The aligned primitives, independent errors and atomistic exact signal aggregation of S1 hold. The inherited law $Q$ and every admissible encoder output used below have finite second moments. For an upgrading type $c$, the nonempty feasible Model-A classes obey $\mathcal M_c(B_c)\subseteq\mathcal M_c(B'_c)$ when $B'_c\succeq B_c$. For every feasible encoder, its affine mean coefficients are freely and jointly optimized, and its positive constant variance is freely optimized; these head optima are feasible and interior. Global population Gaussian-log-score minima over the complete classes are attained under the same $Q$.

The investors' masses, risk preferences and signal qualities, the exogenous state law, and the supply-noise distribution are held fixed. Non-upgrading types retain their deployed Model-A rules. Their private-signal loadings are nonnegative, as holds for the aligned population-optimal heads considered here. The aggregate clearing slopes of both compared profiles are positive. No stationarity or retraining condition is imposed on the market after deployment.
\end{assumption}

The head must attain its unrestricted least-squares and variance optima; binding coefficient penalties or variance bounds would change this step. Attainability concerns population optima, which a particular finite-data training run need not achieve.

For each upgraded type, both capacities solve S2's Gaussian forecasting problem under the same $Q$:
\begin{equation}\label{eq:S4-loss}
 \mathcal L_c(M;Q)=\frac12\E_Q\!\left[
 \log(2\pi v_c)+\frac{(D_{t+1}-\mu_{ci,t}(p_t^Q))^2}{v_c}\right].
\end{equation}
Here and below $\E_Q$ includes the representative type-$c$ private signal, the marginal denoted $Q_c$ in S2. Actual demand then follows from S3.2:
\begin{equation}\label{eq:S4-demand-slope}
 h_{ci,t}(p)=\frac{\alpha_c+b_c^\top z_{c,t}+\beta_cs_{ci,t}-(R_f-\chi_c)p}{\gamma_cv_c},
 \qquad \iota_c\equiv\frac{\partial h_{ci,t}(p)}{\partial s_{ci,t}}
 =\frac{\beta_c}{\gamma_cv_c}.
\end{equation}
The coefficient $\iota_c$ measures how many additional shares an investor wants to hold when her private signal rises by one unit, holding the candidate price and public state fixed. A smaller $\iota_c$ therefore means a weaker trading response to the same favorable private news. The forecast weight $\beta_c$ affects that response through the expected payoff, while risk aversion $\gamma_c$ and fitted risk $v_c$ determine the position taken on that expectation. Thus $\iota_c$ measures sensitivity to private information, not total trading volume. The holding is optimal under the learned Gaussian belief; the proof does not require that belief to be the true conditional distribution in the newly cleared market.

\subsection{How much remains for the private signal to explain?}\label{subsec:S4-projection}

Fix a type and a feasible encoder. Before adding its own signal, the head can use an intercept, the computed public state $z_{c,t}$ and the inherited current price $p_t^Q$. Define the fundamental prediction error from those regressors by
\begin{equation}\label{eq:S4-P}
 P=\Pi_c(z;Q)\equiv\inf_{a,b,d}\E_Q\!\left[
 (\theta_t-a-b^\top z_{c,t}-dp_t^Q)^2\right].
\end{equation}
Here $a,b,d$ are auxiliary projection coefficients; the scalar $a$ is unrelated to the latent-state loading in S1. The index $c$ identifies the encoder class. Including $p_t^Q$ measures what the private clue adds beyond both computed public data and price.

Equation~\eqref{eq:S4-P} is an unconditional linear-projection residual under $Q$. It is a single nonnegative number for this encoder and law. It need not equal $\V(\theta_t\mid z_{c,t},p_t^Q)$, which can vary with the conditioning values, or the minimum error over all nonlinear functions of those values. A nonlinear encoder can create useful nonlinear predictors of the raw history while the final projection remains affine in its output.

\begin{lemma}[Population-optimal affine head]\label{lem:S4-projection}
For a fixed feasible encoder under Assumption~\ref{ass:S4-T1}, let $P=\Pi_c(z;Q)$. Its optimal Model-A head has
\begin{equation}\label{eq:S4-beta-v}
 \beta_c(P)=\frac{\lambda P}{P+\nu_c},\qquad
 v_c(P)=\sigma_\varepsilon^2+\frac{\lambda^2\nu_cP}{P+\nu_c}.
\end{equation}
Consequently,
\begin{equation}\label{eq:S4-iota}
 \boxed{\quad
 \iota_c(P)=\frac{\beta_c(P)}{\gamma_cv_c(P)}
 =\frac{\lambda P}
 {\gamma_c[\sigma_\varepsilon^2\nu_c+(\sigma_\varepsilon^2+\lambda^2\nu_c)P]}.
 \quad}
\end{equation}
These quantities are uniquely determined by $P$, even if redundant public regressors make some individual public coefficients nonunique.
\end{lemma}

\begin{proof}
Suppress $c,i,t$ where no ambiguity arises. Let $\mathcal V_z=\operatorname{span}(1,z,p^Q)$ in $L^2(Q)$, and let $\widehat\theta_z$ be the orthogonal projection of $\theta$ onto this space. Write $r=\theta-\widehat\theta_z$, so $\E_Q[r^2]=P$ and $r$ is orthogonal to every regressor in $\mathcal V_z$. This finite-dimensional projection exists even if the displayed regressors are linearly dependent.

The independent, mean-zero individual noise has zero projection on $\mathcal V_z$. The residual of $s$ after projection is therefore $r+\xi$. Likewise, the residual of $D$ is $\lambda r+\varepsilon$. Their covariance and the signal residual's variance are
\begin{equation}\label{eq:S4-FWL}
 \E_Q[(\lambda r+\varepsilon)(r+\xi)]=\lambda P,
 \qquad \E_Q[(r+\xi)^2]=P+\nu>0.
\end{equation}
Frisch--Waugh--Lovell then gives $\beta=\lambda P/(P+\nu)$. Equivalently, the optimal fitted mean can be represented as
\begin{equation}\label{eq:S4-fitted-mean}
 \mu^*=\bar d+\lambda\widehat\theta_z+\beta(s-\widehat\theta_z).
\end{equation}
Its residual is $(\lambda-\beta)r-\beta\xi+\varepsilon$. Independence of the primitive errors and the projection orthogonality eliminate the cross terms, giving
\begin{align}\label{eq:S4-mse}
 \E_Q[(D-\mu^*)^2]
 &=(\lambda-\beta)^2P+\beta^2\nu+\sigma_\varepsilon^2\\
 &=\sigma_\varepsilon^2+\frac{\lambda^2\nu P}{P+\nu}.\nonumber
\end{align}
For a constant variance, minimizing \eqref{eq:S4-loss} over the mean is least squares. For any mean with residual MSE $R>0$, the variance-dependent objective is $\tfrac12\log(2\pi v)+R/(2v)$. Its derivative is $(v-R)/(2v^2)$, so its unique positive minimum is at $v=R$. Substituting \eqref{eq:S4-mse} establishes the formula for $v$. Division gives \eqref{eq:S4-iota}. The private-signal coefficient is unique because the residual signal variance is strictly positive; the fitted mean is the unique projection onto $\mathcal V_z+\operatorname{span}(s)$.
\end{proof}

The argument uses orthogonality of a linear-projection residual, not independence of the fundamental from the encoder state. It therefore applies to nonlinear public encoders and does not assume that $(\theta,z,p^Q)$ is jointly Gaussian. Gaussianity of the learned belief matters for the demand formula inherited from S3; the projection identities themselves depend on the stated second moments and independence of the primitive errors.

\subsection{Why the reduction in risk does not reverse crowd-out}\label{subsec:S4-mechanism}

Lemma~\ref{lem:S4-projection} makes both effects of improved prediction explicit. When $P$ falls, the private clue receives less weight in the mean forecast, while lower forecast risk encourages a larger position for a given expected excess payoff. Why does the first effect dominate? The two effects are linked by the optimal-head identity
\begin{equation}\label{eq:S4-risk-floor}
 v_c(P)=\sigma_\varepsilon^2+\lambda\nu_c\beta_c(P),\qquad
 \iota_c(P)=\frac1{\gamma_c}
 \frac{\beta_c(P)}{\sigma_\varepsilon^2+\lambda\nu_c\beta_c(P)}.
\end{equation}
Only the second component of forecast risk shrinks with the private-signal coefficient. The independent payoff disturbance remains uncertain even after the current fundamental is learned. It places a floor under risk and prevents the increase in position size from fully offsetting the decline in the signal's predictive weight. Thus a better public-data forecast leads to less trading on a given change in the private clue, even after the investor adjusts her position for the lower estimated risk.

Formally, differentiating the three head quantities gives
\begin{align}\label{eq:S4-derivatives}
 \beta_c'(P)&=\frac{\lambda\nu_c}{(P+\nu_c)^2}>0,\qquad
 v_c'(P)=\frac{\lambda^2\nu_c^2}{(P+\nu_c)^2}>0,\\
 \iota_c'(P)&=\frac{\lambda\sigma_\varepsilon^2\nu_c}
 {\gamma_c[\sigma_\varepsilon^2\nu_c+(\sigma_\varepsilon^2+\lambda^2\nu_c)P]^2}>0.\nonumber
\end{align}
At $P=0$, the derivatives are understood from the right. An encoder whose affine head already predicts $\theta$ exactly has $\beta_c=\iota_c=0$ and $v_c=\sigma_\varepsilon^2$.

For comparison, remove only this payoff disturbance and take $\sigma_\varepsilon^2=0$. For $P>0$, the formulas yield
\begin{equation}\label{eq:S4-zero-noise}
 v_c(P)=\lambda\nu_c\beta_c(P),\qquad
 \iota_c(P)=\frac1{\gamma_c\lambda\nu_c}.
\end{equation}
The two effects then cancel exactly: improving the forecast reduces its private-signal weight but leaves the private-signal demand slope unchanged. This boundary calculation is outside the maintained positive-noise assumptions and excludes $P=0$, where the positive-risk portfolio problem degenerates. It explains why a strictly positive irreducible payoff risk gives strict crowd-out when forecasting improves. The disturbance is the independent payoff shock already present in S1.

\subsection{From a larger feasible class to a less precise price signal}\label{subsec:S4-theorem}

It remains to show that greater computational capacity lowers the relevant $P$, rather than simply assuming that the selected public representation contains more information. After optimizing the head at a fixed encoder, its Gaussian loss is
\begin{equation}\label{eq:S4-profile}
 \mathcal L_c^{\mathrm{head},*}(z;Q)
 =\frac12\log\!\left(2\pi e\,v_c(\Pi_c(z;Q))\right).
\end{equation}
By \eqref{eq:S4-derivatives}, this expression is strictly increasing in $\Pi_c(z;Q)$. Consequently, any global optimum over a feasible complete Model-A class also minimizes this projection residual over the encoders whose unrestricted heads are feasible. A larger budget can reproduce the old complete algorithm, so it cannot have a higher minimum loss or a higher selected residual. Write the latter as $P_c^B$. Then
\begin{equation}\label{eq:S4-order}
 \mathcal M_c(B_c)\subseteq\mathcal M_c(B'_c)
 \quad\Longrightarrow\quad P_c^{B'}\le P_c^B.
\end{equation}
The states selected at the two optima may have different dimensions or summarize different aspects of the record. No inclusion between $\sigma(z_c^B)$ and $\sigma(z_c^{B'})$ is used. Moreover, all global minimizers at a given budget have the same $P_c^B$ and hence the same $\beta_c$, $v_c$ and $\iota_c$, even if their public predictions differ.

Individual private signals share the common component $\theta_t$, while their idiosyncratic errors average to zero within each type. Consequently, the private-signal contribution to aggregate demand is $A\theta_t$, where $A$ is the population-weighted sum of the responses $\iota_c$. S3's competitive clearing result gives, for either deployed profile,
\begin{align}\label{eq:S4-clearing}
 A&=\sum_c\pi_c\frac{\beta_c}{\gamma_cv_c},\qquad
 \Delta=\sum_c\frac{\pi_c(R_f-\chi_c)}{\gamma_cv_c}>0,\\
 K_t&=\sum_c\frac{\pi_c(\alpha_c+b_c^\top z_{c,t})}{\gamma_cv_c},\qquad
 p_t=\frac{K_t+A\theta_t-\bar S-\zeta_t}{\Delta}.\nonumber
\end{align}
Conditional on the inherited pre-auction public record and the deployed policies, $K_t$ is known. Removing this computed public component and rescaling gives
\begin{equation}\label{eq:S4-experiment}
 \Delta p_t-K_t+\bar S=A\theta_t-\zeta_t,
 \qquad J=\frac{A^2}{\sigma_\zeta^2}.
\end{equation}
When $A\ne0$, dividing the adjusted price by $A$ gives the measurement $\theta_t-\zeta_t/A$. Its noise variance is $\sigma_\zeta^2/A^2$, and $J$ is the reciprocal of that variance: the likelihood precision of the price signal about $\theta_t$. At $A=0$, define $J=0$. The clearing denominator cancels from this comparison because it scales both the fundamental and supply-noise terms.

\begin{samepage}
\begin{theorem}[Inherited-law computational crowd-out]\label{thm:one}
Under Assumption~\ref{ass:S4-T1}, each upgraded type's globally optimized Gaussian forecasting loss, fitted constant variance, private-signal forecast loading, and risk-adjusted demand response weakly decrease as its resource budget expands. Writing $\iota_c^B=\beta_c^B/(\gamma_cv_c^B)$ and $A^B=\sum_c\pi_c\iota_c^B$, the \emph{incremental fresh-signal likelihood precision} of the corresponding first-deployment price experiment satisfies
\begin{equation}\label{eq:S4-T1}
 \boxed{\quad B'\succeq B\quad\Longrightarrow\quad
 J^{\rm impact}(Q;B')\equiv\frac{(A^{B'})^2}{\sigma_\zeta^2}
 \le\frac{(A^B)^2}{\sigma_\zeta^2}\equiv J^{\rm impact}(Q;B).\quad}
\end{equation}
The last inequality is strict if a positive-mass upgraded type strictly improves globally optimal forecasting loss. This is a comparison of \emph{new private-signal precision after subtracting the pre-auction predictable price component}; it does not order unconditional price information or equilibrium mutual information after a new market-law adjustment.
\end{theorem}
\end{samepage}

Equation~\eqref{eq:S4-T1} says that unchanged supply noise obscures a weaker fundamental signal. Better public-data prediction makes each upgrading investor's demand less responsive to her private clue. The resulting fall in $A$ reduces the fundamental's contribution to aggregate orders while leaving the supply-noise variance unchanged. Once the computed public component is removed, the price therefore provides a noisier measurement of $\theta_t$. This conclusion allows the public component $K_t$ and clearing slope $\Delta$ to change; it concerns the relative strength of privately informed orders and supply noise.

\begin{proof}
For each upgraded type, nestedness of the complete feasible classes weakly lowers the attained minimum loss. Its freely optimized head satisfies Lemma~\ref{lem:S4-projection}, and the strictly increasing profile \eqref{eq:S4-profile} then gives $P_c^{B'}\le P_c^B$. The derivatives in \eqref{eq:S4-derivatives} imply weak decreases in $\beta_c$, $v_c$ and $\iota_c$. Non-upgrading types keep the same responses, so $A^{B'}\le A^B$.

For upgraded types, \eqref{eq:S4-iota} is nonnegative because $P\ge0$ and all primitive scale parameters are positive. The unchanged types also have nonnegative loadings by the maintained population-optimal aligned specification, or the explicit sign condition in Assumption~\ref{ass:S4-T1}. Hence $0\le A^{B'}\le A^B$, and squaring preserves the inequality. The supply variance is unchanged and positive, which proves \eqref{eq:S4-T1}.

If an upgraded type of mass $\pi_c>0$ has a strictly smaller optimal loss, strict monotonicity of \eqref{eq:S4-profile} implies $P_c^{B'}<P_c^B$. Equation~\eqref{eq:S4-derivatives} then gives $\iota_c^{B'}<\iota_c^B$. No other type's response rises, so $A^{B'}<A^B$; since both are nonnegative, their squared loadings are strictly ordered as well.
\end{proof}

Here $B$ and $B'$ denote the budget profiles, with unchanged coordinates for non-upgrading types. A partial upgrade can be represented by splitting an investor group into low- and high-capacity subtypes, as in S3.5. If group $g$ of mass $\pi_g$ adopts at share $x$, and both candidates remain fitted to $Q$, then
\begin{equation}\label{eq:S4-adoption}
 A(x)=A_{-g}+\pi_g[(1-x)\iota_{gL}+x\iota_{gH}],\qquad
 J'(x)=\frac{2A(x)\pi_g}{\sigma_\zeta^2}(\iota_{gH}-\iota_{gL})\le0.
\end{equation}
This is an adoption comparison within the same inherited-law experiment, wherever clearing has $\Delta(x)>0$. If the candidate improvement is strict, every positive adoption increase strictly lowers $J$; the derivative can equal zero at an endpoint with $A(x)=0$. The formula is not the derivative along a path of newly retrained market laws.

\subsection{A numerical illustration and the equality cases}\label{subsec:S4-example}

For a small illustration of the head identities, normalize $\lambda=\nu_c=\gamma_c=\sigma_\varepsilon^2=1$. Suppose the selected encoder's projection residual falls from $P=1$ to $P=1/4$. The quantities in Table~\ref{tab:S4-example} then follow exactly. These are illustrative forecast-head values, not a calibration or a reproduction of a main-text figure.

\begin{table}[H]\centering\small
\caption{A forecast improvement reduces the private-signal response despite lower risk.}\label{tab:S4-example}
\begin{tabularx}{\linewidth}{@{}X cc@{}}\toprule
 & \textbf{Low capacity} & \textbf{High capacity}\\\midrule
Projection residual $P$ & $1$ & $1/4$\\
Private-signal forecast coefficient $\beta(P)$ & $1/2$ & $1/5$\\
Fitted payoff variance $v(P)$ & $3/2$ & $6/5$\\
Private-signal demand response $\iota(P)$ & $1/3$ & $1/6$\\\bottomrule
\end{tabularx}\end{table}

The signal's forecast coefficient falls by 60\%, while fitted risk falls by 20\%. Their ratio falls by 50\%. Thus an increase of one unit in an investor's private signal, holding her public state and the quoted price fixed, raises her desired holding by $1/6$ rather than $1/3$ of a share. If an entire otherwise homogeneous population underwent this change with fixed supply noise, its aggregate private-signal loading would halve and the corresponding likelihood precision would fall to one quarter of its former value. Other unchanged traders would attenuate that aggregate comparison.

A larger budget need not produce any strict change. It may select the old program, or a different program with exactly the same optimal loss. By the strictly increasing profile, equal optimal losses imply equal $P$, $\beta$, $v$ and $\iota$. Thus under the theorem's conditions, equality in $J$ occurs exactly when every upgraded positive-mass type has unchanged optimal loss. An improvement by a negligible individual has no aggregate effect. The selected public coefficients and the price level can still change when $J$ is unchanged; $J$ is a measure of one signal channel, not a complete description of prices.

\subsection{Why the head restriction matters}\label{subsec:S4-boundaries}

Model A makes both optimized forecast loss and the private-signal response increasing functions of the same residual $P$. Model B lets the signal coefficient $\beta_c(z)$ and fitted risk $v_c(z)$ vary across public states. Its average forecasting loss can improve while trading on private information becomes stronger in some states: the loss averages logarithms of fitted risks, whereas price precision depends on the square of the aggregate demand response.

Appendix~\ref{app:S4-modelB} gives a complete counterexample with aligned payoffs and independent Gaussian primitive shocks, fitting both candidates under one inherited law. The higher-capacity system distinguishes ordinary public releases from unusually large ones and fits a separate head in each state. Its private-signal response falls in the ordinary state and rises in the large-release state. Both its optimized forecasting loss and its average demand response fall, yet its average squared aggregate response, and hence average structural price precision, rise. This shows that the constant-head restriction is economically substantive, without establishing a universal opposite sign for Model B.

\subsection{What the precision comparison leaves for S5}\label{subsec:S4-forward}

The price experiment in \eqref{eq:S4-experiment} conditions on the pre-auction public record and knowledge of the deployed policies, so its public component can be removed. A computationally limited observer may not be able to reproduce that component. An advanced algorithm's trades can therefore make the price more informative to that observer even while $J$ falls. Theorem~\ref{thm:one} isolates the privately informed trading channel; it does not rank the usefulness of the entire price to every observer.

Nor does $P_c^B$ in the learning proof represent the prior variance used in every price-information calculation. It is a training-law linear-projection residual that already includes $p^Q$. In a correctly specified jointly Gaussian price experiment, if a pre-price field $\mathcal U$ contains the public price component and $P_{\mathcal U}=\V(\theta\mid\mathcal U)$ is its prior variance, then
\begin{equation}\label{eq:S4-MI}
 I(\theta;p\mid\mathcal U)=\frac12\log(1+JP_{\mathcal U}).
\end{equation}
Here $P_{\mathcal U}$ is defined before observing the current price. Along changing stationary market laws it may itself change, and joint Gaussianity must be verified. Theorem~\ref{thm:one} does not sign that stationary mutual-information comparison.

Finally, frozen Model-A heads keep $A$ and $J$ fixed as the post-deployment market generates a new history, even though encoder outputs, forecasts and prices change. Retraining can subsequently change the coefficients and the law under which projection errors are evaluated. The theorem does not sign the additional effects of refitting under that changed law. S5 relates price information to feasible public computation and the private value of an upgrade; S6 constructs an equilibrium in which the opposing informational effects coexist. Both use the market and learning distinctions established here.

\clearpage
\begingroup
\renewcommand{\thesubsection}{\thesection.A}
\subsection{A state-dependent-head counterexample}\label{app:S4-modelB}
\setcounter{equation}{0}
\renewcommand{\theequation}{\thesection.A.\arabic{equation}}
\renewcommand{\theHequation}{\arabic{section}.A.\arabic{equation}}

This appendix supplies the construction used in S4.6. It compares globally optimized forecasts under one inherited law and derives the new clearing price after deployment. Its purpose is to show that a better Model-B log score need not reduce average structural price precision.

Consider an i.i.d. specialization of S1 with $\theta\sim N(0,1)$ and public release $m=\rho\theta+\omega$, where $\rho=9/10$ and $\omega\sim N(0,1-\rho^2)$ is independent. Hence $m$ is standard normal but does not reveal $\theta$. Set $D=\theta+\varepsilon$, $s_i=\theta+\xi_i$, $\sigma_\varepsilon^2=\nu=20$, $\gamma=1$, $\sigma_\zeta^2=1$, $\bar d=\bar S=0$ and $R_f=1.05$. All other shocks are independent, with the same continuum signal aggregation as before.

Use a deliberately small pair of nested admissible architecture classes. The low class retains no public state and has a free constant-affine head. The high class may copy that system or retain the single binary feature $z=\mathbf1\{|m|>1\}$ and use a free Model-B affine head separately in each of its two states. Each class includes all coefficients and positive variances for its stated head. This is a specified nested pair of computational architectures, not a claim that an unrestricted high-budget optimizer must choose this feature. Both candidate systems are globally optimized over these classes under the same inherited law.

Take the inherited law to be the all-low Gaussian competitive equilibrium. Write its normalized price as $y^Q=\Delta_0p^Q=\alpha\theta-\zeta$. The optimal low head and clearing give
\begin{equation}\label{eq:S4-B-root}
 \alpha=\frac1{440+400\alpha^2},\qquad
 0<\alpha<\frac1{440}.
\end{equation}
Indeed, its fundamental residual after $y^Q$ is $P=1/(1+\alpha^2)$, and \eqref{eq:S4-iota} gives $A=P/(400+40P)=1/(440+400\alpha^2)$. The polynomial $400\alpha^3+440\alpha-1$ is strictly increasing and has opposite signs at $0$ and $1/440$, so this root is unique. The inherited clearing denominator is positive, as verified below. The resulting law is stationary because the primitive inputs are i.i.d. and the low head uses no lagged state.

The binary feature distinguishes unusually large public releases from ordinary ones. It changes the conditional dispersion of the fundamental without changing its conditional mean. Let $q_j=\Pr(z=j)$ and $T_j=\V(\theta\mid z=j)$. Symmetry gives $\E[\theta\mid z=j]=0$. With $\varphi$ and $\Phi$ the standard normal density and distribution function,
\begin{equation}\label{eq:S4-B-states}
 \begin{split}
 q_0&=2\Phi(1)-1,\qquad q_1=1-q_0,\\
 T_0&=1-\rho^2\frac{2\varphi(1)}{q_0},\qquad
 T_1=1+\rho^2\frac{2\varphi(1)}{q_1}.
 \end{split}
\end{equation}
Thus $0<T_0<1<T_1<3$ and $q_0T_0+q_1T_1=1$. The upper bound follows, for example, from $1-\Phi(1)>\varphi(1)/2$: integrating $\varphi(u)/u$ by parts on $[1,\infty)$ gives $1-\Phi(1)=\varphi(1)-\int_1^\infty\varphi(u)/u^2\,du>\varphi(1)-(1-\Phi(1))$.

Conditional least squares on $(s,y^Q)$ in a state with variance $T$ gives
\begin{equation}\label{eq:S4-B-head}
 \begin{split}
 \beta(T)&=\frac{T}{20+(1+20\alpha^2)T},\qquad
 \psi(T)=20\alpha\beta(T),\\
 \mu(T;s,p)&=\beta(T)s+\psi(T)\Delta_0p,\qquad
 v(T)=20+20\beta(T),\\
 \iota(T)&=\frac{\beta(T)}{v(T)}
 =\frac{T}{400+\kappa_BT},\qquad \kappa_B=40+400\alpha^2<41.
 \end{split}
\end{equation}
To obtain these expressions directly, residualizing $\theta$ on $y^Q$ within the state gives $P(T)=T/(1+\alpha^2T)$; apply the same covariance calculation as in Lemma~\ref{lem:S4-projection}. The conditional distribution is not Gaussian after observing the binary feature, but these are exactly the optimal affine means and constant-within-state variances for the Gaussian log score. The high class can attain them because it has independent head parameters for its two states. The low head is \eqref{eq:S4-B-head} at $T=1$.

The function $\log v(T)$ is strictly concave for $T>0$: $\beta(T)$ is strictly increasing and strictly concave, $v(T)>0$ is strictly concave, and $\log$ is increasing and concave. Since the two $T_j$ are distinct and their mean is one, Jensen's inequality gives
\begin{equation}\label{eq:S4-B-loss}
 \mathcal L_H^*=\frac12\sum_{j=0}^1q_j\log(2\pi e\,v(T_j))
 <\frac12\log(2\pi e\,v(1))=\mathcal L_L^*.
\end{equation}
The state-dependent system is therefore the global optimum in the high class and strictly improves the population forecasting loss.

It also clears a well-defined new market. The inherited denominator is
\begin{equation}\label{eq:S4-B-delta}
 \Delta_0=\frac{R_f}{v(1)(1+20\alpha^2)}>0.
\end{equation}
After all accounts deploy the high head trained on this inherited law, its candidate-price slope in state $j$ is $\chi_j=\psi(T_j)\Delta_0$, and the new clearing denominator is $\Delta_j=(R_f-\chi_j)/v(T_j)$. Because $\Delta_0<R_f/20$, $\beta(T_j)<1$ and $\alpha<1$, we have $\chi_j<R_f\alpha<R_f$. Thus $\Delta_j>0$ in both states, and the actual new price satisfies $\Delta_jp=\iota(T_j)\theta-\zeta$. No historical price is substituted for this new clearing price.

Yet mean likelihood precision rises. For $f(T)=\iota(T)^2$,
\begin{equation}\label{eq:S4-B-convex}
 f''(T)=\frac{800(400-2\kappa_BT)}{(400+\kappa_BT)^4}>0
 \qquad (T_0\le T\le T_1),
\end{equation}
because $\kappa_B<41$ and $T_1<3$. A second application of strict Jensen gives
\begin{equation}\label{eq:S4-B-J}
 \E_Q[J_H]=\sum_{j=0}^1q_j\iota(T_j)^2
 >\iota(1)^2=J_L.
\end{equation}
For orientation, $T_0\simeq0.4258$, $T_1\simeq2.2354$ and $\alpha\simeq0.00227272$. The optimized scores are approximately $2.940065$ and $2.939045$, while mean precisions are $5.16524\times10^{-6}$ and $7.33112\times10^{-6}$, respectively. The strict convexity and concavity arguments establish these signs without relying on numerical rounding. By continuity in the adoption share, the strict precision increase also persists when a sufficiently small positive mass keeps the low policy.

In this example the high-state loading rises, the low-state loading falls, and even the average loading falls because $\iota(T)$ is concave. Its average square nevertheless rises. This demonstrates why neither state-by-state crowd-out nor a decline in average structural precision follows from a lower Model-B log score alone. It establishes no universal sign in the opposite direction and makes no claim about stationary conditional mutual information after retraining.
\endgroup

\clearpage\setcounter{table}{0}\section{Dual price information and the value of computation}\label{sec:S5}

A price can reveal less to an investor who can already perform sophisticated public-data analysis, yet reveal more to someone who cannot. The second investor may learn from the trades of others without repeating their computations. This section makes that distinction precise and asks how it changes the private value of a computational upgrade.

We compare two feasible forecasting capabilities for one negligible investor in the same market. Both use the same current price and private clue; the higher capability can additionally extract information from the public record. We first ask what the same price reveals under each capability, then relate the gain from upgrading to the information it adds beyond price and the private signal. Theorem~\ref{thm:two} shows how adoption by other investors changes this gain and identifies conditions under which it falls. The comparison follows a changing stationary market law, extending the analysis beyond S4's first-auction comparison.

\subsection{One market, two feasible levels of public analysis}\label{subsec:S5-comparison}

Let $x$ denote the adoption share in the investor group whose computational capacity is varied, as in S3, and let $Q_x$ be the resulting stationary market law. We evaluate an investor in economic group $g$, with fixed risk aversion $\gamma_g$ and private-signal noise variance $\nu_g$. Her negligible mass means that her own capability choice changes neither the adoption share nor the clearing price. Changes in $x$ therefore describe adoption by other investors. Maintain the aligned payoff and private clue,
\begin{equation}\label{eq:S5-primitives}
 D_{t+1}=\bar d+\lambda\theta_t+\varepsilon_{t+1},\qquad
 s_{gi,t}=\theta_t+\xi_{gi,t},
 \qquad \lambda,\gamma_g,\nu_g,\sigma_\varepsilon^2>0.
\end{equation}
Here $\V(\xi_{gi,t})=\nu_g$ and $\V(\varepsilon_{t+1})=\sigma_\varepsilon^2$; the independent Gaussian errors have the timing and independence properties of S1.

Following S2.8, let $z_{L,x,t}$ and $z_{H,x,t}$ be the outputs that the two algorithms compute from public records within their respective resource budgets. To describe the information in these outputs, we use a \emph{sigma-algebra}, also called a \emph{sigma-field}: a collection of events containing the full sample space and closed under complements and countable unions. For a real vector $z$, $\sigma(z)$ is the smallest such collection containing $\{z_j\le a\}$ for every coordinate $j$ and real threshold $a$. It represents the information available from observing $z$. With $\mathscr H_t^-$ denoting the full public record available before the auction, we assume
\begin{equation}\label{eq:S5-fields}
 \mathcal F_{L,x,t}=\sigma(z_{L,x,t})
 \subseteq\mathcal U_{H,x,t}=\sigma(z_{H,x,t})
 \subseteq\mathscr H_t^-.
\end{equation}
Thus the higher-capacity algorithm retains all information available from the lower-capacity algorithm. This is an explicit condition on the two outputs; larger resource budgets alone do not guarantee it. Finite capacity refers to the resource budgets used to compute $z_L$ and $z_H$. Neither algorithm uses the current price or the investor's private clue to produce its public-data output. Both add those same observations when forming their final forecasts.

\begin{samepage}
\begin{assumption}[Gaussian comparison of feasible capabilities]\label{ass:S5-T2}
Consider a differentiable family of stationary laws $Q_x$ generated by the market of S1--S3. Under each law, the vector $(\theta_t,D_{t+1},p_t,s_{gi,t},z_{L,x,t},z_{H,x,t})$ is jointly Gaussian. After redundant coordinates are removed, the covariance matrices required for conditioning are nonsingular, and the fundamental residual variances used below are positive and differentiable in $x$. Each algorithm yields the correct Gaussian conditional mean and variance given its public-data output and the same $p_t,s_{gi,t}$, under the same $Q_x$. The primitive payoff and private-signal parameters in \eqref{eq:S5-primitives} are fixed along the comparison.
\end{assumption}
\end{samepage}

The information assigned to the higher-capacity algorithm must be information that the proposed upgrade can actually compute. For example, S2's finite-history benchmark uses
\begin{equation}\label{eq:S5-K3}
 \mathcal F_{L,x,t}=\sigma(m_t),\qquad
 \mathcal U_{H,x,t}=\sigma\!\left(m_t,\{m_{t-j},p_{t-j},D_{t+1-j}\}_{j=1}^{3}\right).
\end{equation}
All entries are available before the auction, including $D_t$ at $j=1$. If the upgrade provides these three periods of history in addition to $m_t$, both its private value and the price information available to its user must be calculated using this same information. Giving the observer an unlimited history would describe a different upgrade.

The law $Q_x$ may arise with frozen incumbent algorithms or with population retraining. In either case, both alternatives for the negligible investor are fitted correctly to that law. Thus incumbent algorithms may remain fixed while the investor's two candidate algorithms are fitted anew to evaluate the upgrade. Assumption~\ref{ass:S5-T2} requires a suitable stationary Gaussian branch; it does not establish its existence. We suppress $g,i,t$ and the subscript $x$ on information sets where their meaning is clear. All information quantities use natural logarithms and are therefore measured in nats.

\subsection{What the price reveals to each observer}\label{subsec:S5-information}

We compare what the same price reveals to an observer using each of the two algorithms. For each capability, measure how much observing price reduces uncertainty about the fundamental after the algorithm has processed the public data:
\begin{equation}\label{eq:S5-SF}
 S(x)=\MI_{Q_x}(\theta;p\mid\mathcal F_L),\qquad
 F(x)=\MI_{Q_x}(\theta;p\mid\mathcal U_H).
\end{equation}
Thus $S$ measures what the current price adds after limited public analysis, and $F$ measures what it adds after advanced feasible public analysis. The observer differs only in computational capability across the two comparisons. Neither conditions on an individual's private clue; that clue enters the trading-value calculation below. In expressions such as $\MI(\theta;\mathcal U_H\mid\mathcal F_L)$, the information set denotes its generating observations: here this means $\MI(\theta;z_H\mid z_L)$.

\begin{lemma}[Two-channel information accounting]\label{lem:S5-chain}
For any two information sets represented by sigma-algebras $\mathcal F\subseteq\mathcal U$, with finite relevant mutual informations,
\begin{equation}\label{eq:S5-chain}
 \boxed{\quad
 \MI(\theta;p\mid\mathcal F)
 =\MI(\theta;p\mid\mathcal U)
 +\underbrace{\MI(\theta;\mathcal U\mid\mathcal F)
 -\MI(\theta;\mathcal U\mid\mathcal F,p)}_{\mathcal T}.
 \quad}
\end{equation}
This identity does not require Gaussianity.
\end{lemma}
\begin{proof}
Expand $\MI(\theta;(p,\mathcal U)\mid\mathcal F)$ first by adding $\mathcal U$ and then $p$, and then in the reverse order. Equating the two expressions and using $\mathcal F\subseteq\mathcal U$ gives \eqref{eq:S5-chain}.
\end{proof}

The term $\mathcal T$ compares the informational advantage of advanced analysis before and after observing price. When $\mathcal T>0$, price reduces that advantage: part of what advanced public analysis could reveal is also conveyed through price. This is the sharing channel in the main text. The term $\mathcal T$ can also be negative. Public analysis and price can instead complement one another, making the advanced analysis more useful after price is observed.

For a simple illustration, let $\theta,u,n$ be mutually independent normal random variables, each with mean zero and variance one, and let $p=\theta+u+n$. An observer with the lower capability has no observations before seeing price; an observer with the higher capability also observes $u$. Formally, $\mathcal F=\{\varnothing,\Omega\}$, where $\Omega$ is the full sample space, and $\mathcal U=\sigma(u)$. The first information set contains no observations, although the observer knows the model and its distributions. Here $S=\tfrac12\log(3/2)$ and $F=\tfrac12\log2$, so $\mathcal T=\tfrac12\log(3/4)<0$. Knowing $u$ alone says nothing about $\theta$, but it lets the observer subtract $u$ from price and use the less noisy signal $p-u=\theta+n$. This illustrates the information identity; the price in this example has not been derived from market clearing. Neither the identity nor greater capacity by itself determines how its terms change as others adopt.

Under Assumption~\ref{ass:S5-T2}, information can be expressed through four conditional variances:
\begin{equation}\label{eq:S5-variances}
 \begin{aligned}
 P_L(x)&=\V_{Q_x}(\theta\mid\mathcal F_L),&
 P_H(x)&=\V_{Q_x}(\theta\mid\mathcal U_H),\\
 L(x)&=\V_{Q_x}(\theta\mid\mathcal F_L,p),&
 H(x)&=\V_{Q_x}(\theta\mid\mathcal U_H,p).
 \end{aligned}
\end{equation}
The $P$'s are uncertainty before seeing the current price; $L$ and $H$ are what remains afterward, but before adding the investor's own signal. Joint Gaussianity makes them deterministic at each $x$, independent of realized conditioning values. Because the higher-capacity algorithm retains all the information in the lower-capacity output, $P_H\le P_L$ and $H\le L$. These are conditional variances, unlike S4's linear-projection residual under a possibly nonlinear encoder.

The entropy of a scalar normal variable of variance $R$ is $\tfrac12\log(2\pi eR)$. Subtracting conditional entropies therefore gives
\begin{equation}\label{eq:S5-entropy}
 S=\frac12\log\frac{P_L}{L},\qquad
 F=\frac12\log\frac{P_H}{H},\qquad
 L=P_Le^{-2S},\quad H=P_He^{-2F}.
\end{equation}
These equalities separate two reasons an investor's remaining uncertainty can change: a change in what she knows before price, and a change in what price adds. Along an endogenous market law, both can matter.

\subsection{The private value of a computational upgrade}\label{subsec:S5-value}

An upgrade is useful to a trader only if it improves decisions beyond what she can already learn from price and her own clue. The upgrade's value comes from the subsequent trading payoff. We evaluate that gain before the auction's public states, price and private signal are realized, when the investor chooses her computational capability. This is the sense in which the valuation is \emph{ex ante}.

To measure this gain, first add the private signal to the information available under each capability. If her fundamental uncertainty after public analysis and price is $R$, Gaussian updating with $s=\theta+\xi$ leaves variance $\nu_gR/(\nu_g+R)$. The payoff forecast variance is consequently
\begin{equation}\label{eq:S5-v}
 v_g(R)=\sigma_\varepsilon^2+\frac{\lambda^2\nu_gR}{\nu_g+R},\qquad
 v_g'(R)=\frac{\lambda^2\nu_g^2}{(\nu_g+R)^2}>0.
\end{equation}
Thus the lower- and higher-capacity algorithms leave payoff forecast variances $v_L=v_g(L)$ and $v_H=v_g(H)$ after all observations. The irreducible payoff shock is retained in both.

We measure the one-period trading payoff relative to cash; adding the same deterministic initial wealth to both policies does not change their difference. Let
\begin{equation}\label{eq:S5-trading-info}
 \mathcal I_L=\mathcal F_L\vee\sigma(p,s),\qquad
 \mathcal I_H=\mathcal U_H\vee\sigma(p,s),\qquad
 r=D-R_fp,
\end{equation}
where $\vee$ means that both sets of observations are available. Correct forecasting and S3's CARA demand formula give $h_k^*=\E[r\mid\mathcal I_k]/(\gamma_gv_k)$ for $k\in\{L,H\}$. Define the ex-ante certainty equivalent of that policy by
\begin{equation}\label{eq:S5-CE-definition}
 \CE_{gk}(Q_x)=-\frac1{\gamma_g}\log\E_{Q_x}
 \!\left[\exp\{-\gamma_gh_k^*r\}\right],\qquad
 \Delta\CE_g(x)=\CE_{gH}(Q_x)-\CE_{gL}(Q_x).
\end{equation}
This averages utility over trading opportunities before converting it to a certain payoff. Averaging the conditional certainty equivalents from S3 would give a different object.

\begin{lemma}[Ex-ante value of feasible public computation]\label{lem:S5-CE}
Under Assumption~\ref{ass:S5-T2},
\begin{equation}\label{eq:S5-CE}
 \boxed{\quad
 \Delta\CE_g(x)=\frac1{2\gamma_g}\log\frac{v_g(L(x))}{v_g(H(x))}
 =\frac1{\gamma_g}\MI_{Q_x}\!\left(D;\mathcal U_H\mid\mathcal F_L,p,s\right).
 \quad}
\end{equation}
The upgrade premium is nonnegative. It is strictly positive if and only if $H<L$, meaning that the upgrade reduces fundamental uncertainty after price has been observed.
\end{lemma}

\begin{proof}
Fix $x$ and write $a_k=\E[r\mid\mathcal I_k]$. Since $p$ is observed under either capability, $\V(r\mid\mathcal I_k)=\V(D\mid\mathcal I_k)=v_k$. Conditional Gaussian integration and the optimal holding imply
\begin{equation}\label{eq:S5-conditional-utility}
 \E[e^{-\gamma_gh_k^*r}\mid\mathcal I_k]
 =\exp\!\left(-\gamma_gh_k^*a_k+
 \frac{\gamma_g^2(h_k^*)^2v_k}{2}\right)
 =\exp\!\left(-\frac{a_k^2}{2v_k}\right).
\end{equation}
Let $\bar r=\E[r]$ and $\tau^2=\V(r)>0$ under this common law. The conditional mean $a_k$ is an affine function of jointly Gaussian observations, so it is Gaussian. Total variance then gives $a_k\sim N(\bar r,\tau^2-v_k)$. For $X\sim N(m,q)$ and $v>0$, completing the square in its density yields
\begin{equation}\label{eq:S5-integral}
 \E[e^{-X^2/(2v)}]
 =\sqrt{\frac{v}{v+q}}\exp\!\left[-\frac{m^2}{2(v+q)}\right].
\end{equation}
The formula also holds at $q=0$ by direct substitution. Applying it with $v=v_k$ and $q=\tau^2-v_k$ gives
\begin{equation}\label{eq:S5-CE-level}
 \CE_{gk}(Q_x)=\frac1{2\gamma_g}\log\frac{\tau^2}{v_k}
 +\frac{\bar r^2}{2\gamma_g\tau^2}.
\end{equation}
The same $\bar r$ and $\tau^2$ enter both policies because the investor takes the same price law as given. Subtraction proves the first equality in \eqref{eq:S5-CE}, even when the unconditional expected excess payoff is nonzero.

For completeness, conditional on public analysis and price, the residual of $\theta$ has variance $R$ and its covariance with the residual of $s$ is $R$. The latter has variance $R+\nu_g$. Gaussian conditioning therefore leaves $R-R^2/(R+\nu_g)=\nu_gR/(R+\nu_g)$; independence of $\varepsilon$ gives \eqref{eq:S5-v}. Finally, \eqref{eq:S5-fields} implies $\mathcal I_L\subseteq\mathcal I_H$. The conditional entropy difference for $D$ is $\tfrac12\log(v_L/v_H)$, which proves the second equality. Since $v_g$ is strictly increasing and $H\le L$, the premium and its strictness follow.
\end{proof}

Equation~\eqref{eq:S5-CE} measures the value of information that the upgrade still adds after both commonly observed price and the investor's private clue have been used. The additional information concerns the asset's payoff, whereas $S$ and $F$ concern the fundamental before adding that private clue. The factor $1/\gamma_g$ converts the payoff information in nats into the investor's certainty-equivalent payoff units. This is a gross one-period benefit before computing costs. A deterministic fee paid before trading would be worth paying up to $\Delta\CE_g/R_f$ in current-date units.

\subsection{How adoption by others changes the incentive to upgrade}\label{subsec:S5-theorem}

Adoption by other investors can either strengthen or weaken the incentive to upgrade. By \eqref{eq:S5-CE}, the upgrade becomes more valuable when the ratio of payoff forecast variances $v_g(L)/v_g(H)$ rises, and less valuable when it falls. Theorem~\ref{thm:two} distinguishes these cases. We then focus on how greater sharing of public analysis through price, together with less information beyond that analysis, can reduce the private gain from upgrading. To express these effects through $S$ and $F$, we must also account for changes in uncertainty before observing price, $P_L$ and $P_H$.

Define the elasticity of payoff risk with respect to remaining fundamental uncertainty by
\begin{equation}\label{eq:S5-weight}
 w_g(R)=\frac{Rv_g'(R)}{v_g(R)}
 =\frac{\lambda^2\nu_g^2R}{(\nu_g+R)^2v_g(R)}>0\qquad(R>0).
\end{equation}
It translates a proportional change in fundamental uncertainty into a proportional change in payoff forecast risk. A prime below denotes differentiation with respect to the adoption share $x$; at a boundary the derivative is one-sided.

\begin{theorem}[Adoption by others and the private value of upgrading]\label{thm:two}
Under Assumption~\ref{ass:S5-T2}, the derivative of the upgrade value with respect to other investors' adoption share satisfies
\begin{equation}\label{eq:S5-bridge}
 \boxed{\quad
 2\gamma_g\Delta\CE_g'
 =w_g(L)\!\left(\frac{P_L'}{P_L}-2S'\right)
 -w_g(H)\!\left(\frac{P_H'}{P_H}-2F'\right).
 \quad}
\end{equation}
In particular, $P_L'=0$, $P_H'\ge0$, $S'>0$ and $F'<0$ imply $\Delta\CE_g'<0$. Conditional on $P_L'=0$, the sharper necessary-and-sufficient criterion is
\begin{equation}\label{eq:S5-sharp}
 \boxed{\quad
 \Delta\CE_g'<0\quad\Longleftrightarrow\quad
 \frac{P_H'}{P_H}>2F'-2\frac{w_g(L)}{w_g(H)}S'.
 \quad}
\end{equation}
More generally, $L'<0$ and $H'\ge0$ suffice, whether or not the two price-information derivatives have been separately signed.
\end{theorem}

\begin{proof}
Differentiate the first equality in \eqref{eq:S5-CE}:
\begin{equation}\label{eq:S5-CE-derivative}
 2\gamma_g\Delta\CE_g'
 =\frac{v_g'(L)}{v_g(L)}L'
 -\frac{v_g'(H)}{v_g(H)}H'
 =w_g(L)\frac{L'}{L}-w_g(H)\frac{H'}{H}.
\end{equation}
Logarithmic differentiation of \eqref{eq:S5-entropy} yields $L'/L=P_L'/P_L-2S'$ and $H'/H=P_H'/P_H-2F'$, proving \eqref{eq:S5-bridge}. Under the stated sufficient conditions, the first bracket is strictly negative and the second strictly positive, and both weights are positive. Setting $P_L'=0$ and rearranging the strict inequality gives \eqref{eq:S5-sharp}. The last conclusion follows directly from \eqref{eq:S5-CE-derivative} and $v_g'>0$.
\end{proof}

The derivative formula covers both directions. Under $P_L'=0$, reversing the strict inequality on the right of \eqref{eq:S5-sharp} gives $\Delta\CE_g'>0$; equality gives $\Delta\CE_g'=0$. For example, if adoption raises $L$ while leaving $H$ unchanged, the lower-capacity forecast deteriorates while the higher-capacity forecast does not, so the upgrade becomes more valuable. This is a sufficient pattern of variance changes; whether it occurs must be established from the market law.

The sufficient conditions emphasized in Theorem~\ref{thm:two} describe the opposite direction. In \eqref{eq:S5-bridge}, the two brackets are the proportional changes in uncertainty after price under the lower and higher capabilities. With unchanged $P_L$, an increase in $S$ reduces $L$: the investor can learn more from price without upgrading. With unchanged or increasing $P_H$, a decrease in $F$ raises $H$. Both changes reduce the gain from upgrading. The sharper criterion also allows $P_H$ to fall, provided that this improvement does not offset these effects. No restriction on the number of public predictors appears in the formula; what matters is their effect on these residual variances.

Table~\ref{tab:S5-illustration} illustrates the mechanism with $\lambda=\nu_g=\gamma_g=1$, $\sigma_\varepsilon^2=0.1$, and unchanged $P_L=2$, $P_H=1$. The entries are illustrative conditional variances and the values implied by the identities, not estimated market equilibria. As $L$ falls and $H$ rises, the upgrade remains useful but its gross premium declines from $0.1627$ to $0.0534$ payoff units.

\begin{table}[H]\centering\small
\caption{A smaller information advantage yields a smaller private upgrade premium.}\label{tab:S5-illustration}
\begin{tabularx}{\linewidth}{@{}X cc@{}}\toprule
 & \textbf{Initial values} & \textbf{Changed values}\\\midrule
Low-capacity uncertainty after price, $L$ & $1$ & $3/4$\\
High-capacity uncertainty after price, $H$ & $1/2$ & $3/5$\\
Price information after limited analysis, $S$ (nats) & $0.3466$ & $0.4904$\\
Price information after advanced analysis, $F$ (nats) & $0.3466$ & $0.2554$\\
Gross upgrade premium, $\Delta\CE_g$ (payoff units) & $0.1627$ & $0.0534$\\\bottomrule
\end{tabularx}\end{table}

Holding the upgrade cost fixed, a falling premium means that others' adoption weakens the investor's incentive to upgrade; a rising premium means that it strengthens that incentive. These are, respectively, \emph{strategic substitution} and \emph{strategic complementarity} in the private benefit of adoption. They describe a force that restrains or reinforces investment in computation. Whether an arms race develops also depends on computing costs, the available upgrades and how investors adjust their choices. S6 establishes the declining-premium case locally along a family of competitive equilibria; the conditions here also identify when the opposite sign would occur.

\subsection{Checking the conditions in the competitive price}\label{subsec:S5-market}

The preceding conditions can be evaluated from the covariance law of the actual clearing price. On a stationary Gaussian branch of S3's market, write the centered price as
\begin{equation}\label{eq:S5-price}
 p_t=\frac{A(x)\theta_t+C(x)^\top Z_t-\zeta_t}{\Delta(x)},\qquad
 \Delta(x)>0,\qquad J(x)=\frac{A(x)^2}{\sigma_\zeta^2},
\end{equation}
where $Z_t$ collects the public predictors appearing in aggregate demand. These coefficients come from investors' schedules and competitive clearing. Suppose the higher-capacity output contains enough information to reconstruct $Z_t$ (formally, $Z_t$ is measurable with respect to $\mathcal U_{H,x,t}$), and current supply noise is independent of $(\theta_t,\mathcal U_{H,x,t})$, as in S1. The high-capacity observer can then subtract the public component. Gaussian updating gives
\begin{equation}\label{eq:S5-high-price}
 F=\frac12\log(1+JP_H),\qquad
 H^{-1}=P_H^{-1}+J.
\end{equation}
The first expression is the information supplied by price; the second adds its likelihood precision to prior precision. In particular, $F'<0$ requires $(JP_H)'<0$, not merely $J'<0$. If the public component cannot be reconstructed from the information retained by the higher-capacity algorithm, \eqref{eq:S5-high-price} is unavailable, while Theorem~\ref{thm:two} still applies using the directly computed $H$ and $F$.

For the low-capacity observer define $c_L=\Cov(\theta,p\mid\mathcal F_L)$ and $u_L=\V(p\mid\mathcal F_L)>0$. Gaussian conditioning gives
\begin{equation}\label{eq:S5-low-price}
 L=P_L-\frac{c_L^2}{u_L},\qquad
 S=-\frac12\log\!\left(1-\frac{c_L^2}{P_Lu_L}\right).
\end{equation}
The ratio $c_L^2/u_L$ is how much observing price reduces her fundamental uncertainty. These covariance expressions allow the public part of price to help an observer who cannot compute it independently.

\begin{samepage}
\begin{corollary}[Sufficient conditions from the stationary clearing law]\label{cor:S5-market}
Maintain Assumption~\ref{ass:S5-T2}, \eqref{eq:S5-price}, and the measurability and independence conditions above. At an adoption share with $P_L'=0$, $A\ne0$ and $c_L\ne0$, all three signs $S'>0$, $F'<0$ and $\Delta\CE_g'<0$ follow from
\begin{equation}\label{eq:S5-market-tests}
 \frac{d}{dx}\!\left(\frac{c_L^2}{u_L}\right)>0,\qquad
 \frac{d}{dx}(JP_H)<0,\qquad
 \frac{d}{dx}(P_H^{-1}+J)\le0.
\end{equation}
The last condition is sufficient but not necessary for a falling premium; \eqref{eq:S5-sharp} gives the sharper criterion.
\end{corollary}
\end{samepage}
\begin{proof}
The first condition and $P_L'=0$ give $L'<0$ and $S'>0$ by \eqref{eq:S5-low-price}. The second gives $F'<0$ by \eqref{eq:S5-high-price}. If $G=P_H^{-1}+J=H^{-1}>0$, the third gives $H'=-G'/G^2\ge0$. Apply the last conclusion of Theorem~\ref{thm:two}.
\end{proof}

This test also shows the limit of importing S4's result into a stationary comparison. S4 orders $J$ at first deployment under a common inherited training law. Here adoption can change the pre-price information and the law used to fit policies. Even with fixed Model-A coefficients, a change in the distribution of public histories can change $P_H$ and hence $F$, while $J$ stays fixed. Retraining can change both.

\subsection{Economic interpretation and the next step}\label{subsec:S5-conclusion}

The private premium in \eqref{eq:S5-CE} compares two choices by the same investor under one prevailing law. Its derivative asks how that advantage changes as others adopt advanced computation and the market changes. A falling premium does not say that the high-capacity investor's total utility falls, or that the low-capacity investor's total utility rises: the common terms in \eqref{eq:S5-CE-level} can also change with $x$.

Information can nevertheless have a direct benefit for a specified recipient. For example, a passive observer who chooses an estimate $a$ to minimize expected squared error $(a-\theta)^2$, using $\mathcal F_L$ and price, chooses $a=\E[\theta\mid\mathcal F_L,p]$ and has minimum expected loss $L$. Thus $L'<0$ benefits this particular decision maker. It is not an aggregate-welfare comparison, which would also require the costs of computation and the other participants' objectives and payoffs.

The information accounting identity is general; the conversion to an ex-ante trading premium and Theorem~\ref{thm:two} require correct forecasts under a jointly Gaussian law, the inclusion of feasible public information in \eqref{eq:S5-fields}, and the same market law for both capabilities. The theorem supplies an exact relationship and conditions for its sign. S6 takes the next step: it derives a branch of competitive equilibria from economic primitives and shows that more price information for the limited observer, less information beyond advanced analysis, and a lower private upgrade premium coexist locally, under both population retraining and frozen incumbent policies.

\clearpage\setcounter{table}{0}\section{Coexistence in competitive equilibrium}\label{sec:S6}

S5 identifies conditions under which adoption by other investors lowers the private value of a computational upgrade. We now show that optimal forecasting, trading and market clearing can generate those conditions. In the equilibrium we construct, adoption makes price more informative to an investor using limited public analysis, less informative to one who can already perform advanced analysis, and reduces the gain from upgrading.

For an analytic proof, we specialize S1--S3 to a fundamental with two independent components. One is already public but usable only by the higher-capacity algorithm; the other is inferred from private clues and price. We use this economy throughout S6. Theorem~\ref{thm:three} shows that all three effects coexist for small adoption shares, with and without population retraining, and persist under small changes in the economic parameters.

The numerical analysis tests these mechanisms in richer settings. Main-text Fig.~3 shows all three effects across the reported adoption range in an eight-state dynamic economy, where algorithms use public histories containing past prices, under both frozen and retrained policies. Across 180 parameter settings, the population-equilibrium calculations converge in 147; all three effects coexist in those 147 cases (Fig.~\ref{fig:S7-mechanisms-summary}C). Separate finite-data experiments with linear state-space and gated recurrent unit (GRU) models find improved forecasts and weaker discovery on average (main-text Fig.~2B,C). Thus the numerical evidence extends beyond the economy used for the proof.

\subsection{Two computational capabilities in a two-state economy}\label{subsec:S6-economy}

We make the primitive shocks independent across dates so that forecasts and clearing prices can be derived without solving for changing price histories. The parameter values below simplify the algebra; the result persists under small changes in those values. All variables are centered, and mean net supply is zero. At each date,
\begin{equation}\label{eq:S6-primitives}
 \theta_t=m_t+e_t,\qquad D_{t+1}=\theta_t+\varepsilon_{t+1},\qquad
 s_{i,t}=\theta_t+\xi_{i,t},\qquad S_t=\zeta_t.
\end{equation}
The variables $m_t,e_t,\xi_{i,t}$ and $\zeta_t$ are mutually independent normal variables with mean zero and variance one. The independent payoff shock $\varepsilon_{t+1}$ is normal with mean zero and variance $1/10$. As in S1, individual signal errors aggregate to zero in each investor type. All investors have CARA risk aversion $\gamma=1$, and the gross risk-free return is $R_f=1.05$.

There are two investor groups, each of mass $1/2$. A fraction $x$ of group 1 uses the higher computational capability; everyone else uses the lower capability. Thus the total mass using the higher capability is $q=x/2$. The groups have the same risk preferences and private-signal precision, so varying $x$ changes only the computational capabilities in use.

The public record contains $m_t$ before the auction. The lower-capacity algorithm does not process it, whereas the higher-capacity algorithm can retain it as an input. In the notation defined in S5,
\begin{equation}\label{eq:S6-information-sets}
 z_{L,t}=0,\qquad z_{H,t}=m_t,\qquad
 \mathcal F_{L,t}=\{\varnothing,\Omega\}\subseteq
 \mathcal U_{H,t}=\sigma(m_t).
\end{equation}
The lower capability therefore supplies no public observation to the forecast. The higher capability can use $m_t$ or ignore it, so its feasible forecast class contains the lower class. Both algorithms use the same private clue and current price. This is a difference in the ability to process already public data, rather than a difference in access to private data.

Each algorithm has an unrestricted affine forecasting head and a positive constant variance, as in Model A. Whenever an algorithm is fitted to a market law, it attains the correct Gaussian conditional mean and variance given its inputs. Such a fit also minimizes the population Gaussian forecast loss. Historical observations contain no additional predictive information in this independent-date example, and neither algorithm uses a history state. For the derivations below, suppress the date and individual subscripts.

\subsection{Optimal forecasts and the clearing price}\label{subsec:S6-clearing}

We now derive optimal forecasts and the clearing price in the economy defined in S6.1. With a linear forecast, individual demand is affine in the private clue, the available public input and the proposed current price. After private errors aggregate out, the numerator of the clearing price is therefore a linear combination of $\theta$, $m$ and supply noise. Write its candidate form as
\begin{equation}\label{eq:S6-price}
 p=\frac{Ae+Bm-\zeta}{\Delta},\qquad
 y\equiv\Delta p=Ae+Bm-\zeta,\qquad \Delta>0.
\end{equation}
Here $y$ is simply price multiplied by a positive constant. Observing $p$ or $y$ provides the same information. The coefficients $A,B,\Delta$ will be determined by optimal forecasts and clearing; they are not imposed price weights. In S3's notation the numerator is $A\theta+(B-A)m-\zeta$. Because supply noise has coefficient $-1$ in $y$, $A$ and $B$ measure each fundamental component's weight relative to that noise.

First fix a candidate pair $(A,B)$. For an investor who cannot use $m$, define
\begin{equation}\label{eq:S6-L}
 W\equiv\V(y)=A^2+B^2+1,\qquad
 U\equiv\Cov(\theta,y)=A+B,\qquad
 L\equiv\V(\theta\mid y)=2-\frac{U^2}{W}.
\end{equation}
The conditional mean before the private clue is $(U/W)y$. Updating with $s=\theta+\xi$, whose independent noise has variance one, gives
\begin{equation}\label{eq:S6-low-forecast}
 \mu_L=\beta_Ls+\psi_Ly,\qquad
 \beta_L=\frac{L}{1+L},\quad
 \psi_L=\frac{U/W}{1+L},\quad
 v_L=\frac1{10}+\frac{L}{1+L}.
\end{equation}
The irreducible payoff noise contributes $1/10$ to the forecast variance.

The investor who can use $m$ can subtract $Bm$ from $y$ and $m$ from $s$. The resulting observations are $Ae-\zeta$ and $e+\xi$. Combining these with the unit-variance prior for $e$ yields
\begin{equation}\label{eq:S6-high-update}
 \E[e\mid m,y,s]=\frac{s-m+A(y-Bm)}{2+A^2},\qquad
 \V(e\mid m,y,s)=\frac1{2+A^2}.
\end{equation}
Consequently,
\begin{equation}\label{eq:S6-high-forecast}
 \begin{aligned}
 \mu_H&=\beta_Hs+\psi_Hy+d_Hm,\qquad
 v_H=\frac1{10}+\frac1{2+A^2},\\
 \beta_H&=\frac1{2+A^2},\qquad
 \psi_H=\frac{A}{2+A^2},\qquad
 d_H=1-\frac{1+AB}{2+A^2}.
 \end{aligned}
\end{equation}
These are conditional payoff forecasts. In particular, $1/(2+A^2)$ includes the private clue and differs from the uncertainty $H$ defined in S5 before that clue is added.

Under either capability, the CARA demand formula from S3 is
\begin{equation}\label{eq:S6-demand}
 h_i(p)=\frac{\mu_i(p)-R_fp}{v_i}.
\end{equation}
To aggregate these demands, define
\begin{equation}\label{eq:S6-aggregates}
 \begin{aligned}
 \bar A&=(1-q)\frac{\beta_L}{v_L}+q\frac{\beta_H}{v_H},&
 D_m&=q\frac{d_H}{v_H},\\
 W_d&=\frac{1-q}{v_L}+\frac{q}{v_H},&
 \Psi&=(1-q)\frac{\psi_L}{v_L}+q\frac{\psi_H}{v_H}.
 \end{aligned}
\end{equation}
Because the average private clue is $\theta$, aggregate demand is
$\bar A\theta+D_mm+(\Delta\Psi-R_fW_d)p$.
Equating this to $\zeta$ and matching \eqref{eq:S6-price} gives the equilibrium restrictions
\begin{equation}\label{eq:S6-equilibrium}
 \boxed{\quad
 A=(1-q)\frac{\beta_L}{v_L}+q\frac{\beta_H}{v_H},\qquad
 B=A+q\frac{d_H}{v_H},\qquad
 \Delta=\frac{R_fW_d}{1+\Psi},\quad q=\frac{x}{2}.
 \quad}
\end{equation}
The expression for $\Delta$ follows from $\Delta=R_fW_d-\Delta\Psi$. Thus the actual slope of aggregate demand with respect to $p$ is $-\Delta$. A solution with $\Delta>0$ gives a unique clearing price at every realized state.

Equations~\eqref{eq:S6-low-forecast}--\eqref{eq:S6-equilibrium} express the consistency requirement under population retraining: forecasts must be correct for the price law that their own trading generates. The first two restrictions determine $(A,B)$; the third recovers the scale of price. In particular, multiplying price by $\Delta$ has not removed a clearing condition.

\subsection{How the retrained equilibrium responds to adoption}\label{subsec:S6-retrained}

At $x=0$, every investor uses the lower capability. The second restriction in \eqref{eq:S6-equilibrium} then gives $B=A$. Write their common value as $\alpha$. Substituting $L=2/(2\alpha^2+1)$ into the first restriction gives
\begin{equation}\label{eq:S6-baseline}
 \alpha=\frac{20}{2\alpha^2+23},\qquad
 2\alpha^3+23\alpha-20=0.
\end{equation}
The polynomial is strictly increasing, is negative at zero and positive at one. It therefore has exactly one root $\alpha\in(0,1)$, numerically $\alpha\simeq0.821378$. At this root, $v_L>0$, $\psi_L>0$ and
$\Delta_0=R_f/(v_L+\psi_L)>0$. This establishes the all-low equilibrium. The higher-capacity forecast can also be fitted under this law even though no positive mass uses it yet.

To determine what happens when a small share upgrades, write
$g_L(A,B)=\beta_L/v_L$, $g_H(A)=\beta_H/v_H$ and
$b_H(A,B)=d_H/v_H$. The two equations for price weights are
\begin{equation}\label{eq:S6-E}
 E_1=A-(1-x/2)g_L-(x/2)g_H=0,\qquad
 E_2=B-A-(x/2)b_H=0.
\end{equation}
These functions are smooth near $(A,B,x)=(\alpha,\alpha,0)$. Since
$g_L=10L/(1+11L)$ and
$\partial_A L=\partial_B L=-4\alpha/(2\alpha^2+1)^2$ at that point, both partial derivatives of $g_L$ equal
\begin{equation}\label{eq:S6-k}
 k\equiv-\frac{40\alpha}{(2\alpha^2+23)^2}<0.
\end{equation}
The Jacobian with respect to the two price weights is therefore
\begin{equation}\label{eq:S6-Jacobian}
 \frac{\partial(E_1,E_2)}{\partial(A,B)}
 =\begin{pmatrix}1-k&-k\\-1&1\end{pmatrix},\qquad
 \det\frac{\partial(E_1,E_2)}{\partial(A,B)}=1-2k>0.
\end{equation}
The implicit function theorem now gives a locally unique continuously differentiable pair $A(x),B(x)$ satisfying both clearing restrictions. Recovering $\Delta(x)$ from \eqref{eq:S6-equilibrium} gives a positive scale for $x$ sufficiently close to zero. The primitive shocks have the same distribution at every date, so these forecasts and prices generate a stationary Gaussian market law. This establishes equilibrium existence locally; it does not require an iterative training algorithm to converge.

A prime in the remaining derivations denotes the right derivative with respect to $x$ at zero. Define the two effects of changing population weights while holding the baseline forecasts fixed:
\begin{equation}\label{eq:S6-direct}
 \delta\equiv\frac12(g_H-g_L)\big|_0
 =-\frac{\alpha}{4(\alpha^2+12)}<0,\qquad
 d\equiv\frac12 b_H\big|_0=\frac5{\alpha^2+12}>0.
\end{equation}
For these simplifications, use $g_L|_0=\alpha$ and
$g_H|_0=b_H|_0=10/(\alpha^2+12)$, together with \eqref{eq:S6-baseline}. Differentiating \eqref{eq:S6-E} gives
\begin{equation}\label{eq:S6-linear-system}
 (1-k)A'-kB'=\delta,\qquad B'-A'=d.
\end{equation}
Solving,
\begin{equation}\label{eq:S6-retrained-slopes}
 \boxed{\quad
 A'=\frac{\delta+kd}{1-2k}<0,\qquad
 B'=A'+d,\qquad
 A'+B'=\frac{5-\alpha/2}{(\alpha^2+12)(1-2k)}>0.
 \quad}
\end{equation}
The last inequality uses $\alpha<1$. It also implies $B'>0$ because $A'<0$.

The weight on $e$ relative to supply noise falls, while the corresponding weight on the public component $m$ rises. Their combined weight $A+B$ also rises. The equation $A'=\delta+k(A'+B')$ separates two reasons for the first change. Replacing some lower-capacity algorithms directly reduces the aggregate private-signal response by $\delta<0$. The improved information in price also leads the remaining lower-capacity investors to put less weight on their own clues: $k(A'+B')<0$. At the all-low baseline, these investors are the only positive-mass users whose refitting contributes at first order.

\subsection{What each observer learns, and the value of upgrading}\label{subsec:S6-effects}

The same changes in price have different implications for the two observers. An investor who already knows $m$ subtracts $Bm$ from $y$, leaving $Ae-\zeta$. Because $A'<0$, this remaining signal becomes less informative. An investor who cannot use $m$ observes the whole price; its stronger response to the public component can improve her forecast of $\theta=m+e$.

Precisely, the uncertainties before price are constant: $P_L=2$ and $P_H=1$. After price, but before the private clue,
\begin{equation}\label{eq:S6-information}
 H=\frac1{1+A^2},\qquad
 F=\MI(\theta;p\mid m)=\frac12\log(1+A^2),\qquad
 S=\MI(\theta;p)=\frac12\log\frac2L.
\end{equation}
At $A=B=\alpha$, differentiating \eqref{eq:S6-L} and \eqref{eq:S6-information} yields
\begin{equation}\label{eq:S6-information-slopes}
 \begin{aligned}
 L'&=-\frac{4\alpha(A'+B')}{(2\alpha^2+1)^2}<0,&
 H'&=-\frac{2\alpha A'}{(1+\alpha^2)^2}>0,\\
 S'&=-\frac{L'}{2L}>0,&
 F'&=\frac{\alpha A'}{1+\alpha^2}<0.
 \end{aligned}
\end{equation}
For the limited observer, the increase in $A+B$ dominates the accompanying change in price variance at this equilibrium. The sign of $S'$ is therefore derived from the full conditional variance, not from the increase in $B$ alone.

For a negligible investor in group 1 choosing between the two capabilities, both alternatives are fitted to this same price law. The S5 certainty-equivalent gain is
\begin{equation}\label{eq:S6-premium}
 \Delta\CE(x)=\frac12\log\frac{v_L(x)}{v_H(x)}
 =\frac12\log\frac{\frac1{10}+\frac{L(x)}{1+L(x)}}
 {\frac1{10}+\frac1{2+A(x)^2}}.
\end{equation}
It is evaluated before the observations are realized and is a gross benefit before computation costs. Since $v(R)=1/10+R/(1+R)$ increases in $R$, \eqref{eq:S6-information-slopes} makes the lower-capacity payoff forecast more precise and the higher-capacity forecast less precise. Hence
\begin{equation}\label{eq:S6-premium-slope}
 \Delta\CE'
 =\frac12\left[\frac{L'}{(1+L)^2v_L}
 -\frac{H'}{(1+H)^2v_H}\right]<0.
\end{equation}
This also verifies the sufficient conditions in Theorem 2 directly: both pre-price variances are unchanged, while $S'>0$ and $F'<0$.

\subsection{The same signs with frozen trading algorithms}\label{subsec:S6-frozen}

The three effects do not require the incumbent algorithms to be retrained. Fit both capabilities under the all-low law just constructed, and denote those coefficients by a subscript $0$. Keep their entire forecasting rules fixed as $x$ rises. In particular, the coefficient on the actual price is fixed at $\chi_{k0}=\Delta_0\psi_{k0}$. The deployed forecasts are
\begin{equation}\label{eq:S6-frozen-heads}
 \mu_L^F=\beta_{L0}s+\chi_{L0}p,\qquad
 \mu_H^F=\beta_{H0}s+\chi_{H0}p+d_{H0}m,
 \qquad v_k^F=v_{k0}.
\end{equation}
Freezing $\chi_{k0}$ matters: keeping only the coefficient on a newly rescaled price fixed would change the underlying trading rule.

Aggregate these demands directly, using the fixed forecast variances. Define
\begin{equation}\label{eq:S6-frozen-price}
 \begin{aligned}
 A_F(x)&=(1-x/2)\frac{\beta_{L0}}{v_{L0}}
              +(x/2)\frac{\beta_{H0}}{v_{H0}},\\
 B_F(x)&=A_F(x)+(x/2)\frac{d_{H0}}{v_{H0}},\\
 \Delta_F(x)&=(1-x/2)\frac{R_f-\chi_{L0}}{v_{L0}}
                   +(x/2)\frac{R_f-\chi_{H0}}{v_{H0}},\\
 p_x^F&=\frac{A_F(x)e+B_F(x)m-\zeta}{\Delta_F(x)}.
 \end{aligned}
\end{equation}
At zero, $A_F=B_F=\alpha$ and $\Delta_F=\Delta_0>0$. Continuity gives a unique competitive clearing price for small positive $x$. The rules maximize utility under their fixed fitted beliefs; after adoption, those beliefs need not be correct under the new law. No claim of population refitting is made for these incumbent policies.

The price weights in \eqref{eq:S6-frozen-price} are affine in $x$, so their derivatives follow immediately:
\begin{equation}\label{eq:S6-frozen-slopes}
 \boxed{\quad
 A_F'=\delta=-\frac{\alpha}{4(\alpha^2+12)}<0,\qquad
 B_F'=\delta+d,\qquad
 A_F'+B_F'=\frac{5-\alpha/2}{\alpha^2+12}>0.
 \quad}
\end{equation}
Thus the sign calculations in \eqref{eq:S6-information-slopes} apply to the frozen-policy price as well. The lower-capacity observer learns more from price, and the observer who can use $m$ learns less beyond that public input.

For the upgrade-value comparison, S5 fits both alternatives for a negligible investor to the \emph{current} law of $p_x^F$. Her forecasts therefore use the current $A_F,B_F$ in \eqref{eq:S6-low-forecast} and \eqref{eq:S6-high-forecast}; they need not equal the incumbents' frozen forecasts. The private premium is \eqref{eq:S6-premium} evaluated at $A_F,B_F$, and its derivative is negative by the same argument. This compares two feasible choices for the investor without changing anyone else's deployed rule.

Table~\ref{tab:S6-local-values} reports the baseline and the derivatives derived above. All derivatives are per unit of $x$, the adoption share \emph{within group 1}, whose mass is one half. Both protocols start from the same market. Retraining changes the magnitudes, but the three signs agree.

\begin{table}[H]\centering\small
\caption{The three effects at the all-low equilibrium. Derivatives are evaluated at $x=0$.}\label{tab:S6-local-values}
\begin{tabularx}{\linewidth}{@{}X r r r@{}}\toprule
 & \textbf{Baseline} & \textbf{Retrained derivative} & \textbf{Frozen derivative}\\\midrule
Weight on $e$ in $y$, $A$ & $0.821378$ & $-0.034264$ & $-0.016201$\\
Weight on $m$ in $y$, $B$ & $0.821378$ & $0.360224$ & $0.378287$\\
Price information given $m$, $F$ (nats) & $0.257806$ & $-0.016806$ & $-0.007946$\\
Price information without $m$, $S$ (nats) & $0.427064$ & $0.113963$ & $0.126593$\\
Upgrade premium, $\Delta\CE$ (payoff units) & $0.083351$ & $-0.058864$ & $-0.060092$\\\bottomrule
\end{tabularx}\end{table}

Because the primitive states are independent across dates and the algorithms use no history state, the first-deployment price law is already stationary when the fitted rules are frozen. There is no separate adjustment of the input history in this example. S3 allows that additional adjustment in markets whose algorithms use past prices; the present construction isolates the effects of adoption and refitting.

\begin{samepage}
\subsection{The local result and its scope}\label{subsec:S6-theorem}

The following statement collects the implications of the construction. A local result concerns adoption shares sufficiently close to zero; it does not require the signs to hold throughout $x\in[0,1]$.

\begin{theorem}[Coexistence with and without population retraining]\label{thm:three}
In the economy of S6.1, the all-low equilibrium has the unique positive normalized price weights $A(0)=B(0)=\alpha\in(0,1)$ satisfying \eqref{eq:S6-baseline}. There is a locally unique continuously differentiable family $(A(x),B(x),\Delta(x))$ of competitive equilibria with correctly fitted Gaussian forecasts, for $x\ge0$ sufficiently close to zero. At zero,
\begin{equation}\label{eq:S6-three-signs}
 \boxed{\qquad
 \left.\frac{dF}{dx}\right|_0<0,\qquad
 \left.\frac{dS}{dx}\right|_0>0,\qquad
 \left.\frac{d\Delta\CE}{dx}\right|_0<0.
 \qquad}
\end{equation}
The same three strict signs hold for the stationary competitive price generated by the frozen forecasts in \eqref{eq:S6-frozen-heads}, when both alternatives in the negligible investor's value comparison are fitted to the resulting law. For each protocol, the signs hold on a sufficiently small adoption interval and persist in an open neighborhood of the stated positive economic parameters, within the same Gaussian model and information structure.
\end{theorem}
\end{samepage}

\begin{proof}
Equation~\eqref{eq:S6-baseline} establishes the baseline. The nonsingular Jacobian in \eqref{eq:S6-Jacobian} gives local existence and uniqueness of the retrained price coefficients; the positive clearing slope gives a unique auction price. Equations~\eqref{eq:S6-retrained-slopes}, \eqref{eq:S6-information-slopes} and \eqref{eq:S6-premium-slope} prove the three strict signs. Equations~\eqref{eq:S6-frozen-price} and \eqref{eq:S6-frozen-slopes} establish the frozen-policy counterpart.

For robustness, allow small changes in the positive state, signal, payoff and supply variances, risk aversions, payoff loading, risk-free return and group masses, while retaining the two feasible public inputs and the independent Gaussian timing. Gaussian regression coefficients depend smoothly on these parameters because the conditioning covariance matrices are nonsingular. If group parameters differ, their separate population weights enter the demand aggregates, as in S3. The same two price weights still determine their Gaussian forecasts. Consequently, the two clearing equations and their derivatives are smooth in the parameters as well as $(A,B,x)$. The nonzero Jacobian determinant at the stated economy permits the implicit function theorem with these parameters included. The clearing slope and all variances remain positive nearby, and the three derivatives vary continuously. Their strict signs therefore hold for a small parameter neighborhood and adoption interval. Baseline-fitted frozen coefficients and their clearing prices have the same continuity property, establishing the corresponding conclusion for that protocol.
\end{proof}

Correct population refitting therefore need not eliminate the effects. Convergence of a particular training procedure, comparative statics over the full adoption range, and the increasing-premium equilibrium case discussed in S5.4 require separate analysis.

The welfare interpretation remains the one in S5. A passive observer estimating $\theta$ from price has a lower minimum squared-error loss because $L$ falls. The private gain from a computational upgrade also falls. Neither statement by itself ranks total trader utility or aggregate welfare, which would require the other payoff components and computing costs. These distinctions allow the same price to share more useful public analysis while revealing less beyond what an advanced investor can already compute.

\clearpage\setcounter{table}{0}\setcounter{figure}{0}\setcounter{section}{6}
\renewcommand{\thetable}{S7.\arabic{table}}\renewcommand{\theHtable}{S.7.\arabic{table}}
\renewcommand{\thefigure}{S7.\arabic{figure}}\renewcommand{\theHfigure}{S.7.\arabic{figure}}
\section{Population Gaussian markets: information, learning, and computation}\label{sec:S7}

A stronger forecasting algorithm can improve an investor's predictions without improving what the market learns from investors' private observations. At the same time, prices can become more useful to someone who cannot perform the same public-data analysis. The preceding sections explain why these effects are possible. Here we calculate their magnitudes in a dynamic market where investors forecast payoffs, choose their holdings, and jointly determine prices.

Our starting point is the eight-state economy used in the numerical results of the main text. Economic conditions evolve over time, two noisy macroeconomic indicators are announced before each auction, and each investor receives a private clue about the fundamental that determines the next payout. We vary one group's ability to use earlier prices, payouts, and macroeconomic releases in its forecast. The economic shocks, private-signal quality, and investor preferences remain unchanged. Because the economy and forecasting rules in this section are linear and Gaussian, we can calculate the market's stationary distribution and the optimal forecasting coefficients directly from population covariances. The results therefore do not depend on finite training samples or neural-network optimization.

We examine the initial change in trading, the new price distribution generated by fixed algorithms, and the outcome after the algorithms have been fitted again to that distribution. We then measure how much the resulting price tells two observers with different public-data processing capabilities, and how this affects the incentive to acquire additional computation. Finally, we distinguish changes in forecasts of expected payoffs from changes in forecast risk, and examine the role of investors who did not upgrade.

\subsection{The market and the two forecasting capabilities}\label{subsec:S7-environment}

We use the economic environment and within-date timing of S1--S3. A latent state $f_t\in\mathbb R^8$ follows a stationary Gaussian transition, and two public macroeconomic statistics reveal noisy functions of that state. The scalar $\theta_t=a^\top f_t$ is the fundamental about which investors receive private clues. The asset pays $D_{t+1}$ at the following date:
\begin{align}\label{eq:S7-economy}
 f_{t+1}&=Ff_t+G\eta_{t+1},&m_t&=Hf_t+\omega_t,\\
 \theta_t&=a^\top f_t,&D_{t+1}&=1.15\theta_t+\varepsilon_{t+1},\nonumber\\
 s_{ci,t}&=\theta_t+\xi_{ci,t},&S_t&=\zeta_t.\nonumber
\end{align}
All innovations in \eqref{eq:S7-economy} are Gaussian and mutually independent across their specified primitive streams. As in S1, the errors in individual signals integrate to zero within each continuum of investors. The supply disturbance $\zeta_t$ is not known before the auction. The claim's payout is aligned with the privately signaled fundamental, while the eight latent factors and their persistence make public-history forecasting nontrivial.

There are two investor groups. Group 1 has mass $0.55$ and group 2 has mass $0.45$; their risk-aversion coefficients and private-signal precisions differ. We measure computational capacity by the number of previous public observations the forecasting algorithm can process. At date $t$ all investors can use the current released vector $m_t$. A low-capacity forecast uses that vector alone, whereas a high-capacity forecast also uses the three preceding public records. Let
\begin{equation}\label{eq:S7-history}
 o_t=(m_t,p_{t-1},D_t),\qquad
 z_{k,t}=\big(m_t,\{m_{t-j},p_{t-j},D_{t+1-j}\}_{j=1}^{k}\big),\qquad k\in\{0,3\}.
\end{equation}
For $k=0$, $z_{0,t}=m_t$ has two entries. For $k=3$, the fourteen entries include three past prices, three already paid dividends, and six past macroeconomic observations. The current auction price $p_t$ is not in this vector; it enters the individual forecast later as an argument of the demand schedule. Thus, even at $k=3$, there is no use of the current auction's outcome before it has been determined.

\begin{table}[htbp]\centering\small
\caption{The eight-state population economy. The full state transition is constructed once from the fixed generator described in S7.9; no primitive is changed when computational adoption changes.}\label{tab:S7-primitives}
\begin{tabular}{@{}p{.50\linewidth}p{.39\linewidth}@{}}\toprule
Quantity & Value or construction\\\midrule
Latent state; largest transition eigenvalue modulus & 8 factors; $\rho(F)=0.94$\\
Public macro releases & 2 noisy indicators; noise SDs $(0.15,0.22)$\\
Payout coefficient and independent shock SD & $\lambda=1.15$, $\sigma_\varepsilon=0.43$\\
Net supply shock SD; risk-free gross return & $\sigma_\zeta=0.90$, $R_f=1.02$\\
Group masses $(\pi_1,\pi_2)$ & $(0.55,0.45)$\\
Risk aversion $(\gamma_1,\gamma_2)$ & $(0.9,1.2)$\\
Private-clue noise SDs & $(1.0,0.3)$\\
Low and high pre-auction history & $k=0$ and $k=3$ public lag blocks\\
Adoption share $x$ & Fraction of group 1 using $k=3$\\\bottomrule
\end{tabular}
\end{table}

We begin with both groups using $k=0$. We then allow a fraction $x\in[0,1]$ of group 1 to adopt $k=3$. Group 2 continues to trade with $k=0$ throughout this adoption exercise. The technology used by an individual investor can change without changing her risk aversion or the noise in her private signal. Later, when we calculate the private value of computation, we consider a negligible \emph{candidate} from group 2 who compares $k=0$ with $k=3$ under the same market law. The candidate does not alter market prices.

\subsection{From optimal forecasts to a stationary market price}\label{subsec:S7-population-map}

For a specified market law $Q$, each investor type chooses the best Gaussian forecast available to its capacity. The forecast mean is affine in its processed public history, private clue, and a proposed auction price. Since the resulting data law is jointly Gaussian, the population-optimal coefficients are exactly the coefficients of linear regression. For an active forecasting type $c$ with public-history capacity $k_c$, write
\begin{equation}\label{eq:S7-head}
 \mu_{ci,t}(p)=b_c^\top z_{k_c,t}+\beta_c s_{ci,t}+\chi_c p,
 \qquad v_c>0.
\end{equation}
The numerical economy is centered, so the intercept is zero. The type's capacity class allows the coefficients in \eqref{eq:S7-head} and the positive forecast variance to be selected jointly. In particular, a high-capacity forecast can ignore additional lagged observations by setting their coefficients to zero.

To calculate its population fit, let $X_{ci,t}=(z_{k_c,t}^\top,s_{ci,t},p_t^Q)^\top$, where $p_t^Q$ is the price realized in a historical observation drawn from $Q$. With covariance matrices computed under this \emph{same} training law,
\begin{align}\label{eq:S7-fit}
 \widehat b_c^{\mathrm{all}}(Q)&=\operatorname{Var}_Q(X_{ci,t})^{-1}\operatorname{Cov}_Q(X_{ci,t},D_{t+1}),\\
 v_c(Q)&=\operatorname{Var}_Q(D_{t+1})-
 \operatorname{Cov}_Q(D_{t+1},X_{ci,t})\operatorname{Var}_Q(X_{ci,t})^{-1}
 \operatorname{Cov}_Q(X_{ci,t},D_{t+1}).\nonumber
\end{align}
Here $\widehat b_c^{\mathrm{all}}$ stacks the public, private, and price coefficients $(b_c,\beta_c,\chi_c)$. The fitted variance is the population mean squared forecast error. These are the Gaussian-log-score minimizers described in S2. Each individual then chooses a CARA-optimal holding from her fitted belief, as derived in S3:
\begin{equation}\label{eq:S7-demand}
 h_{ci,t}(p)=\frac{b_c^\top z_{k_c,t}+\beta_c s_{ci,t}-(R_f-\chi_c)p}{\gamma_c v_c}.
\end{equation}

Individual clue errors cancel when orders are aggregated. Write every public coefficient vector $b_c$ in the common fourteen-dimensional feature space by padding unused lags with zeros. For the active population shares $\pi_c$, define
\begin{equation}\label{eq:S7-clearing-coefficients}
 A=\sum_c\frac{\pi_c\beta_c}{\gamma_c v_c},\qquad
 C=\sum_c\frac{\pi_c b_c}{\gamma_c v_c},\qquad
 \Delta=\sum_c\frac{\pi_c(R_f-\chi_c)}{\gamma_c v_c}.
\end{equation}
When $\Delta>0$, the same competitive clearing rule used throughout the paper gives
\begin{equation}\label{eq:S7-clearing}
 \boxed{p_t=\frac{A\theta_t+C^\top z_{3,t}-\zeta_t}{\Delta}}.
\end{equation}
This is not a price equation assigned independently of investors' choices: $A,C,\Delta$ are constructed from the learned forecasting rules and the resulting optimal demands.

The price has two informational components. The term $C^\top z_{3,t}$ carries the results of processing public history. The term $A\theta_t$ conveys the common fundamental embedded in many investors' private clues, obscured by supply noise. An observer who can reconstruct the public term can transform the observed price into $A\theta_t-\zeta_t$. Its structural likelihood precision about $\theta_t$ is
\begin{equation}\label{eq:S7-precision}
 J=\frac{A^2}{\sigma_\zeta^2}.
\end{equation}
This is the fresh-price precision shown in Fig.~2A of the main text and Fig.~\ref{fig:S7-mechanisms-summary}A--B. It is different from the conditional mutual information introduced in S5 and measured below.

\paragraph{Computing the market law.}
With the forecasting coefficients fixed, combine the eight latent factors and the three lagged public records into a state $h_t$. The clearing price and the macroeconomic and dividend observations generate a linear stochastic recursion,
\begin{equation}\label{eq:S7-fastlaw}
 h_{t+1}=M(q)h_t+N(q)e_t,\qquad
 \Sigma_h=M(q)\Sigma_h M(q)^\top+N(q)N(q)^\top,
 \qquad q=(A,C^\top,\Delta)^\top.
\end{equation}
The state has twenty entries and $q$ has sixteen. When $\rho(M(q))<1$, the discrete Lyapunov equation gives the unique stationary covariance $\Sigma_h$. This covariance also determines the regression moments in \eqref{eq:S7-fit}. We therefore calculate the market law and the population-optimal forecasting policies from the same system of economic primitives, without drawing a finite training sample.

At a fixed adoption share $x$, the calculation defines a mapping $q_{n+1}=\Phi_x(q_n)$: given the current stationary price law, fit every permitted forecasting policy, form its demand schedule, and clear the new market. A population-retrained equilibrium satisfies $q^*(x)=\Phi_x(q^*(x))$. The calculation uses S3's learning and market operators; $n$ denotes the learning iteration, not the trading date. We check the positive clearing denominator, the stationary fast law, and the equilibrium residual separately. Local stability of the learning iteration is assessed by the eigenvalues of $D_q\Phi_x$.

\subsection{Comparing adoption before and after the market learns}\label{subsec:S7-stages}

An algorithm trained on yesterday's market can change the next auction even before its trades have generated new price histories. To isolate that initial effect, we first fit the low- and high-capacity alternatives using the \emph{same} stationary low-adoption law $Q_0$. We change the share using the higher-capacity forecast and clear the next auction, leaving the non-upgraders' learned coefficients unchanged. This is the first-deployment comparison from S3--S4.

We also allow the altered market to run with all forecasting coefficients fixed. The new prices then become inputs to subsequent pre-auction histories, so forecasts may change even though their coefficients do not. In Model A the structural precision $J$ remains constant throughout this frozen-policy adjustment because $A$ and $\sigma_\zeta$ are fixed. Conditional information measures, however, can change as the public-history distribution changes. Finally, we let both groups refit their forecast rules to the stationary law generated after adoption and find a population-retrained fixed point.

\begin{table}[htbp]\centering\small
\caption{Price discovery in the eight-state population market when group 1 upgrades from no lagged records to three. The first-deployment and retrained columns use the same exogenous economy.}\label{tab:S7-precision-stages}
\begin{tabularx}{\linewidth}{@{}p{.40\linewidth}>{\centering\arraybackslash}X>{\centering\arraybackslash}X>{\centering\arraybackslash}X@{}}\toprule
 & Before adoption & First deployment & Retrained equilibrium\\\midrule
High-capacity share within group 1 & 0 & 1 & 1\\
Fresh-price precision $J$ & 2.469507 & 2.262529 & 2.073514\\
Change relative to baseline & --- & $-8.38\%$ & $-16.04\%$\\
Local retraining Jacobian radius & 0.6871 & --- & 0.7236\\
Fast-market transition radius & 0.9400 & --- & 0.9400\\\bottomrule
\end{tabularx}
\end{table}

\begin{figure}[htbp]\centering
\includegraphics[width=.86\linewidth]{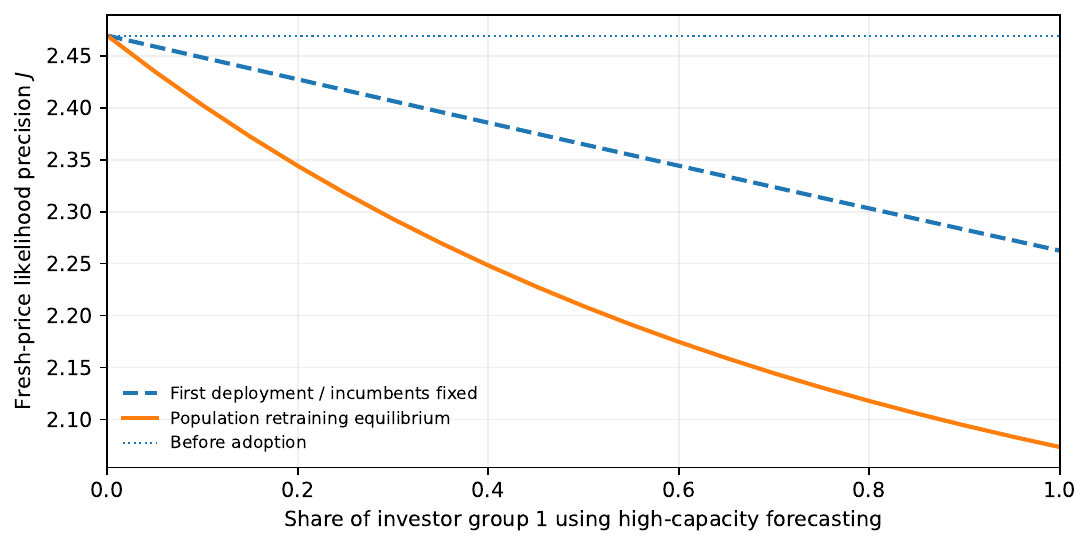}
\caption{Fresh-price precision as the share of group 1 using three public lag blocks increases from zero to one. The dashed curve compares algorithms fitted to the original market law and then deployed with their coefficients fixed. The solid curve is calculated after all active types have been refitted to their own stationary price law. A stationary frozen-policy adjustment does not separately change $J$ under the constant-affine final head. These are exact Gaussian population calculations, not neural training results.}\label{fig:S7-J-adoption}
\end{figure}

The first upgrade reduces $J$ from $2.4695$ to $2.2625$, a loss of about $8.4\%$. After every active forecasting type learns from the new stationary market law, precision falls further to $2.0735$, a total decline of about $16.0\%$ (Table~\ref{tab:S7-precision-stages}). Figure~\ref{fig:S7-J-adoption} shows the comparison at 21 adoption shares. Thus, in this economy, the initial reduction in private-information aggregation survives and becomes larger after population retraining.

The first decline can be traced directly to the upgrading group's forecast. Under the inherited market law, its coefficient on the private clue falls from $0.19696$ to $0.12459$ when the three lag blocks become usable. Its optimal forecast variance also falls, from $0.41141$ to $0.32818$. The Gaussian forecast loss for the upgrading group falls from $0.97485$ to $0.86185$ (population negative log score). The lower variance makes the investor more willing to bear risk, while the smaller private-signal coefficient makes her forecast respond less to the private clue. With both changes included, the group's risk-adjusted individual clue response $\beta_1/(\gamma_1v_1)$ falls from approximately $0.532$ to $0.422$. This is the mechanism behind the aggregate reduction in $A$ and therefore $J$.

The reported baseline and high-capacity population fixed points are locally attracting for the learning operator, with radii below one; their stationary fast-market transition matrices have spectral radii approximately $0.94$. These are checks for this calibration, not a general claim that every computational market converges. The difference between the initial and retrained $J$ values is also specific to this calibration; Theorem 1 supplies the first-deployment sign under its conditions, not a universal sign for retraining feedback.

\subsection{How useful is the same price to different observers?}\label{subsec:S7-information}

A price can lose fresh information yet become more useful to someone who has less computational capacity. To see this in the same economy, we evaluate two hypothetical observers who both see the \emph{same current competitive price} generated at each adoption share. The limited observer processes only the two current macroeconomic indicators. The more capable observer also processes the three previous public records:
\begin{equation}\label{eq:S7-fields}
 \mathcal F_{L,t}=\sigma(m_t),\qquad
 \mathcal U_{H,t}=\sigma\big(m_t,\{m_{t-j},p_{t-j},D_{t+1-j}\}_{j=1}^{3}\big),
 \qquad \mathcal F_{L,t}\subseteq\mathcal U_{H,t}.
\end{equation}
These are the feasible \emph{three-lag} information fields used in the numerical comparison. In particular, the high-capacity observer is not an unlimited-history or forty-lag oracle.

Let $P_L=\operatorname{Var}(\theta_t\mid\mathcal F_{L,t})$ and $P_H=\operatorname{Var}(\theta_t\mid\mathcal U_{H,t})$ be the information remaining before the current price is observed. Define the corresponding variances after price as $L=\operatorname{Var}(\theta_t\mid\mathcal F_{L,t},p_t)$ and $H=\operatorname{Var}(\theta_t\mid\mathcal U_{H,t},p_t)$. Because all variables are jointly Gaussian, the additional information provided by price to each observer is
\begin{equation}\label{eq:S7-two-values}
 S=\tfrac12\log\frac{P_L}{L},\qquad
 F_{\!\mathrm{info}}=\tfrac12\log\frac{P_H}{H}.
\end{equation}
Here $S$ is the price information for the limited observer and $F_{\!\mathrm{info}}$ is the \emph{fresh information} remaining for the capable observer, both measured in nats. The symbol $F_{\!\mathrm{info}}$ distinguishes this informational outcome from the state transition matrix $F$ in \eqref{eq:S7-economy}; it is the $F$ of S5's information-accounting identity.

The more capable observer can remove the computed public component $C^\top z_{3,t}$ from price. Its incremental information therefore obeys
\begin{equation}\label{eq:S7-fresh-mi}
 F_{\!\mathrm{info}}=\tfrac12\log(1+JP_H).
\end{equation}
The limited observer cannot generally perform that subtraction because the full three-lag information vector is not available to its processing rule. We compute its $S$ directly from the stationary covariance of $(\theta_t,m_t,p_t)$. This keeps the information comparison faithful to the two observers' different processing abilities.

\begin{figure}[htbp]\centering
\includegraphics[width=.86\linewidth]{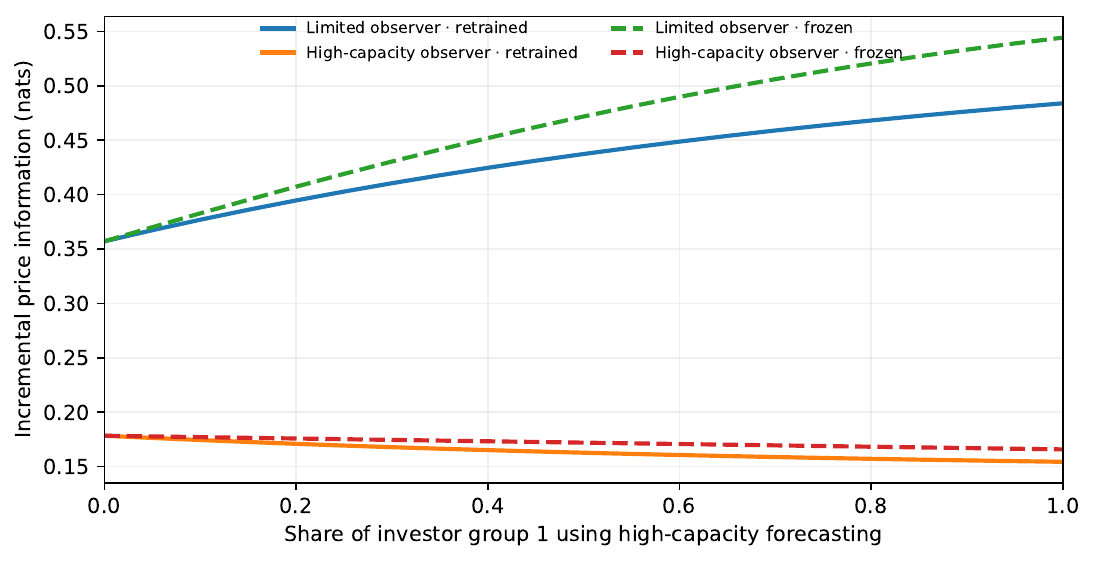}
\caption{Information carried by the same current price under two finite public-data processing capabilities. $S$ measures information gained after the current macro releases alone; $F_{\!\mathrm{info}}$ measures information gained after those releases and three past public records. Solid curves use population-retrained stationary markets; dashed curves keep incumbent forecasting policies fixed after deployment, while market histories adjust. All information measures are in nats.}\label{fig:S7-price-info}
\end{figure}

\begin{table}[htbp]\centering\small
\caption{Information available from price and the marginal value of computation in the same population market. Values at $x=1$ are measured under two different market-adjustment protocols.}\label{tab:S7-information-moments}
\begin{tabular}{@{}p{.48\linewidth}rrr@{}}\toprule
 & $x=0$ & $x=1$, frozen & $x=1$, retrained\\\midrule
$P_L$, prior variance after $m_t$ & 0.422095 & 0.422095 & 0.422095\\
$P_H$, prior variance after 3 lags & 0.173595 & 0.173802 & 0.174489\\
$L$, posterior variance after price (limited) & 0.206670 & 0.142132 & 0.160367\\
$H$, posterior variance after price (high) & 0.121506 & 0.124747 & 0.128131\\
Price information $S$ (nats) & 0.357055 & 0.544236 & 0.483883\\
Fresh price information $F_{\!\mathrm{info}}$ (nats) & 0.178380 & 0.165813 & 0.154406\\
Candidate upgrade value $\Delta\CE_2$ (model units) & 0.023257 & 0.006083 & 0.010213\\\bottomrule
\end{tabular}
\end{table}

Under full retraining, price's fresh information for the capable observer falls from $0.17838$ to $0.15441$ nats, about a $13\%$ reduction. At the same time, the information received by the limited observer rises from $0.35705$ to $0.48388$ nats, about a $36\%$ increase. The first effect means there is less information in price \emph{beyond} the public analysis available to a capable observer. The second means that price transmits more of the market's accumulated analysis to an observer who cannot reproduce it. The two statements concern different conditioning information; they are not competing estimates of one scalar notion of market efficiency.

The two outcomes have the same directions when incumbent policies remain fixed but new histories are allowed to form (Table~\ref{tab:S7-information-moments}). Retraining makes the loss of fresh information larger while reducing some of the additional information received by the limited observer in this example. Their difference is determined by the market law generated by investors' forecasts and trades, not by an assumed price response.

\subsection{Why the private benefit of more computation can fall}\label{subsec:S7-CE}

The price can also reduce an investor's reason to acquire a more capable algorithm. We evaluate this effect for a single negligible investor from group 2. The candidate compares a forecast based on current macroeconomic information with another forecast that also processes the same three lagged public records. She observes the same current price and the same private signal in either case. At each adoption level, both of her hypothetical forecasts are fitted anew to that \emph{same} market law. Her choice has no effect on prices.

For this aligned-payoff Gaussian economy, an investor whose posterior variance about $\theta$ is $R$ after observing price and the relevant public information has optimally fitted payout variance
\begin{equation}\label{eq:S7-risk-of-state}
 v_2(R)=\sigma_\varepsilon^2+\lambda^2\frac{\nu_2R}{\nu_2+R},
 \qquad \sigma_\varepsilon^2=0.43^2,\quad \nu_2=0.30^2,\quad\lambda=1.15.
\end{equation}
The candidate's ex-ante gain from upgrading, measured in CARA certainty-equivalent units before deducting computational costs, is
\begin{equation}\label{eq:S7-compute-value}
 \Delta\CE_2(x)=\frac{1}{2\gamma_2}\log\frac{v_2(L_x)}{v_2(H_x)},
 \qquad \gamma_2=1.2.
\end{equation}
This is the Gaussian information-to-utility expression proved in S5. Its two residual variances are calculated under the same adoption-dependent market law; it measures the value of additional feasible processing for the candidate, rather than any equilibrium welfare effect.

\begin{figure}[htbp]\centering
\includegraphics[width=.86\linewidth]{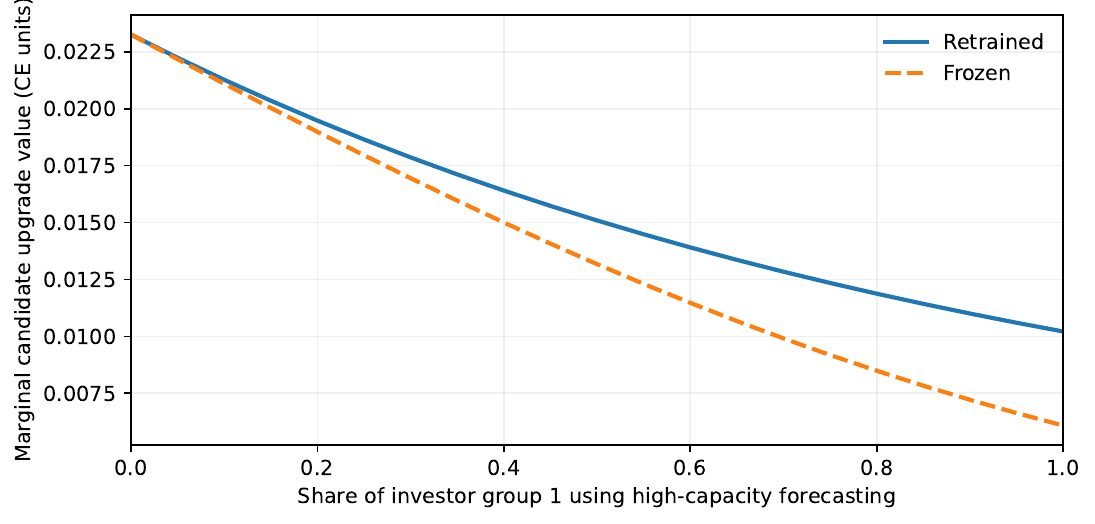}
\caption{The ex-ante certainty-equivalent value to a negligible group-2 investor of processing three additional public lag records. The two candidate strategies both observe the same current price and private clue and refit under the stationary law at each adoption share. Model units have no direct monetary calibration.}\label{fig:S7-CE}
\end{figure}

The candidate's benefit falls from $0.02326$ to $0.01021$ after full adoption and population retraining, a decline of about $56\%$. Prices are increasingly helpful to the limited observer, so a separate computation of the more detailed public history adds less. This is the third sign reported in the main-text Fig.~3C.

The exact comparative-static identity in S5 also lets us separate the channels locally. At adoption share $x=0.5$ along the retrained branch, implicit differentiation of the population learning map gives
\begin{equation}\label{eq:S7-derivative}
 \frac{dS}{dx}=0.120021,\quad
 \frac{dF_{\!\mathrm{info}}}{dx}=-0.022380,\quad
 \frac{dP_H}{dx}=0.000851,\quad
 \frac{d\Delta\CE_2}{dx}=-0.012462.
\end{equation}
Applying the identity from S5 to these derivatives yields a change of $-0.0101075$ from the increasing information supplied by price to the limited observer, $-0.0021225$ from the decreasing fresh information available to the capable observer, and $-0.0002317$ from the change in the capable observer's pre-price uncertainty. These contributions sum to $-0.0124617$ certainty-equivalent units per unit increase in $x$. Their relative sizes characterize this calibration. The derivative of $P_H$ is accounted for rather than held fixed; the high-capacity public history contains past prices whose law changes with adoption.

\subsection{Mean forecasts, risk forecasts, and who retrains}\label{subsec:S7-mechanisms}

The same model allows us to see why a more capable forecast can have two opposing effects on private-information aggregation. Under the inherited low-capacity market law, we fit the upgrading group's low- and high-capacity Gaussian heads. We then substitute either the new mean coefficients or the new variance into its old forecasting head while leaving the other component unchanged. At every substitution we derive the first-auction clearing price anew, with all other investor policies held fixed. These comparisons isolate mechanical contributions of two parts of a \emph{jointly fitted} forecasting system; they do not correspond to separately reoptimized mean-only or variance-only algorithms.

\begin{table}[htbp]\centering\small
\caption{The two opposing components of an upgrade at first deployment. All four cases use the inherited low-adoption market law and the same policies for non-upgrading investors. The outcome is the structural precision $J$.}\label{tab:S7-factorial}
\begin{tabular}{@{}p{.57\linewidth}rr@{}}\toprule
Forecasting head deployed for upgrading group & $J$ & Change from original\\\midrule
Old mean, old variance & 2.469507 & ---\\
New mean, old variance & 2.108368 & $-14.62\%$\\
Old mean, new variance & 2.735404 & $+10.77\%$\\
New mean, new variance & 2.262529 & $-8.38\%$\\\bottomrule
\end{tabular}
\end{table}

Replacing the mean coefficients alone reduces private-price precision, because investors' holdings become less sensitive to their private clues. Replacing the variance alone increases precision, because lower forecast risk raises the risk-adjusted response to those clues. The joint upgrade lowers precision in this calibration because the first force dominates the second. Averaging the two possible orders of substitution assigns $-0.417007$ units of $J$ to the mean change and $+0.210029$ to the variance change. Their sum is the observed first-deployment change, $-0.206977$. The decomposition illustrates how the terms in the optimal demand rule \eqref{eq:S7-demand} interact; it is neither an additive welfare calculation nor a decomposition into two independently trained algorithms.

Retraining can amplify these effects through investors who never acquire the more capable algorithm. After the full group-1 upgrade, we solve four population markets. In each one we permit a selected set of investors to update their fitted forecasts under the new stationary law; all other fitted coefficients remain fixed. This keeps the information sources and economic primitives unchanged while isolating the role of the learning decisions.

\begin{table}[htbp]\centering\small
\caption{Which investors relearn after a full group-1 upgrade? Every entry is calculated with the competitive clearing price and the corresponding stationary Gaussian law.}\label{tab:S7-group-update}
\begin{tabular}{@{}p{.60\linewidth}r@{}}\toprule
Groups permitted to refit their forecasts & Fresh-price $J$\\\midrule
Neither (initial deployment, coefficients fixed) & 2.262529\\
Only upgrading group 1 & 2.272432\\
Only non-upgrading group 2 & 1.994346\\
Both groups & 2.073514\\\bottomrule
\end{tabular}
\end{table}

In this example, refitting only the non-upgrading group has a much larger effect on $J$ than refitting only the upgrading group. The stronger algorithm changes prices, and those prices become part of the public record from which other investors learn. Investors who retain their old computational capacity can therefore change the way they trade on private information. The four outcomes are separate counterfactual fixed points; differences between them should not be added as though they were independent causal effects.

\subsection{How often do the three informational signs coexist?}\label{subsec:S7-screen}

The baseline economy shows that reduced fresh information, increased information to a limited observer, and lower private value of computation can occur together. We also assess these signs in a set of nearby aligned Gaussian economies. We examine 180 calibrations that vary economic quantities such as supply noise, payout noise, risk aversion, signal noise, investor masses, and persistence. Of these, 147 passed the previously specified requirements for a converged population equilibrium with admissible market clearing. The other 33 could not be included as verified stationary equilibria under that screen. They were not treated as failed sign tests.

For each of the 147 retained calibrations, we compute the frozen-incumbent and fully retrained stationary laws at group-1 adoption shares $x=0.25$ and $x=0.75$. The limited and high-capacity observers process exactly the fields in \eqref{eq:S7-fields}, and the negligible candidate's upgrade value is re-evaluated at each law.

\begin{table}[htbp]\centering\small
\caption{Signs in the 147 preselected converged aligned economies between adoption shares $0.25$ and $0.75$. Counts are conditional on the convergence and clearing screen; they are not success rates among all 180 draws.}\label{tab:S7-robustness}
\begin{tabular}{@{}p{.59\linewidth}rr@{}}\toprule
Condition satisfied & Frozen policies & Population retraining\\\midrule
Price information to limited observer increases & 147/147 & 147/147\\
Fresh information beyond three public lags decreases & 147/147 & 147/147\\
Marginal candidate's computation value decreases & 147/147 & 147/147\\
High-capacity pre-price variance $P_H$ nondecreases & 76/147 & 108/147\\
Limited posterior variance falls and high posterior variance rises & 140/147 & 145/147\\\bottomrule
\end{tabular}
\end{table}

\begin{figure}[htbp]
\centering
\includegraphics[width=.96\linewidth]{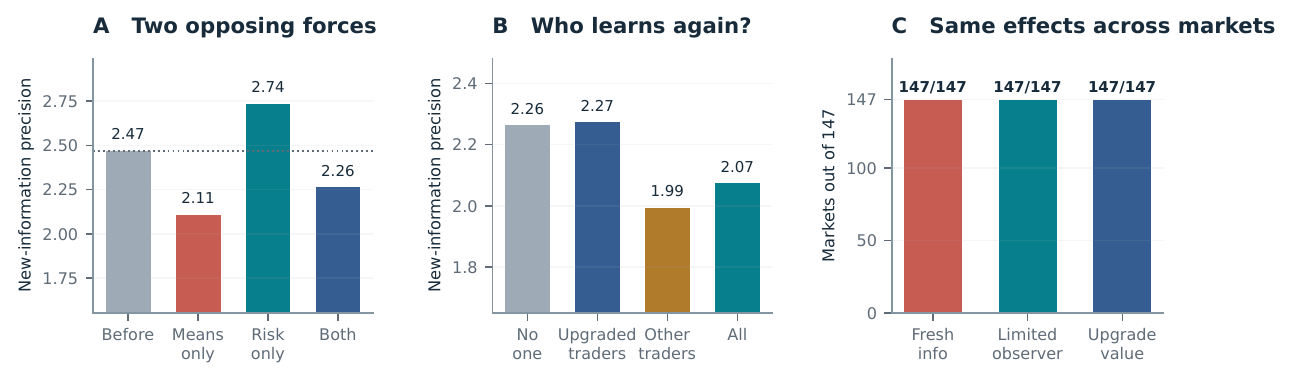}
\caption{\textbf{Forecasting mechanisms and robustness in the population market.} (A) Replacing the predicted mean alone reduces fresh-price precision, while replacing fitted risk alone raises it. These are substitutions of already fitted forecasting coefficients, not separately optimized models. (B) Updating forecasts for investors who retained the low-capacity technology contributes substantially to the decline in fresh-price precision. (C) The three principal directional results hold in all 147 calibrations satisfying the convergence and market-clearing criteria among 180 candidates.}
\label{fig:S7-mechanisms-summary}
\end{figure}

All three headline signs appear in every one of the 147 economies retained by the convergence screen under both protocols (Fig.~\ref{fig:S7-mechanisms-summary}C). The stronger restriction that high-capacity pre-price uncertainty must rise holds in fewer cases. S5's sharper condition for a falling computation premium is therefore useful: the premium can fall even when the advanced public record becomes more informative, provided the price provides sufficiently greater gains to the limited observer. An additional local calculation at $x=0.5$ confirms the sharp derivative condition in all 147 retained calibrations under each protocol. This is a numerical audit of selected stable economies, not an unrestricted theorem for all economic parameters or nonlinear learning systems.

\subsection{Numerical checks and interpretation}\label{subsec:S7-validation}

The population results are determined by covariance matrices rather than simulated training observations. Nevertheless, several separate calculations were used to check that the reported precision and information measures come from the same clearing price. The fast-market covariance is solved from the Lyapunov equation; the population Gaussian regression is calculated from that covariance; the fitted demand schedules determine $A,C,\Delta$; and the next market law is recalculated from those coefficients. For the main benchmark, the population fixed-point residual is below $7\times 10^{-12}$ at both endpoints. The training-map Jacobian derived from Lyapunov and regression sensitivities agrees with an independent centered finite-difference calculation to approximately $3.6\times10^{-10}$ in maximum absolute entrywise difference.

An independent trading simulation of the same market mechanism also verified competitive clearing, with maximum demand residual approximately $2.7\times10^{-15}$ over 26,000 observations. A separate 85,000-period Gaussian simulation checked stationary moments against the analytic covariance. These stochastic checks verify implementation and moment calculations; the results in Tables~\ref{tab:S7-precision-stages}--\ref{tab:S7-robustness} themselves come from the population solution rather than estimated Monte Carlo moments.

The reported comparisons distinguish three objects. The first-auction result uses forecasts fitted to a common inherited law; the frozen-policy results allow price histories to change without refitting coefficients; and the retrained equilibrium refits all designated investor types to a self-consistent stationary law. All use the constant-affine Gaussian head, even when discussing how more complex neural markets differ in S8--S10. In particular, the mechanical substitutions in Table~\ref{tab:S7-factorial} do not establish a theorem for state-dependent forecast variances, and the retained calibration screen does not establish universal monotonicity at arbitrary adoption shares. A population-market equilibrium can also be dynamically unstable under repeated retraining, a distinction examined separately in the numerical dynamics reported with the later experiments.

Taken together, the calculations show why measuring price quality requires specifying both the information available to an observer and the stage of market adjustment. In the same competitive market, a larger population of advanced processors reduces the price's fresh private-information content but improves its usefulness to a limited observer. Because the price shares some of the advanced analysis, the marginal value of undertaking that analysis independently also falls. The finite-model experiments in S8--S10 ask how these relationships appear when algorithms are trained on limited data and the market need not reach a population learning equilibrium.

\subsection{Exact implementation and reproduction record}\label{subsec:S7-reproduction}

The eight-factor system is generated by a fixed procedure based on NumPy random draws. It uses NumPy's \texttt{default\_rng} with seed $20261015+8$ to construct a random orthogonal basis $Q$ by QR factorization of an $8\times8$ standard-normal matrix, followed by a deterministic sign convention for the diagonal of $R$. The transition matrix is $F=Q\operatorname{diag}(\rho)Q^\top$, with eigenvalues
\begin{equation}\label{eq:S7-primitive-recipe}
 \rho=(0.94,0.88,0.78,0.68,0.53,-0.43,0.32,0.15).
\end{equation}
The innovation loading matrix is $G=Q\operatorname{diag}(g)$, with scale vector
\[g=(0.35,0.34,0.36,0.40,0.42,0.35,0.40,0.32).\] The normalized target loading has nonzero raw entries $a_1=1$ and $a_4=0.25$ before normalization. The two rows of the public observation matrix have raw nonzero entries $(H_{1,2},H_{1,5})=(1,0.3)$ and $(H_{2,3},H_{2,6})=(1,0.4)$. These details, together with Table~\ref{tab:S7-primitives}, specify the economy; the stored generator fixes the precise floating-point realization of $Q$.

The implementation stacks the eight latent factors and three lagged blocks of $(p_t,D_{t+1},m_t)$ into the state of \eqref{eq:S7-fastlaw}. It solves the stationary Lyapunov equation, fits Gaussian forecast coefficients from population covariances, and reconstructs the market-clearing coefficients in \eqref{eq:S7-clearing-coefficients}. The population calculations use stationary covariance solutions, Gaussian forecast fitting, and the market-clearing rules described above. The figures are constructed from those population quantities using the fixed economy specified here.

\clearpage\setcounter{table}{0}\setcounter{figure}{0}
\renewcommand{\thetable}{S8.\arabic{table}}\renewcommand{\theHtable}{S.8.\arabic{table}}
\renewcommand{\thefigure}{S8.\arabic{figure}}\renewcommand{\theHfigure}{S.8.\arabic{figure}}
\section{Finite learning, neural trading, and the response to retraining}\label{sec:S8}

The population calculations in S7 describe what happens when every forecasting system is fitted exactly to the market distribution it faces. Actual learning algorithms have finite training data and may reach different fitted models even when their available public information is the same. This section asks how the market effects change when the public record is processed by trained state-space and recurrent neural networks. We first compare small and large algorithms on a common inherited market. We then let trained algorithms trade repeatedly, changing the prices that will enter their next training samples. Finally, we examine which investors' subsequent learning matters most and whether changes in forecasts of expected payoffs or in forecasts of risk drive the response.

The experiments use the investor decisions and competitive clearing rule derived in S3. An algorithm summarizes past public observations before the auction, combines that summary with the investor's own private clue and a proposed current price, and produces a forecast of next period's payout. The investor chooses her holding using that forecast. All new prices are determined by clearing the resulting orders, rather than being supplied by an independent forecasting benchmark. Keeping this sequence intact makes it possible to distinguish better statistical prediction from more information incorporated into the price.

\subsection{Two experiments with trained forecasting systems}\label{subsec:S8-design}

We use two related sets of computations, each with a different purpose. In the first, a small and a large algorithm are fitted to the \emph{same} historical market and are evaluated when their trading rules are deployed in its next auction. This isolates the change in the forecasts and orders associated with a capacity upgrade before the market generates a new price history. The second experiment starts from saved trained algorithms, runs markets with their own price histories, and allows specified investor groups to train again on the observations that those markets generate. The first experiment measures an initial impact; the second measures a finite sequence of market and learning adjustments.

\begin{table}[htbp]\centering\small
\caption{The two finite-learning experiments. The first is an inherited-market comparison; the second follows a market through repeated trading and a finite number of model updates. The numerical economies and training regimes are different, so values of their information statistics are not pooled.}\label{tab:S8-studies}
\begin{tabularx}{\linewidth}{@{}p{31mm}XX@{}}\toprule
 & Initial deployment & Endogenous-history and retraining experiment\\\midrule
Public-history encoder & Linear state-space model (SSM) or gated recurrent unit (GRU) & GRU for group 1; no public-history encoder for group 2\\
Model sizes & 4 and 12 hidden coordinates within each architecture & 4 and 32 hidden coordinates in group 1\\
Historical window & 48 observations & 16 observations\\
Forecasting head & Constant affine Gaussian (Model A); selected Model-B comparisons & Constant affine Gaussian (Model A)\\
Primary comparison & Prediction loss and fresh-price precision at the first auction & Precision after endogenous price-history adjustment and up to 12 or 6 model updates\\
Independent learning runs & Three matched initial fitting seeds in the aligned comparison & Three inherited trained initial seeds, reused in paired adoption and retraining comparisons\\\bottomrule
\end{tabularx}
\end{table}

An increase in hidden-state dimension makes a richer representation feasible \emph{within} each architecture. The SSM and GRU are not interchangeable measures of computational cost: a hidden dimension in one architecture requires a different number of parameters and operations from a hidden dimension in the other. Moreover, finite training does not generally find the population-optimal predictor assumed in Theorem 1. A small model may outperform a larger model in an individual fit even though the larger feasible class can reproduce the smaller model.

\subsection{How neural forecasts enter the auction}\label{subsec:S8-neural-clearing}

At date $t$, a public input consists of the information newly released before trading, the previous auction's price, and the payout that has just matured. For a model that processes $L$ dates, write its input as $U_{t-L+1:t}=(o_{t-L+1},\ldots,o_t)$, where $o_t=(m_t,p_{t-1},D_t)$ as in S2. The encoder calculates $z_{c,t}=T_{\phi_c}(U_{t-L+1:t})$. It sees neither the current clearing price $p_t$ nor the next payout $D_{t+1}$. The investor's private observation enters the final forecasting head separately.

For the experiments using Model A, the forecast of the next payout and its constant reported variance take the form
\begin{equation}\label{eq:S8-forecast}
 \mu_{ci,t}(p)=\alpha_c+b_c^\top z_{c,t}+\beta_c s_{ci,t}+\chi_c p,
 \qquad v_c>0.
\end{equation}
An encoder can be nonlinear even though the final coefficients $\beta_c$, $\chi_c$ and $v_c$ are constant. Model B, also tested in the initial-deployment study, permits the mean coefficients and variance to depend on $z_{c,t}$. All models are trained to forecast the payout using a Gaussian negative log score. Portfolio positions are subsequently computed from those fitted beliefs; the training objective is not trading profit.

When the Model-A rules trade, their fitted private-clue coefficients and variances determine the aggregate loading
\begin{equation}\label{eq:S8-loading}
 \iota_c=\frac{\beta_c}{\gamma_cv_c},\qquad
 A=\sum_c\pi_c\iota_c,\qquad
 J=\frac{A^2}{\sigma_\zeta^2}.
\end{equation}
The last quantity is the structural likelihood precision of the fresh private-signal channel, after an observer removes the pre-auction predictable public component of the price. It is not a mutual-information calculation and need not order information received by a less capable observer. In the general state-dependent Model-B head, $A_t$ can vary with public histories; we do not extend a Model-A sign claim to that case.

During market generation, all encoder and head parameters are frozen. At each auction the model calculates its public state, price-contingent individual holdings are formed, and the unique price is obtained from their aggregate demand. The realized price is added to the \emph{next} date's public record. When retraining is permitted, a block of such trading dates is simulated before any fitted parameter is updated. In particular, a fixed model may issue different predictions when the price history changes; this does not require the model to retrain. The implementation reproduces the clearing equation in S3, including the risk-adjusted forecast coefficients, rather than assuming a response of price to AI adoption.

\subsection{Initial deployment: better forecasts and weaker fresh-price signals}\label{subsec:S8-initial}

We begin with an eight-factor stochastic economy in which the next payout is aligned with the scalar fundamental observed imperfectly by individual investors. A previously prevailing set of diagnostic trading policies generates a common historical training law. These incumbent policies are used to produce training observations, not as the small-model benchmark. We fit the small and large SSM or GRU separately to that same law, holding the 48-date public input window and final head specification fixed within each architecture. The newly fitted model then replaces the selected investor type's forecasting rule at the first auction. The underlying fundamental, supply shock, inherited public history, and other investors' schedules are held fixed. Thus the current-price argument of the new head is the newly clearing price, while its encoder still processes the inherited \emph{past} prices.

In the aligned economy, both architecture families show lower held-out forecast loss with greater capacity on average. At the common 48-date input window, the width-4 and width-12 Model-A SSMs have 28 and 76 trainable parameters, respectively, while the corresponding GRUs have 128 and 664. Approximate per-date inference operation counts are 1,156 and 3,468 for the SSMs and 5,188 and 29,388 for the GRUs. These are implementation-specific resource measures, not hardware timing measurements. They also show a lower average $J$ in the first deployed auction (Table~\ref{tab:S8-initial}). The two changes need not move together: $J$ is determined by the sensitivity of trading to fresh private observations after adjusting for fitted risk, whereas forecast loss assesses the accuracy of the entire prediction.

\begin{table}[htbp]\centering\small
\caption{Initial-deployment results in the aligned eight-factor economy. Each entry averages three matched training seeds. Smaller negative log score is better; smaller $J$ means less fresh private information reaches price through the structural supply-noise channel. The large and small models share an inherited training law.}\label{tab:S8-initial}
\begin{tabular}{@{}lrrr@{}}\toprule
Architecture & Hidden dimension & Held-out negative log score & Fresh-price $J$\\\midrule
Linear SSM & 4 & 0.94043 & 0.27184\\
Linear SSM & 12 & 0.93946 & 0.26559\\
GRU & 4 & 0.95134 & 0.27605\\
GRU & 12 & 0.94584 & 0.27392\\\bottomrule
\end{tabular}
\end{table}

\begin{figure}[htbp]\centering
 \includegraphics[width=.97\linewidth]{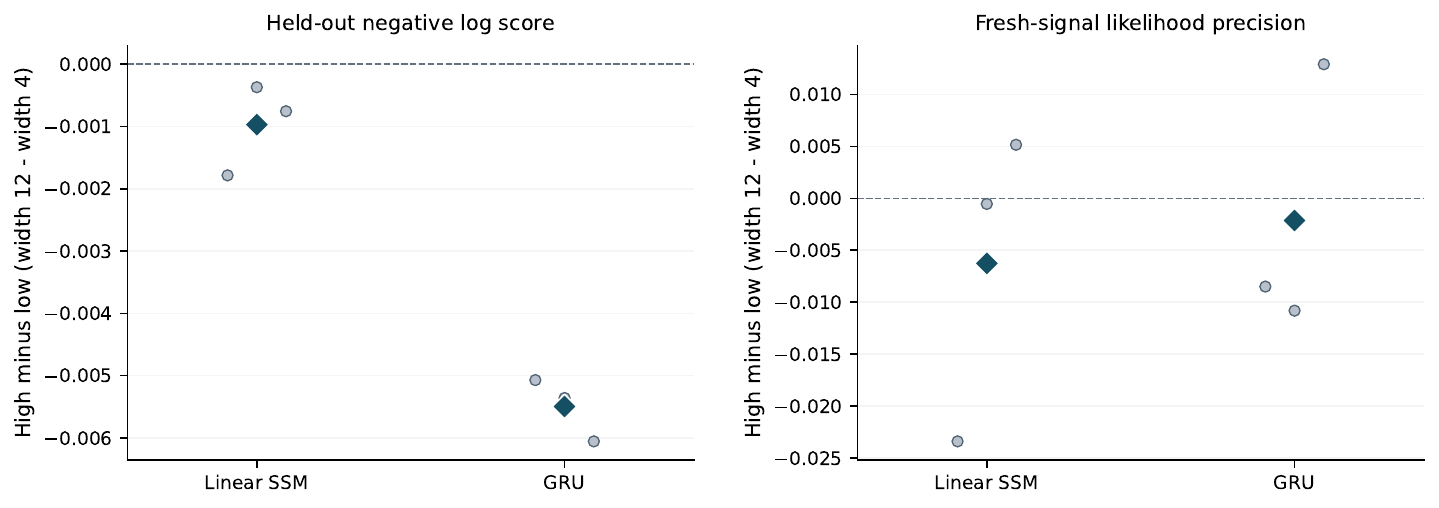}
 \caption{Changes from a 4-dimensional to a 12-dimensional public encoder, paired by training seed within architecture. Dots represent the three individual seed differences; diamonds are the means. The negative mean changes in held-out loss and $J$ underlie the separate finite-learning panels of main-text Fig.~2. The positive $J$ differences for some seeds are retained in the display.}\label{fig:S8-initial}
\end{figure}

The average response is not uniform across the six matched pairs. Under both SSM and GRU, two of the three seeds produce a lower $J$ when capacity increases, while one produces a higher $J$. Re-evaluating the trained checkpoints on a new sample of shocks generated by the inherited market preserves the sign for each pair. The seed dependence therefore reflects differences between fitted algorithms rather than only randomness in the price-evaluation paths. The results support the possibility identified by Theorem 1 without claiming its population-optimal comparative static for stochastic training.

Two additional initial-deployment settings illustrate the scope of this finding. When the next payout also depends on a factor outside the privately signaled target, the simple scalar $J$ in \eqref{eq:S8-loading} no longer measures the complete information about that target. We instead assess the improvement of an independent linear decoder when price is added to an inherited public record. Its price-added MSE gain is not uniformly monotone with model size, and the larger SSM has slightly worse average payout log score in this nonaligned case. 

A second setting allows state-dependent payoff risk and a Model-B forecasting head. For width-12 models, the average held-out negative log score improves from $1.03551$ to $1.03053$ for the SSM and from $1.04616$ to $1.03900$ for the GRU. This comparison changes mean and variance flexibility together; it is not a variance-only experiment. These extensions provide descriptive boundaries for the aligned, constant-head result rather than further theorem claims.

\subsection{A market that learns from the prices it generates}\label{subsec:S8-dynamic}

The first experiment fits models on inherited prices and ends after their initial deployment. We next run a market in which learned algorithms generate future prices, those prices enter the public information available on later dates, and the algorithms may eventually be refitted. The economic environment remains based on eight latent factors, but this is a different numerical calibration from the population Gaussian market of S7. At each date, eight public indicators are noisy nonlinear functions of the current latent state. The fundamental combines a persistent component with a newly drawn component. The latter contributes 20\% of fundamental variance in this calibration. The payoff is $D_{t+1}=1.15\theta_t+\varepsilon_{t+1}$, with payout-noise standard deviation $0.43$, and the independent net-supply noise has standard deviation $0.9$. Individual private-signal quality and investor preferences remain fixed when computational adoption changes.

There are two economic groups. Group 1 comprises 55\% of the population, has risk aversion $\gamma_1=0.9$ and private-signal noise variance $\nu_1=0.25$, and can use a GRU with hidden width 4 or 32. Both GRUs receive the same 16-date input window, containing eight current and historical public indicators together with lagged prices and matured payouts. Group 2 comprises the remaining 45\%, has $\gamma_2=1.2$ and $\nu_2=0.09$, and forecasts from its own private clue and the current candidate price without a public-history encoder. The adoption share $x$ is the fraction of \emph{group 1} using the width-32 GRU. It is not the fraction of the whole market using advanced AI. Group 2's feasible information class does not change; its fitted coefficients can change through subsequent retraining.

\begin{table}[htbp]\centering\small
\caption{Core settings for the nonlinear neural-market experiments. Each training epoch uses separate independent simulation paths, with model weights frozen during the generation of each bank.}\label{tab:S8-dynamic-design}
\begin{tabularx}{\linewidth}{@{}p{.45\linewidth}X@{}}\toprule
Object & Implemented value\\\midrule
Investor masses; risk aversion; private noise variance & $(\pi_1,\pi_2)=(0.55,0.45)$; $(\gamma_1,\gamma_2)=(0.9,1.2)$; $(\nu_1,\nu_2)=(0.25,0.09)$\\
Group-1 public-data models & GRU widths 4 and 32; Model-A constant-affine mean and variance head; 16-date window\\
Group-2 model & Affine mean in private signal and current price; constant fitted variance; no public encoder\\
Adoption $x$ and initial random seeds & $x\in\{0,0.5,1\}$; seeds $18013$, $19031$, $20047$\\
Training observations per update & 512 independent paths, 2,048 usable dates each; 1,048,576 observations\\
Validation and held-out evaluation & 65,536 and 131,072 observations, respectively; independent shocks\\
Fast-market and training timing & 512-date burn-in per path; no information window crosses path boundaries\\
Optimizer & Adam, one pass through eight disjoint training chunks, initial learning rate $0.001$ followed by decay; no profit-based training\\\bottomrule
\end{tabularx}
\end{table}

For each slow learning update, we first simulate the entire market with the current models frozen. The model then sees its own generated price histories as historical inputs. Once the training, validation, and evaluation banks have been formed, each eligible GRU is updated using the payout log score. Its previously fitted normalization is retained when the network is warm-started; the optimizer is reset for the new epoch. The second group's affine Gaussian head is refitted by least squares on the generated training law. The new models are deployed simultaneously \emph{after} training, and the next block of trading begins. Training paths are independent of evaluation paths; primitive shock streams are coordinated across adoption comparisons, but every market recalculates its own prices recursively.

The fitted heads retain constant $\beta_c$, $\chi_c$ and $v_c$ within each deployment epoch. Consequently, their structural $J$ remains the same as histories adjust with those coefficients fixed, even though realized prices, public encoder states, forecast errors, and prices' usefulness to observers can all change. Across epochs, the learned private-signal weights and fitted variances change, and hence $J$ can change. This distinction is useful for interpreting the finite retraining paths.

\subsection{What twelve learning updates establish}\label{subsec:S8-continuation}

In the first series of neural-market updates, the three initial fitted model pairs are deployed at adoption shares $x=0$, $0.5$, and $1$. Each of the nine market paths is followed for twelve simultaneous learning updates. The end-of-path averages in Table~\ref{tab:S8-E5} show less fresh-price precision at higher adoption. The high-capacity GRU remains more accurate when both candidate algorithms are evaluated on the same market, and its risk-adjusted response to a private clue remains smaller. Even at $x=0$, the inactive high-capacity system is trained and evaluated as a candidate; it does not affect the market price. These paths exhibit the crowd-out pattern after multiple changes in market histories and fitted models.

\begin{table}[htbp]\centering\small
\caption{Twelve-update neural-market paths. Entries average the three inherited seeds at each adoption share. The three paths for each adoption rate begin from the same inherited model profile, but retrain on different market histories. MSE gain is the percentage reduction for the high-capacity candidate relative to the low-capacity candidate, evaluated on the same market; private-response difference is the percentage difference in $\iota$. The 12-update endpoints are not certified retraining fixed points.}\label{tab:S8-E5}
\begin{tabular}{@{}lrrr@{}}\toprule
Group-1 adoption & End-point $J$ & High-model MSE gain & High versus low $\iota$\\\midrule
$0\%$ & 3.3489 & $3.73\%$ & $-8.36\%$\\
$50\%$ & 3.1737 & $2.78\%$ & $-6.40\%$\\
$100\%$ & 3.0647 & $2.08\%$ & $-4.90\%$\\\bottomrule
\end{tabular}
\end{table}

The time paths require a separate assessment of whether retraining has settled. The strict convergence diagnostic requires changes in $J$, prediction error, private-signal loading, and forecast variance to remain within a $0.5\%$ relative tolerance over three successive updates. None of the nine twelve-update paths passed that test. Moreover, an independent check took the three saved all-low update-12 profiles, generated fresh data \emph{without changing their deployed rules}, and fitted four independent next-step updates for each seed. The mean implied change in $J$ was $-0.383\%$, with a three-seed 95\% interval of $[-1.886\%,1.120\%]$. That interval does not fit inside the additional $\pm1\%$ practical stability band. Thus, even though late averages of some trading statistics change little, the saved all-low profiles do not pass the stated convergence diagnostics.

We retain these profiles because they provide fully reproducible starting models for a controlled comparison of how different investor groups respond to the same adoption shock. Their role is a \emph{common finite-learning baseline}, not an equilibrium selected by the population best-response fixed-point conditions of S3. There is no need to infer convergence to use a fixed saved profile as the starting point for a paired finite-horizon experiment.

\subsection{Which investors' retraining changes private-information aggregation?}\label{subsec:S8-groups}

Starting from each saved all-low update-12 profile, we now impose a new adoption share of $0$, $0.5$, or $1$ in group 1 and let prices and public histories adjust while \emph{all fitted parameters remain fixed}. This is the common stage-zero market for four subsequent comparisons. In one comparison nobody refits. In a second, group 1 updates both its GRUs while group 2's forecasting head remains frozen. In a third, only group 2 refits. In the fourth, both groups refit. For every active updating rule, we apply six simultaneous model updates and retain the seven deployed states from step zero through step six. Each arm starts with identical saved policies within a seed and adoption level; a non-updating group retains its exact parameter values even though the price histories it subsequently observes can differ.

A straightforward comparison between a refitted market at $x=1$ and its saved baseline would combine two effects: adjustment caused by adoption and ordinary residual learning that would have occurred even if adoption remained zero. Because the starting profiles are not exact fixed points, the second effect cannot be assumed away. We therefore measure how much \emph{additional} retraining is associated with the adoption change. For a market statistic $Y$ and a choice $a$ of groups permitted to refit, define
\begin{equation}\label{eq:S8-adoption-adjustment}
 R_a(x,n)=Y_a(x,n)-Y_{\varnothing}(x,0),\qquad
 T_a(x,n)=R_a(x,n)-R_a(0,n),
\end{equation}
where $\varnothing$ means that no group updates, $n$ is the number of slow learning updates, and $Y_a(x,0)$ is measured after the fixed-policy market has adjusted to adoption $x$. Subtracting the corresponding $x=0$ updating path removes the ordinary baseline drift under the same rule and horizon. It is an ordered finite-path comparison, not an assertion that the baseline is a retraining equilibrium.

The two groups can affect each other's learning through the endogenous prices. To retain that interaction rather than assign it arbitrarily, define
\begin{equation}\label{eq:S8-interaction}
 \mathcal I(x,n)=Y_{\mathrm{both}}(x,n)-Y_{1}(x,n)
                    -Y_{2}(x,n)+Y_{\varnothing}(x,n).
\end{equation}
The adoption-specific retraining effect when both groups update satisfies the exact accounting identity
\begin{equation}\label{eq:S8-group-identity}
 T_{\mathrm{both}}(x,n)=T_1(x,n)+T_2(x,n)
                  +\big[\mathcal I(x,n)-\mathcal I(0,n)\big].
\end{equation}
For $Y=J$, we express each term as a percentage of the same seed's original all-low $J$, then average across the three seeds. This common denominator makes the pieces additive and avoids treating the correlated arms as independent markets.

\begin{table}[htbp]\centering\small
\caption{Decomposition of the adoption effect on fresh-price precision after six updates. All entries are percentages of the same seed's original all-low $J$, averaged over three paired initial seeds. The interaction is the additional nonadditivity after subtracting its $x=0$ counterpart. The last row equals the sum of direct adoption, the two group effects, and their interaction.}\label{tab:S8-E6}
\begin{tabular}{@{}p{.61\linewidth}rr@{}}\toprule
Contribution & Group-1 adoption 50\% & 100\%\\\midrule
Direct adoption, coefficients fixed & $-3.40$ & $-6.75$\\
Additional retraining by group 1 alone & $-0.48$ & $+0.63$\\
Additional retraining by group 2 alone & $-1.83$ & $-3.00$\\
Additional interaction between the groups & $+0.48$ & $+0.65$\\\midrule
Both-group additional retraining & $-1.83$ & $-1.72$\\
Total: adoption and both-group retraining & $-5.23$ & $-8.47$\\\bottomrule
\end{tabular}
\end{table}

\begin{figure}[htbp]\centering
 \includegraphics[width=.98\linewidth]{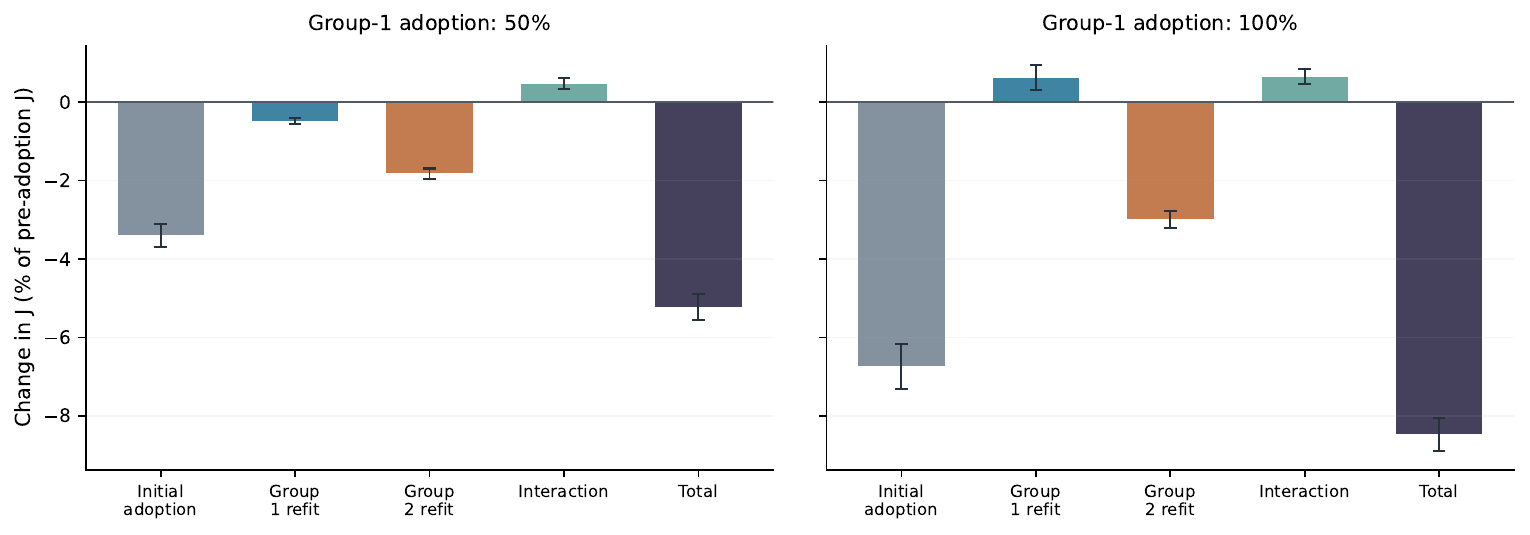}
 \caption{The adoption-related reduction in fresh-price precision and its group-level accounting. Each error bar is a descriptive two-sided 95\% Student-$t$ interval from three paired seed-level contrasts ($2$ degrees of freedom). The first four bars form an additive decomposition of the total; the fifth bar shows that total. Finite learning and the shared starting profiles, rather than a certified equilibrium, underlie these comparisons.}\label{fig:S8-groups}
\end{figure}

With no updating, adopting the wider GRU for all of group 1 reduces $J$ by $6.75\%$. Allowing only group 2 to refit adds another $3.00\%$ decline, after removing that group's ordinary $x=0$ learning. Allowing only group 1 to refit instead offsets the decline by $0.63\%$. When both groups learn, their interaction offsets part of group 2's effect; the total decline is $8.47\%$. At 50\% adoption, the total decline is $5.23\%$. The numbers depend on group sizes, signal quality, and finite optimization in this calibration, but they demonstrate a concrete feedback channel: investors whose computational capacity remains unchanged can alter their private-signal trading response because they refit to prices influenced by other investors' algorithms.

The interaction has two possible origins. The statistic $J=A^2/\sigma_\zeta^2$ is nonlinear even if changes in aggregate private-signal loading $A$ from the two groups are additive. In addition, when groups refit sequentially across learning epochs, the price histories on which they train may themselves differ. The original implementation separates the algebraic square-term from nonadditivity in the fitted aggregate $A$. The first contribution is small in this calibration, so the reported interaction mostly reflects the learning response, rather than merely squaring the sum of independent loadings. It should nonetheless remain an explicit term in the accounting identity.

\subsection{Mean forecasts, forecast risk, and the trading response}\label{subsec:S8-mean-risk}

Why does adoption or retraining reduce fresh-price precision? A trained investor responds to a private clue through $\iota_c=\beta_c/(\gamma_cv_c)$. A better public forecast can reduce the predictive coefficient $\beta_c$ on the private clue, making trades less responsive to it. The same improvement can reduce fitted forecast variance $v_c$, causing a risk-averse investor to take a larger position for a given signal. The first channel tends to lower $J$; the second tends to raise it. The net effect depends on the size of both changes.

We measure these contributions by replacing parts of two \emph{already fitted} forecasting systems while holding the inherited economic state, supply shock, and public history fixed. Let $Y_{ab}$ denote the resulting $J$, where $a=1$ replaces the full forecast-mean rule (including encoder and all mean coefficients), $b=1$ replaces the constant reported variance, and zero retains the old fitted component. The four combinations are $Y_{00}$, $Y_{10}$, $Y_{01}$, and $Y_{11}$. We calculate
\begin{align}\label{eq:S8-mean-risk}
 \mathcal M_Y&=\tfrac12\big[(Y_{10}-Y_{00})+(Y_{11}-Y_{01})\big],\\
 \mathcal V_Y&=\tfrac12\big[(Y_{01}-Y_{00})+(Y_{11}-Y_{10})\big],\nonumber\\[-4pt]
 \mathcal M_Y+\mathcal V_Y&=Y_{11}-Y_{00}.\nonumber
\end{align}
The two terms $\mathcal M_Y$ and $\mathcal V_Y$ are the mean and variance contributions. Averaging the two replacement orders allocates their interaction symmetrically. These quantities are mechanical \emph{post-fit} substitutions: they do not represent models trained under separate mean-only and variance-only restrictions, and the mixed fitted profiles are not alternative retraining equilibria.

\begin{table}[htbp]\centering\small
\caption{Mean and variance contributions to the effect on $J$ at update six. Entries are percentages of pre-adoption $J$ and averages across three seeds. Adoption rows refer to the first changed auction; additional-retraining rows subtract the same-arm $x=0$ path. Each pair sums to its net effect up to rounding.}\label{tab:S8-E7}
\begin{tabular}{@{}p{.52\linewidth}rrr@{}}\toprule
Group-1 adoption and comparison & Mean channel & Variance channel & Net effect\\\midrule
50\%: initial adoption & $-4.90$ & $+1.50$ & $-3.40$\\
50\%: both groups' additional retraining & $-2.60$ & $+0.77$ & $-1.83$\\
100\%: initial adoption & $-9.72$ & $+2.97$ & $-6.75$\\
100\%: only group 2 refits, additional & $-4.09$ & $+1.09$ & $-3.00$\\
100\%: both groups' additional retraining & $-2.40$ & $+0.68$ & $-1.72$\\\bottomrule
\end{tabular}
\end{table}

In both adoption comparisons, mean learning exerts the larger force and reduces private-information precision; the fall in predicted risk partly offsets it. Retraining adds contributions of the same kinds, with different sizes depending on which investors update. For Model A, structural loadings $A$, individual private-response coefficients, and $J$ depend on fitted coefficients rather than the distribution of historical encoder inputs. Consequently their changes can be decomposed at each update and summed across successive updates. The same calculation cannot be used to sum changes in \emph{stationary price variance}: changes in the law of past prices also affect that statistic. Price-variance substitutions are therefore treated only as one-auction diagnostics.

\subsection{Validation, statistical scope, and reproduction}\label{subsec:S8-validation}

The computation uses initial policies, market generators, training configurations, saved fitted models, and seed-indexed shock streams. Their validation checks address the order of events in the model. The current auction price and future payout never enter the encoder prematurely; individual signal noise cancels only at aggregate clearing; public-history windows remain inside independent paths; and a group's parameters do not change when its updates are disabled. For the six-update group experiment, all 198 distinct simulated states pass the saved-profile and generation-before-learning checks. The maximum clearing residual is $2.66\times10^{-15}$, the minimum aggregate downward clearing slope is $2.4142$, and the retained initial-price coupling error after burn-in is zero at numerical precision. Six engine checks and four additional accounting checks pass. The initial-deployment calculations satisfy the numerical checks for competitive clearing and consistent information timing.

The comparisons use three inherited initial learning seeds. We preserve the seed-level results, pair the same seeds across adoption rates and updating rules, and report descriptive intervals using a Student-$t$ reference distribution with two degrees of freedom. The number of simulated trading dates, training observations, model-update rounds, and correlated experimental arms does not enlarge the number of independent initial learning runs. We do not interpret small fixed-model Monte Carlo uncertainty as uncertainty about how a different random sample of learned algorithms would behave.

The model classes and numerical outcomes must also be kept distinct. The initial 48-date SSM/GRU experiment changes model width at an inherited market and reports both Model-A fresh-price precision and selected Model-B forecasting scores. The subsequent 16-date GRU experiment changes adoption shares and refits policies on endogenous market histories. Its twelve-update paths and six-update group comparisons remain finite-learning results, since the twelve-update checkpoints do not meet the strict settling and independent refit criteria. A correctly specified stationary Gaussian population equilibrium is constructed and analyzed in S7; its convergence and information formulas do not automatically carry over to these finite nonlinear forecasting markets.

Reproduction requires the stated SSM and GRU architectures, training splits, common random seeds, market generator, learning budgets, and simulated-price clearing rule. The group comparisons use matched initial models and independent evaluation paths, with all four retraining conditions fitted under the same design. Mean--variance substitutions are computed from the fitted heads at each update. Figure intervals are obtained from paired differences across the three initial training seeds.

The finite-learning studies show two complementary things. Better neural forecasts can coexist with a weaker fresh-price signal at first deployment, although not in every fitted seed. And the subsequent loss of private-signal aggregation can be amplified by retraining among investors who did not acquire the more capable encoder. These results concern the structural fresh-price channel and the sequence of fitted market rules. S9 uses the resulting neural markets to ask a different question: how valuable is the price to observers with different public-data processing capabilities, and how much does an individual investor gain by performing additional computation herself?

\clearpage\setcounter{table}{0}\setcounter{figure}{0}
\renewcommand{\thetable}{S9.\arabic{table}}\renewcommand{\theHtable}{S.9.\arabic{table}}
\renewcommand{\thefigure}{S9.\arabic{figure}}\renewcommand{\theHfigure}{S.9.\arabic{figure}}
\section{What neural-market prices tell investors, and the value of computing again}\label{sec:S9}

When more investors use powerful forecasting algorithms, the resulting prices contain a different mix of information. A trader who already runs an advanced algorithm may learn little from the current price, while a trader with a simpler algorithm may learn more. The price can also make a further upgrade less worthwhile: why pay to compute information that competitors have already partly conveyed through their trades? S5 establishes exact relationships for correct Gaussian forecasts, and S7 calculates them in a population Gaussian market. Here we examine both questions in the neural markets constructed in S8, without requiring the neural predictions or the resulting price distribution to be jointly Gaussian.

We take nine markets from S8: three initial training seeds, each evaluated at three adoption shares. In each market, the trading algorithms have already undergone six updates using prices generated by the market itself. We then hold all those algorithms fixed. Two additional investors, one with a small forecasting network and one with a larger network, learn separately from that market's data. Their holdings are too small to change its price. We can therefore ask what the \emph{same} observed price tells each investor and how much an investor gains by using the larger network, without changing the market being evaluated. The answers are measured with held-out dividend predictions and the investor's actual expected utility under the simulated economy.

\subsection{The nine markets and the investors observing them}\label{subsec:S9-markets}

The market used here is the nonlinear eight-factor economy defined in S8. Each date brings eight public economic indicators, a new noisy clue to each investor, and an uncertain next-period dividend. The public record available before trading also contains past prices and matured dividends. Investors in group 1, which has population mass $0.55$, use a gated recurrent unit (GRU) to summarize the preceding 16 public observations. Its smaller version has hidden width 4, and its larger version has width 32. Group 2, with mass $0.45$, trades using a private clue and the current quoted price but no public-history network. Their forecasts use Model-A constant-affine Gaussian heads after nonlinear public encoding. Both groups choose CARA-optimal positions from these forecasts, and their aggregate orders determine the clearing price as in S3.

Let $x$ denote the fraction of \emph{group 1} using the width-32 GRU. We study $x\in\{0,0.5,1\}$ under the three inherited seeds $18013$, $19031$, and $20047$. For every $x$ and seed, we use the market profile in which both groups have been allowed to refit for six learning updates (S8.6). These nine profiles are the endpoints of a finite learning procedure. Each is used as a \emph{fixed} generator of new prices and payouts during the measurement here; we do not claim that the learning procedure has reached a population fixed point.

A separate, negligible candidate investor belongs to group 1. She has the same risk aversion $\gamma_1=0.9$, private-signal noise variance $\nu_1=0.25$, and access to the prevailing current price as other members of that group. Her low-capacity option runs the width-4 public encoder; her high-capacity option runs the width-32 encoder. Both process the same 16-date public window. The options differ in what the trained networks can compute from that window, not in the raw financial observations they are permitted to see. The candidate's hypothetical orders are not inserted into clearing. Thus, within a market, her two computing choices face identical prices, fundamentals, private-clue realizations, and supply disturbances.

A new pair of candidate networks is fitted under \emph{each} of the nine market laws. The candidates start from the same seed-specific pre-adoption checkpoints across the three adoption shares, rather than inheriting the already updated networks used by incumbent traders. For each market, the low- and high-capacity candidates both receive $1{,}048{,}576$ independently simulated training observations and $4{,}096$ optimizer updates, using the same schedule and Gaussian negative log score. The inherited input normalization and the 16-date window are retained. A separate validation set is used during training. Once training ends, neither candidate's parameters change during the measurement. This design compares algorithms of different capacity after giving them a common opportunity to learn about the prevailing market, without attributing a candidate's behavior to a different incumbent retraining history.

The larger model has $4{,}260$ trainable parameters and approximately $66{,}080$ multiply--accumulate operations per forecast, compared with $200$ parameters and $2{,}884$ operations for the smaller model. These counts describe the specified implementation, not measured running time. Candidate calibration and outcome evaluation use additional, independent simulated paths. Table~\ref{tab:S9-protocol} records the principal distinctions.

\begin{table}[htbp]\centering\small
\caption{Market and candidate evaluation design. Each market is generated by its own frozen, six-update group-1/group-2 profile. The candidate investor never enters price clearing. The same primitive shocks and current prices are used when comparing candidates within a market.}\label{tab:S9-protocol}
\begin{tabularx}{\linewidth}{@{}p{.42\linewidth}X@{}}\toprule
Item & Implemented choice\\\midrule
Market profiles & Three inherited seeds $\times$ group-1 adoption $0\%,50\%,100\%$; both groups refitted for six updates\\
Candidate information & Width-4 or width-32 GRU of 16 pre-auction public observations; current price added at the forecasting stage; own private clue added for utility evaluation\\
Candidate learning & $512$ paths $\times$ $2{,}048$ observations; $4{,}096$ Adam updates for each capacity; independent validation paths\\
Affine-head calibration & $128$ paths $\times$ $2{,}048$ observations, separate from neural training and evaluation\\
Held-out evaluation & Four independent banks, each $256$ paths $\times$ $1{,}024$ dates; $512$ burn-in dates per path\\
Statistical comparisons & Three paired initial seeds; descriptive Student-$t$ intervals ($2$ degrees of freedom); additional path-level diagnostics\\\bottomrule
\end{tabularx}
\end{table}

\subsection{How much does the current price improve a dividend forecast?}\label{subsec:S9-E3-method}

Imagine a trader who is trying to forecast the dividend paid next period. She has already processed the previous 16 public observations using her GRU, obtaining a vector $z_{k,t}$, where $k\in\{L,H\}$ denotes the smaller or larger network. She can forecast using that vector alone, or she can also examine today's clearing price $p_t$. To isolate the extra predictive value of price, we fit two new, affine forecasting rules to her learned public representation: one uses only $z_{k,t}$, and the other adds $p_t$. Neither rule uses the trader's own private clue in this comparison. Both rules are fitted on the same independent calibration data and evaluated on the same held-out dividends.

Write $\widehat D_{k,t}^{\,0}$ for the prediction based on the public representation alone and $\widehat D_{k,t}^{\,p}$ for the prediction that also uses price. Specifically, the calibrated heads have the forms
\begin{equation}\label{eq:S9-E3-heads}
 \widehat D_{k,t}^{\,0}=a_{k,0}+b_{k,0}^{\top}z_{k,t},\qquad
 \widehat D_{k,t}^{\,p}=a_{k,p}+b_{k,p}^{\top}z_{k,t}+c_{k,p}p_t.
\end{equation}
The separate coefficients allow the best fitted affine readout to change when price is available. For each market $x$ and capacity $k$, let $M_{k}^{0}(x)$ and $M_{k}^{p}(x)$ denote the corresponding held-out mean squared errors for $D_{t+1}$. We measure the additional predictive value of seeing price as
\begin{equation}\label{eq:S9-E3-gain}
 G_k(x)=M_k^{0}(x)-M_k^{p}(x),\qquad
 T(x)=G_L(x)-G_H(x).
\end{equation}
We use \emph{absolute} dividend MSE reductions, not percentage reductions with different denominators. A positive $G_k$ means that the price helps that particular predictor. A positive $T$ means it helps the smaller predictor more. It has a useful exact interpretation:
\begin{equation}\label{eq:S9-E3-identity}
 T(x)=\big[M_L^{0}(x)-M_H^{0}(x)\big]
       -\big[M_L^{p}(x)-M_H^{p}(x)\big].
\end{equation}
Thus, $T>0$ says that observing price narrows the high--low prediction-error gap. Both terms on the right are calculated on the same market and the same observations. This is an empirical prediction-loss identity, not the conditional-mutual-information identity proved in S5.

The calibrated readouts here can be interpreted as two specified decision makers who can extract different information from a common price. An observer with the width-4 GRU may see in price some consequences of computations performed by other traders that her own network cannot reproduce. The width-32 observer may have already processed more of those public patterns and may therefore gain less from the same price. The price also contains information aggregated from private trading and random supply orders. Our MSE comparison measures its \emph{total} additional predictive usefulness to these particular processors; it does not separately identify the part due to shared public analysis and the part due to new private observations.

\subsection{Prices narrow the predictive advantage of a larger network}\label{subsec:S9-E3-results}

Table~\ref{tab:S9-E3} gives the price-related MSE reductions at the three adoption shares. At $x=0$, adding price lowers the smaller model's dividend MSE by $0.04795$, compared with $0.03212$ for the larger model. The price therefore closes $0.01583$ of the prediction-error gap. For example, without price the two held-out MSEs are $0.34887$ and $0.31509$; with price they fall to $0.30092$ and $0.28297$. The advanced model retains a predictive advantage, but the advantage is smaller when both observe the price.

\begin{table}[htbp]\centering\small
\caption{Additional value of the current price for dividend prediction. Each estimate averages three matched initial seeds; brackets give descriptive two-sided 95\% Student-$t$ intervals across those seeds. All entries are \emph{absolute reductions in dividend MSE}. The last column is the difference between the first two, not a separate information estimator.}\label{tab:S9-E3}
\begin{tabularx}{\linewidth}{@{}p{.18\linewidth}*{3}{>{\centering\arraybackslash}X}@{}}\toprule
Adoption by group 1 & Smaller $G_L$ & Larger $G_H$ & Difference $T=G_L-G_H$\\\midrule
$0\%$ & $0.04795$ & $0.03212$ & $0.01583$\\
 & $[0.04637,0.04953]$ & $[0.03175,0.03249]$ & $[0.01453,0.01713]$\\[2pt]
$50\%$ & $0.05136$ & $0.03081$ & $0.02056$\\
 & $[0.04935,0.05337]$ & $[0.03049,0.03113]$ & $[0.01876,0.02235]$\\[2pt]
$100\%$ & $0.05392$ & $0.02994$ & $0.02398$\\
 & $[0.05155,0.05629]$ & $[0.02973,0.03015]$ & $[0.02172,0.02623]$\\\bottomrule
\end{tabularx}
\end{table}

\begin{figure}[htbp]\centering
\includegraphics[width=.98\linewidth]{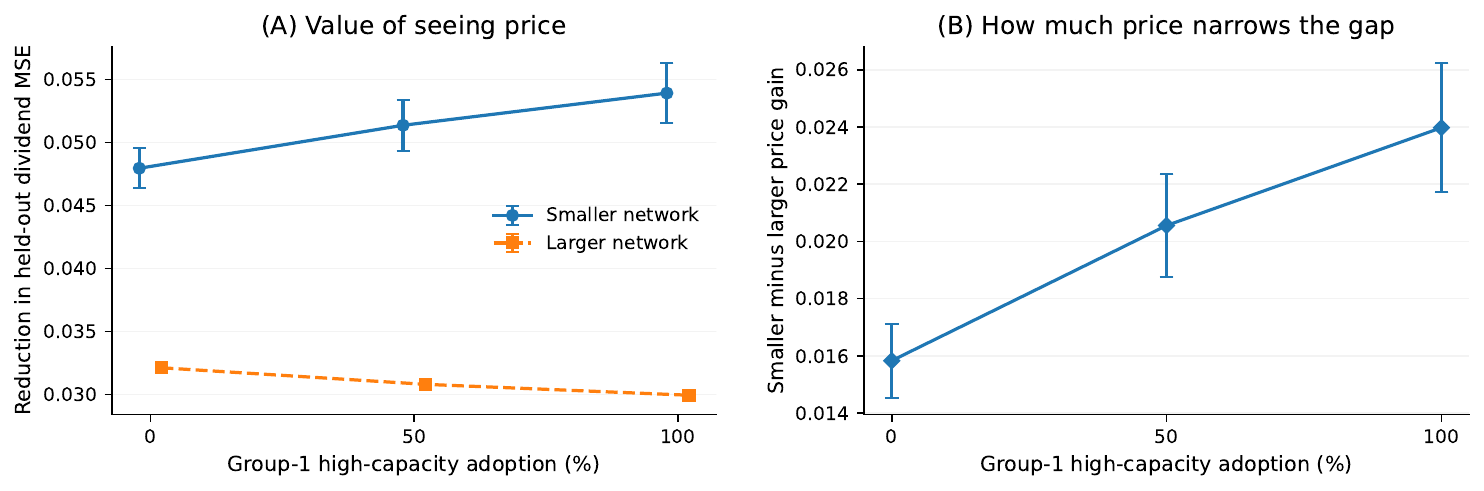}
\caption{What price adds to a smaller and a larger neural observer's forecast. \textbf{(A)} Held-out reduction in next-dividend MSE from including the current price. \textbf{(B)} The difference between the two reductions, which is also the amount by which price closes the difference in predictive MSE. Points are means across three training seeds; error bars are descriptive 95\% Student-$t$ intervals with two degrees of freedom. The underlying prices are generated by the same frozen six-update neural markets studied in S8.}\label{fig:S9-price-gain}
\end{figure}

As the share of group 1 using the wider GRU rises from zero to one, the price becomes more useful to the smaller observer: $G_L$ increases from $0.04795$ to $0.05392$, a $12.4\%$ increase. For the larger observer, $G_H$ instead falls from $0.03212$ to $0.02994$, a $6.8\%$ decrease. Consequently, the price-induced narrowing of the prediction gap rises to $0.02398$. The increase in $T$ between zero and full group-1 adoption is $0.00815$, with a paired-seed interval of $[0.00718,0.00911]$. Each of the three seed-level comparisons has the same sign. The intermediate adoption share also lies between the two endpoints (Fig.~\ref{fig:S9-price-gain}).

To check that the result is not solely an artifact of the larger encoder discarding something the smaller encoder retains, we also build a predictor from the concatenated vector $(z_L,z_H)$. It must run \emph{both} networks and then fit a new affine head; it is therefore more expensive than simply running width 32. The price-related difference between the smaller observer and this nested representation rises from $0.01603$ at $x=0$ to $0.02402$ at $x=1$. This confirms the reported pattern for a specifically implemented representation containing both public summaries. It does not make the separately trained width-32 representation automatically nested in the width-4 representation or establish the unrestricted value of all possible computations on the record.

\subsection{How much is an investor willing to pay for the larger network?}\label{subsec:S9-E4-method}

The previous test asks how well a price helps an observer forecast the dividend. The incentive to buy a stronger forecasting system is a different economic question. An investor choosing between two systems already sees the current price and also receives her own private signal. What she gains from the upgrade is not just lower prediction error: a change in perceived risk affects how aggressively she trades. We therefore evaluate the investment positions produced by the trained neural forecasts under the \emph{true} simulated payout process, rather than converting MSE reductions into utility with a Gaussian information formula.

For each capacity $k$, the investor uses the network's learned Gaussian mean $\widehat\mu_{k,t}(z_{k,t},s_{i,t},p_t)$ and constant fitted dividend variance $\widehat v_k>0$. Her price-taking risky holding is the CARA-optimal holding under that fitted belief,
\begin{equation}\label{eq:S9-E4-holding}
 h_{k,t}=\frac{\widehat\mu_{k,t}-R_fp_t}{\gamma_1\widehat v_k},
 \qquad \gamma_1=0.9.
\end{equation}
Both capacities see the same price and the same realization of the investor's private signal. For each $k$, define the ex-ante certainty equivalent of the resulting holding, normalizing initial wealth to zero, by
\begin{equation}\label{eq:S9-E4-CE}
 \CE_k(x)=-\frac{1}{\gamma_1}\log
 \E_{Q_x}\!\left[\exp\{-\gamma_1 h_{k,t}(D_{t+1}-R_fp_t)\}\right].
\end{equation}
Here $Q_x$ is the distribution generated by the given fixed neural-market profile, not a population-retrained Gaussian equilibrium. The expectation in \eqref{eq:S9-E4-CE} uses the true dividend outcomes under that market and the investor's \emph{fitted} holdings. In particular, it evaluates what a candidate actually gains from using a prediction-oriented learning rule, without assuming the rule maximizes true expected utility among all feasible algorithms.

In the simulation, $D_{t+1}=1.15\theta_t+\varepsilon_{t+1}$, where the independent dividend innovation has standard deviation $0.43$. To reduce evaluation noise, we integrate this innovation exactly while leaving the fitted holdings unchanged. Conditional on the pre-innovation simulated variables,
\begin{align}\label{eq:S9-E4-integration}
 &\log\E_{\varepsilon}\!\left[\exp\{-\gamma_1h_{k,t}(D_{t+1}-R_fp_t)\}\mid \theta_t,p_t,h_{k,t}\right]\\
 &\qquad=-\gamma_1 h_{k,t}(1.15\theta_t-R_fp_t)
           +\frac{\gamma_1^2h_{k,t}^2(0.43)^2}{2}.\nonumber
\end{align}
The remaining expectation across simulated public states, private clues, and prices is evaluated by a stable log-sum-exp calculation. The latent $\theta_t$ is available to the \emph{researcher} for this scoring operation only; it is never an input supplied to either candidate. We do not presume that the nonlinear neural-market price and payoff are jointly Gaussian.

Our measure of the gross private value of an upgrade is
\begin{equation}\label{eq:S9-E4-upgrade}
 U(x)=\CE_H(x)-\CE_L(x).
\end{equation}
This is the payoff-equivalent advantage of the fitted larger network to a negligible group-1 investor under one prevailing market law. It is evaluated before any charge for extra inference, training, or equipment. We compare $U(x)$ \emph{across} the nine fixed market profiles to measure how that advantage changes as other group-1 traders adopt larger networks and the incumbent models subsequently undergo six updates.

\subsection{The benefit of upgrading declines as larger networks spread}\label{subsec:S9-E4-results}

The larger network has a positive gross certainty-equivalent advantage in all three adoption environments (Table~\ref{tab:S9-E4}). At zero adoption among incumbent group-1 traders, $U(0)=0.02128$ in the model's one-period payoff units. At full adoption, $U(1)=0.01188$. The decline is $0.00940$, or $44.2\%$ of the initial benefit. The paired-seed interval for this change is $[-0.01051,-0.00829]$, and all three seed-level changes are negative. The mid-adoption result, $0.01595$, lies between the endpoints.

\begin{table}[htbp]\centering\small
\caption{Private value of a higher-capacity predictor under the fixed neural-market laws. The main column compares the learned neural heads and reports the high-minus-low true exponential-utility certainty equivalent. The other columns use separately fitted affine heads as checks. Means are across three initial seeds; bracketed intervals are descriptive 95\% Student-$t$ intervals across those seeds. All values are gross of computation costs.}\label{tab:S9-E4}
\begin{tabularx}{\linewidth}{@{}p{.18\linewidth}*{3}{>{\centering\arraybackslash}X}@{}}\toprule
Adoption by group 1 & Learned neural head & Recalibrated affine head & Both encoders and affine head\\\midrule
$0\%$ & $0.02128$ & $0.02129$ & $0.02145$\\
 & $[0.01900,0.02356]$ & $[0.01906,0.02352]$ & $[0.01915,0.02375]$\\[2pt]
$50\%$ & $0.01595$ & $0.01596$ & $0.01597$\\
 & $[0.01428,0.01763]$ & $[0.01434,0.01759]$ & $[0.01433,0.01761]$\\[2pt]
$100\%$ & $0.01188$ & $0.01189$ & $0.01192$\\
 & $[0.01069,0.01307]$ & $[0.01075,0.01303]$ & $[0.01078,0.01305]$\\\bottomrule
\end{tabularx}
\end{table}

\begin{figure}[htbp]\centering
\includegraphics[width=.83\linewidth]{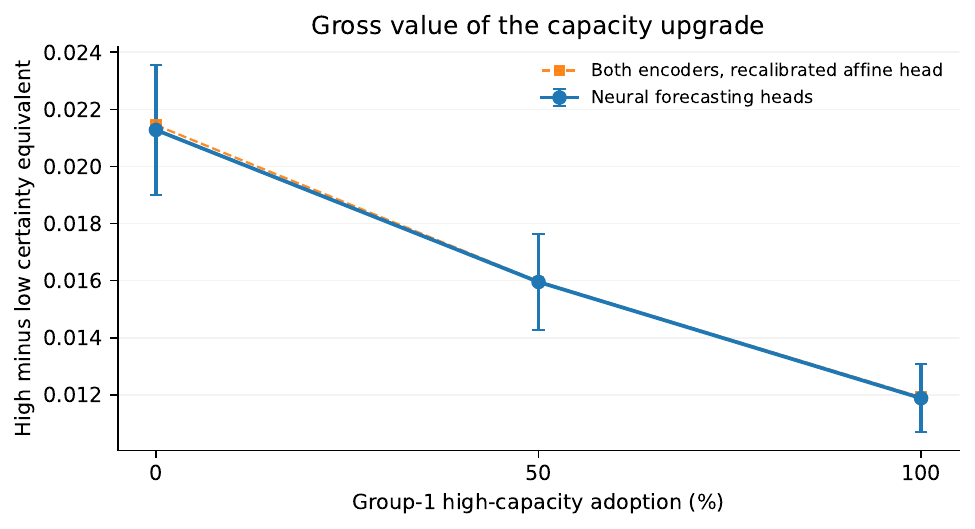}
\caption{Gross certainty-equivalent benefit from using the larger GRU, evaluated by true ex-ante exponential utility under each market. Candidates are fitted separately to the same market law at each adoption rate, see the same current price and private clue, and do not affect clearing. Points are means of three paired initial seeds; error bars are descriptive 95\% Student-$t$ intervals ($2$ degrees of freedom). This numerical utility comparison is not derived by inserting neural forecast errors into the Gaussian information identity of S5.}\label{fig:S9-upgrade-value}
\end{figure}

The reduction in the value of computation fits the economic interpretation developed in S5: when more trading is based on sophisticated public-data processing, the price can substitute for some calculations that another investor would otherwise perform herself. The current numerical exercise also shows that the limited observer obtains \emph{more} incremental predictive value from price as adoption rises, even though the incremental price value for the high-capacity observer falls. Both observations are made within the same nine neural markets. The two measures thus offer complementary evidence about how other investors' computing choices alter the information that a price provides and the incentive to compute again.

We repeat the utility comparison after refitting an affine forecast mean and constant residual variance on a separate calibration set, instead of using the original learned neural head. The recalibrated high-minus-low benefit falls from $0.02129$ to $0.01189$. When the higher alternative additionally processes both public representations, the corresponding benefit falls from $0.02145$ to $0.01192$. These checks show that the pattern is not sensitive to modest recalibration of the final readout or to using a public representation that explicitly retains the smaller model's information. They require their own additional fitting and computation. Neither is the primary width-32-versus-width-4 neural comparison.

\subsection{Independent evaluation and statistical checks}\label{subsec:S9-validation}

Every market is evaluated using four new banks of simulated paths that are independent of candidate training, validation, and head calibration. Each bank contains 256 paths and 1,024 usable trading dates after burn-in, yielding $1{,}048{,}576$ evaluated observations per market. Market price histories are recomputed from frozen incumbent rules and a seed-controlled stream of exogenous shocks. Across adoption comparisons, the same primitive draws are used while the market price is allowed to respond endogenously. Within each market, the low- and high-capacity candidates receive the same evaluation observations, which makes their differences paired rather than comparisons of different macroeconomic realizations.

We distinguish uncertainty about the initial learned algorithms from uncertainty in evaluating a fixed pair. The displayed intervals first compute the candidate difference for each of three inherited training seeds, then report a two-sided Student-$t$ interval with two degrees of freedom. These are descriptive intervals based on only three independent initial fits. Separately, we resample entire evaluation paths within each of the four banks, pairing low- and high-capacity outcomes and using 1,000 bootstrap repetitions. Those path-level intervals quantify Monte Carlo evaluation variability \emph{conditional on the fitted models}; they do not create additional independent training seeds. Independent-bank and first-half/second-half evaluations are also retained to check that the reported utility advantage is not concentrated in a small time segment.

The exponential-utility calculation is sensitive to tail observations because it averages exponentials rather than payoffs themselves. We therefore inspect the normalized contribution of each evaluation outcome to the exponential average. Across the 18 primary neural capacity--market evaluations, the smallest effective weight count is $947{,}703$ out of $1{,}048{,}576$ observations. The largest contribution from one observation is below $0.001\%$; the largest combined share from the top $0.1\%$ of observations is approximately $0.30\%$. These checks show limited tail concentration in the simulated banks. They do not prove that the exponential moment is finite in every possible continuation of a nonlinear market.

The source-code tests validate the measurement pipeline independently of the market's substantive result. They check the dividend-noise integration against numerical Gaussian quadrature, recover a known Gaussian benchmark, return zero utility for zero holdings, verify numerical stability of the exponential averaging, and confirm the forecast-error difference identity and joint-encoder fitting. The measurement checks confirm the forecast-error difference identity, numerical stability of exponential utility, the separation of learning and evaluation data, and the absence of candidate effects on market clearing. The maximum saved market-clearing residual across the nine evaluation profiles is $1.78\times10^{-15}$. The calculations use the nine market profiles, separately fitted candidate predictors, and independent evaluation banks described in this section.

\subsection{What these measurements establish}\label{subsec:S9-scope}

The results show a concrete pattern in markets whose prices arise from fitted neural forecasts and competitive trading. As wider algorithms become more common, the current price adds more to a limited observer's dividend forecast and less to the forecast of an advanced observer. Under those same market laws, a negligible investor's gross payoff-equivalent gain from adopting the advanced algorithm falls. The additional value of price and the value of an upgrade are measured separately, but on a common set of market profiles and independent evaluation data.

Three distinctions locate these findings within the paper. First, the predictive measure $T$ is an MSE-based comparison of specified implemented forecasts. It is not the conditional mutual information of S5, and the low- and high-capacity neural summaries are not automatically nested. Second, $U$ is a true-utility evaluation of trading rules \emph{trained to predict payouts}; it is not the highest possible utility over all feasible computational programs, and it excludes computing costs. Third, the six-update profiles are fixed finite-learning markets, not certified population-retrained equilibria. The theoretical Gaussian results in S5--S7 establish exact information relationships and equilibrium examples under their stated assumptions; the present experiment shows analogous response patterns for nonlinear public encoders and finite learning in a market whose prices still clear investors' orders.

S10 examines how these finite-market patterns respond to changes in training budgets, learning dynamics, and economic specifications, and records the computational details needed to reproduce the underlying experiments.

\clearpage\setcounter{table}{0}\setcounter{figure}{0}
\renewcommand{\thetable}{S10.\arabic{table}}\renewcommand{\theHtable}{S.10.\arabic{table}}
\renewcommand{\thefigure}{S10.\arabic{figure}}\renewcommand{\theHfigure}{S.10.\arabic{figure}}
\section[Robustness, learning dynamics, and reproducibility]{Robustness, learning dynamics, and computational reproducibility}\label{sec:S10}

The neural-market experiments in S8--S9 show that more capable forecasting algorithms can change both the information that investors extract from prices and their incentive to upgrade their own models. These results come from trained models in an endogenously generated market, so their interpretation also depends on how the models were trained and evaluated. New training samples may produce different forecasts; alternative prediction rules may extract more from the same price; and repeated retraining may continue to change the market.

This section examines those questions in the same computational economy. We first vary how the two candidate forecasting models are trained and how their predictions are read out, while leaving the nine markets of S9 unchanged. We then extend the market's own retraining process and distinguish fluctuations caused by new training data from the behavior obtained when the same random shocks are reused. Finally, we change two assumptions about the asset's payout and repeat the immediate-deployment and frozen-policy comparisons. The experiments share the CARA trading rule and competitive clearing developed in S3, but they change different parts of the economy; Table~\ref{tab:S10-map} shows exactly what stays fixed in each one.

\begin{table}[htbp]\centering\small
\caption{The three additional numerical experiments. The adoption rate $x$ is the fraction of group 1 using the width-32 rather than width-4 GRU; group 1 represents $55\%$ of investors.}\label{tab:S10-map}
\begin{tabularx}{\linewidth}{@{}p{.15\linewidth}X X@{}}\toprule
Experiment & What changes & What remains fixed\\\midrule
Training and prediction & Training length and random training samples for negligible candidates; supervised readout used to evaluate the price & Nine six-update markets (three seeds $\times$ $x=0,0.5,1$), incumbent trading rules, and economic primitives\\
Market learning & More rounds of refitting by market participants; newly drawn versus reused primitive shocks & Three original market seeds, underlying economy, model classes, and per-update learning budget\\
Payout laws & Whether payouts are aligned with private clues and whether payoff noise varies with a public state & Market clearing, model classes, matched primitive draws, and a common first-deployment/frozen-history protocol\\\bottomrule
\end{tabularx}
\end{table}

\subsection{Do the price-information results depend on the candidate's training procedure?}\label{subsec:S10-learning}

Recall the two measurements in S9. In a fixed market, a negligible candidate investor processes public history with either the small or large GRU. For the prediction experiment, let $G_L(x)$ and $G_H(x)$ be the reductions in held-out dividend mean squared error when today's price is added to the respective forecasts. Their difference is
\begin{equation}\label{eq:S10-T}
 T(x)=G_L(x)-G_H(x).
\end{equation}
A larger $T$ means that observing price closes more of the predictive gap between the two public-data processors. In the separate investment experiment, the candidate sees the same price and the same private signal under either computing choice. Let $U(x)=\CE_H(x)-\CE_L(x)$ be the gross certainty-equivalent benefit of the larger network, evaluated under the actual simulated payoff distribution. The endpoints of interest are
\begin{equation}\label{eq:S10-changes}
 \Delta T=T(1)-T(0),\qquad \Delta U=U(1)-U(0).
\end{equation}
These are changes across adoption shares within group 1, not information quantities measured in nats. $T$ is a difference in dividend MSE reductions; $U$ is a monetary certainty equivalent in the units of the simulated claim.

We begin with the same nine fixed markets measured in S9: three initial market seeds and adoption shares of $0\%$, $50\%$, and $100\%$. In those markets, both investor groups have already completed six policy updates. For each market we separately train two small/large \emph{candidate} networks. These candidates observe but do not change the prevailing price. We then vary the candidates' training while generating fresh, common evaluation paths for every comparison. The market participants' parameters, their prices, and the economic calibration never change within this experiment.

The original candidate training uses $1{,}048{,}576$ windows and $4{,}096$ Adam updates for each capacity, starting from the common seed-specific inherited checkpoints described in S9. Two conditions add one or three further training blocks of the same size to these candidates. Each block uses new samples from the \emph{same fixed market law} and restarts the original optimizer schedule. Two more conditions repeat the original training budget with different samples, validation observations, and optimizer shuffling, again starting from the inherited checkpoints. Thus the comparisons change the implemented finite-learning procedure; additional blocks do not guarantee a global predictive optimum.

The original forecasting readouts are recalibrated on an independent set of $262{,}144$ observations. All trained choices are evaluated on four new banks, each containing $256$ paths and $1{,}024$ usable dates after burn-in. The $1{,}048{,}576$ held-out observations per market are generated from identical primitive shock streams for the matched candidate comparisons. For the utility experiment, the forecasts' learned means and variances determine CARA positions, and expected exponential utility is evaluated under the true payout law, integrating the independent Gaussian dividend shock analytically. This is the same measurement principle as in S9.

\begin{table}[htbp]\centering\small
\caption{Changing the learning procedure does not change the direction of the adoption comparison in the nine fixed neural markets. Entries are means and descriptive $95\%$ Student-$t$ intervals from the three original market seeds. Candidate training variants share the same market laws and are not independent market replications.}\label{tab:S10-learning}
\begin{tabularx}{\linewidth}{@{}p{.23\linewidth}X X@{}}\toprule
Candidate training & $\Delta T$: low-minus-high price value & $\Delta U$: upgrade value\\\midrule
Original training & $0.00814\ [0.00710,0.00918]$ & $-0.00943\ [-0.01056,-0.00831]$\\
One extra block & $0.00813\ [0.00704,0.00922]$ & $-0.00942\ [-0.01054,-0.00830]$\\
Three extra blocks & $0.00809\ [0.00707,0.00912]$ & $-0.00943\ [-0.01056,-0.00830]$\\
New sample, first refit & $0.00810\ [0.00715,0.00905]$ & $-0.00942\ [-0.01053,-0.00832]$\\
New sample, second refit & $0.00809\ [0.00721,0.00898]$ & $-0.00941\ [-0.01051,-0.00832]$\\\bottomrule
\end{tabularx}
\end{table}

Across these five training conditions, the difference in the price's predictive value increases by approximately $0.0081$ between zero and full adoption within group 1, while the gross private value of a high-capacity upgrade falls by about $0.0094$. All fifteen seed-by-training-condition endpoint comparisons show those two signs. These are correlated comparisons based on three initial markets; repeating candidate fits does not turn them into fifteen independent draws of an economy. The stability nevertheless indicates that the reported effects are not especially sensitive to the particular continuation length or the two additional training samples examined here.

\subsection{Can a different forecasting readout change the price comparison?}\label{subsec:S10-decoders}

The forecasting experiment in S9 evaluates what a price adds to a forecast after each candidate GRU has processed the public history. Its principal comparison uses separately fitted affine readouts of the GRU representation, with and without the current price. This gives a clear prediction-loss measure, but the readout is itself a limited forecasting procedure. We therefore keep the trained encoders fixed and allow the final forecasting rule to use nonlinear transformations of their outputs.

The alternative rules use random Fourier features (RFFs), a finite collection of sinusoidal transformations of the input representation. We fit versions with $64$ and $128$ random features and with corresponding price-interaction features. In each case, the price-augmented predictor retains all the features available without price. A ridge penalty is selected from $10^{-6},10^{-5},10^{-4},10^{-3},10^{-2},0.1,1$ using an independent validation bank. The feature draws and the set of penalties are fixed before evaluation. To compare readouts on equal terms, these additional models and an affine-ridge control are calibrated using the same first $65{,}536$ calibration observations. The full-data affine least-squares readout of S9 remains the primary result.

There are two comparisons of public processing. The first uses the original small and large GRU representations separately. The second gives the high-capacity option the \emph{combined} small and large representations, so that it explicitly retains the smaller model's computed features. This latter option consumes both encoders and extra readout computation. It is a useful implementation of nested candidate information, but it is not the same computational technology as the large GRU alone.

\begin{table}[htbp]\centering\small
\caption{Robustness of the price's relative predictive value to the readout. Each entry is the mean change in $T$ from $0\%$ to $100\%$ adoption, with a descriptive $95\%$ interval over the three inherited seeds. For this matched fitting-budget exercise the affine row is a ridge control, not the full-calibration OLS baseline in S9.}\label{tab:S10-decoder}
\begin{tabularx}{\linewidth}{@{}p{.25\linewidth}X X@{}}\toprule
Forecast readout & Small GRU versus large GRU & Small GRU versus combined representations\\\midrule
Affine ridge & $0.00814\ [0.00708,0.00920]$ & $0.00793\ [0.00680,0.00906]$\\
64 random features & $0.00811\ [0.00686,0.00936]$ & $0.00791\ [0.00653,0.00930]$\\
128 random features & $0.00805\ [0.00680,0.00929]$ & $0.00777\ [0.00650,0.00904]$\\\bottomrule
\end{tabularx}
\end{table}

The price reduces the high--low forecast gap more at higher adoption in all nine seed-by-decoder comparisons for the separate encoders, and in all nine comparisons for the combined representations. The size varies modestly with decoder flexibility. This strengthens the interpretation that the result is not specific to a single affine readout. The readouts remain finite supervised prediction rules; their errors are not an estimate of the unrestricted Bayes risk of the encoded information.

\subsection{What happens when the market keeps retraining?}\label{subsec:S10-dynamics}

So far, the nine neural markets have been held fixed while negligible candidates learn from them. We now return to the \emph{market participants} themselves. S8 follows the initial deployment of the width-32 GRU and six rounds in which both investor groups refit forecasts using prices that their trading has generated. Here we extend those same nine market-learning histories to twelve updates. Each new forecast changes investors' demand schedules, the new schedules determine prices, and the resulting public histories become the next training sample. The per-update training budget and market primitives stay fixed.

Two sources of variation matter. In the first set of continuations, fresh random training data are generated at each update. This resembles an implemented learning procedure that repeatedly draws observations from its current market. In a second set, conducted at $50\%$ adoption for each of the three seeds, the same primitive random shocks and the same optimizer ordering are reused at each update. Prices and public histories are \emph{still recomputed} after every policy change. Reusing the shocks removes one source of sampling variation; it does not freeze the market record or replace competitive clearing with a prescribed price equation.

To assess the resulting market behavior we examine the fresh-price likelihood precision $J=A^2/\sigma_\zeta^2$ and the risk-adjusted private-clue response of each active forecasting rule, denoted $\iota_L,\iota_H,\iota_O$ for the small GRU, large GRU, and other investor group. These coefficients are derived from fitted demand heads as in S3 and S8. Our operational convergence screen requires the relative changes in $J$ and all three $\iota$ coefficients to be below $0.5\%$ on \emph{each} of the final three updates. At the endpoint adoption shares, an inactive type still has a stored candidate rule; we also check active-rule versions of the screen.

\begin{table}[htbp]\centering\small
\caption{Behavior between the sixth and twelfth market-learning updates. Changes in $J$ are relative to update 6. Intervals are descriptive $95\%$ Student-$t$ intervals from three initial market seeds. The settling screen requires all final three relative changes in $J$ and private-response coefficients to be below $0.5\%$.}\label{tab:S10-dynamics}
\begin{tabularx}{\linewidth}{@{}p{.26\linewidth}X p{.21\linewidth}p{.15\linewidth}@{}}\toprule
Learning samples / adoption & Change in $J$ at update 12 ($\%$) & Passes settling screen & Period-two flags\\\midrule
Fresh samples, $x=0$ & $0.352\ [-0.790,1.495]$ & $0/3$ & $0/3$\\
Fresh samples, $x=0.5$ & $0.388\ [-0.800,1.577]$ & $0/3$ & $0/3$\\
Fresh samples, $x=1$ & $0.424\ [-0.654,1.502]$ & $0/3$ & $0/3$\\
Reused shocks, $x=0.5$ & $0.169\ [-0.823,1.160]$ & $3/3$ & $0/3$\\\bottomrule
\end{tabularx}
\end{table}

The mean change in $J$ from update 6 to update 12 is small in all four comparisons. Yet none of the nine fresh-sample continuations passes the stricter final-update test. All three reused-shock continuations pass. The difference tells us that the continuing variation in fitted coefficients is sensitive to the random training samples. It does not establish that the fresh-sample procedures would diverge indefinitely or that the reused-shock procedures converge to the population-optimal market law.

\begin{figure}[htbp]\centering
\includegraphics[width=.96\linewidth]{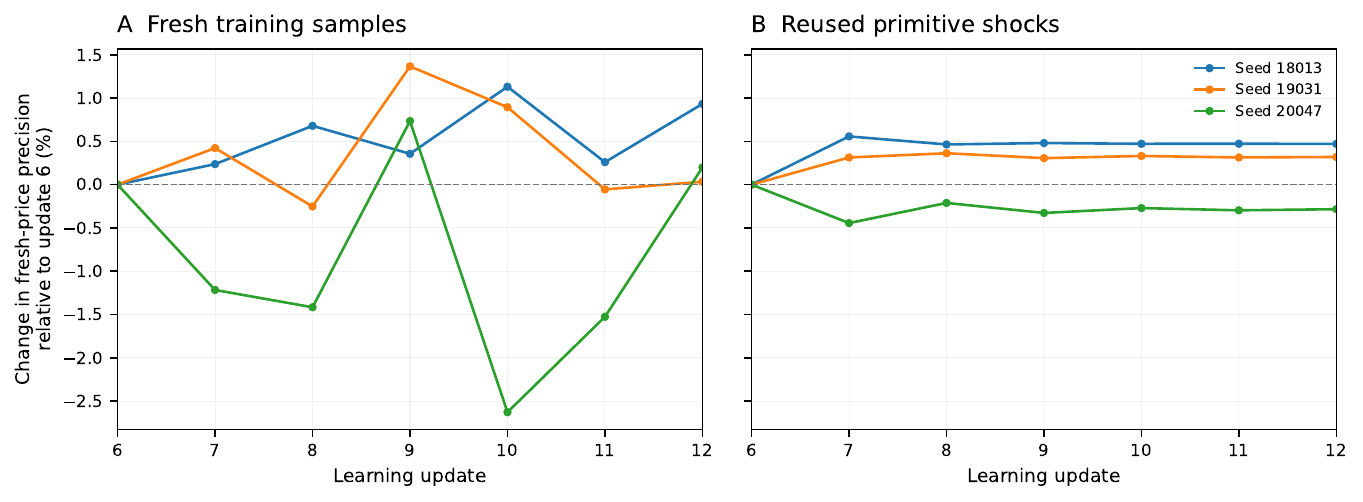}
\caption{Continued market learning with $50\%$ adoption within group 1. Each line is one of the three inherited market seeds, expressed as the change in fresh-price precision $J$ relative to that seed's value at update 6. In panel A, each learning update draws new training observations. In panel B, primitive shock streams and optimizer order are reused, while price histories are recomputed under the updated trading rules. The smaller fluctuations in panel B do not by themselves establish population-equilibrium convergence.}\label{fig:S10-dynamics}
\end{figure}

We also checked for a stable alternation between two patterns of learning, rather than convergence to one. The screen requires at least four sign reversals in the six observed steps of $J$, a two-step return distance below one quarter of the average one-step change in the normalized vector $(J,\iota_L,\iota_H,\iota_O)$, and a median one-step change exceeding $0.5\%$. None of the twelve trajectories passes this period-two screen. This is evidence against a clear two-cycle in the \emph{observed twelve-update trajectories}; it does not exclude cycles with longer periods or later onset.

\subsection{Do small changes in aggregate prices mean that the forecasting rules have stopped changing?}\label{subsec:S10-functional}

Market-level precision may appear nearly constant even while the individual forecasting rules change. To investigate that possibility, we take the saved twelve-update profiles at $50\%$ adoption and fit each one \emph{four more times}, independently, using new draws while holding the inherited market law fixed. Each refit begins from the same stored profile. It is therefore a one-step diagnostic of variation around that profile, not the next step of a single evolving market trajectory.

For each seed we construct intervals for the expected one-step changes in $J$ and the three private-response coefficients. Under a tolerance requiring the full $95\%$ interval to lie inside $\pm1\%$ of its starting value, only four of the twelve seed-by-metric intervals pass. None of the three twelve-update profiles passes the criterion for all four metrics. In particular, the other investor group's private-response coefficient continues to drift downward in all three seed-specific intervals. This provides a more demanding check than comparing the average values of $J$ across the last few updates.

We separately evaluate the learned \emph{functions}, not just their aggregate demand slopes. Each saved policy is applied to the same held-out histories, prices, and private signals drawn under the sixth-update law. This isolates how the fitted forecasts and submitted orders change when model parameters are updated. Table~\ref{tab:S10-functional} reports the size of those changes for the last three updates at $50\%$ adoption, averaged over the three initial seeds. Forecast changes are scaled by the standard deviation of the dividend; order changes are scaled by each group's initial root mean squared holding.

\begin{table}[htbp]\centering\small
\caption{Changes in implemented forecasts and orders across the last three learning updates, evaluated on the same held-out price histories. Values are root mean squared changes in percent of the indicated reference scale, averaged over three seeds at $50\%$ adoption.}\label{tab:S10-functional}
\begin{tabularx}{\linewidth}{@{}p{.25\linewidth}X X X X@{}}\toprule
 & \multicolumn{2}{c}{Fresh samples} & \multicolumn{2}{c}{Reused shocks}\\
Forecasting rule & Forecast & Holding & Forecast & Holding\\\midrule
Small GRU & $1.3832\%$ & $4.4078\%$ & $0.0523\%$ & $0.1672\%$\\
Large GRU & $1.6467\%$ & $5.0188\%$ & $0.1387\%$ & $0.4235\%$\\
Other group & $0.2584\%$ & $0.7229\%$ & $0.0253\%$ & $0.0730\%$\\\bottomrule
\end{tabularx}
\end{table}

Under new training samples, fitted order schedules continue changing by several percent for the investors using the GRUs. These changes are much smaller when shocks are reused. The calculation evaluates every policy on one specified anchor distribution; it does not establish equality or stability of the rules on every possible history. Together, the trajectory, fixed-law, and functional checks show why finite neural retraining results should be reported as outcomes of specified learning procedures rather than certified population equilibria.

\subsection{Do the results survive changes in what determines the asset payout?}\label{subsec:S10-payoffs}

Our final experiment changes the process generating dividends. The aligned-payoff assumption is especially useful in S4 because the private clue and the conditional expected payout concern the same scalar fundamental. Actual assets can also depend on other economic forces, and the risk of an asset may change with conditions investors can observe. We examine both departures while retaining the competitive market and limited forecasting technologies.

The comparison includes three economies. The \emph{matched control} pays a multiple of the privately signaled fundamental $\theta_t$, together with independent dividend noise of variance $0.43^2$. In the \emph{nonaligned} economy, an additional, unit-variance latent factor $\psi_t$, contemporaneously orthogonal to $\theta_t$, accounts for $20\%$ of the unconditional fundamental contribution to dividend variance. If $P_\theta=\V(\theta_t)$, the conditional payout mean is
\begin{equation}\label{eq:S10-nonaligned}
 \E[D_{t+1}\mid f_t]=\lambda\!\left(\sqrt{0.8}\,\theta_t+\sqrt{0.2P_\theta}\,\psi_t\right).
\end{equation}
The total unconditional variance of the fundamental payout component is matched to the control. Investors' individual clues remain noisy observations of $\theta_t$, so the extra factor affects dividends without being added to those clues.

In the \emph{variable-risk} economy, the payoff mean remains $\lambda\theta_t$, but the conditional dividend-noise variance is
\begin{equation}\label{eq:S10-risk-law}
 \V(\varepsilon_{t+1}\mid m_t)=0.43^2\left[1+0.6\tanh\!\left(\frac{m_{0,t}}{\operatorname{SD}(m_{0,t})}\right)\right].
\end{equation}
Here $m_{0,t}$ is the first publicly released macroeconomic indicator. The multiplier stays between $0.4$ and $1.6$ and has unconditional mean one by symmetry. Investors can therefore observe a variable related to true dividend risk, but the deployed forecasting heads still use Model A's single fitted variance per type. This is an experiment with misspecified constant risk forecasts; it is not an experiment with the state-dependent forecasting heads of Model B.

We run the \emph{same sequence of market comparisons} in all three economies. In each, small- and large-capacity models and the second investor group's model are fitted using a common inherited all-low market. We then change the fraction of group 1 using the larger model. The first auction reflects the new mix of fitted demand schedules. For the subsequent price-information and investment measurements, we keep the deployed policies fixed while their generated price histories adjust to a new market law. At each of these resulting laws, we refit negligible candidate low- and high-capacity models and evaluate them using the same procedures as S9. Market participants do not jointly retrain again in this experiment. The matched control uses this same timing, so its results can be compared to the two changes in the dividend law, but not directly to the market levels after six rounds of joint retraining.

Table~\ref{tab:S10-extensions} displays three endpoint responses. The first is the percentage change in first-auction fresh-price precision $J$ when all of group 1 upgrades. The other two are the changes, after fixed-policy history adjustment, in the price-induced prediction-error narrowing $T$ and the gross capacity value $U$ of a negligible candidate. The three statistics consequently refer to \emph{different stages} of the comparison, even though they share the same economic primitives and adoption levels.

\begin{table}[htbp]\centering\small
\caption{Relaxing two dividend assumptions. The control and extensions use the same inherited-law fitting and frozen-policy adjustment protocol. Each cell shows the mean $0\%$-to-$100\%$ adoption change, followed by a descriptive $95\%$ Student-$t$ interval across three paired seeds. $J$ concerns $\theta$, not the entire dividend in the nonaligned economy.}\label{tab:S10-extensions}
\begin{tabularx}{\linewidth}{@{}p{.255\linewidth}>{\centering\arraybackslash}X>{\centering\arraybackslash}X>{\centering\arraybackslash}X@{}}\toprule
Economy & Initial $J$ change ($\%$) & $\Delta T$ (dividend MSE) & $\Delta U$ (certainty equivalent)\\\midrule
Matched aligned control & \shortstack{$-6.645$\\$[-7.497,-5.793]$} & \shortstack{$0.00962$\\$[0.00883,0.01041]$} & \shortstack{$-0.01113$\\$[-0.01217,-0.01008]$}\\[3pt]
Nonaligned payout & \shortstack{$-4.386$\\$[-10.599,1.827]$} & \shortstack{$0.01135$\\$[0.01036,0.01233]$} & \shortstack{$-0.01140$\\$[-0.01789,-0.00491]$}\\[3pt]
Public-state-dependent risk & \shortstack{$-6.637$\\$[-7.564,-5.709]$} & \shortstack{$0.00968$\\$[0.00904,0.01032]$} & \shortstack{$-0.01095$\\$[-0.01192,-0.00998]$}\\\bottomrule
\end{tabularx}
\end{table}

Both extended economies retain the positive change in $T$ and the decline in the gross private upgrade value in these three-seed comparisons. The precision response $J$ also falls in every nonaligned seed, but its magnitude varies enough that the interval for the mean extends across zero. More fundamentally, because private signals remain signals about $\theta$, $J$ continues to quantify the price's fresh likelihood precision about $\theta$; it no longer describes the price's information about the \emph{whole dividend}. By contrast, The forecasting exercise predicts the full dividend, while the utility exercise evaluates the true payoff distribution. These outcomes are therefore useful even when dividend alignment is relaxed.

The gross value of an upgrade falls between $0\%$ and $100\%$ group-1 adoption from $0.02134$ to $0.01021$ in the matched control, from $0.02835$ to $0.01695$ with nonaligned payoffs, and from $0.02110$ to $0.01015$ with state-dependent payoff noise. These are model-unit certainty equivalents, net of neither hardware costs nor other implementation expenses. A three-seed simulation of two departures from the payoff model cannot establish a universal sign for all nonaligned or nonlinear trading environments; it shows that these particular effects do not rely on the two restrictions in these calibrations.

\subsection{Training, evaluation, and numerical checks}\label{subsec:S10-validation}

For the learning and payout extensions, we preserve the same within-date information timing as in S1--S3. The GRU processes a 16-date public window, excluding the current clearing price and future dividend. The final affine forecast head adds a candidate current price and the investor's own private clue to form the demand schedule. Idiosyncratic private-clue errors cancel in the aggregate continuum clearing rule. Once the market price is selected, it becomes part of future public histories. In experiments evaluating a negligible candidate, that candidate's hypothetical holdings are excluded from the market-clearing equation.

The additional payout experiments use $512$ training paths of $2{,}048$ windows per refit, $4{,}096$ Adam updates, an independent $262{,}144$-observation calibration sample, and four evaluation banks totaling $1{,}048{,}576$ held-out observations per frozen market law. Primitive shock streams are paired across dividend laws and adoption levels. The optimization, normalization, and data-window rules remain those of S8--S9. The nonaligned factor enters evaluation of the true dividend, not a forecaster's feature vector. For the variable-risk economy, the independent Gaussian payout innovation is integrated analytically with its \emph{state-dependent true variance} when calculating exponential utility. No clipping is applied to the exponential average.

Finite-sample certainty equivalents can be sensitive to a small number of extreme paths. We therefore record the normalized exponential weights and the effective number of contributing observations. In the candidate-learning robustness evaluations, the smallest exponential-weight count is $946{,}850$ out of $1{,}048{,}576$, the largest single normalized weight is $0.0010\%$, and the largest combined share of the top $0.1\%$ of weights is $0.31\%$. For the payout extensions, the corresponding minimum is $918{,}493$, the largest single weight is $0.0012\%$, and the largest top-$0.1\%$ share is $0.33\%$. These figures assess concentration in the retained evaluation sample. They are not time-series effective sample sizes or guarantees about unobserved tail realizations. Separate $500$-draw, path-paired bootstraps stratified by evaluation bank quantify Monte Carlo variability conditional on the saved fitted policies.

We use descriptive Student-$t$ intervals over the three initial training seeds for market-level comparisons. Additional candidate refits, learning rounds, and evaluation observations are correlated within those seeds and are not treated as independent economies. For the one-step fixed-law diagnostic, its four independent refits per fixed seed instead give seed-specific $t$ intervals with three degrees of freedom. The numerical code also verifies that market rules and new training banks are applied in the specified order and that saved simulated paths satisfy competitive clearing. The maximum retained clearing residual is $1.33\times10^{-15}$ in the candidate-learning tests, $2.66\times10^{-15}$ in continued market retraining, and $2.22\times10^{-15}$ in the payout extensions. These are numerical identities at the saved simulation states, not estimates of global convergence.

\subsection{Reproducing the numerical evidence}\label{subsec:S10-repro}

Numerical replication requires the market generator, learning and evaluation settings, seed-specific market checkpoints, and the corresponding run summaries. The principal software environment used for the original calculations was Python 3.12 and CPU PyTorch 2.5.1, with additional dependencies recorded in each experiment's requirements file. The implementation specifies primitive-noise keys, normalization, training lengths, and evaluation-bank conventions. Different hardware or library versions may change floating-point details, especially when neural networks are retrained, so the saved outputs and their SHA-256 manifests provide the primary point of comparison for an independent reproduction.

The learning-dynamics and payout-extension calculations use the drivers \texttt{run\_e9.py}, \texttt{functional\_check.py}, \texttt{run\_e10.py}, and \texttt{audit\_inherited\_laws.py}, plus economy specifications and seed-level diagnostic files. The model specifications, training parameters, and seed-level results are used to regenerate the tables and figures. The learning and evaluation algorithms follow the timing in S8--S9.

The candidate-training robustness comparisons are calculated on the nine markets held fixed after six learning updates. Across these markets we vary the number of candidate training steps and the supervised readout, then evaluate the forecast and utility measures on independent generated histories. The sample construction, training budgets, and statistical summaries are reported above.

The preceding sections specify the population, initial-deployment, and candidate-valuation calculations. For the present experiments, Table~\ref{tab:S10-repro} summarizes the distinct objects to be regenerated; each uses different points in the market-learning timeline. Seed-level tables are calculated before plotting and constructing the descriptive confidence intervals.

\begin{table}[htbp]\centering\small
\caption{Computational steps for reproducing the additional numerical experiments.}\label{tab:S10-repro}
\begin{tabularx}{\linewidth}{@{}p{.14\linewidth}X X@{}}\toprule
Exercise & Computational procedure & Materials required\\\midrule
Candidate training & Hold the nine neural markets fixed; refit candidate predictors or change the readouts; calculate $T$ and $U$ on new evaluation paths & Saved market policies, training settings, candidate fits, evaluation samples\\
Market retraining & Continue both investor groups from update six; evaluate $J$, the $\iota$ coefficients, and one-step policy changes & Market checkpoints, shock streams, learning configurations, seed-level statistics\\
Alternative payoffs & Refit the inherited laws for each payout specification; clear new adoption mixes; evaluate candidates on fixed-policy laws & Economic parameters, inherited model fits, paired shocks, evaluation summaries\\\bottomrule
\end{tabularx}
\end{table}

\subsection{How these experiments qualify the main results}\label{subsec:S10-conclusion}

The numerical experiments show that two of the paper's central informational effects survive several meaningful changes to the finite-learning exercise. In the same nine neural markets, the price closes more of a limited observer's predictive disadvantage when advanced trading becomes common, and the gross private value of an additional computational upgrade falls. The signs remain when candidates are trained longer, trained on different data, or evaluated with richer supervised readouts. They also appear in the two examined economies with nonaligned dividend components or publicly varying true payoff risk, under the separate frozen-policy protocol used for those extensions.

The learning-dynamics results place a clear boundary on equilibrium language. Twelve rounds of neural refitting can leave the aggregate price statistic close to its earlier level without bringing every trading rule to rest. Fresh-sample continuations fail the strict settling screen, and the short observed paths do not show a stable two-cycle. These facts support the interpretation of S8--S9 as finite-learning numerical experiments; the population equilibrium and exact information identities established for the Gaussian subeconomies in S3--S7 remain analytically distinct. With the theory, population calculations, and neural-market experiments stated on their respective domains, the numerical evidence can be reproduced and interpreted without treating one model's robustness checks as a universal claim about all AI-mediated markets.

\clearpage
\setcounter{section}{0}\setcounter{table}{0}\setcounter{figure}{0}
\renewcommand{\thesection}{E\arabic{section}}
\renewcommand{\theHsection}{E.\arabic{section}}
\renewcommand{\thetable}{E\arabic{section}.\arabic{table}}
\renewcommand{\theHtable}{E.\arabic{section}.\arabic{table}}
\renewcommand{\thefigure}{E\arabic{section}.\arabic{figure}}
\renewcommand{\theHfigure}{E.\arabic{section}.\arabic{figure}}
\section{The value of market prices under different public-data processing capabilities}\label{sec:E1-market}

A stock price brings together information from many sources. Some investors trade on observations that others do not possess. Others study company accounts, economic reports, and historical market data that are publicly available, but difficult to analyze in full. Their trades can convey the results of that analysis through the price. Consequently, the same price may tell two observers different things: someone who has already extracted much of the information in the public record may learn less from the price than someone who has processed only a small part of that record. This section examines that possibility using U.S. stock-market and earnings data.

We assemble financial information for U.S. listed firms at the end of each calendar quarter and use it to forecast the surprise in their next reported quarterly earnings. We compare four forecasts for each firm. Two use a relatively simple statistical model and a limited set of public accounting and past-market variables; two use a more flexible model and a longer public history. Within each pair, we ask how the forecast changes when the model also receives the firm's latest stock price, market capitalization, and one-month return. We call these three observations the \emph{current-market snapshot}. All four predictions are evaluated on the same 58,807 out-of-sample firm-quarter observations from 2018 through 2025. The comparison tells us how much the same recent market information improves prediction at each level of public-data processing.

We then follow the firms to their next earnings announcements. An earnings announcement can move the stock price when the newly reported results differ from what investors expected. If a richer analysis of previously available financial data already helps anticipate some of the reported earnings, that predicted component may elicit a smaller price response at the announcement than the part the model could not predict. We examine this distinction using announcement-period stock returns. Together, the two exercises study the additional value of market information for different processors and the subsequent market response to the earnings components they forecast.

\subsection{Measuring the value of the same market information to different processors}\label{subsec:E1-question}

Section S5 distinguishes the information a market price provides to observers with different abilities to analyze an already public record. The observers see the same price. Their prior knowledge differs because one can compute a richer summary of the history of releases, payouts, and past prices. In the correctly specified Gaussian model, the additional information provided by the price is measured by $I(\theta;p\mid\mathcal F_L)$ for a limited public-information set $\mathcal F_L$ and by $I(\theta;p\mid\mathcal U_H)$ for a richer set $\mathcal U_H$. The empirical exercise makes a corresponding comparison using the improvement in forecasts of an observable future earnings outcome.

Let $Y$ denote the next standardized quarterly earnings surprise. A limited forecasting technology $L$ uses a small set of public accounting and past-market variables, while a richer technology $H$ processes additional past information with a more flexible statistical method. For each technology, we make a forecast before and after providing the current-market snapshot defined above. We call the four resulting predictions $\widehat Y_{L0}$, $\widehat Y_{L1}$, $\widehat Y_{H0}$, and $\widehat Y_{H1}$, where $0$ means that the current snapshot is absent and $1$ means that it has been added. Every forecast is evaluated against the same realized $Y$ for the same firm and quarter. Comparing $L0$ with $L1$ tells us what the limited technology gains from the current snapshot; comparing $H0$ with $H1$ gives the corresponding gain for the richer technology.

We measure gains in predictive accuracy with an out-of-sample loss $\ell$, for which smaller values indicate better forecasts. The four average loss reductions are
\begin{align}\label{eq:E1-moments}
 S_{\ell}&\equiv\mathbb E[\ell(L0)-\ell(L1)], &
 F_{\ell}&\equiv\mathbb E[\ell(H0)-\ell(H1)],\\
 E_{\ell}&\equiv\mathbb E[\ell(L0)-\ell(H0)], &
 V_{\ell}&\equiv\mathbb E[\ell(L1)-\ell(H1)].\nonumber
\end{align}
Here $S_{\ell}$ and $F_{\ell}$ are the respective gains from adding current market information. The other two quantities measure the forecasting advantage of richer processing before ($E_{\ell}$) and after ($V_{\ell}$) that information is added. Our main comparison is $S_{\ell}-F_{\ell}$. A positive difference means that the same current-market observations are more useful to the limited processor. Because the four forecasts concern the same outcome, they also satisfy the accounting identity
\begin{equation}\label{eq:E1-identity}
 \boxed{\qquad S_{\ell}-F_{\ell}=E_{\ell}-V_{\ell}.\qquad}
\end{equation}
Thus, if the current snapshot closes part of the forecasting gap between the two technologies, it must also have a larger incremental value to the limited technology. The identity follows directly from the loss definitions and holds for each observation.

The empirical question is therefore quite specific: after each technology has used the financial records it can process, how much does its forecast improve when the same three current market variables are added? We use the loss notation $S_{\ell}$ and $F_{\ell}$ to distinguish this measured prediction gain from the information quantities in the theoretical model.

\subsection{Firms, earnings outcomes, and available information}\label{subsec:E1-data}

\paragraph{The firm-quarter observations.}
For each listed common stock, we collect a public-information record at the end of a calendar quarter. We refer to that quarter-end as the \emph{forecast date}; it is called the formation date in the underlying data construction. Starting with CRSP monthly data from March 2011 through December 2025, we retain eligible U.S.-incorporated common stocks traded on the NYSE, NYSE American, or Nasdaq. CRSP identifies each security by PERMNO and supplies its monthly price, shares outstanding, and returns. Financial statements from Compustat are linked using the CRSP--Compustat Merged (CCM) history, and earnings records from I/B/E/S are matched using the security identifiers valid at the forecast date. Each security-quarter combination appears once in the constructed panel.

We predict the I/B/E/S standardized unexpected earnings-per-share score (\texttt{suescore}) at the firm's next U.S. quarterly earnings announcement, provided that the announcement occurs after the forecast date and within 180 calendar days. The score describes realized earnings relative to analysts' prior expectations, in the standardized units supplied by I/B/E/S. We denote it by $Y_{i,t+1}$. The median interval from the forecast date to the announcement is 36 days. For example, an observation assembled at the end of March uses information then available to predict the first qualifying later quarterly earnings report. The realized earnings, analysts' corresponding ex post surprise score, and the subsequent announcement are the evaluation outcome; they are not inputs to the four forecasts.

\paragraph{Financial statements and their release dates.}
The baseline accounting record comes from Compustat annual statements for U.S. industrial firms that report standard, consolidated financial accounts in U.S. dollars. We use each statement only after its recorded filing date. When a filing date is missing or falls before the fiscal year-end date, we assign a conservative availability date six months after fiscal year-end. The statement must also be connected to the stock by a CCM link that was valid at the forecast date. These rules yield a panel of 163,912 firm-quarter records, with accounting information matched for 91.41\% of them. A future standardized EPS surprise is available for 58.26\% of the panel. Table~\ref{tab:E1-samples} shows the sample construction and the observations entering the forecasting and event-time analyses.

\begin{table}[htbp]\centering\small
\caption{Construction and timing of the empirical samples. The out-of-sample observations are a selected subset with a valid future I/B/E/S surprise.}\label{tab:E1-samples}
\begin{tabularx}{\linewidth}{@{}Xr@{}}\toprule
\textbf{Diagnostic} & \textbf{Count or share}\\\midrule
Eligible CRSP monthly common-stock observations & 872,009\\
Unique quarter-end PERMNO--forecast-date rows, 2011--2025 & 163,912\\
Distinct PERMNOs at forecast dates & 7,807\\
Annual-accounting match rate at forecast dates & 91.41\%\\
I/B/E/S first future announcement within 180 days & 67.26\%\\
Rows with a nonmissing future standardized EPS surprise & 58.26\%\\
Common out-of-sample forecasts, 2018--2025 & 58,807\\
Distinct firms (PERMNO) in the out-of-sample comparison & 4,183\\
Observed forecast-quarter clusters in the out-of-sample comparison & 22\\
Next-announcement events linked for Step 2 & 55,841\\
Complete principal announcement-return regressions & 55,740\\\bottomrule
\end{tabularx}
\end{table}

\paragraph{Historical information and the current market observations.}
At each forecast date, the models can use eligible financial statements and stock-market histories available by the preceding month. Those histories include past returns, past prices, and measures of market volatility; the richer model retains a longer record. We separately identify three market variables observed at the forecast date itself: the log share price, the log of market capitalization (share price times shares outstanding), and the return over the most recent month. This three-variable block is the current-market snapshot. The two forecasts labeled $0$ use past market information but exclude this current block. The forecasts labeled $1$ append exactly the same three variables within each technology. This timing lets us ask what is gained from up-to-date market information after different degrees of processing of the older public record.

\subsection{Four forecasting specifications}\label{subsec:E1-technologies}

The limited technology uses ridge regression, which predicts the earnings surprise from a small number of accounting ratios and past-market variables. The richer technology uses histogram gradient boosting, which can capture nonlinear patterns and interactions, together with more observations from the firm's earlier financial and price histories. In each case we estimate a forecast with and without the same current-market block. Table~\ref{tab:E1-tech} summarizes the four specifications. The design holds the firm-quarter and prediction target fixed while changing the available history and the method used to process it.

\begin{table}[htbp]\centering\small
\caption{The four counterfactual forecasting technologies. An identical current-market snapshot is appended within each processing family.}\label{tab:E1-tech}
\begin{tabularx}{\linewidth}{@{}l l c X@{}}\toprule
\textbf{Model} & \textbf{Estimator} & \textbf{Inputs} & \textbf{Information available}\\\midrule
$L0$ & Ridge & 14 & Limited accounting variables and lagged market data; no current snapshot\\
$L1$ & Ridge & 17 & Exactly $L0$ plus current log price, log market equity, and current one-month return\\
$H0$ & Gradient boosting & 54 & All $L0$ input families plus longer lagged return, price, volatility, and financial-statement histories\\
$H1$ & Gradient boosting & 57 & Exactly $H0$ plus the same three current-snapshot fields as $L1$\\\bottomrule
\end{tabularx}
\end{table}

The limited model receives ten accounting-related variables, including assets, profitability, leverage, sales growth, and the age of the latest financial statement, together with four measures of previous stock-market performance. The richer model receives these 14 fields plus 40 further variables describing longer return and price histories, volatility at several horizons, monthly return lags, and earlier accounting statements. These 40 additional inputs are all dated before the current-market block. Section~\ref{subsec:E1-featureappendix} records the exact feature names for reproduction.

Missing predictor values are replaced using medians calculated from the training observations, and indicators retain information about which values were missing. The ridge model also standardizes its inputs and uses a regularization parameter of $\alpha=10$. The gradient-boosting model uses a learning rate of $0.05$, 250 iterations, at most 15 leaf nodes, at least 40 observations per leaf, and $L_2$ regularization of $1.0$ (random seed 20261029). We keep these specifications fixed across the out-of-sample comparisons.

The richer model thus differs from the limited model in two ways: it has access to more historical information and it has a more flexible forecasting method. The historical inputs are nested, since the richer specification includes every predictor available to the limited one. This provides a concrete comparison between a relatively simple processor and one that can make more extensive use of the same public financial record. The models are not intended to isolate the contribution of any single forecasting technique.

\subsection{Out-of-sample forecasting and statistical inference}\label{subsec:E1-loss}

We assess forecasts in calendar years 2018--2025, one year at a time. For each evaluation year, all four models are fitted using firm-quarter observations whose forecast dates fall in preceding calendar years. For example, the forecasts evaluated in 2018 are estimated using observations dated from 2011 through 2017; the 2019 forecasts can also use observations dated in 2018. We then generate the four predictions for firms observed in the evaluation year and compare them with their subsequently announced earnings surprises. The final evaluation uses 58,807 firm-quarter observations for which every model has a prediction and the future earnings outcome is observed. Within each training set, the earnings target is clipped at its first and ninety-ninth percentiles; the same training-derived cutoffs are applied to that year's evaluation outcomes. Missing-value imputation and other preprocessing are fitted on the training records.

All four forecasts for a given evaluation year are scored on the same scale. Let $\sigma_y$ be the standard deviation of the earnings-surprise target calculated from the training data for year $y$, $\widetilde Y_{it}$ the realized outcome after applying that training set's clipping cutoffs, and $\widehat Y_{M,it}$ the prediction from model $M$. We use the common-scale Gaussian pseudo-negative-log-likelihood
\begin{equation}\label{eq:E1-loss}
 \ell_{it}(M)=\frac12\log(2\pi\sigma_y^2)
 +\frac{(\widetilde Y_{it}-\widehat Y_{M,it})^2}{2\sigma_y^2},
 \qquad y=\operatorname{year}(t).
\end{equation}
The common scale makes the loss difference between two models equal to their squared-error difference divided by $2\sigma_y^2$. Accordingly, the estimated gains describe improvements in the prediction of earnings-surprise means. They can be compared across the four models without attributing differences to model-specific variance forecasts.

We average the firm-quarter loss differences in equation~\eqref{eq:E1-moments}. Confidence intervals allow for repeated observations of the same stock and for shocks shared by stocks in the same quarter. The principal standard errors use two-way clustering by CRSP PERMNO and forecast quarter, following the Cameron--Gelbach--Miller estimator. We also obtain 95\% percentile intervals by resampling entire forecast-quarter blocks 1,999 times. The evaluation sample contains 22 distinct forecast-quarter clusters. Annual patterns and comparisons by firm size and analyst coverage are reported as additional descriptive checks.

\subsection{Results: how much does the current market information add?}\label{subsec:E1-results}

The richer technology predicts the next earnings surprise more accurately on average, with or without the current-market block. The mean losses of $L0$, $L1$, $H0$, and $H1$ are $2.946876$, $2.943544$, $2.930524$, and $2.929735$, respectively; their corresponding root mean squared errors are $4.591450$, $4.577614$, $4.524215$, and $4.521013$. Having established these differences in overall predictive performance, we compare the improvements attributable to the three current-market variables within each model family.

\begin{figure}[htbp]\centering
 \includegraphics[width=0.94\linewidth]{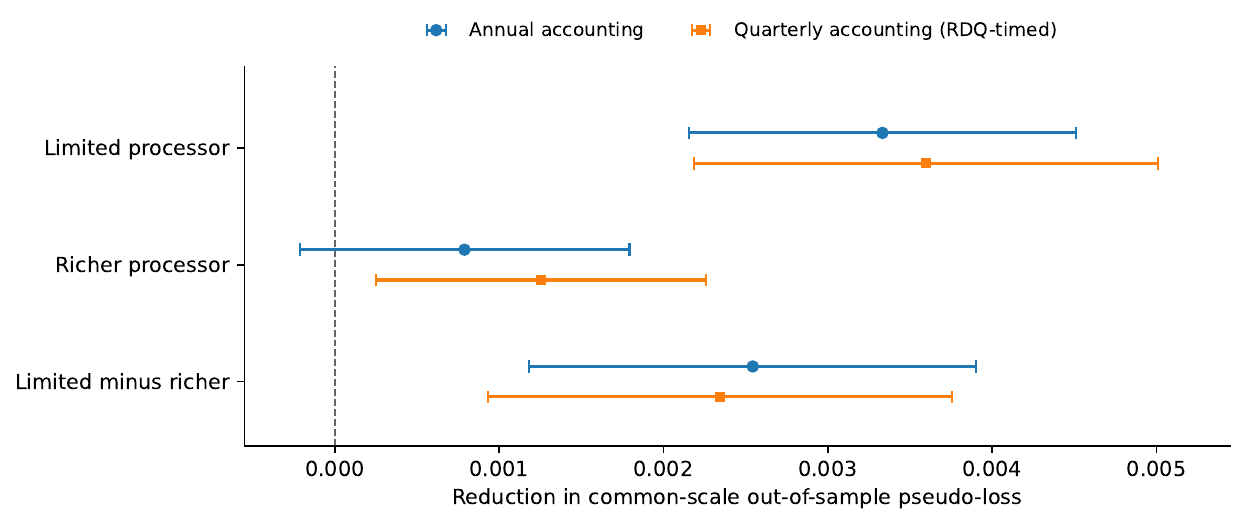}
 \caption{Current market-information snapshot: incremental out-of-sample loss reductions for limited and richer processors, with the difference $S_{\ell}-F_{\ell}$. Circles are the annual-accounting baseline; squares replace public accounting inputs with RDQ-timed Compustat Quarterly data on the same 58,807 observations. Bars are two-way firm/forecast-quarter clustered 95\% confidence intervals. These are pseudo-loss improvements, not mutual information or an observed change in prices.}\label{fig:E1-primary}
\end{figure}

The limited processor gains more from the current market information (Table~\ref{tab:E1-primary} and Figure~\ref{fig:E1-primary}). Its average loss falls by $S_{\ell}=0.003332$ when the three current variables are added, with a two-way clustered 95\% confidence interval of $[0.002155,0.004509]$. For the richer processor, the corresponding gain is $F_{\ell}=0.000789$, with a confidence interval of $[-0.000214,0.001793]$. The difference between these gains is $S_{\ell}-F_{\ell}=0.002543$ (95\% CI $[0.001181,0.003904]$). The confidence interval from quarter-block resampling also lies entirely above zero. Thus the same recent market observations improve prediction more for the limited model, even though the richer model has better forecasts overall.

\begin{table}[htbp]\centering\small
\caption{Main out-of-sample loss contrasts, 2018--2025 (58,807 common firm-quarters). All four models use the same fold-specific scale; intervals use the indicated clustering or block resampling.}\label{tab:E1-primary}
\begin{tabularx}{\linewidth}{@{}l r r X X@{}}\toprule
\textbf{Contrast} & \textbf{Estimate} & \textbf{Two-way SE} & \textbf{Two-way 95\% CI} & \textbf{Quarter-block 95\% CI}\\\midrule
$S_{\ell}=L0-L1$ & $0.003332$ & $0.000601$ & $[0.002155,0.004509]$ & $[0.002483,0.004359]$\\
$F_{\ell}=H0-H1$ & $0.000789$ & $0.000512$ & $[-0.000214,0.001793]$ & $[-0.000162,0.001739]$\\
$V_{\ell}=L1-H1$ & $0.013809$ & $0.002007$ & $[0.009876,0.017743]$ & $[0.010575,0.017534]$\\
$E_{\ell}=L0-H0$ & $0.016352$ & $0.002144$ & $[0.012149,0.020555]$ & $[0.012926,0.020260]$\\
$S_{\ell}-F_{\ell}$ & $0.002543$ & $0.000695$ & $[0.001181,0.003904]$ & $[0.001352,0.003748]$\\\bottomrule
\end{tabularx}
\end{table}

The identity in equation~\eqref{eq:E1-identity} offers another interpretation. Before seeing the current-market block, the richer model has a loss advantage of $E_{\ell}=0.016352$ over the limited one. Once both see that block, the advantage falls to $V_{\ell}=0.013809$. The reduction of $0.002543$ amounts to about 15.5\% of the original technology gap in measured forecast loss. In this sense the current market observations substitute for part of the additional analysis performed by the richer model.

\paragraph{Using more recent financial statements.}
Annual accounts can be several months old when a quarterly forecast is made. We therefore repeat the prediction exercise using Compustat Quarterly statements and keep the same stocks, forecast dates, current-market variables, future earnings outcomes, and four forecasting specifications. The quarterly reconstruction dates each statement using its recorded earnings-announcement date (\texttt{rdq}), discards invalid or missing dates, sums the relevant flows over the preceding four quarters, and uses properly spaced observations for annual growth. Under this availability convention, 94.9\% of firm-quarter records receive accounting data, and all attached records have dates no later than the corresponding forecast date. This comparison allows the models to process more recent financial information while preserving the original evaluation sample.

The main result remains. With quarterly accounting, the limited processor gains $0.003598$ from the current market observations and the richer processor gains $0.001256$; the difference is $0.002342$ (95\% CI $[0.000930,0.003754]$). The richer processor's gain is now itself estimated precisely above zero, but the limited processor continues to gain more. Table~\ref{tab:E1-quarterly} compares all four loss contrasts under annual and quarterly accounting. The stability of the main difference shows that it is not driven solely by the use of older annual statements.

\begin{table}[htbp]\centering\small
\caption{Accounting-frequency robustness for common out-of-sample observations. Values are mean common-scale loss improvements with two-way clustered 95\% intervals.}\label{tab:E1-quarterly}
\begin{tabularx}{\linewidth}{@{}l X X@{}}\toprule
\textbf{Moment} & \textbf{Annual accounting} & \textbf{RDQ-timed quarterly accounting}\\\midrule
$S_{\ell}$ & $0.003332\ [0.002155,0.004509]$ & $0.003598\ [0.002187,0.005008]$\\
$F_{\ell}$ & $0.000789\ [-0.000214,0.001793]$ & $0.001256\ [0.000251,0.002261]$\\
$E_{\ell}$ & $0.016352\ [0.012149,0.020555]$ & $0.022087\ [0.012134,0.032040]$\\
$V_{\ell}$ & $0.013809\ [0.009876,0.017743]$ & $0.019745\ [0.010747,0.028743]$\\
$S_{\ell}-F_{\ell}$ & $0.002543\ [0.001181,0.003904]$ & $0.002342\ [0.000930,0.003754]$\\\bottomrule
\end{tabularx}
\end{table}

\paragraph{Variation across years and firms.}
The difference $S_{\ell}-F_{\ell}$ is positive in seven of the eight evaluation years; 2019 is the exception. The difference is also positive in each of three groups formed by firm size and in each of three groups formed by the number of analysts following the firm. Table~\ref{tab:E1-stability} gives the underlying estimates. These patterns suggest that the average result is not confined to a single year or a narrow class of firms.

These estimates establish a straightforward empirical pattern. When a forecaster has already processed a longer record of public financial information with a richer statistical method, the three current market variables add less to its earnings prediction. We next examine what kinds of earnings variation the richer forecasts capture and how those components relate to price movements when firms announce their results.

\subsection{What do the richer forecasts predict before earnings announcements?}\label{subsec:E1-events}

The earnings-surprise score compares a firm's reported earnings with analysts' expectations, while the stock-price reaction reflects what market participants learn when the announcement arrives. The two need not move together. Consider a firm whose publicly available financial history suggests that earnings will be higher than analysts expect. A forecasting model that detects this pattern can predict part of the eventual earnings surprise. If investors have already taken the same information into account when trading, the predictable part may provoke little additional price movement at the announcement. By contrast, the part that neither the model nor investors anticipated may move the price more strongly. We use the next earnings announcement to compare how these two components relate to short-window stock returns.

\paragraph{Earnings events.}
We link the four forecasts made at each quarter-end to the firm's next I/B/E/S quarterly earnings announcement. Occasionally, more than one prior quarter-end observation points to the same announcement. In those cases we keep the latest forecast date that precedes it, so each firm-announcement pair enters once. This produces 55,841 distinct events. The principal return analysis has complete observations for 55,740 events, representing 4,128 firms and 1,249 announcement trading dates.

For each earnings release, we match the I/B/E/S calendar announcement date to the first CRSP trading day on or after that date, denoted $\tau$. We then measure the firm's stock return relative to the value-weighted market return over the event day and the following trading day:
\begin{equation}\label{eq:E1-car}
 CAR_{i,[0,+1]}=\sum_{d=\tau}^{\tau+1}(R_{id}-R^{VW}_{d}),
\end{equation}
where $R_{id}$ is the firm's daily return and $R^{VW}_{d}$ is the market return. This two-day interval allows for earnings releases made after the market closes. Earnings surprises and event returns are separately winsorized at the first and ninety-ninth percentiles within the announcement year; model forecasts and their differences are left unchanged. The I/B/E/S date agrees with Compustat Quarterly's earnings-release date (\texttt{rdq}) on the same calendar day for 97.82\% of linked events and within one day for 98.50\%. The one-day-matched subsample contains 54,929 complete return observations.

\paragraph{Separating expected earnings from the remaining surprise.}
Suppose a model forecasts an earnings-surprise score of $\widehat Y_M$ and the realized score is $Y$. We call $u_M=Y-\widehat Y_M$ the earnings component the model did not forecast. The difference between two model predictions measures the earnings component that becomes predictable when a particular piece of information or processing technology is added. We use the four forecasts to construct
\begin{align}\label{eq:E1-event-signals}
 p_L&=\widehat Y_{L1}-\widehat Y_{L0}, &
 p_H&=\widehat Y_{H1}-\widehat Y_{H0},\\
 q_{H,\mathrm{pub}}&=\widehat Y_{H0}-\widehat Y_{L0}, &
 q_{H,\mathrm{post}}&=\widehat Y_{H1}-\widehat Y_{L1}.\nonumber
\end{align}
The first two differences isolate what the current market observations contribute to each forecasting technology. The last two isolate the additional predictions generated by richer processing, first without and then with the current-market block. For example, $q_{H,\mathrm{pub}}$ measures the part of the predicted earnings surprise that is added by the richer model's analysis of previously available records before it sees the current snapshot.

We estimate how the event-window return is associated with each incremental forecast component $x$, while also accounting for the remaining surprise $u_{M(x)}$:
\begin{equation}\label{eq:E1-eventreg}
 CAR_{i,[0,+1]}=\beta_x u_{M(x),i}+\delta_x x_i
 +X_i^\top\Gamma_x+\alpha_{\tau(i)}+\varepsilon_i.
\end{equation}
For $x=p_L,p_H,q_{H,\mathrm{pub}},q_{H,\mathrm{post}}$, the corresponding residual comes from $L1,H1,H0,H1$, respectively. The controls $X_i$ describe firm size and analyst coverage at the earlier forecast date, including their tercile groups; $\alpha_{\tau(i)}$ absorbs variation shared by firms announcing on the same trading day. Standard errors are clustered by firm and event date. In each regression, $\beta_x$ measures the association between the model's unpredicted earnings component and the announcement return, while $\delta_x$ measures the association for the component added to the forecast. Comparing $\delta_x$ with $\beta_x$ asks whether the two earnings components have similarly sized event-time return responses.

\begin{table}[htbp]\centering\small
\caption{Earnings-announcement market-adjusted CAR[0,+1] and incrementally predicted next-SUE components. Two-way firm/event-date clustered standard errors in parentheses; $N=55{,}740$ throughout.}\label{tab:E1-events}
\begin{tabularx}{\linewidth}{@{}X r r r r@{}}\toprule
\textbf{Increment $x$} & \textbf{$\beta_x$ (SE)} & \textbf{$\delta_x$ (SE)} & \textbf{$p:\delta_x=0$} & \textbf{$p:\delta_x=\beta_x$}\\\midrule
Limited snapshot, $p_L$ & $0.004266\ (0.000125)$ & $0.003696\ (0.002007)$ & $0.0655$ & $0.7769$\\
Rich snapshot, $p_H$ & $0.004342\ (0.000124)$ & $0.000453\ (0.002053)$ & $0.8255$ & $0.0574$\\
Rich public, $q_{H,\mathrm{pub}}$ & $0.004334\ (0.000124)$ & $0.001218\ (0.000686)$ & $0.0761$ & $<0.0001$\\
Rich after snapshot, $q_{H,\mathrm{post}}$ & $0.004341\ (0.000124)$ & $0.000660\ (0.000704)$ & $0.3487$ & $<0.0001$\\\bottomrule
\end{tabularx}
\end{table}

The differences are largest for the two increments generated by richer public-data processing (Table~\ref{tab:E1-events}). The event-return coefficients on $q_{H,\mathrm{pub}}$ and $q_{H,\mathrm{post}}$ are $0.001218$ and $0.000660$, respectively, while the coefficients on the corresponding remaining earnings surprises are approximately $0.00434$. The hypothesis that the predicted and residual components have equal return coefficients is rejected at $p<0.0001$ in both cases. Thus the extra earnings variation forecast by richer processing is associated with a much smaller stock-price reaction at the eventual earnings announcement than the earnings variation the model leaves unexplained.

\paragraph{Accounting for overlap among the forecast components.}
All four increments come from the same set of predictions and are linked by the exact identity
\begin{equation}\label{eq:E1-eventidentity}
 q_{H,\mathrm{post}}=q_{H,\mathrm{pub}}+p_H-p_L.
\end{equation}
We therefore also estimate one regression with the high-model residual $u_{H1}$ and the three increments $p_L$, $p_H$, and $q_{H,\mathrm{pub}}$ entered together. In this joint comparison, the coefficient on the rich-public component is $0.000735$ (SE $0.000717$), compared with $0.004339$ (SE $0.000124$) for the remaining earnings surprise. Their difference is statistically clear ($p=4.496\times10^{-7}$). Replacing $q_{H,\mathrm{pub}}$ with the algebraically equivalent $q_{H,\mathrm{post}}$ representation produces the same fitted returns. Table~\ref{tab:E1-joint} reports all three coefficients from the joint regression.

\begin{table}[htbp]\centering\small
\caption{Joint announcement-return decomposition using $u_{H1}$ and three independent forecast increments. The coefficient on $u_{H1}$ is $0.004339$ (SE $0.000124$).}\label{tab:E1-joint}
\begin{tabularx}{\linewidth}{@{}X r r r@{}}\toprule
\textbf{Increment} & \textbf{Coefficient (SE)} & \textbf{$p:\delta=0$} & \textbf{$p:\delta=\beta$}\\\midrule
$p_L$ & $0.002361\ (0.002057)$ & $0.2511$ & $0.3387$\\
$p_H$ & $0.000645\ (0.002087)$ & $0.7573$ & $0.0754$\\
$q_{H,\mathrm{pub}}$ & $0.000735\ (0.000717)$ & $0.3056$ & $4.496\times10^{-7}$\\\bottomrule
\end{tabularx}
\end{table}

\subsection{Checking the event-time result and its interpretation}\label{subsec:E1-eventlimits}

The smaller event-time responses to richer forecast components persist when we vary how earnings announcements and abnormal returns are dated. Restricting the sample to announcements whose I/B/E/S date agrees with the Compustat earnings-release date within one calendar day produces almost the same estimates. Replacing the two-day market-adjusted cumulative return with either a three-day return or a two-day buy-and-hold abnormal return also preserves the difference. Re-estimating the models using the quarterly accounting records described above leads to the same broad pattern (Table~\ref{tab:E1-eventrobust}). The event response is therefore not specific to one earnings-date match, return window, or frequency of financial-statement inputs.

\begin{table}[htbp]\centering\small
\caption{Event-time robustness: high-processing increments versus residual SUE. Entries are incremental coefficient $\delta$ (clustered SE), followed by the test $p$-value for equality with the residual-SUE loading $\beta$.}\label{tab:E1-eventrobust}
\begin{tabularx}{\linewidth}{@{}X X X@{}}\toprule
\textbf{Sample or outcome} & \textbf{$q_{H,\mathrm{pub}}=H0-L0$} & \textbf{$q_{H,\mathrm{post}}=H1-L1$}\\\midrule
Annual, CAR$[0,+1]$ & $0.001218\ (0.000686);\ p<0.0001$ & $0.000660\ (0.000704);\ p<0.0001$\\
Annual, RDQ agreement within 1 day & $0.001187\ (0.000692);\ p<0.0001$ & $0.000624\ (0.000716);\ p<0.0001$\\
Annual, CAR$[-1,+1]$ & $0.001178\ (0.000741);\ p<0.0001$ & $0.000750\ (0.000758);\ p<0.0001$\\
Annual, BHAR$[0,+1]$ & $0.001233\ (0.000690);\ p<0.0001$ & $0.000660\ (0.000709);\ p<0.0001$\\
Quarterly, CAR$[0,+1]$ & $0.001173\ (0.000961);\ p=0.0010$ & $0.000784\ (0.000980);\ p=0.0003$\\
Quarterly, RDQ agreement within 1 day & $0.001118\ (0.000970);\ p=0.0008$ & $0.000687\ (0.000996);\ p=0.0002$\\\bottomrule
\end{tabularx}
\end{table}

\paragraph{Stock-price movements before the earnings release.}
The forecast is made before the earnings announcement, leaving time for investors to respond to information in the interim. To look for gradual incorporation, we relate the richer forecast components to market-adjusted returns between the forecast date and the day before the release. Across 55,696 complete intervals, the median length is 24 trading days. The estimated associations are small and statistically indistinguishable from zero, both over the full pre-announcement interval and in the first and last 20 trading days where those windows are available. Table~\ref{tab:E1-pre} reports the coefficients. These data do not show a detectable average drift associated with the richer forecasts in the weeks leading up to the announcement.

\paragraph{What happens to earnings in later quarters?}
The market may respond differently to an earnings component that reverses quickly than to one that changes expectations about long-run profits. We investigate this possibility by relating the high-processing forecast increments to standardized earnings surprises at the current and next four announcements. With fixed effects for the original forecast quarter, these components remain positively associated with subsequent surprises through the fourth later announcement. When we additionally include firm fixed effects, their within-firm association declines sharply: it is near zero by the second later announcement and negative at the fourth. For example, at that fourth horizon, the coefficients are $-0.170$ (SE $0.067$) for $q_{H,\mathrm{pub}}$ and $-0.159$ (SE $0.074$) for $q_{H,\mathrm{post}}$. The pattern also appears in the quarterly-accounting version. Table~\ref{tab:E1-persistence} reports the annual-accounting estimates. These findings point to substantial between-firm differences and mean reversion within firms in the predictable earnings-surprise component.

The event results indicate that richer processing predicts earnings components that are less closely associated with immediate announcement returns than the remaining surprise. The timing and later-earnings exercises help interpret this difference. In particular, the absence of pre-announcement drift leaves open whether the information reached prices earlier or had less influence on valuation, and the within-firm reversal shows that the future-SUE association is not a simple measure of permanent cash-flow news. We discuss these boundaries in relation to the theory below.

\subsection{Interpretation and relation to the theory}\label{subsec:E1-interpretation}

The first empirical result shows that the value of a price observation depends on how much related public information a forecaster has already processed. Two technologies predict the same earnings outcome for the same stocks on the same dates, but adding the current price, market capitalization, and monthly return improves the limited forecast more than the richer forecast. This is the central link to the information-sharing mechanism in S5: prices can be especially useful to an observer who cannot reproduce a more extensive analysis of the public record. The subsequent earnings-event analysis adds that the extra predictions generated by richer processing have a smaller association with announcement-day price changes than earnings news left unexplained by the model.

These comparisons measure forecasting and event-return relationships in an existing market. The three current-market variables combine price and quantity information, and the two technologies differ in both the histories they process and the forecasting methods they use. Their loss differences provide a real-data measure of predictive substitution rather than the conditional mutual information or private value of computation in S5. The event regressions leave open whether smaller return responses reflect earlier incorporation of information, a different relation to firm value, or both. Relatedly, standardized earnings surprises depend on analysts' forecasts and do not directly measure permanent cash-flow shocks. For quarterly accounting, the $\texttt{rdq}$ convention records earnings-release dates and may precede the disclosure of every item in a full financial statement.

The empirical data contain neither an intervention that changes investors' AI adoption nor observations of traders' private clues and demand schedules. The discovery and equilibrium comparative statics in S4--S6 are therefore established by the model and market simulations, while the present results document the heterogeneous usefulness of market information and the behavior of more extensively processed earnings forecasts. The next empirical section studies a complementary question by experimentally varying a synthetic private clue supplied to language models.

\subsection{Additional measurement details and reproducibility}\label{subsec:E1-reproduction}

\paragraph{Forecasting features.}\label{subsec:E1-featureappendix}
The 14 inputs of the limited technology are \path{log_assets}, \path{operating_profitability}, \texttt{roa}, \path{asset_growth}, \texttt{leverage}, \path{cash_holdings}, \texttt{accruals}, \path{sales_growth}, \texttt{tangibility}, \path{statement_age_days}, \path{return_1m_lag1}, \path{return_12m_lag1}, \path{return_volatility_12m_lag1}, and \path{log_market_equity_lag1}. The richer technology adds 3-, 6-, and 24-month historical returns and their volatility measures; lagged log prices over 1, 3, 6, 12, and 24 months; monthly \path{return_lag2} through \path{return_lag12}; and one- and two-statement lags of nine accounting ratios. The current block adds only \path{current_log_price}, \path{current_log_market_equity}, and \path{current_return_1m}. The full construction code defines all transformations and handles missing values within the training pipeline.

\paragraph{Yearly and cross-sectional comparisons.}
Table~\ref{tab:E1-stability} provides the mean loss differences for each evaluation year and for groups defined by firm size and analyst coverage. The size and coverage groups are formed separately within each forecast quarter, using observations available before the current market block is added. These comparisons describe the variation in measured forecast gains across the sample.

\begin{table}[htbp]\centering\small
\caption{Descriptive stability of incremental snapshot value. Values are means of the common-scale forecast-loss contrasts, not separate causal effects.}\label{tab:E1-stability}
\begin{tabularx}{\linewidth}{@{}l r r r r@{}}\toprule
\textbf{OOS year or split} & $N$ & $S_{\ell}$ & $F_{\ell}$ & $S_{\ell}-F_{\ell}$\\\midrule
2018 & 2,333 & 0.000728 & -0.002317 & 0.003046\\
2019 & 4,783 & 0.001106 & 0.001427 & -0.000321\\
2020 & 9,820 & 0.003786 & 0.002325 & 0.001461\\
2021 & 11,311 & 0.004592 & 0.001446 & 0.003146\\
2022 & 8,921 & 0.003237 & -0.001424 & 0.004661\\
2023 & 5,661 & 0.002938 & 0.001043 & 0.001895\\
2024 & 5,432 & 0.003389 & 0.002139 & 0.001250\\
2025 & 10,546 & 0.003408 & 0.000095 & 0.003313\\\midrule
Analyst coverage: high & 19,579 & 0.002332 & 0.000134 & 0.002198\\
Analyst coverage: middle & 19,574 & 0.003452 & 0.000660 & 0.002791\\
Analyst coverage: low & 19,588 & 0.004240 & 0.001519 & 0.002721\\\midrule
Firm size: high & 19,593 & 0.003313 & 0.000107 & 0.003206\\
Firm size: middle & 19,586 & 0.003829 & 0.000561 & 0.003268\\
Firm size: low & 19,600 & 0.002854 & 0.001667 & 0.001187\\\bottomrule
\end{tabularx}
\end{table}

\paragraph{Returns before the earnings release.}
Table~\ref{tab:E1-pre} reports the pre-announcement return estimates. Returns are measured on trading dates after the forecast date and before the next earnings-release trading day. The full-interval outcome is the average daily market-adjusted return; the first- and last-20-day outcomes are cumulative abnormal returns, so their coefficients have different units.

\begin{table}[htbp]\centering\small
\caption{High-processing forecast increments and pre-announcement market-adjusted returns. Two-way firm/forecast-date clustered standard errors. The first row is average daily return; the others are cumulative over the indicated 20-trading-day period.}\label{tab:E1-pre}
\begin{tabularx}{\linewidth}{@{}X r X X@{}}\toprule
\textbf{Pre-event outcome} & $N$ & $q_{H,\mathrm{pub}}$ coefficient (SE) & $q_{H,\mathrm{post}}$ coefficient (SE)\\\midrule
Full interval, average daily & 55,696 & $0.000091\ (0.000101)$ & $0.000130\ (0.000101)$\\
First 20 trading days & 42,073 & $0.000654\ (0.002302)$ & $0.000879\ (0.002712)$\\
Last 20 trading days & 42,071 & $0.001622\ (0.002002)$ & $0.001632\ (0.002411)$\\\bottomrule
\end{tabularx}
\end{table}

\paragraph{Earnings surprises at later announcements.}
Table~\ref{tab:E1-persistence} follows forecast components from the first predicted earnings announcement ($h=0$) through four subsequent announcements. The intervals between successive announcements are required to be 45--180 days. The first specification includes fixed effects for the original forecast quarter; the second adds firm fixed effects to focus on changes within the same firm over time. Since the outcome remains standardized earnings relative to analysts' expectations, these regressions describe the persistence of predictable earnings surprises rather than the persistence of the firm's underlying cash flows.

\begin{table}[htbp]\centering\small
\caption{Associations between high-processing forecast increments and standardized earnings surprise at later announcements (annual-accounting specification). Entries are coefficients (two-way clustered SE).}\label{tab:E1-persistence}
\begin{tabularx}{\linewidth}{@{}c r X X X X@{}}\toprule
 & & \multicolumn{2}{c}{\textbf{Formation-date FE}} & \multicolumn{2}{c}{\textbf{Firm + formation-date FE}}\\
\cmidrule(lr){3-4}\cmidrule(l){5-6}
\textbf{$h$} & \textbf{$N$} & $q_{H,\mathrm{pub}}$ & $q_{H,\mathrm{post}}$ & $q_{H,\mathrm{pub}}$ & $q_{H,\mathrm{post}}$\\\midrule
0 & 55,771 & $0.771\ (0.059)$ & $0.723\ (0.064)$ & $0.673\ (0.065)$ & $0.663\ (0.068)$\\
1 & 43,557 & $0.719\ (0.092)$ & $0.635\ (0.096)$ & $0.120\ (0.053)$ & $0.161\ (0.069)$\\
2 & 34,671 & $0.655\ (0.095)$ & $0.598\ (0.096)$ & $0.005\ (0.063)$ & $0.048\ (0.069)$\\
3 & 28,404 & $0.615\ (0.083)$ & $0.549\ (0.085)$ & $-0.090\ (0.060)$ & $-0.087\ (0.067)$\\
4 & 22,657 & $0.566\ (0.078)$ & $0.515\ (0.081)$ & $-0.170\ (0.067)$ & $-0.159\ (0.074)$\\\bottomrule
\end{tabularx}
\end{table}

\paragraph{Data and reproducibility.}
The data are drawn from CRSP, Compustat, and I/B/E/S under institutional licenses. The sample restrictions, timing conventions, forecasting features, fitted models, and inference procedures are described above. The source datasets are available to authorized users through their respective providers.

\clearpage\setcounter{table}{0}\setcounter{figure}{0}
\section{How AI forecasts respond to private information when public financial histories are available}\label{sec:E2-behavior}

An investor trying to predict a firm's next earnings announcement can examine information that everyone can obtain: recent financial statements and the firm's past stock-market performance. The investor might also purchase a research estimate based on a separate investigation. How much should that additional estimate change the forecast? A forecaster that has already learned a great deal from public information may have less reason to change its prediction than one that has extracted less from the same record. This is the information-substitution mechanism studied in the theoretical model.

We examine how two large language models respond to this situation. For 50 historical U.S. firm-quarter observations, we prepare an anonymous collection of public financial data. Each model first receives the firm's accounting information. In a second condition, it also receives a long sequence of past stock-market returns. We then add a hypothetical independent research estimate of the firm's next earnings surprise. Holding the public information fixed, we change that estimate from unfavorable to favorable and ask how much the model changes its forecast. The difference in forecasts gives a direct measure of the model's response to the additional information.

The later model, GPT-6.1 Sol, responds less to the independent estimate than a fixed GPT-5.2 model when past market returns are included. Moreover, adding the market history makes the difference between the two models larger. The effect appears in most of the 50 company-quarter cases: with past returns included, GPT-6.1 Sol is less responsive in 42 of them. We next explain exactly what the models saw, how we measured their responses, and what the differences tell us about public and private information.

\subsection{The question: does public analysis change the value of a private clue?}\label{subsec:E2-question}

The experiment isolates a single forecasting decision. Imagine that an AI has already examined a company's public financial data. It is then told that an independent research group expects a negative earnings surprise. In a second, otherwise identical prompt, that group expects a positive surprise instead. The AI produces a forecast in both cases. We measure how far its forecast moves as the independent estimate changes. For a forecaster that has extracted more useful information from the public data, a given additional estimate may change the answer less. This is the specific response that we compare across models.

In the Gaussian economy of Section S4, this relationship can be calculated exactly. Let $\theta$ be the payoff-relevant state. Suppose the public information available to a forecaster leaves uncertainty $P=\operatorname{Var}(\theta\mid\mathcal U)$, and the forecaster receives an independent noisy signal $s=\theta+\xi$, with $\operatorname{Var}(\xi)=\nu>0$. The coefficient on $s$ in the correctly specified Gaussian forecast of $\theta$ is
\begin{equation}\label{eq:E2-privateweight}
 \beta(P,\nu)=\frac{P}{P+\nu}.
\end{equation}
A more informative analysis of public records reduces $P$. With the private signal's noise held fixed, its weight $\beta$ then falls. For the aligned payoff $D=\bar d+\lambda\theta+\varepsilon$, the corresponding coefficient in the payoff forecast is $\lambda\beta$. Section S4 additionally examines how forecast risk changes the response of \emph{trading demand}. Our experiment studies the simpler, observable counterpart: the response of an AI's \emph{forecast mean} to an additional research estimate.

In the real-data experiment, the extra estimate is not an actual confidential report. We write a short, fictional independent-research note and supply it as part of the prompt. The note reports a forecast of the next standardized quarterly earnings surprise and states that its error has mean zero and a standard deviation of $0.7$ in the same units. We alternate the reported estimate between $-0.8$ and $+0.8$. These are controlled inputs, rather than simulated draws around the firm's actual subsequent earnings. This lets us compare the models' treatment of exactly the same stated information.

\subsection{What the models were given}\label{subsec:E2-inputs}

\paragraph{Public company information.}
We start with 50 U.S. company-quarter observations from 2017--2021, selecting ten from each year before making any calls to the language models. For each observation, we collect publicly reported accounting numbers that would have been available at the forecast date. We remove the company's name, ticker, identifying codes, actual calendar and forecast dates, and the subsequently realized earnings surprise. We call each resulting set of numbers a \emph{company-quarter case}: the information for one firm at one point in time. The models forecast the next standardized unexpected earnings (SUE), a measure of how reported earnings per share differ from analysts' prior forecasts. The task concerns a future earnings measure, not a forecast of the stock's future return.

\paragraph{Past stock-market information.}
In one condition, the company record contains the accounting numbers alone. In the other, it also contains a numerical account of the company's past stock-market performance. This includes 52 observations of weekly returns and 60 recent daily observations, along with an index reconstructed from returns and rescaled to a common starting value. The record gives a historical sequence, not the firm's actual dollar share price, a ticker symbol, or a current market-clearing quote. Both models receive the same values in the same order. The difference between conditions is whether this additional public financial history is provided.

\paragraph{The independent research estimate.}
Finally, each model receives the fictional research note described above. The wording and stated reliability of the note are the same for both models. For every case, we run one prompt in which the note says $-0.8$ and another in which it says $+0.8$. Nothing else in the company record changes between those prompts. Table~\ref{tab:E2-inputs} summarizes the information available in each comparison.

\begin{table}[htbp]\centering\small
\caption{Information presented to each language model. The historical market series is omitted or added as a whole; the research estimate changes while all public inputs stay fixed.}\label{tab:E2-inputs}
\begin{tabularx}{\linewidth}{@{}p{34mm}X X@{}}\toprule
\textbf{Input} & \textbf{What it contains} & \textbf{How it is used}\\\midrule
Company accounts & Publicly reported financial numbers from before the forecast & Included in every prompt\\
Past market returns & 52 weekly and 60 recent daily observations, plus a return-based index & Included in the historical-price condition only\\
Independent research note & Hypothetical next-earnings-surprise estimate of $-0.8$ or $+0.8$; stated error SD $0.7$ & Varied between paired prompts\\
Future earnings and identity & Realized next earnings surprise, company name and ticker, dates, actual dollar price & Withheld from every prompt\\\bottomrule
\end{tabularx}
\end{table}

The two models are a dated GPT-5.2 snapshot (\texttt{gpt-5.2-2025-12-11}) and GPT-6.1 Sol. Both are instructed to return a numerical forecast using the same response format and the same nominal high reasoning-effort setting. They are evaluated in separate calls; one response is never shown to the other model, and earlier answers do not enter subsequent prompts. The model comparison changes an entire trained forecasting system, including its training history, rather than only the number of inference operations available to an otherwise identical algorithm.

\subsection{How we measure the response}\label{subsec:E2-measure}

For every case, model, and public-information condition, we submit the favorable and unfavorable research notes in fresh calls. Each combination is repeated three times to average over variation in the models' responses. Let $\widehat\mu_{jmc}^{(+)}$ and $\widehat\mu_{jmc}^{(-)}$ be the average reported earnings-surprise forecasts for case $j$, model $m$, and condition $c$, with the research note set to $+0.8$ and $-0.8$, respectively. We measure the response to a one-unit change in the research estimate by
\begin{equation}\label{eq:E2-response}
 \widehat\beta_{jmc}=\frac{\widehat\mu_{jmc}^{(+)}-\widehat\mu_{jmc}^{(-)}}{1.6}.
\end{equation}
For example, if changing the research estimate by $1.6$ units moves the earnings forecast by $0.96$ units, the measured response is $0.60$. A smaller response means that the model changes its forecast less when the research estimate is altered by exactly the same amount. This comparison does not require the research estimate to have been observed by a real investor.

We first compare the responses of GPT-6.1 Sol and GPT-5.2 under each public-information condition. We then compare the difference between models when the historical market returns are added. Write
\begin{equation}\label{eq:E2-gaps}
 \begin{aligned}
 G_{\mathrm{accounts}}&=\mathbb E_j[\widehat\beta_{j,6,\mathrm{accounts}}-\widehat\beta_{j,5,\mathrm{accounts}}],\\
 G_{\mathrm{history}}&=\mathbb E_j[\widehat\beta_{j,6,\mathrm{history}}-\widehat\beta_{j,5,\mathrm{history}}],\\
 \Delta G&=G_{\mathrm{history}}-G_{\mathrm{accounts}}.
 \end{aligned}
\end{equation}
A negative $G_{\mathrm{history}}$ means that GPT-6.1 Sol responds less to the research note when both models see the longer public record. A negative $\Delta G$ means that the difference between the models becomes larger after that record is added. The same company-quarter case and the same research-note values appear in both public-information conditions, so the final contrast compares changes in response to the identical input.

There are $50\times2\times2\times2\times3=1{,}200$ independent model calls: 50 cases, two models, two public-information conditions, two research-note values, and three repetitions. All 1,200 returned valid structured responses. We average repetitions within each case before estimating the differences in equation~\eqref{eq:E2-gaps}. Confidence intervals are obtained by resampling the 50 cases 10,000 times. The unit of statistical inference is a company-quarter case, not an individual API call.

\subsection{Results: the difference between models grows with public history}\label{subsec:E2-main}

Table~\ref{tab:E2-main} reports the main results. When the models receive only accounting information, GPT-5.2 has an average response of $0.7050$ to the independent research estimate, compared with $0.6835$ for GPT-6.1 Sol. Their difference is $-0.0215$ (95\% confidence interval $[-0.0365,-0.0058]$). With the historical market returns added, the average responses are $0.7492$ and $0.7072$, respectively. The model difference is now $-0.0420$ (95\% interval $[-0.0521,-0.0315]$), with GPT-6.1 Sol less responsive in 42 of the 50 company-quarter cases.

\begin{table}[H]\centering\small
\caption{Private research-estimate response in 50 independently selected company-quarter cases. Each response is the change in the earnings-surprise forecast per unit change in the hypothetical research estimate. Confidence intervals for model differences resample cases, not API calls.}\label{tab:E2-main}
\begin{tabularx}{\linewidth}{@{}X r r r X@{}}\toprule
\textbf{Public information} & \textbf{GPT-5.2} & \textbf{GPT-6.1 Sol} & \textbf{Difference} & \textbf{95\% interval}\\\midrule
Accounts only & 0.7050 & 0.6835 & $-0.0215$ & $[-0.0365,-0.0058]$\\
Accounts and past returns & 0.7492 & 0.7072 & $-0.0420$ & $[-0.0521,-0.0315]$\\\midrule
\multicolumn{3}{@{}l}{\textbf{Change in the model difference, $\Delta G$}} & $-0.0205$ & $[-0.0317,-0.0093]$\\\bottomrule
\end{tabularx}
\end{table}

\begin{figure}[H]\centering
 \includegraphics[width=0.94\linewidth]{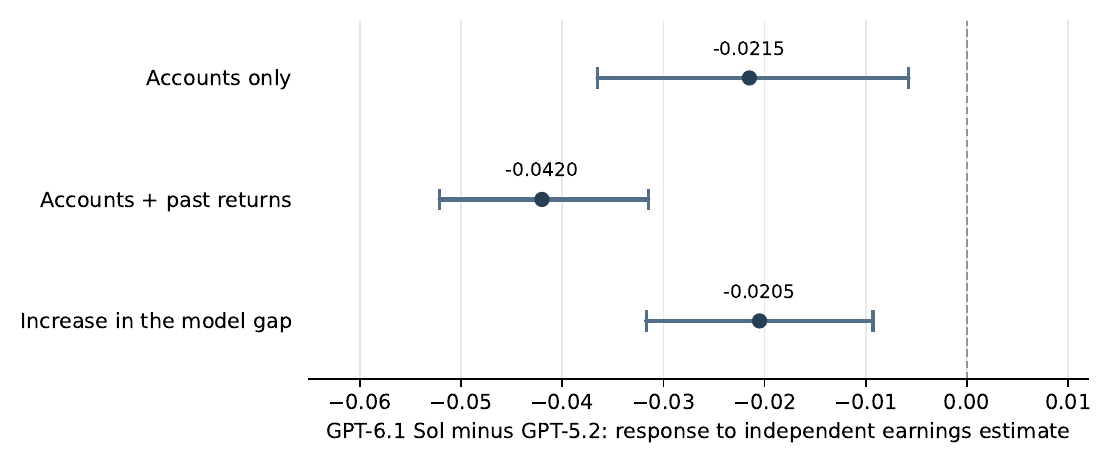}
 \caption{\textbf{How a historical market series changes model differences in the use of an independent earnings estimate.} Points show the average GPT-6.1 Sol response minus the average GPT-5.2 response for the 50 new company-quarter cases. Horizontal lines are 95\% confidence intervals from resampling cases. The third row is the change in that model difference when past market returns are included. Negative values indicate a smaller response by GPT-6.1 Sol or a widening of the difference.}\label{fig:E2-main}
\end{figure}

The important comparison is what happens to the \emph{difference between models} after the market history is added. That difference becomes more negative by $0.0205$ (95\% interval $[-0.0317,-0.0093]$). The change is negative in 33 of the 50 company-quarter cases. Figure~\ref{fig:E2-main} displays the two model differences and their change, including uncertainty intervals. These 50 cases are separate from the earlier cases used to set up the task.

The separate responses also matter for interpretation. Adding the historical market series raises the GPT-5.2 response from $0.7050$ to $0.7492$, an increase of $0.0442$. It raises the GPT-6.1 Sol response from $0.6835$ to $0.7072$, an increase of $0.0236$ when calculated from unrounded values. Thus both models respond \emph{more}, not less, to the independent estimate when market history is available. The increase is smaller for GPT-6.1 Sol, producing the larger difference between models. The experiment establishes this relative pattern; it does not show that adding the price series reduces either model's absolute use of the independent estimate.

\subsection{Checks on how the models interpret the information}\label{subsec:E2-checks}

Before the 50-case evaluation, a separate set of 20 earlier company records from 2012--2016 was used to develop the task. These records are not included in the 50-case estimates above. Two additional comparisons on those 20 records are useful for understanding what the model responses mean.

First, we changed the stated quality of the independent research estimate. Its error standard deviation was raised from $0.7$ to $1.4$, while the accounting numbers, market history, and reported estimate were held fixed. Under the true-history condition, the average response fell by $0.2019$ for GPT-5.2 (95\% interval $[-0.2263,-0.1769]$) and by $0.3147$ for GPT-6.1 Sol (95\% interval $[-0.3243,-0.3053]$). The response fell in all 20 records for both models; similar decreases occurred when no history or an alternative history was supplied. The models therefore pay attention to the stated reliability of the extra estimate. GPT-6.1 Sol's lower average response in the main experiment is not simply an instance of ignoring that estimate altogether.

Second, we replaced the company-specific part of the past return series with returns from another company, while retaining the original sequence of market-wide returns and the same numerical format. The models were not told that a substitution had been made. The model difference was $-0.0517$ under this alternative series, compared with $-0.0612$ under the target company's original history. The difference between these effects was $-0.0095$ (95\% interval $[-0.0233,0.0043]$). Much of the effect of presenting a long financial series therefore persists without the target company's own sequence of firm-specific returns. This points toward common market information or general features of financial time series as plausible influences on model behavior, rather than establishing extraction of the target firm's private fundamentals from its past prices.

We also checked that the numerical inputs were understood at a basic level. Across 80 accounting and return-arithmetic questions per model in these 20 records, both models answered all 160 questions correctly. This confirms that they could read and calculate from the supplied data. It does not demonstrate that one model made better forecasts from the public record than the other.

\subsection{A calculation with known public information}\label{subsec:E2-gaussian}

In the historical company records, we do not know the information that each model has actually extracted from the return series. A separate Gaussian exercise removes this uncertainty by supplying public observations with a known relationship to the outcome. It provides a direct check of the economic prediction in equation~\eqref{eq:E2-privateweight} and of the numerical procedure used to measure responses.

Let $\theta\sim N(0,1)$. Each informative public observation is $z_k=\theta+\eta_k$, where the observation errors are independent with standard deviation $1.5$. The private research estimate is $s=\theta+\xi$, with independent error SD $0.7$. If the record contains $n$ such public observations, the remaining variance of $\theta$ and the optimal weight on $s$ are
\begin{equation}\label{eq:E2-knownprecision}
 P_n=\left(1+\frac{n}{1.5^2}\right)^{-1},
 \qquad \beta_n=\frac{P_n}{P_n+0.7^2}.
\end{equation}
The optimal private-signal weight is $0.6711$ when there are no informative public observations, $0.5193$ with two, and $0.2437$ with twelve. The full synthetic records in this exercise all have twelve rows; the remaining rows are unrelated market observations, so the change is in information about $\theta$ rather than the length of the input. Both models reproduce these calculated responses essentially exactly. In a second exercise involving 48 observations with different factor loadings and noise levels, two GPT-6.1 Sol reasoning-effort settings also recover the known posterior to numerical precision. This is a check on the mathematical mechanism and the prompt's interpretation, not a separate estimate of the comparative ability of GPT-5.2 and GPT-6.1 Sol in the historical company task.

\subsection{What the results imply for the market model}\label{subsec:E2-interpretation}

The main experiment shows that the two AI models make different use of an identical, stated research estimate, and that their difference becomes larger when a longer public financial history is supplied. The Gaussian example describes a mechanism that could generate such a pattern: better public information reduces the uncertainty left for an independent private signal to resolve. The real-data experiment, however, measures models' reported responses rather than the precision of the information they extracted. The two observations are consistent with information substitution, but the observed model differences need not be caused by that mechanism alone.

In particular, the historical price series were not shown to add verified, firm-specific predictive information about the next earnings surprise in a separate researcher-side out-of-sample check on the 50 cases. Mean squared error was $4.7670$ using the accounting information alone and $4.7782$ after adding the target's price features. Replacing the firm-specific return sequence with another firm's data also preserved most of the model-response difference in the 20-case check. The evidence therefore supports a statement about model behavior in the presence of a rich public financial record. It does not establish that GPT-6.1 Sol extracts more information from the particular firm's past prices or that the temporal ordering of those prices causes the result.

The theory's first theorem concerns how optimally fitted forecasts and their risks jointly determine the demand placed on individual private signals. Here we measure the response of the forecast \emph{mean}. The models' self-reported forecast uncertainty was not well calibrated in the earlier company-record exercise: nominal 95\% intervals covered realized outcomes only 45--60\% of the time. We therefore do not convert the responses into empirical CARA demand coefficients. The experiment also does not alter actual investors' orders or the market price. It is a complementary observation about AI forecasting behavior, not a test of an adoption-driven market equilibrium.

Together with Section E1, the results illustrate two closely related ideas using different data. The market-data exercise in E1 finds that recent market information adds less to a richer statistical forecast than to a limited one. Here, the response of an AI forecast to a hypothetical private research estimate depends on both the model used and the amount of public financial history presented. The theoretical and simulated markets in Sections S3--S6 explain how these individual information choices can affect price discovery and the transmission of information among investors.

\subsection{Methods and reproducibility}\label{subsec:E2-replication}

The 50 company-quarter cases were selected before any calls on those cases, using seed \texttt{step4-holdout-v1} and ten observations from each year 2017--2021. The questions, input fields, model settings, and response format were fixed before this behavioral test. Each of the 1,200 requests used a fresh Responses API call, a strict JSON response schema, no earlier conversation, and \texttt{store: false}. All requests returned valid responses. Three repeated forecasts were averaged within each case and condition; the reported intervals use 10,000 case-level bootstrap resamples. The earlier 20-case checks do not contribute to these estimates.

The GPT-5.2 model was fixed to the dated snapshot stated above. GPT-6.1 Sol was accessed through a model alias, and the resolved model identifier was recorded for each call. Reproduction requires the anonymized company information, the experimental prompts, model versions and settings, and the case-level analysis procedure. The underlying firm-level financial records are subject to data-provider licenses.

\par\medskip
\noindent\textbf{Code availability.}
The source code and reproduction instructions for the numerical simulations and empirical analyses are available in the GitHub repository \emph{ai-market-information}: \url{https://github.com/shimaohajime/ai-market-information}.
The repository excludes licensed financial data, raw language-model responses, and neural-model checkpoints; its documentation describes the scope of the released materials.

\end{document}